\documentclass[%
preprint,
onecolumn,
amsmath,
amssymb,
aps,%
superscriptaddress
]{revtex4-2}

\usepackage{array}
\usepackage{ragged2e} 
\usepackage{graphicx} 
\usepackage{siunitx}
\usepackage{subcaption}
\usepackage{newtxtext,newtxmath}
\usepackage{textgreek}
\usepackage[draft]{hyperref}
\usepackage{placeins} 
\usepackage{cleveref}
\crefname{figure}{Fig.}{Figs.}
\Crefname{figure}{Figure}{Figures}

\crefname{table}{Table}{Tables}
\Crefname{table}{Table}{Tables}

\crefname{equation}{Eq.}{Eqs.}
\Crefname{equation}{Equation}{Equations}

\crefname{section}{Sec.}{Secs.}
\Crefname{section}{Section}{Sections}

\crefname{subsection}{Subsec.}{Subsecs.}
\Crefname{subsection}{Subsection}{Subsections}

\DeclareSIUnit{\wtpercent}{wt\%}

\newcommand{\DeNum}{\mathrm{De}}

\begin{document}

\title{Airway Mucus Rheology: Physical Insights for Navigating through Health to Pathology and Clinical Applications}
\author{Zhiwei Liu}
\email{zhiweiliu@ucas.ac.cn}
\affiliation{%
Wenzhou Institute, University of Chinese Academy of Sciences, Wenzhou, Zhejiang 325000, China}%
\affiliation{%
Oujiang Laboratory (Zhejiang Lab for Regenerative Medicine, Vision and Brain Health), Wenzhou, Zhejiang 325000, China}%

\author{Bo Che}
\affiliation{Changzhou Key Laboratory of Respiratory Medical Engineering, Institute of Biomedical Engineering and Health Sciences, School of Medical and Health Engineering, Changzhou University, Changzhou, 213164, China}

\author{Hailin Zhang}
\email{zhanghailin@wmu.edu.cn}
\affiliation{Department of Children’s Respiratory Disease, the Second Affiliated Hospital \& Yuying Children's Hospital, Wenzhou Medical University, Wenzhou 325000, China}

\author{Linhong Deng}
\email{dlh@cczu.edu.cn}
\affiliation{Changzhou Key Laboratory of Respiratory Medical Engineering, Institute of Biomedical Engineering and Health Sciences, School of Medical and Health Engineering, Changzhou University, Changzhou, 213164, China}

\date{\today}

\begin{abstract}
Airway mucus is a complex gel with an anisotropic three-dimensional network structure. As a crucial component of the respiratory defense barrier, it plays a vital role in maintaining airway hydration and supporting the function of airway epithelial cells. Through linear and nonlinear rheological mechanisms such as ciliary motion and coughing, airway mucus expels foreign pathogens and toxic nano- and microparticles while selectively allowing the passage of specific nutrients and proteins. These protective and clearance functions depend on the proper rheological properties of mucus under normal physiological conditions. However, in respiratory disease such as cystic fibrosis (CF), chronic obstructive pulmonary disease (COPD), asthma, and COVID-19, excessive mucus secretion is often accompanied by abnormal rheological behaviors. This leads to impaired mucus flow, airway obstruction, and potentially life-threatening conditions. Therefore, this review examines the rheological behaviors of airway mucus in relation to health and disease, focusing on both macrorheology and microrheology. Macrorheology provides insights into the overall viscoelastic behavior of mucus, revealing its general mechanical properties but often overlooks local variations. Microrheology addresses this limitation by examining heterogeneity, local mechanical properties, and the diffusion characteristics of drugs and viruses within the mucus gel network at nano- and microscale levels. The review highlights those changes in the chemical composition and microstructure of airway mucus, especially under pathological conditions, that can significantly alter its rheological behavior. Rheological parameters can also serve as biological indicators to study the role of mucus in clearance functions and aid in developing pulmonary drug delivery systems. By integrating findings from both macro- and microrheological studies, this review aims to enhance our understanding of the complex behavior of airway mucus, supporting better diagnosis, treatment, and management of chronic respiratory diseases.
\end{abstract}

  \maketitle
\tableofcontents

\section{Introduction}

The mucociliary system of the pulmonary airways, comprising cilia, the periciliary fluid layer (PCL), and the mucus layer, is a crucial innate defense mechanism essential for maintaining lung health \citep{Fischer_2006,Cone_2009,King_2009,Lai_2009,Voynow_2009,Fahy_2010,Rubin_2010,Cohen_2012,Hartl_2012,Tuder_2012,Broedersz_2014,Spagnolie_2015,Duncan_2016a,Kesimer_2017,Witten_2018,Boucher_2019,Nawroth_2020,McShane_2021,Meldrum_2021,Hill_2022,Kavishvar_2023,Pangeni_2023,Abrami_2024}. 
Normally, cilia extend from the surface of bronchial epithelial cells, through the PCL, and into the mucus layer, where they beat in a coordinated fashion to clear the airway lumen of excess mucus \citep{King_2009,Fahy_2010,Rubin_2010,Tuder_2012,Wagner_2017,Boucher_2019,Vanaki_2020,Nawroth_2020,McShane_2021,Huck_2022,Kavishvar_2023,Pangeni_2023,Abrami_2024}. 
This constant secretion and clearance of mucus are vital for several physiological functions, including nutrient exchange, airflow lubrication, and the entrapment and removal of inhaled hazardous particles such as pathogens and airborne pollutants \citep{Cone_2009,Fahy_2010,Cohen_2012,Kesimer_2017,Boucher_2019,Hill_2022}. 

\begin{figure*}[htbp]
   \centering
    \includegraphics[width=0.6\textwidth]{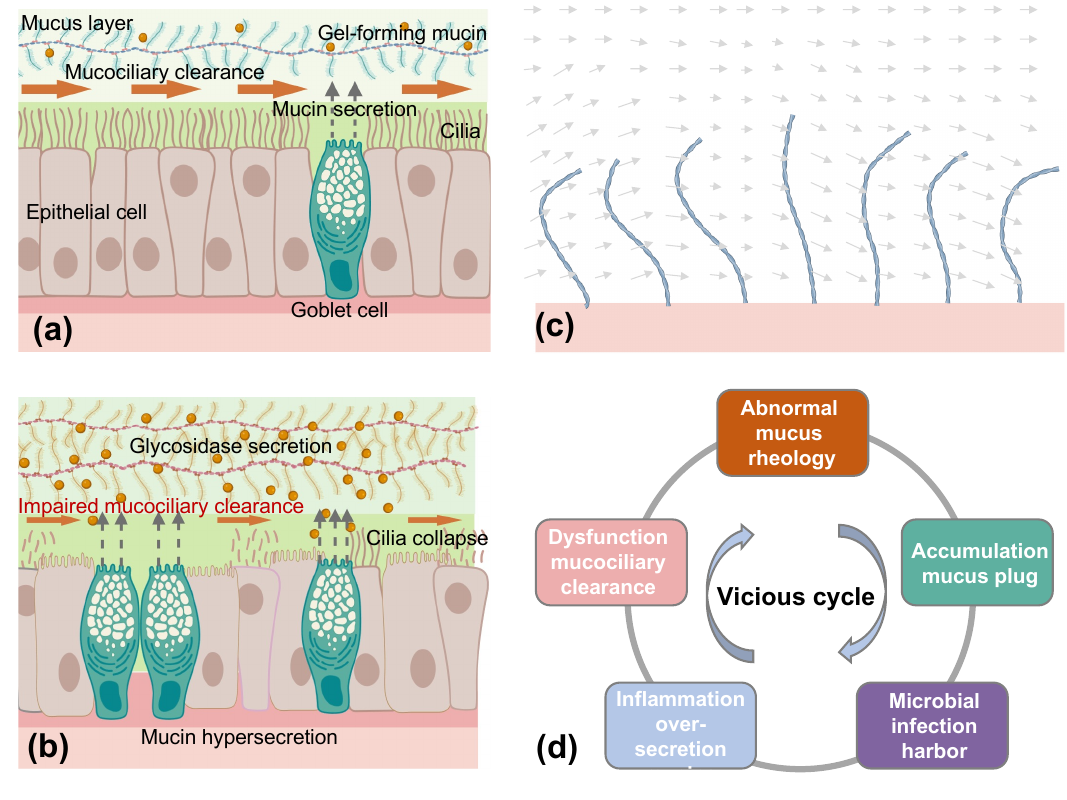}
  \caption{(a) Healthy mucociliary clearance system,
  (b) impaired mucociliary clearance system,
  (c) synchronous metachronal movement of cilia, (d) vicious cycle during pathological progress.}
  \label{fig:1}
\end{figure*}

In the large airways, mucus is often cleared effectively by coughing-induced high-velocity airflow. Conversely, in the distal small airways, mucus clearance is primarily driven by the upward mobilization due to ciliary beating \citep{Fahy_2010,Nawroth_2020,Vanaki_2020,Hill_2022}. 
The effectiveness of mucociliary clearance is highly dependent on the physical, specifically rheological, properties of the mucus \citep{Cone_2009,King_2009,Lai_2009,Grotberg_2011,Kesimer_2017,Wang_2017,Boucher_2019,Hill_2022,Kavishvar_2023}. 
Healthy airway mucus, which contains \SI{97}{\percent} water and \SI{3}{\percent} solids (mucins, nonmucinous proteins, salts, lipids, and cellular debris), exhibits a solid-like state with relatively low viscosity and elasticity, allowing easy transport by ciliary action.


However, in conditions such as simple dehydration or diseases like asthma, chronic obstructive pulmonary disease (COPD), and cystic fibrosis, mucus can become highly viscoelastic, making it difficult to be mobilized and cleared by both cough and ciliary action \citep{Fischer_2006,Cone_2009,King_2009,Lai_2009,Voynow_2009,Fahy_2010,Rubin_2010,Cohen_2012,Hartl_2012,Tuder_2012,Broedersz_2014,Spagnolie_2015,Duncan_2016a,Kesimer_2017,Witten_2018,Boucher_2019,Nawroth_2020,McShane_2021,Meldrum_2021,Hill_2022,Kavishvar_2023,Pangeni_2023,Abrami_2024}. 
Chronic impairment of mucus clearance leads to mucus accumulation or even mucus plugging, which promotes chronic bacterial infection. 
This results in a cascade of events, including persistent inflammation and airway obstruction, disease progression, and increased morbidity and mortality, ultimately creating a vicious cycle as shown in \cref{fig:1} \citep{Fischer_2006,Cone_2009,King_2009,Lai_2009,Voynow_2009,Fahy_2010,Rubin_2010,Cohen_2012,Hartl_2012,Tuder_2012,Broedersz_2014,Spagnolie_2015,Duncan_2016a,Kesimer_2017,Witten_2018,Boucher_2019,Nawroth_2020,McShane_2021,Meldrum_2021,Hill_2022,Kavishvar_2023,Pangeni_2023,Abrami_2024}. 
This phenomenon has been widely observed in COVID-19 patients during the recent global pandemic \citep{Wang_2020,Kratochvil_2022,Bessonov_2023}. 
From this point of view, understanding the rheology of airway mucus is essential not only for comprehending mucus clearance dysfunction associated with respiratory diseases but also for evaluating mucolytic-based drugs designed to promote mucus clearance for disease treatment.

\begin{figure*}[htbp]
   \centering
    \includegraphics[width=0.99\textwidth]{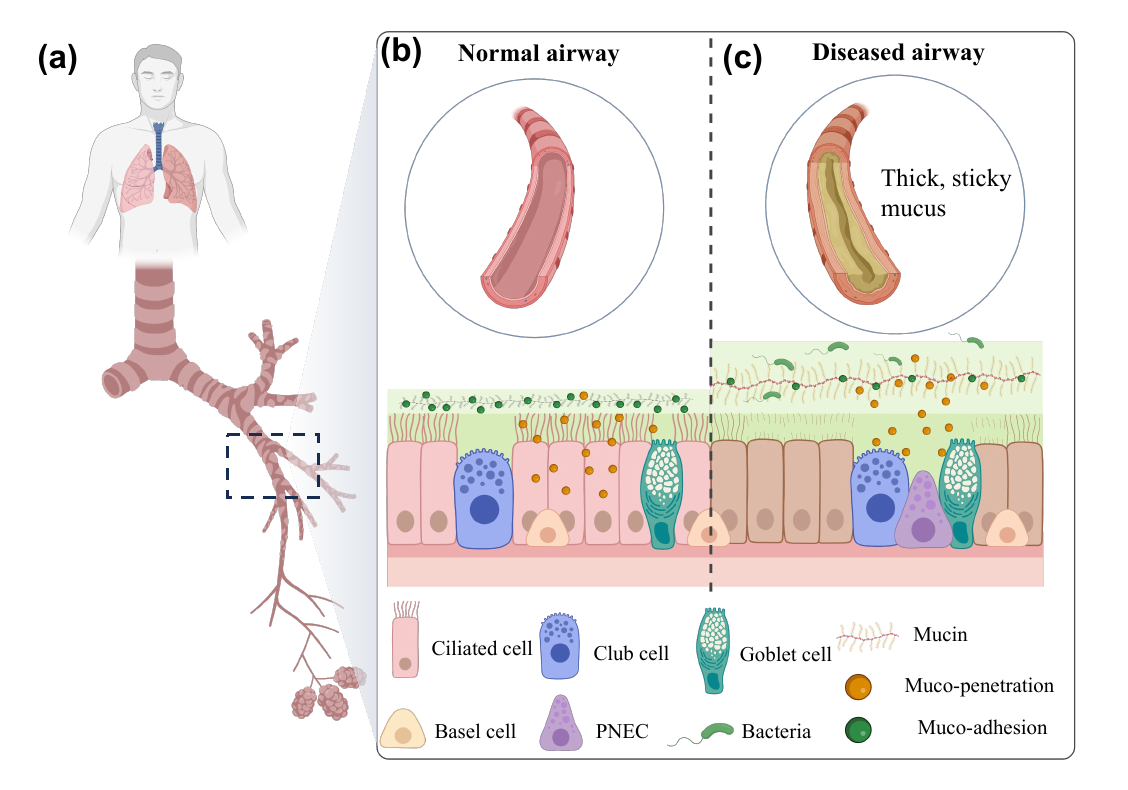}
  \caption{Schematic of airway and mucus in normal and pathological state. 
  (a) Macro to micro airway structure,
  (b) mucus in normal airway state, 
  (c) mucus in pathological airway state.
  }
  \label{fig:AirwayStructure}
\end{figure*}

\section{Sources and collection of native human airway mucus from healthy and pathology individuals}

Despite the significant impact of changes in mucus quantity and composition on the mortality and morbidity of airway diseases, accurately measuring the rheological properties of mucus  remains challenging \citep{Atanasova_2019,McCarron_2021,Huck_2022,Abrami_2018}.  One major obstacle is obtaining representative samples directly from the trachea, bronchi, and bronchioles. However, due to the fact that mucus differs widely in composition, pH value and rheological properties across different tissue origins depending on its specific barrier function, it may not be suitable to use mucus samples that are more easily obtained in quantity such as the nasal and salivary mucus as surrogates of airway mucus for every type of research. \citep{KARNAD_1990,Fischer_2006,Shah_2016,Meldrum_2018}

For this reason, researchers have developed various methods for collecting native mucus from human airways. These techniques often require specialized equipment, trained personnel, and ethical approvals, limiting sample availability \citep{Araujo_2018,Huck_2022,Hill_2022,Atanasova_2019}. Common approaches include collecting mucus from healthy volunteers or sputum from patients with obstructive lung diseases. Healthy individuals rarely produce sputum naturally, so methods like hypertonic saline inhalation are used to induce sputum production. However, these techniques may alter the solid content and rheology of the mucus. 
For instance, \citet{Serisier_2009} demonstrated that the viscoelastic properties of sputum differ between healthy individuals and those with obstructive diseases, suggesting its potential as a disease biomarker. And  when rapidly analyzed, the fresh sputum from healthy individuals and those with asthma, COPD, and CF exhibits a progressive increase in viscoelasticity, reflecting an increasing difficulty of mucus clearance and thus greater tendency of airway obstruction. Therefore, it requires multiple techniques and strategies to assess mucus properties to ensure data accuracy and relevance to pathological states. Below, we enlist the currently available airway mucus collection methods, and outline their advantages and disadvantages with highlight on the operational challenges and sample volume requirements. For a detailed comparison, refer to \cref{table:source_mucus}.
\subsection{Bronchoscopy}
Bronchoscopy is a widely used method for direct collection of airway mucus from the lungs and provides essential diagnostic information \citep{Weynants_1985,Zayas_1990,Pizzichini_1996,Kamin_2006,Innes_2009,Andreasson_2016,Fernandez-Petty_2019}. This technique was first applied by \citet{Jeanneret-Grosjean_1988} in the late 1980s, when a microbiology specimen brush was inserted into the trachea to obtain fresh, undiluted mucus directly from the bronchial mucosa. Although the mucus yield is small, it is sufficient for rheological experiments. Additionally, bronchoscopy facilitates the study and management of mucus plugs, as mucolytics can be applied during the procedure \citep{Kamin_2006}. Despite these advantages, the method requires general anesthesia and specialized medical personnel \citep{Atanasova_2019,Bej_2022,Huck_2022}.
%
%
\subsection{Endotracheal tubes}
Mucus can also be collected using endotracheal (ET) tubes during elective surgeries in healthy patients, as described by \citet{Rubin_1990}. ET tubes, primarily used for airway management during general anesthesia \citep{Lapinsky_2015}, can accumulate mucus, which sometimes forms plugs \citep{Mietto_2014,Berra_2012,Atanasova_2019}. This method has the advantage of allowing direct collection from healthy patients, yielding large sample volumes \citep{Atanasova_2019}. However, variations in hydration levels between mucus inside and outside the tubes may affect experimental reproducibility~\citep{Rubin_1990,Atanasova_2019}. 
Recent studies by \citet{Markovetz_2022} have shown that even isotonic mucus can be adjusted to solid contents between \SI{2}{\percent} and \SI{4.6}{\percent}, providing ample material for analysis. Moreover, \citet{Radiom_2021} utilized magnetic rotational spectroscopy to evaluate the viscoelastic moduli of mucus collected from healthy bronchial tubes and cultured bronchial resections, further advancing our understanding of mucus \citep{Song_2021,Radiom_2021,Markovetz_2019,Duncan_2016,Lapinsky_2015} .
%
%
\subsection{Induced and spontaneous airway mucus}
Sputum, expelled from the lower airways via coughing, provides vital information on mucus and mucins. 
For simplicity, this review uses ``mucus'' to refer to all types of mucus samples, including sputum. 
Sputum can be spontaneous or induced \citep{Hall_1971,Pizzichini_1996,Bartoli_2002,Sagel_2007,Serisier_2009,Suk_2009,Vieira_2011,Weiszhar_2013,Duncan_2016,Patarin_2020}. Spontaneous sputum is naturally coughed out by patients, commonly seen in respiratory diseases like CF, COPD, or asthma \citep{Patarin_2020,Duncan_2016}.
Its advantage is that it can be collected without medical intervention, but its limitations include the inability to collect mucus from healthy individuals and potential contamination from saliva and unpredictable sputum yield \citep{Huck_2022,Hall_1971,Atanasova_2019}. 
The severity of lung inflammation can also affect the composition and quantity of mucus in the samples, necessitating careful collection to minimize contamination \citep{Huck_2022,Atanasova_2019}.
Induced sputum is produced by introducing hypertonic saline aerosols into the patient's airways, following standardized protocols to enhance reproducibility and rigor across studies \citep{Weiszhar_2013,Pizzichini_1996}. 
Unlike spontaneous sputum, induced sputum is non-invasive and can be collected from healthy volunteers, providing a broader sample base for research. 
However, the introduction of saline aerosols may have some dilution effects to alter the mucus rheological and barrier properties. 
Studies found that induced samples had reduced total solid content compared to the undiluted samples obtained via bronchoscopy \citep{Hall_1971}. 
The success of sputum induction can also be influenced by the patient's level of inflammation, and caution on potential adverse effect of this procedure is needed for asthmatic patients as hypertonic saline can cause bronchoconstriction \citep{Bartoli_2002,Vieira_2011}. 
Despite the disadvantages of spontaneous and induced sputum, both are crucial for understanding airway mucus and inflammation \citep{Atanasova_2019,Sagel_2007}.

\subsection{Mucus derived from cultured human bronchial epithelial cells} 

Human bronchial epithelial cells (HBE) cultured at the air-liquid interface (ALI) provide an ideal platform for studying airway mucus without external infection and inflammation interference. 
This model collects mucus through apical washes, effectively simulating and evaluating the impact of various interventions on mucus secretion \citep{Davis_2002,Kemp_2004,Matsui_2006,Kesimer_2009,Sears_2011,Abdullah_2012,Hill_2014,Vasquez_2016,Jory_2019,Jory_2022,Markovetz_2019,Lin_2020,Radiom_2021,Liegeois_2024}.
However, ensuring complete removal of mucins requires strict adherence to standardized procedures \citep{Huck_2022,Atanasova_2019}. 
While this method can replicate the viscoelastic properties of native airway mucus at physiological concentrations, multiple washes may complicate mucus and cell separation, leading to cell contamination and difficulty to distinguish mucus layers from cell layers \citep{Cai_2024,Liegeois_2024}. 
Moreover, these samples may not be suitable for studying mucus barrier properties in patients with obstructive lung diseases, as key components like DNA and actin found in airway secretions of these patients are often missing in culture-derived mucus samples \citep{Abdullah_2012,Kemp_2004,Kesimer_2009}.


\subsection{Mucus from animal models}

Animal models are indispensable for respiratory research, particularly when human mucus samples are difficult to obtain \citep{Lai_2009,Vasquez_2014,Murgia_2016,Gross_2017,Schneider_2017,Morgan_2021,Atanasova_2019}. Species with airway structures and mucus properties similar to humans—such as pigs and dogs—are preferred for comparative studies. Genetically engineered pigs deficient of cystic fibrosis transmembrane conductance regulator (CFTR), for example, spontaneously develop cystic fibrosis (CF)-like mucus, making them invaluable for CF research \citep{Tomaiuolo_2014,McCarron_2021}. Normal pig mucus closely matches human mucus in its rheological properties. Similarly, tracheal mucus from dogs exhibits viscoelasticity within ±2\% of human mucus, providing a reliable model for airway mucus studies \citep{Murgia_2016,King_1977,King_1989,Khan_1976,King_1997}. In contrast, rodents and rabbits, while convenient, exhibit significantly higher mucus viscoelasticity compared to humans, limiting their translational relevance \citep{Tomaiuolo_2014,Lai_2009}.
Mucus is typically collected post-mortem via tracheal scraping or using cytology brushes in anesthetized animals. However, differences in airway anatomy—such as submucosal gland distribution must be carefully considered to ensure the relevance of findings to human physiology\citep{Huck_2022,Lai_2009,Atanasova_2019,Kavishvar_2023}.

\subsection{Artificial synthetic airway mucus}

In the case that natural airway mucus samples are unavailable, artificial synthetic airway mucus model samples are often used as alternatives. These models are typically fabricated to comprise commercial components such as mucin, DNA, actin, lipids, and albumin \citep{Yang_2011,Craparo_2016,Lafforgue_2017}. However, replicating the viscoelastic gel properties of natural mucus at physiological concentrations is challenging \citep{Meldrum_2018,Huck_2019,Carpenter_2021}. Although the mucin concentrations in synthetic models are matched to natural levels, diffusion studies using carboxyl-modified tracer particles have demonstrated that particle migration is significantly slower in synthetic mucus than in native mucus by \citet{Huck_2019}. To address this, researchers have introduced crosslinking or entangling polymers (e.g., polyacrylic acid, xanthan gum, glutaraldehyde) to better mimic the rheological properties of natural mucus  \citep{Tan_2020,Hamed_2014,Sharma_2021,Song_2021}. Recent advancements include chemically synthesized mucin mimetics based on glycoproteins, which exhibit high similarity to natural mucins and offer promising directions for future research \citep{Bej_2022,Wagner_2023,Liu_2024,Degen_2025}.


\begin{table*}[p]
\rotatebox{90}{
  \begin{minipage}{\textheight} 
  \linespread{0.9}\selectfont
\centering
\caption{Advantages and disadvantages of airway mucus collection methods}
\label{table:source_mucus}

\begin{tabular}{
|>{\raggedright\arraybackslash}p{3cm}
|>{\raggedright\arraybackslash}p{6cm}
|>{\raggedright\arraybackslash}p{6cm}
|p{3cm}
|p{3cm}
|>{\raggedright\arraybackslash}p{3cm}|}

\hline
\textbf{Method} & \textbf{Advantages} & \textbf{Disadvantages} & \textbf{Difficulty Level} & 
\textbf{Sample Volume} & \textbf{References} \\
\hline
Bronchoscopy & Direct collection from trachea and bronchi; provides diagnostic information; suitable for rheological experiments; effectively removes mucus plugs. & Requires general anesthesia; needs professional medical personnel. & High & medium to low & [121, 122, 123, 77] \\
\hline
Endotracheal tube (ET) & Direct collection from healthy patients during surgery; collects large samples. & Requires general anesthesia; variability in hydration levels of internal and external mucus can affect experimental outcomes. & Medium & high & [128, 85, 86, 80, 129] \\
\hline
Spontaneous sputum & Collection process does not involve medical intervention; simple process. & Cannot collect from healthy volunteers; potential saliva contamination; unpredictable sputum volume. & Low & medium & [123, 76, 126, 127, 58, 59] \\
\hline
Induced sputum & Non-invasive; can be collected from healthy volunteers; enhances reproducibility and rigor between studies. & Dilution effect can alter mucus barrier properties; patient inflammation status may affect collection success. & Low & Medium to low & [73, 74, 75, 122, 83, 10, 84] \\
\hline
HBE cells & Free from external infection and inflammation; effectively assesses interventions on mucus secretion. & Complex operation; difficulty in separating mucins from cells; unsuitable for studies on obstructive pulmonary disease patients. & High & low & \citep{Kemp_2004,Matsui_2006,Kesimer_2009,Sears_2011,Abdullah_2012,Hill_2014,Vasquez_2016,Lin_2020,Liegeois_2024,Jory_2022} \\
\hline
Animal models & Convenient mucus sampling methods; mucus characteristics similar to humans. & Anatomical structure and mucus properties may differ significantly, such as richness of submucosal glands. & Medium & High & \citep{Vasquez_2014,Murgia_2016,Gross_2017,Schuster_2017,Morgan_2021} \\
\hline
Artificial mucus models & Easy to obtain; controlled composition; adjustable rheological properties. & Difficult to replicate the viscoelastic properties of natural mucus; significant differences in barrier properties. & Low & Adjustable (Low to High) & \citep{Yang_2011,Craparo_2016,Lafforgue_2017,Meldrum_2018,Huck_2019,Kruger_2021,Sharma_2021,Song_2021,Bej_2022,Wagner_2023,Liu_2024,Milian_2024} \\
\hline
\end{tabular}
  \end{minipage}
  }
\end{table*}

\section{CHARACTERISTICS OF AIRWAY MUCUS IN HEALTHY AND PATHOLOGICAL STATES} 
\subsection{Composition and function of healthy airway mucus}

Airway mucus is a critical frontline defense in the respiratory system, trapping inhaled pathogens and particles. Its primary structural component is mucin—a family of high molecular weight, heavily glycosylated proteins constituting \SI{1}{\percent} to \SI{5}{\percent} of mucus by weight. Mucins feature hydrophobic polypeptide backbones with hydrophilic oligosaccharide side chains, enabling them to bind both charged and hydrophobic particles \citep{Spagnolie_2015,Hill_2022,Lai_2009,Wagner_2018,McShane_2021,Pangeni_2023}. These proteins form a mesh-like network stabilized by disulfide bonds and physical entanglement, working alongside other components (proteins, antibodies, salts, lipids, and \SI{95}{\percent} water) to create a viscoelastic gel. This gel balances fluidity and rigidity, ensuring efficient mucociliary and cough clearance while protecting airway surfaces \citep{Murgia_2018,Lai_2009,Boucher_2019,Fahy_2010,Huckaby_2018}.


\subsection{Airway mucins and their roles}
The respiratory tract expresses 14 mucin (MUC) genes categorized into three functional groups.
Monomeric mucins (e.g., MUC7, MUC8) are small, soluble glycoproteins. Gel-forming mucins (e.g., MUC5AC, MUC5B, MUC2) polymerize via cysteine-rich domains to create mucus’s viscoelastic scaffold.
Surface-bound mucins (e.g., MUC1, MUC4) anchor to epithelial cells, forming a protective glycocalyx. Among these, MUC5AC (secreted by goblet cells) and MUC5B (produced by submucosal glands) dominate mucus rheology \citep{Pangeni_2023,Atanasova_2019,Ma_2018,Voynow_2009,Fahy_2010,Thornton_2008,Carpenter_2021,Song_2022,Caughman_2024}. 
In health, their balanced expression ensures mucus permeability to small molecules while trapping pathogens for clearance \citep{Araujo_2018,Pednekar_2022,Garcia-Diaz_2018} .


\subsection{Alterations of airway mucus in respiratory diseases}

Chronic respiratory diseases—COPD, cystic fibrosis (CF), asthma, and bronchiectasis—share mucus dysfunction as a hallmark. Pathological mucus is characterized by hypersecretion, abnormal composition (e.g., elevated DNA/actin), and impaired clearance, leading to airway obstruction and infection \citep{Hill_2022,Lai_2009,Cohen_2012,Kesimer_2017,McShane_2021,Huck_2022}.  
In COPD, mucus hypersecretion stems from goblet cell hyperplasia and submucosal gland hypertrophy, producing thick mucus rich in MUC5AC (\cref{tab:airway_mucus_characteristics}) \citep{Lopez_2006,Decramer_2012,Tuder_2012}. CF mucus, dominated by MUC5B and neutrophil-derived DNA/actin, forms dense plaques that resist clearance \citep{PILEWSKI_1999,Collawn_2014,Rowe_2005,Hauser_2011,Cohen_2012,Hartl_2012}. Asthma involves eosinophilic inflammation and MUC5AC overproduction, while bronchiectasis features neutrophilic inflammation and bacterial colonization \citep{Morcillo_2006,Evans_2009a,Izuhara_2009}.  Neutrophil dysfunction exacerbates pathology: proteases like neutrophil elastase degrade antimicrobial peptides (e.g., α-defensin, LL-37) and damage tissues, while acidic pH and ionic imbalances further impair host defense. These changes create a vicious cycle of mucus stasis, infection, and inflammation. \citet{Fahy_2010} described how mucus transitions from a healthy to a pathological state and summarizes these mechanisms and their impact in \cref{tab:airway_mucus_characteristics}.

\subsection{Implications of mucus management in respiratory medicine}

Balanced viscoelasticity of airway mucus in healthy state ensures efficient pathogen clearance. In diseased state, the mucus is often hyperconcentrated with skewed mucin ratios (e.g., elevated MUC5AC in COPD, and dominant MUC5B in CF), and elevated DNA/actin, which disrupts its both chemical and physical equilibrium. Therapeutic strategies must address these abnormalities, such as using mucolytics like recombinant human DNase (rhDNase) to degrade DNA and thus reduce DNA/actin in CF, and hydration therapies to restore ion-water balance. Anti-inflammatory agents targeting proteases or neutrophil activity have been reported for breaking the cycle of mucus stasis and infection \citep{Suh_2005,Lai_2009,Rubin_2010,Duncan_2016,Araujo_2018,Nordgard_2018,Boucher_2019,Huck_2022}.



\begin{table*}[thbp]
\centering
\begin{minipage}{\textwidth}
\caption{Characteristics of airway mucus in healthy and respiratory disease states}
\label{tab:airway_mucus_characteristics}
\linespread{0.9}\selectfont 
\begin{tabular}{|>{\raggedright\arraybackslash}p{3.2cm}
|>{\raggedright\arraybackslash}p{3cm}
|>{\raggedright\arraybackslash}p{3cm}
|>{\raggedright\arraybackslash}p{3cm}
|>{\raggedright\arraybackslash}p{3cm}|}
\hline
\textbf{Characteristic} &
\textbf{Healthy} &
\textbf{COPD} & 
\textbf{Asthma} & 
\textbf{CF} \\ 
\hline
Mucin concentration & 
\SIrange{20}{30}{\text{mg/mL}} & 
\SIrange{30}{50}{\text{mg/mL}} & 
\SIrange{30}{50}{\text{mg/mL}} & 
\SIrange{50}{100}{\text{mg/mL}} \\ 
\hline
Mucin composition & 
Balanced MUC5AC, MUC5B; Low MUC2 &
Increased MUC5AC; Low MUC2 &
Increased MUC5AC; Low MUC2 & 
Predominantly MUC5B; Increased MUC2 \\ \hline
Other components (DNA and Actin) & Low & 
Increased (due to cell death/damage) & 
Increased (due to cell death/damage) & 
Highly increased (due to cell death/damage and infection) \\ 
\hline
Inflammatory and immune cells & 
Low & 
Increased (neutrophils, macrophages) & 
Increased (eosinophils, macrophages) & 
Increased (neutrophils, macrophages) \\ 
\hline
Cellular debris & Low & Increased & Increased & 
Highly increased \\ 
\hline
Goblet cell and epithelial changes & Normal & 
Hypertrophy and hyperplasia of goblet cells, surface epithelial mucous metaplasia & 
Hypertrophy and hyperplasia of goblet cells, surface epithelial mucous metaplasia & 
Hypertrophy and hyperplasia of goblet cells, surface epithelial mucous metaplasia \\ \hline
Solid content & 
\SIrange{2}{3}{\percent} & 
\SIrange{3}{5}{\percent}& 
\SIrange{3}{5}{\percent} & 
\SIrange{7}{10}{\percent}
 \\ \hline
Rheological properties & 
Low viscosity and elasticity &
Increased viscosity and elasticity &
Increased viscosity and elasticity & 
Highly increased viscosity and elasticity \\ \hline
pH & 
Neutral (6.5--7.0) & 
Slightly acidic (6.0--6.5) &
Neutral (6.5--7.0) & 
Acidic (6.0--6.5) \\ 
\hline
Mucociliary clearance efficiency & Efficient & Impaired & Impaired & Severely impaired \\ \hline
Cough clearance efficiency & Efficient & 
Dependent on severity, often impaired & 
Dependent on severity, often impaired & 
Severely impaired \\ 
\hline
References &
\citep{Voynow_2009,Fahy_2010,Kesimer_2017,McShane_2021,Hill_2022,Abrami_2024,Pangeni_2023,Jory_2019,Serisier_2009}
& 
\citep{Boucher_2019,Lopez_2006,Decramer_2012,Tuder_2012,Woodruff_2015,Raju_2016,Solomon_2017,Chisholm_2019,Thulborn_2019,Lin_2020,Linssen_2021,Volpato_2022}
&
\citep{Morcillo_2006,Evans_2009a,Innes_2009,Izuhara_2009,Lambrecht_2015,Schulz_2016,Solomon_2017,Hammad_2021,Morgan_2021,Song_2021,Volpato_2022}
&
\citep{Suh_2005,Livraghi_2007,Duncan_2016,Abrami_2018,Button_2018,Abrami_2020,Abrami_2021,Abrami_2022,Batson_2022}
\\ 
\hline
\end{tabular}
\end{minipage}
\end{table*}

\section{MACRORHEOLOGY OF AIRWAY MUCUS} 

Macrorheology examines the bulk rheological properties of airway mucus, providing critical insights into its response to stress and strain under physiological and pathological conditions \citep{Lai_2009,Hill_2022,Wagner_2018,Kavishvar_2023,Spagnolie_2015}. This analysis is essential for understanding mucus behavior in health and disease, guiding the development of therapies to improve mucociliary clearance and manage chronic respiratory disorders.

\subsection{Correspondence to physiological and pathological processes}

Airway mucus is a dynamic and functional material propelled by ciliary motion, breathing, coughing, and surfactant gradients combined. Effective clearance requires a delicate balance between mucus rheology (e.g., viscoelasticity, yield stress), periciliary liquid layer properties, and ciliary activity \citep{Voynow_2009,Fahy_2010,Lai_2009,Hill_2022,Wagner_2018,Kavishvar_2023,Spagnolie_2015}. Rheological experiments can be tailored to replicate specific physiological scenarios, bridging laboratory findings to real-world mucus behavior.



During glandular extrusion, mucus experiences oscillatory shear at $\sim$\SI{6}{\radian \per \second} under low stress. 
This process is simulated experimentally through shear stress ramps at low stress levels, creep-recovery tests, steady-state shear at low shear rates and frequency sweeps at low frequencies \citep{Leith_1968,King_1985,Leith_1985,King_1987,Zahm_1986,Agarwal_1994,Mahajan_1994,Lauga_2006,Gupta_2009,Voynow_2009,Grotberg_2011}. 
These experiments capture mucus behavior during secretion, where weak intermolecular bonds dominate its response.
Ciliary beating subjects mucus to higher shear frequencies (\(\omega \leq \SI{60}{\radian \per \second}\)) while maintaining low stress. Relevant experiments include frequency sweeps at intermediate frequencies, steady-state shear at moderate shear rates. 
These conditions reflect mucus behavior during normal mucociliary transport, where elasticity and viscosity jointly govern flow.
Coughing, percussive therapy, or high-frequency ventilation impose extreme stresses and shear rates \citep{Voynow_2009,Grotberg_2011a,Gupta_2017,Montenegro-Johnson_2013,Yang_2017,Dudalski_2020,Feng_2020,Ren_2022,Yi_2021,Seung_2020,Spagnolie_2015}.
These scenarios are replicated using large amplitude oscillatory shear (LAOS), steady-state shear at high shear rates. 
Such studies reveal mucus yielding, fracture, and recovery under pathological stress, informing therapies for obstructive respiratory airway disease.

By aligning rheological experiments with physiological processes, researchers gain actionable insights into mucus behavior across health and disease. 
This approach not only enhances diagnostic accuracy but also accelerates the development of targeted therapies, such as mucolytics or surfactant-based treatments, to restore normal mucus rheology in conditions like asthma, cystic fibrosis and COPD.

\subsection{Steady state shear rheology of airway mucus}

\begin{figure*}[htbp]
   \centering
    \includegraphics[width=0.98\textwidth]{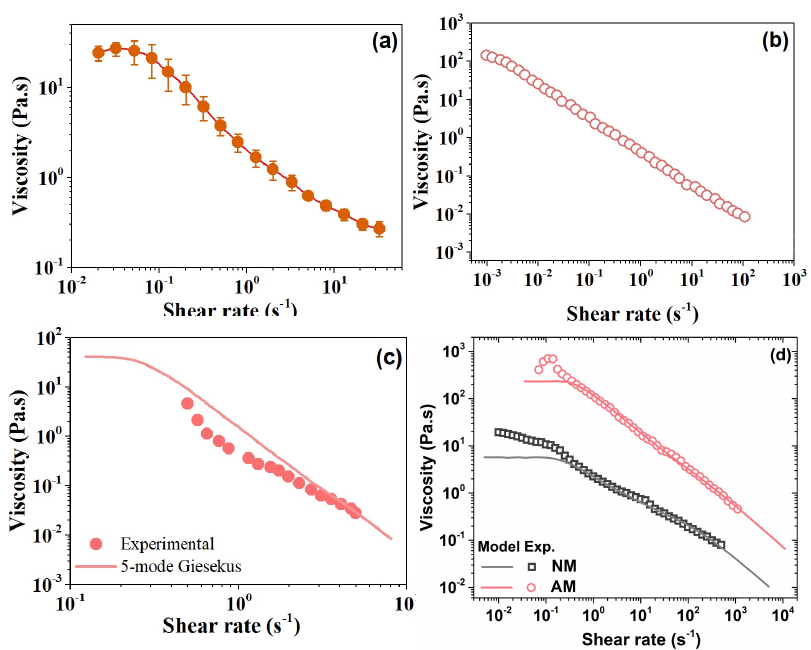}
  \caption{Shear rate dependence of airway mucus viscosity in various conditions. (a) Undiluted CF sputum, adapted from \citet{Suk_2009}. (b) COPD mucus, adapted from \citet{Jory_2022}. (c) \SI{2.5}{\wtpercent} HBE mucus, adapted from \citet{Vasquez_2016}. (d) Simulated normal (NM) and asthmatic (AM) mucus with 5-mode Giesekus fitting, adapted from \citet{Liu_2024}.}
  \label{fig:Steady state}
\end{figure*}

Steady-state shear rheology characterizes the flow behavior of airway mucus under varying shear rates or stresses, offering critical insights into its non-Newtonian nature.
This method measures the steady shear viscosity of the airway mucus as a function of shear rate or stress, revealing a transition between solid-like and fluid-like states for the mucus---a key to understanding its physiological and pathological roles.

Healthy mucus, though challenging to collect (e.g., via endotracheal tubes or bronchial brushes), typically exhibits a viscosity of \SIrange{12}{15}{\pascal\second}, a relaxation time of $\sim$\SI{40}{\second}, and an elastic modulus of $\sim$\SI{1}{\pascal}.
These properties ensure a balance between stability and fluidity for effective mucociliary clearance \citep{Spagnolie_2015,Lai_2009,Puchelle_1987a}. 
Mucus rheology also varies anatomically in the airway system, with lower elasticity in small airways due to reduced solid content compared to tracheal mucus \citep{Lai_2009,Spagnolie_2015,App_1993}.


Mucus from healthy airway exhibits shear-thinning behavior, where viscosity decreases as the shear rate increases. 
This property facilitates mucociliary and cough clearance by reducing viscosity under high shear (e.g., coughing).
The shear-thinning arises from the mucus's complex form of viscoelastic network being disrupted by shear.  
In diseases like cystic fibrosis (CF), COPD, and asthma, mucus becomes hyperviscoelastic due to dehydration and elevated glycoprotein content.
CF sputum, for instance, exhibited pronounced shear-thinning, with its viscosity dropping from $\sim$\SI{110}{\pascal\second} at \SI{0.1}{\per\second} to $\sim$\SI{14}{\pascal\second} at \SI{1}{\per\second}, in a power-law relationship (\(\eta \propto \dot{\gamma}^{\beta}\), where the shear thinning exponent \(\beta=-0.85\) )\citep{Dawson_2003}.
Synthetic normal and asthmatic mucus exhibits similar flow behavior with (\(\beta \approx -0.76\), and -0.56, respectively) despite their biochemical differences \citep{Liu_2024}.
The shear-thinning exponent changes from \(-0.9\) to \(-0.5\) as the mucus changes from healthy to pathological state (e.g., the mucus derived from normal HBE culture vs. sputum from CF patient), a critical behavior for clearance during coughing, sneezing, or therapeutic interventions such as mechanical ventilation and percussive therapies, which can generate shear rates ranging from 100 to \SI{10000}{\per \second}\citep{Voynow_2009,Montenegro-Johnson_2013,Yi_2021,Grotberg_2011a,Ren_2022}.



Interestingly, at low shear rates, mucus can exhibit shear-thickening, where viscosity temporarily increases before transitioning to shear-thinning~\citep{Puchelle_1985,Girod_1992,Banerjee_2001,Dawson_2003,Lai_2009,Vasquez_2014,Kavishvar_2023}. 
CF sputum, for example, shows this behavior up to \SI{1}{\per\second}, likely due to transient mucin network resistance~\citep{Dawson_2003}.
Physiological shear rates in healthy lungs \SI{0.25}{\per \second} in small bronchi and \SI{0.91}{\per \second} in large bronchi may induce similar effects, though disease progression (e.g., reduced ciliary beating in CF) alters this response. 
Shear-thickening is also observed in horse lung mucus~\citep{Vasquez_2014}, HBE mucus~\citep{Vasquez_2016}, and synthetic asthmatic mucus~\citep{Liu_2024}, where condensed mucin chains form temporary networks.

At very low shear rates (\(\dot{\gamma} \rightarrow 0\)), mucus exhibits zero shear viscosity (\(\eta_0\)), behaving like a solid to maintain airway surface protection \citep{Puchelle_1981,Puchelle_1983,Baconnais_1999,Dawson_2003,Matsui_2006,Tang_2021}. 
Reported \(\eta_0\) values vary widely: from \SI{0.01}{\pascal\second} (cell cultures) to \SI{10000}{\pascal\second} (HBE mucus), influenced by organic solid content.
\citet{Hill_2022} summarized those zero-shear viscosity in their recent review. 
These variations are largely due to differences in measurement techniques, mucus sources, and the lack of consistent reporting and control over mucus concentration during measurements.


Airway mucus also displays yield stress (\(\sigma_{\mathrm{y}}\)), a critical shear stress threshold for flow initiation \citep{Kavishvar_2023}. 
For healthy mucus, \(\sigma_{\mathrm{y}}\) typically ranges between 0.05 to \SI{3}{\pascal}, ensuring surface adhesion without impeding clearance \citep{Wong_1977,Naylor_2006,Carpenter_2018}. 
In contrast, in conditions like CF, the cohesive mucus network—composed of mucins, actin-DNA bundles, and bacterial biofilms—leads to significantly higher yield stress, with \( \sigma_{\mathrm{y}} \) ranging from 0.1 to \SI{100}{\pascal}. 
The yield stress of mucus in COPD and asthmatic mucus varies, ranging from 1 to \SI{40}{\pascal} for COPD and 0.2 to \SI{35}{\pascal} for asthmatic mucus, respectively~\citep{Ghanem_2021,Broughton-Head_2007,Tomaiuolo_2014,Lai_2007}.
\citet{Liu_2024} reported that the yield stress of synthetic asthmatic mucus (\SI{32}{\pascal}) far exceeds the synthetic normal mucus (\SI{3}{\pascal}) , aligning with clinical clearance challenges.
However, \citet{Jory_2022} found no significant \(\sigma_{\mathrm{y}}\) differences between healthy, smoker, and COPD mucus collected from human bronchial epithelial (HBE) cultures, suggesting other factors influencing clearance. 
CF sputum from severe patients shows \(\sigma_{\mathrm{y}} = \SI{9}{\pascal}\), four to five times higher than the mild cases, correlating with disease progression as reported by \citet{Tomaiuolo_2014}.

Therapeutic strategies targeting \( \sigma_{\mathrm{y}} \) include 
N-acetylcysteine (NAC) that has been reported to reduce CF sputum \( \sigma_{\mathrm{y}} \) from \SI{16}{\pascal} to \SI{8}{\pascal} by cleaving mucin disulfide bonds, though clinical efficacy is limited due to over-thinning, and 
the recombinant human DNase (rhDNase) capable of lowering \( \sigma_{\mathrm{y}} \) to \SI{1.5}{\pascal} via DNA degradation with proven clinical effectiveness, as well as
heparin known to reduce \( \sigma_{\mathrm{y}} \) from \SI{14}{\pascal} to \SI{8}{\pascal} by disrupting DNA-actin bundles~\citep{Suh_2005,Lai_2009,Rubin_2010,Duncan_2016,Araujo_2018,Nordgard_2018,Boucher_2019,Huck_2022,Pangeni_2023}. 


\begin{figure}[htbp]
   \centering
    \includegraphics[width=0.98\textwidth]{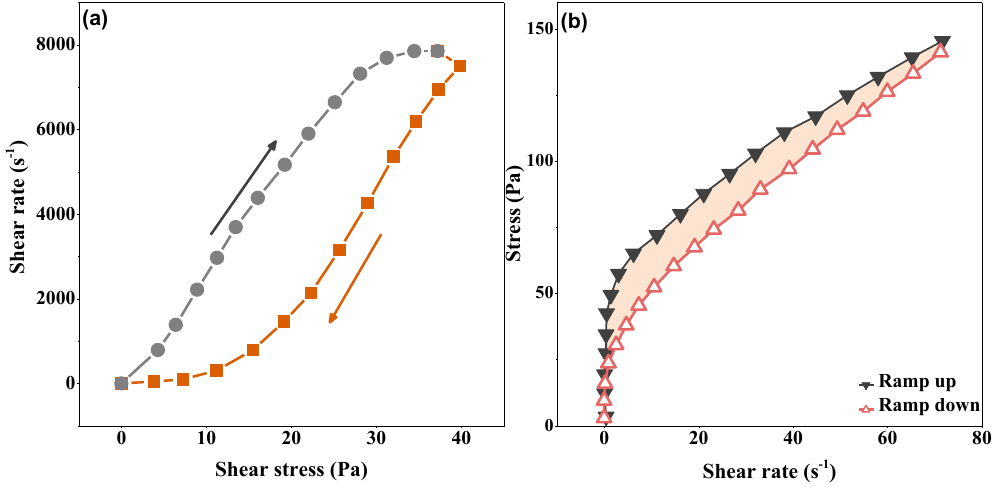}
  \caption{Hysteresis loop of shear rate and shear stress obtained through stress ramp-up and -down methods. (a) CF sputum adapted from \citet{Tomaiuolo_2014}. (b) \SI{2}{\wtpercent} Actigum\texttrademark\ solution, as simulated airway mucus adapted from \citet{Lafforgue_2017a}.
 }
  \label{fig:Loop}
\end{figure}

\subsection{Step shear of airway mucus}

Creep-recovery experiments are essential for investigating the time-dependent rheological properties of airway mucus, particularly under conditions with prolonged relaxation times, such as those found in chronic respiratory diseases \citep{Tang_2021,Hill_2022,Jory_2022}. 
This method provides a more efficient characterization of the terminal region of the linear viscoelastic spectrum compared to traditional dynamic rheological techniques, which can be more time intensive. 
By analyzing the elastic and viscoelastic responses of airway mucus, creep-recovery experiments offer crucial insights into its structural dynamics. During these tests, compliance (\(J(t) = \gamma(t)/\sigma_0\)) is plotted over time, enabling researchers to examine the material’s behavior during both the creep phase (when constant stress is applied) and the recovery phase (after stress is removed).


For example, \citet{Nielsen_2004} demonstrated that CF sputum exhibits non-linear viscoelastic behavior, particularly at higher stress levels (0.7 to \SI{6}{\pascal}), indicating that compliance is not entirely stress independent. 
Initially, compliance increases rapidly, reflecting the elastic response, followed by a slower rise characteristic of viscoelastic behavior. 
The early elastic response corresponds to compliance values around 
\SI{0.08}{\per \pascal}, which equates to an elastic modulus of approximately \SI{12.5}{\pascal}. 
This finding aligns with oscillatory rheology data reporting a storage modulus of \SI{12}{\pascal} at \SI{0.1}{\hertz} under a stress amplitude of \SI{0.7}{\pascal}. 
During the recovery phase, the material exhibits a quick elastic retraction, followed by a slower viscoelastic response. 
At lower stress levels, more than \SI{75}{\percent} of the strain is recovered, underscoring the strong elastic component in CF sputum. 
Conversely, at higher stress levels, compliance tends to increase almost linearly over time, suggesting a shift toward more viscous behavior.
This transition allows the calculation of viscosity from the shear rate, illustrating the complex balance between elasticity and viscosity in mucus.


Comparative analyses of simulated normal (NM) and asthmatic mucus (AM) under constant shear stress, such as those conducted by \citet{Liu_2024}, reveal significant differences in their creep-recovery responses. Normal mucus shows a more pronounced viscous response, characterized by a rapid, linear increase in compliance followed by gradual viscoelastic changes. In contrast, asthmatic mucus exhibits a reduced amplitude of compliance, indicating a stronger elastic component. These differences highlight the critical role of elasticity in maintaining the structural integrity of airway mucus, with pathological changes like those seen in asthma potentially compromising its protective functions.

Integrating frequency sweep data with creep-recovery experiments provides a more comprehensive understanding of mucus behavior. \citet{Evans_2009,Evans_2009a} demonstrated that combining these methods enables the calculation of storage and loss moduli across a wide frequency range. 
The inclusion of long-time scale, low-frequency creep-recovery data from \citet{Tang_2021} facilitated detailed analysis of human bronchial epithelial (HBE) mucus at \SI{1}{\percent} organic solids over five decades of frequency, providing a robust understanding of its viscoelastic properties. Furthermore, creep testing plays a crucial role in determining the yield stress of airway mucus, a key factor in understanding its flow behavior. 
\citet{Jory_2022} found that mucus begins to flow only above a yield stress threshold of approximately \SI{0.05}{\pascal}, which is consistent with steady-state shear rheology findings for COPD mucus derived from HBE cells. 
This yield stress is essential for maintaining the mucus layer on airway surfaces, preventing it from being displaced by gravity and ensuring continued protection of the respiratory tract.

\begin{figure}[htbp]
   \centering
    \includegraphics[width=0.99\textwidth]{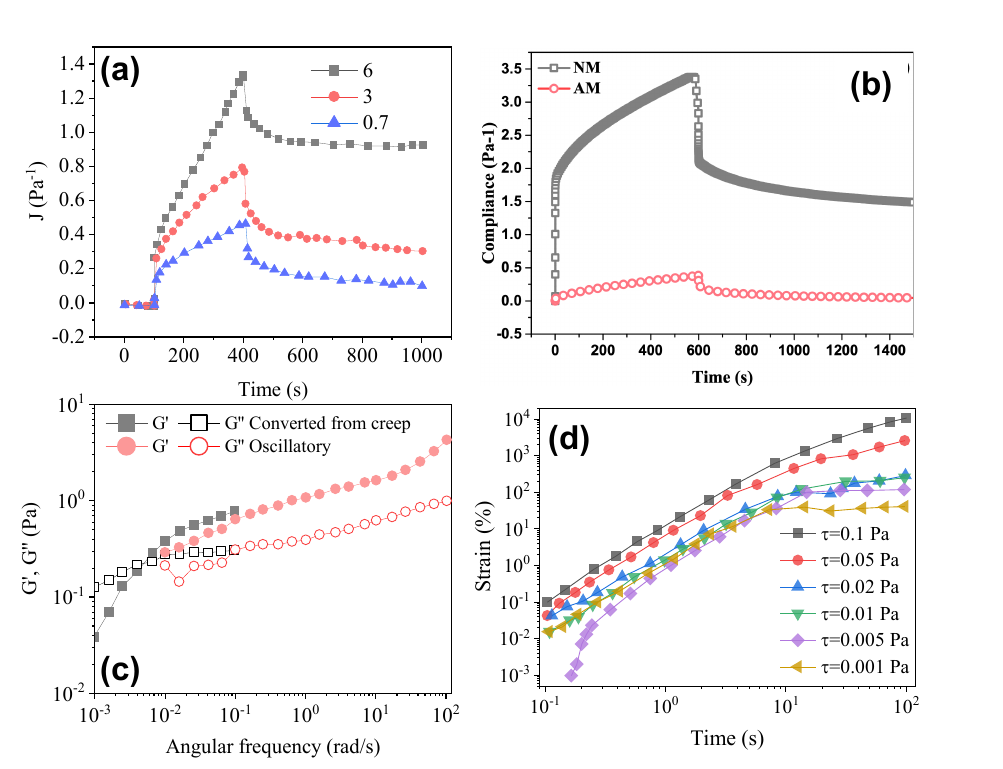}
  \caption{Compliance of airway mucus during creep and recovery experiments. (a) CF sputum, a constant stress (0.7, 3, 6 Pa) applied at \SI{100}{\second} and removed 
  at \SI{400}{\second}, with strain recovery monitored over the subsequent \SI{600}{\second}
  , adapted from \citet{Nielsen_2004}. 
  (b) Synthetic normal and asthmatic airway mucus (NM, AM), \SI{1}{\pascal} stress applied at \SI{0}{\second}, and removed at \SI{600}{\second}, with strain recovery monitored over the subsequent \SI{900}{\second}, adapted from \citet{Liu_2024}.
  (c) Frequency-dependent elastic and viscous moduli (\(G’, G''\)) for human bronchial epithelial (HBE) mucus, adapted from \citet{Tang_2021}. 
  (d) Yield stress of mucus from healthy individual in creep test, illustrating strain as a function of time under applied stress, adapted from \citet{Jory_2022}.
}
  \label{fig:Creep}
\end{figure}


In addition to creep-recovery tests, step shear rate experiments provide important insights into the breakdown and recovery of mucus structure under varying physiological conditions. This process, often referred to as thixotropy, is vital for understanding the mechanical properties of airway mucus. Step shear rate experiments typically consist of three phases: first, a low shear rate is applied to establish the baseline structure of the mucus in its undisturbed state; next, a higher shear rate is imposed, causing disruption to the internal structure; finally, the shear rate is reduced, allowing the mucus to recover. The recovery process is monitored over time to determine how quickly and to what extent the mucus restores its original structure after mechanical deformation \citep{Lafforgue_2017,Lafforgue_2018}.


\citet{Lafforgue_2017} applied this approach to \SI{2}{wt.\percent} airway mucus simulants.
Under moderate shear (\SI{1.6}{\per \second}, comparable to normal breathing), the mucus recovered \SI{90}{\percent} of its structure within \SI{2.7}{\second}, with full recovery taking \SI{75}{\second}. 
However, when exposed to a higher shear rate (\SI{100}{\per \second}, simulating the forces experienced during coughing), recovery times were significantly longer: \SI{90}{\percent} recovery took \SI{17}{\second}, while full recovery required \SI{917}{\second}.
These findings illustrate that greater shear stress leads to more substantial structural damage, resulting in slower regeneration. Further research is required to fully understand the thixotropic properties of airway mucus. 
A time-dependent rheological model could capture the complex behavior of mucus and be used to simulate its response in a model trachea under different air pressure conditions, such as those generated by clearance-assisting devices.

Together, step shear rate tests and creep-recovery experiments provide a comprehensive understanding of how airway mucus behaves under different mechanical stresses. 
Creep-recovery testing is especially valuable for elucidating the complex interplay between the elastic and viscoelastic properties of airway mucus. 
Lower shear rates, such as those encountered during normal breathing, facilitate quicker recovery of mucus structure, which is critical for maintaining respiratory function. In contrast, higher shear rates, such as those experienced during coughing, result in prolonged recovery due to more severe structural disruption. These insights enhance our understanding of mucus behavior in both healthy and pathological conditions and serve as a foundation for developing therapeutic strategies to manage respiratory diseases more effectively.

\begin{figure}[htbp]
   \centering    \includegraphics[width=0.99\textwidth]{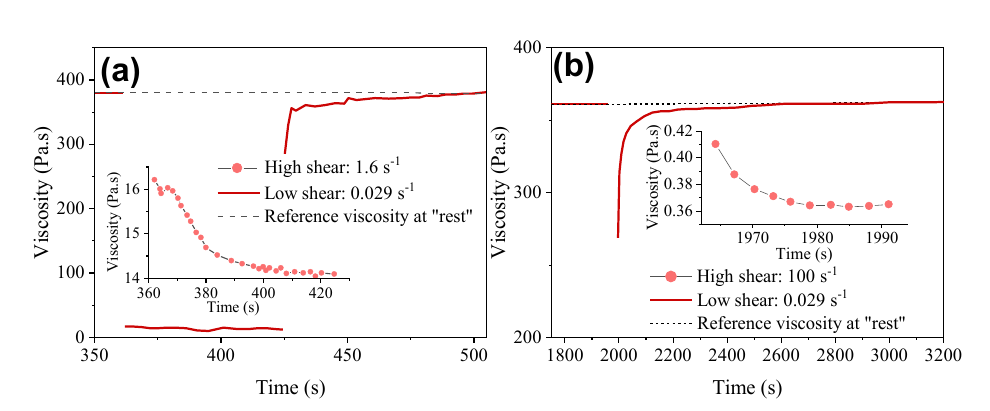}
  \caption{Three-step shear rate experiments for evaluating the thixotropic behavior of synthetic airway mucus composed of \SI{2}{\wtpercent} scleroglucan. 
  (a) Protocol 1: step 1 at 
  \SI{0.029}{\per \second}, 
  step 2 at 
  \SI{1.6}{\per \second}, 
  and step 3 at 
  \SI{0.029}{\per \second}. 
  (b) Protocol 2: step 1 at \SI{0.029}{\per \second}, 
  step 2 at 
  \SI{100}{\per \second}, 
  and step 3 at 
  \SI{0.029}{\per \second}, adapted from \citet{Lafforgue_2018}}
  \label{fig:Step}
\end{figure}

\subsection{Small amplitude oscillatory shear of airway mucus}

\subsubsection{Frequency sweep}

Frequency sweep experiments measure the viscoelastic properties of airway mucus as functions of oscillatory frequency (\(\omega\)). These experiments are typically conducted using small amplitude oscillatory shear (SAOS), a technique that applies a sinusoidal strain small enough to preserve mucus microstructure while quantifying its linear viscoelastic response \citep{hill_2018,Ma_2018,Ahonen_2019,Atanasova_2019,Rouillard_2020}. The complex modulus \(G^*(\omega) = G'(\omega) + iG''(\omega)\) captures mucus behavior: \(G'\) reflects elastic energy storage in the mucin network, while \(G''\) represents energy dissipation through molecular friction during flow and deformation \citep{Celli_2007,Lai_2009,Kavishvar_2023,Milian_2024}. 



Airway mucus exhibits multiscale viscoelasticity due to its hierarchical structure, governed by weak physical bonds (e.g., hydrogen bonds) and strong chemical bonds (e.g., disulfide crosslinks) \citep{Fahy_2010,Voynow_2009,Boucher_2019}. SAOS reveals a power-law frequency dependence in \(G'\) and \(G''\) (\(G' \sim \omega^{m}\)), indicative of the broad relaxation timescales inherent to airway mucus \citep{Mellnik_2014,Wagner_2017,Wagner_2018,Broedersz_2014}. This behavior enables mucus to adapt dynamically to mechanical stresses, balancing solid-like rigidity and fluid-like flow.  

\begin{figure}[htbp]
   \centering
    \includegraphics[width=0.6\textwidth]{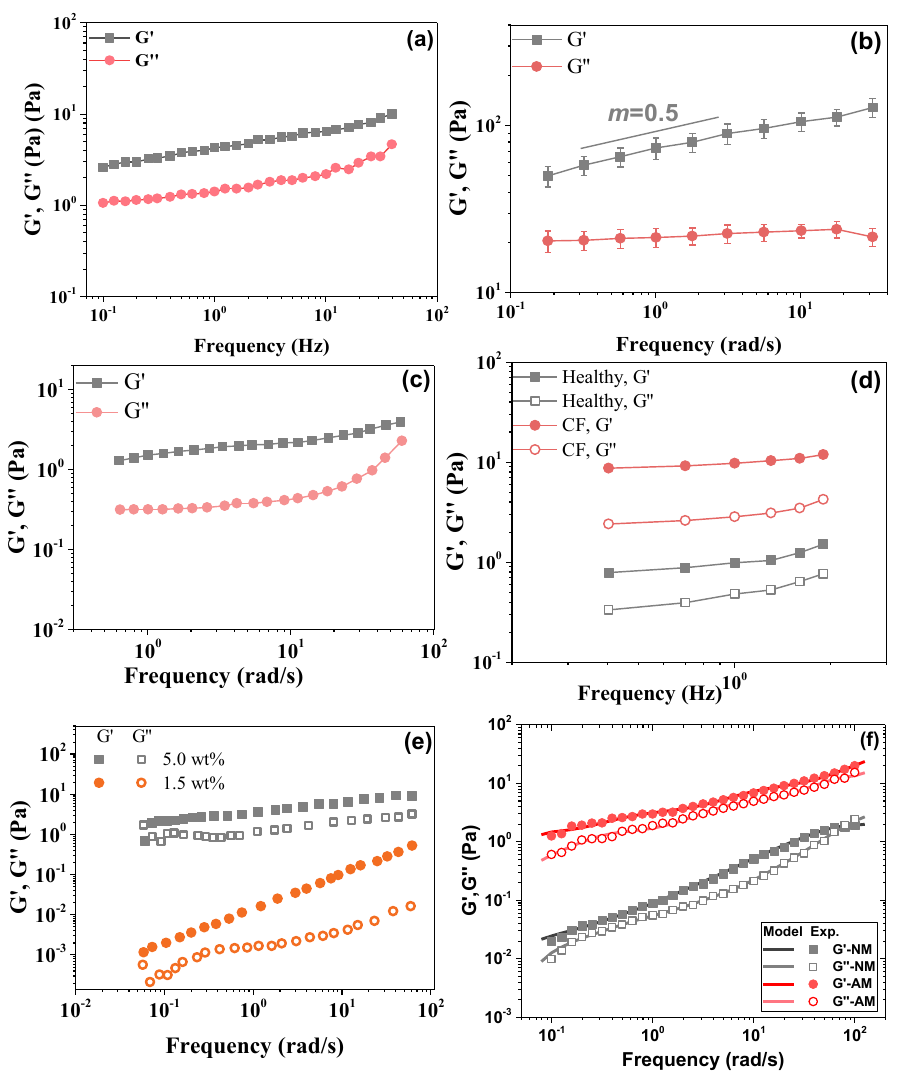}
  \caption{Frequency dependent complex modulus of airway mucus. (a) Healthy sputum: behavior as a cross-linked gel, with elastic modulus (\(G'\)) dominating viscous modulus (\(G''\)) across the physiological frequency range, and the plateau enabling direct determination of both density and molecular weight of entanglements \citep{Innes_2009}. (b-f) CF mucus \citep {Dawson_2003}, COPD mucus \citep{Jory_2022},, Healthy vs. CF mucus \citep{Yuan_2015}, HBE cell mucus with 3.0 to 5.0 wt\% solids \citep{Vasquez_2014}, and synthetic normal versus asthmatic airway mucus (NM vs. AM) \citep{Liu_2024}, respectively.}
\label{fig:FS}
\end{figure}

Healthy mucus predominantly behaves as a gel (\(G' > G''\)) across physiological shear rates \ (0.1–\SI{100}{\radian \per \second}), critical for effective mucociliary clearance as shown in \cref{fig:FS} \citep{Innes_2009,Caicedo_2015,Hill_2014}. At the crossover frequency (\(\omega_c\)), \(G'\) equals \(G''\), marking the transition from gel-like (\(G' > G''\)) to liquid-like (\(G'' > G'\)) behavior. However, this transition does not equate to yielding—permanent structural failure requiring nonlinear techniques like LAOS to reveal \citep{Wagner_2018,Hill_2014,Liu_2024}. For human bronchial epithelial (HBE) mucus, the gel point (GP), where \(G'\) surpasses \(G''\), occurs at $\sim$\SI{4}{wt.\%} mucin concentration \citep {Hill_2014,Vasquez_2016}.

Pathological conditions, such as asthma, CF and COPD, significantly alter mucus rheology \citep{Innes_2009,Patarin_2020,Liu_2024,Hill_2014,Lin_2020,Song_2022,Caughman_2024,Ghanem_2021}. In CF, the presence of DNA and cellular debris forms a secondary network within the mucin matrix, increasing mucus stiffness as shown in \cref{fig:FS}(a) \citep{Innes_2009}. \citet{Dawson_2003} reported that CF sputum exhibits a higher elastic modulus than viscous modulus across all tested frequencies, behaving as a viscoelastic solid. At low and intermediate frequencies, \(G'(\omega)\) follows a power-law relationship (\(G'(\omega) \sim \omega^{1/2}\)), suggesting flexible polymer behavior as shown in \cref{fig:FS}(b). Recent studies by \citet{Liu_2024} found that synthetic asthmatic mucus (AM) exhibits a more pronounced solid-like behavior than synthetic normal mucus (NM), attributed to stronger mucin crosslinking and the presence of pathological materials and highlighting rheological differences between AM and NM (\cref{fig:FS}(f)). These findings align with other studies reporting similar power-law behaviors in mucus as shown in \cref{fig:FS}(c-d) \citep{Vasquez_2016,Philippe_2017,Jory_2022,Wagner_2023}. 
Confocal microscopy has revealed actin-DNA bundles and bacterial biofilms that hinder mucus clearance by cilia, and neutrophil extracellular traps (NET) lung diseases contribute similarly to increasing mucus stiffness
\citep{Linssen_2021,Brinkmann_2004,Thulborn_2019,Kavishvar_2023,Pangeni_2023}. 

The viscoelastic properties revealed by SAOS have significant therapeutic implications. Mucolytic agents, which aim to reduce mucus stiffness and increase mesh size, can be evaluated using SAOS. For example, \citet{Yuan_2015} demonstrated that oxidative agents like dimethyl sulfoxide increase mucus stiffness by forming additional disulfide bonds, mimicking the effects of oxidative stress in CF. Conversely, surfactants such as 1,2-hexanediol reduce viscoelastic moduli by disrupting weak hydrophobic interactions reported by \citet{Wagner_2017}. These findings underscore the potential of SAOS for developing targeted therapies to restore normal mucus rheology in diseases like CF and COPD. Furthermore, inflammatory mediators in CF sputum also appear to influence mucus structure and modulus, although the underlying mechanisms remain unresolved \citep{Serisier_2009}. Finally, SAOS provides critical insights into mucus viscoelasticity, revealing frequency-dependent behavior and pathological deviations. By linking rheological properties to clinical outcomes, this technique guides the development of therapies to improve mucociliary clearance in respiratory diseases.  



\subsubsection{Time sweep}

Time sweep experiments using small-amplitude oscillatory shear (SAOS) are crucial to investigating the time-dependent viscoelastic behavior of airway mucus. These tests measure changes in elastic and viscous properties over time, providing critical insights into how biological factors such as purulence and environmental conditions influence mucus after collection \citep{Innes_2009}. Such information is essential to establishing standardized protocols for airway mucus handling in rheological studies.

\begin{figure}[htbp]
   \centering
    \includegraphics[width=0.9\textwidth]{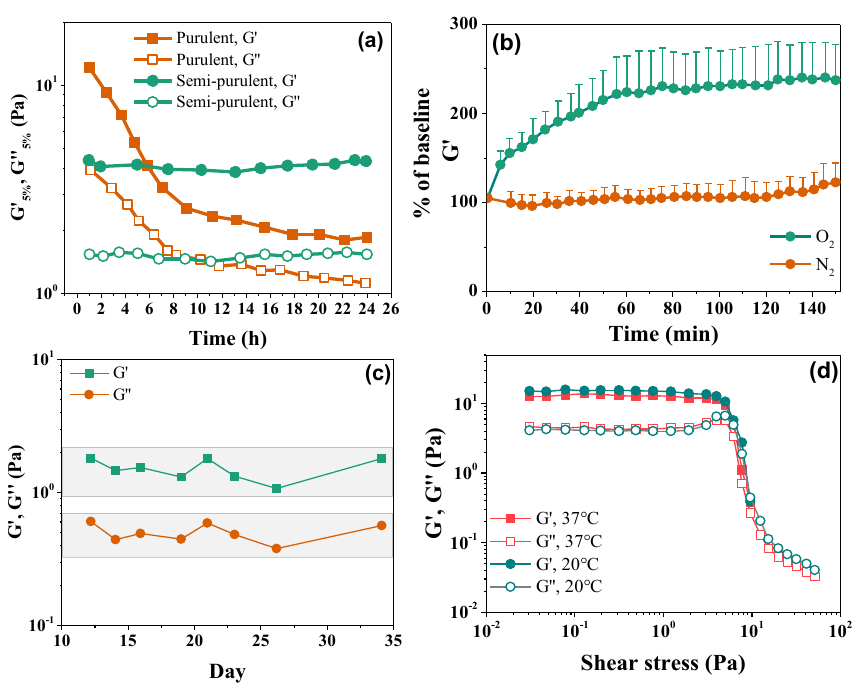}
  \caption{Time and temperature dependent complex modulus of airway mucus under SAOS. 
(a) Purulent and semi-purulent mucus samples: elastic modulus (\(G'\)) and viscous modulus (\(G''\)) evolving over 24-hour, adapted from \citet{Esteban-Enjuto_2023}. (b) Healthy sputum samples: oxygen exposure (O2) increasing elastic modulus over time, but not nitrogen (N2), adapted from \citet{Yuan_2015}. (c) HBE mucus derived from smokers on eight separate days:
\(G', G''\) measured at \SI{1}{\hertz} with \SI{1}{\percent} strain, adapted from \citet{Jory_2022}. (d) \SI{0.75}{\wtpercent} gel as airway mucus simulant: \(G', G''\) measured at 1 Hz and different temperatures as a function of stress amplitude, adapted from \citet{Lafforgue_2018}.} 
  \label{fig:TimeSweep}
\end{figure}

In a study by \citet{Esteban-Enjuto_2023}, the rheological evolution of CF sputum was tracked over a 24-hour period (\cref{fig:TimeSweep}). The purulent samples showed a substantial decrease in both elastic and viscous moduli---approximately \SI{50}{\percent}---within the first six hours, while semi-purulent samples remained stable. This finding indicates that purulence, which reflects biological load and immune activity, plays an important role in sputum degradation, more so than oxidation. Rapid reduction in viscoelasticity underscores the need for prompt analysis, especially for the purulent samples, to avoid structural changes that could compromise the accuracy of rheological measurements. Similar findings were observed by \citet{Innes_2009} in sputum samples from patients with acute asthma. Rapid breakdown of the mucus gel network, particularly in samples with higher biological activity, suggests a connection between mucus stability and its immune response. Further investigations comparing rheological changes with neutrophil and bacterial data could provide deeper insights into the biological processes that influence mucus stability. Additionally, the subjective nature of current purulence assessment methods calls for more objective, quantitative approaches.


Environmental factors also affect mucus rheology. 
\citet{Yuan_2015} demonstrated that exposure to 100\SI{}{\percent} oxygen significantly increased the elastic modulus of healthy airway mucus over time, while nitrogen exposure had no measurable effect. This suggests that oxygen enhances mucus elasticity, which has implications for respiratory therapy in conditions such as cystic fibrosis. Furthermore, the study examined the mucolytic effects of thiol-saccharides, specifically methyl 6-thio-6-deoxy-\( \alpha \)-D-galactopyranoside (TDG), compared to N-acetylcysteine (NAC). 
TDG was found to have stronger and faster mucolytic activity than NAC. In just 2 minutes of exposure, TDG demonstrated a much larger reduction in mucus elasticity than NAC. By 12 minutes, both agents showed similar effects, but the earlier onset of mucolytic action with TDG highlights its potency as a reducing agent. These time-dependent differences in mucolytic activity suggest that thiol-saccharides like TDG may offer more rapid therapeutic benefits in reducing mucus viscosity compared to traditional treatments such as NAC.


Conversely, \citet{Jory_2022} conducted time sweep experiments on mucus from air-liquid interface (ALI) cultures to assess how its viscoelastic properties evolve as the culture matures. Their findings showed that the storage and loss moduli remained consistent from 10 to 30 days, indicating that the maturity of the ALI cultures during this time frame did not significantly affect the rheological properties of the mucus.
This stability makes ALI cultures a reliable model for long-term studies of mucus rheology. Similarly, \citet{Lafforgue_2017} investigated the effect of aging on an airway mucus simulant and found no significant differences in rheological measurements between ``fresh'' simulant and that stored at 4°C for four days, further supporting the robustness of the mucus simulant in such experiments.


In summary, time sweep experiments provide essential insight into the viscoelastic properties of airway mucus and how factors such as purulence, environmental exposure, and culture conditions influence its rheological behavior. The studies highlight the importance of rapid sample analysis, objective purulence classification, and the role of oxygen in modulating mucus properties. Time-dependent differences in mucolytic activity, particularly the faster action of thiol-saccharides such as TDG, suggest potential improvements in therapeutic strategies for managing mucus viscosity. By refining these methodologies, researchers can better understand the rheological changes associated with both healthy and pathological mucus, to improve diagnostic and treatment approaches.

\subsubsection{Temperature sweep}

In addition to time-sweep experiments, temperature-sweep experiments under SAOS are also crucial to understanding how temperature affects the rheological properties and structural integrity of airway mucus. In healthy individuals, airway mucus is prone to degradation by proteases, a process that is significantly influenced by temperature.


\citet{Innes_2009} demonstrated that sputum from healthy subjects incubated at \SI{37}{\celsius} for 24 hours exhibited a marked reduction in both elastic (\(G'\)) and viscous (\(G''\)) moduli compared to that stored at \SI{4}{\celsius}. 
This suggests that higher temperatures accelerate the breakdown of mucus structure, further supported by a reduction in entanglement density---a key indicator of the integrity of the mucus network \citep{Innes_2009,Broedersz_2014,Sharma_2021}. 
Additionally, temperature-dependent changes in the size profile of mucin polymers were observed in healthy sputum. 
After incubation at \SI{4}{\celsius} and \SI{37}{\celsius}, sputum samples were subjected to rate-zonal centrifugation and stained with MAN5ACI (MUC5AC polyclonal antiserum). 
The size profiles differed significantly, with the \SI{37}{\celsius} sample showing predominantly smaller mucins, indicating degradation of the mucus structure at higher temperatures. 
Similarly, \citet{Esteban-Enjuto_2023} demonstrated that heating was highly destructive to CF sputum, whereas freezing at \SI{-80}{\celsius} had no discernible effect on its rheology. Macromolecular colloidal gels, used as simulants of bronchial mucus, also exhibited gel-like behavior across the entire temperature sweep range 
(10--\SI{45}{\celsius}), with a linear decrease in moduli as temperature increased, as shown by \citet{Lafforgue_2017,Lafforgue_2017a}.

In contrast, \citet{Sharma_2021} observed a significant increase in complex modulus of the PEG-based mucus simulant as temperature increased from \SI{25}{\celsius} to \SI{37}{\celsius}
, with \(G'\) increasing by a factor of 30--200 and \(G''\) by a factor of only 3--5. 
This suggests that at lower temperatures, viscous properties dominate, but at \SI{37}{\celsius}, 
elastic properties become dominant. 
Notably, at \SI{37}{\celsius}, 
both \( G' \) and \( G'' \) scale approximately with \(\omega^{\text{0.75--0.8}}\).

\citet{Lafforgue_2017} further measured the temperature dependence of the linear viscoelastic (LVE) range of airway mucus simulants 
(with \SI{0.75}{wt.\percent} Actigum concentration) by comparing stress amplitude sweeps at 
\SI{20}{\celsius} and \SI{37}{\celsius} as shown in \cref{fig:TimeSweep}(d). 
The resulting curves were nearly superimposed. Similarly, \citet{Taylor_2005,Taylor_2005a} found a limited influence of temperature on the rheological properties of mucin–alginate gels. They observed no significant differences in gel behavior across a broader temperature range (10--\SI{60}{\celsius}) during frequency sweep tests. The rheological properties of these aqueous polysaccharide solutions are known to be influenced by macromolecular conformation. For instance, Viscogum (galactomannan chains cross-linked in the presence of sodium tetraborate) and Actigum (an extracellular polysaccharide) can adopt either a rigid helicoidal or a random entangled conformation, depending on temperature. The measurements suggest that Actigum adopts a rigid helicoidal conformation in the mucus simulants tested here, contributing to the observed gel-like behavior, and this conformation remained stable across the temperature range of \SI{20}{\celsius} to \SI{37}{\celsius}. 
Recently, \citet{Jory_2022} conducted a detailed comparison of rheological experiments at \SI{37}{\celsius} and \SI{20}{\celsius} using mucus collected from healthy human bronchial epithelial (HBE) cultures, COPD patients, and smokers without COPD. Their results were consistent across all groups.
These findings underscore the critical role of temperature in modulating the viscoelastic properties of airway mucus, providing valuable insight into how mucus behaves under various physiological and pathological conditions.

\subsection{Large amplitude oscillatory shear of airway mucus}

Large amplitude oscillatory shear (LAOS) has emerged as a vital technique for probing nonlinear viscoelastic properties and yielding transitions of airway mucus \citep{Kamkar_2022,Kavishvar_2023,Liu_2020,Liu_2024}. Unlike the traditional rheological methods confined to the linear viscoelastic region, LAOS enables the study of mucus under large deformations, closely mimicking physiological conditions such as coughing and high-frequency ventilation \citep{Spagnolie_2015}. This technique has been applied to diverse mucus samples, including those from healthy individuals \citep{Schuster_2013,Patarin_2020}, patients with respiratory diseases \citep{Duvivier_1984,Dawson_2003,Nielsen_2004,Lai_2007,Radtke_2018,Esteban-Enjuto_2023},  synthetic simulants \citep{Taylor_2003,Celli_2007,Murgia_2016,Lafforgue_2017,Lafforgue_2018,Tan_2020,Larobina_2021,Liu_2024,Milian_2024}, animal-derived mucus (e.g., horse lung mucus \citep{Vasquez_2014,Gross_2017}, and HBE cell cultures \citep{Vasquez_2016,Jory_2019,Jory_2022}). By independently controlling time scale (frequency) and strain amplitude, LAOS provides simultaneous quantification of elastic and viscous components of the complex modulus, offering critical insights into mucus behavior that are inaccessible via conventional methods like steady shear or creep tests.


Researchers employ two primary approaches to interpret LAOS data of airway mucus. The first, Fourier Transform Rheology (FTR), offers a straightforward approach to analyzing nonlinear output signals by decomposing them into a series of harmonics.  The second approach uses Lissajous–Bowditch curves (stress vs.\ strain or stress vs.\ strain rate) to identify nonlinearities \citep{Vasquez_2016,Kamkar_2022,Liu_2020,Liu_2024}.
In oscillatory shear, the sinusoidal strain input is:
\begin{equation}
\gamma(t) = \gamma_0 \sin(\omega t),    
\end{equation}
where \(\gamma_0\) is the strain amplitude, and \(\omega\) is the angular frequency. Within the linear viscoelastic regime, the stress response remains sinusoidal:
\begin{equation}
\sigma(t) = \sigma_0 \sin(\omega t + \delta),
\end{equation}
where \(\sigma_0\) is the stress amplitude, and \(\delta\) is the phase shift. 

The in-phase component of the stress defines the elastic/storage modulus \(G'\):
\begin{equation}
G' = \frac{\sigma_0}{\gamma_0} \cos \delta,
\end{equation}
while the out-of-phase component defines the viscous/loss modulus \(G''\):
\begin{equation}
G'' = \frac{\sigma_0}{\gamma_0} \sin \delta.    
\end{equation}


As the strain amplitude increases beyond a certain point, the output stress \(\sigma(t)\) becomes nonlinear, and the stress response can no longer be described by a simple sinusoidal function. 
Instead, it can be decomposed using Fourier Transform Rheology (FT-Rheology), expressed as a series of harmonics:
\begin{equation}
\sigma(t) = \sum_{n=1,\mathrm{odd}}^{\infty} \sigma_n \sin(n \omega t + \delta_n),
\end{equation}
where \(n\) is an odd integer representing the harmonic number, \(\sigma_n\) is the amplitude of the \(n\)th harmonic, and \(\delta_n\) is the phase angle. 


The presence of higher harmonics (\(n > 1\)) indicates nonlinear behavior. Nonlinear moduli are then defined as:
\begin{equation}
G'_n(\gamma_0, \omega) = \frac{\sigma_n \cos \delta_n}{\gamma_0}, \quad
G''_n(\gamma_0, \omega) = \frac{\sigma_n \sin \delta_n}{\gamma_0}.
\end{equation}

\begin{figure*}[htbp]
   \centering
    \includegraphics[width=0.85\textwidth]{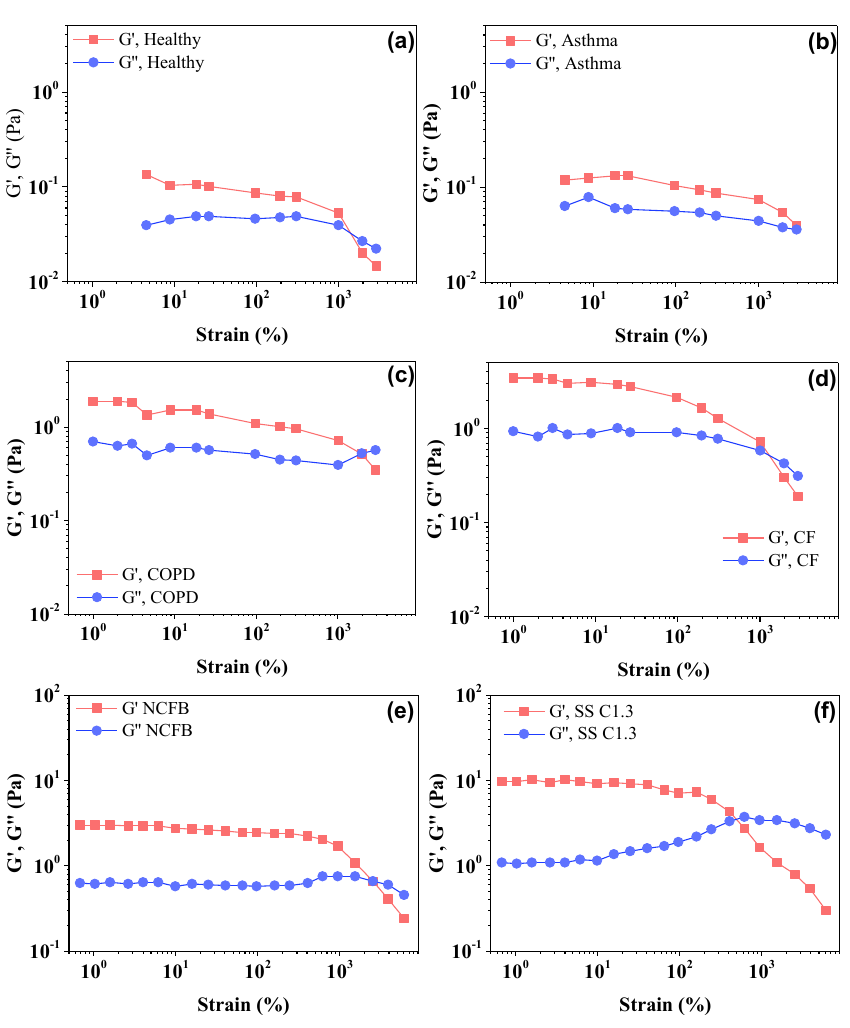}
  \caption{Strain dependent complex moduli (\(G'\), \(G''\)) of airway mucus under LAOS. (a) Healthy sputum, (b-d) mucus samples from asthmatic, COPD, or CF patients \citep{Patarin_2020}. (e, f) 1.3 wt\% PEG-4SH slime-based gel, and sputum from non-CF bronchiectasis patient \citep{Milian_2024}.}
  \label{fig:LAOS}
\end{figure*}

Healthy mucus exhibits a linear viscoelastic plateau (\(G' > G''\)) at low strain amplitudes, transitioning to nonlinear behavior as strain increases as shown in (\cref{fig:LAOS}). This transition is marked by a decrease in \(G'\), an increase in \(G''\), and eventual crossover (\(G'' > G'\)), reflecting microtructural yielding within the mucin network. CF mucus shows a higher yield strain threshold, indicating a denser mucin network (\cref{fig:LAOS} b--d). Similar trends are observed in mucus from COPD and asthmatic patients, where pathological remodeling amplifies nonlinear responses and elevates yield stress or strain as reported by \citet{Patarin_2020}. Notably, the slight overshoot in \(G''\) observed in these systems aligns with the behavior of Type III complex fluids, as classified by \citet{Hyun_2002}. This behavior, also seen in other types of airway mucus, reflects both strain resistance and heterogeneous structural rearrangements within the mucus matrix \citep{Dawson_2003,Celli_2007,Lai_2007,Vasquez_2014,Tan_2020,Jory_2022}. The overshoot is not merely a transient artifact of energy dissipation but rather a signature of localized yielding events within the mucin network.
 This strain resistance is critical for understanding the mucus ability to respond to mechanical forces such as coughing and airway clearance.



\begin{figure}[htbp]
   \centering
    \includegraphics[width=0.95\textwidth]{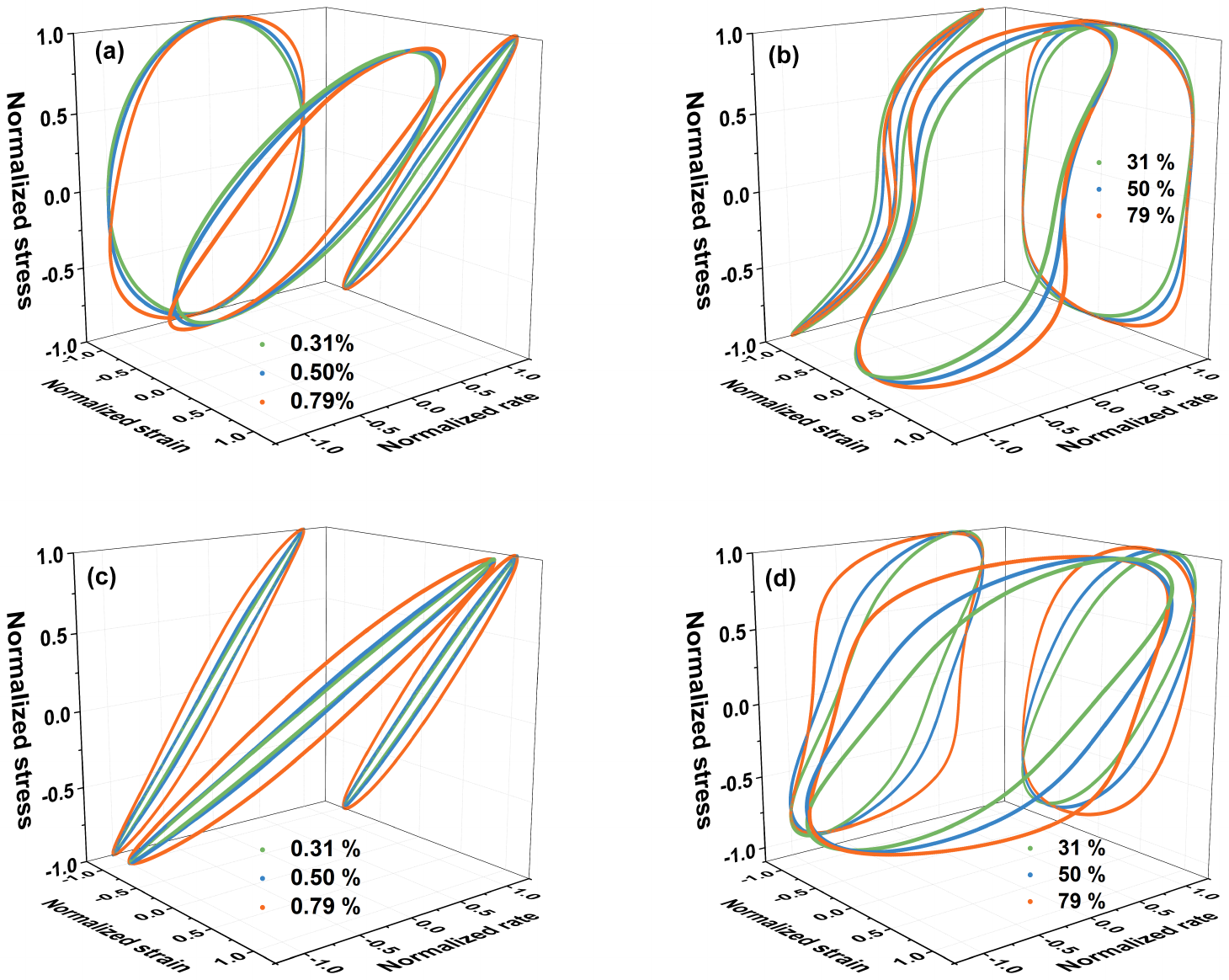}
  \caption{Lissajous curves of synthetic normal and asthmatic airway mucus (NM and AM) under different strain amplitude at \SI{6.28}{\radian \per \second}. 
(a--b) Results of NM. (c--d) Results of AM. 
The three-dimensional curves show measured stress as a function of the orthogonal oscillatory input strain (elastic component) or strain rate (viscous component), respectively \citep{Liu_2024}}
  \label{fig:3DLissajous}
\end{figure}

Furthermore, Lissajous curves provide intuitive visualizations of nonlinearity. Under small strains (\(\leq \SI{1}{\%}\)), the elastic (\(\sigma\)-\(\gamma\)) and viscous (\(\sigma\)-\(\dot{\gamma}\)) curves remain elliptical (linear regime). At large strains (\(\geq \SI{30}{\%}\)), the distortion from ellipticity signifies nonlinearity (\cref{fig:3DLissajous}). For example, \citet{Liu_2024} demonstrated that synthetic normal mucus (NM) exhibits pronounced non-elliptical Lissajous curves at high strains, reflecting stronger nonlinear viscoelasticity compared to synthetic asthmatic mucus (AM) as shown in \cref{fig:3DLissajous}. Similar results were also observed in mucus samples from HBE cell cultures derived from healthy subjects or COPD patients \citep{Jory_2022}, horse lung mucus \citep{Vasquez_2014,Gross_2017}, and simulants \citep{Taylor_2003,Ewoldt_2007}.

To further analyze the nonlinear stress response, the stress signal can be decomposed into elastic and viscous components using Chebyshev polynomials of the first kind. 
These components are represented as:
\begin{gather}
\sigma_{\mathrm{elastic}}(x) = \gamma_0 \sum_{n \, \mathrm{odd}} e_n(\omega, \gamma_0) T_n(x), \\
\sigma_{\mathrm{viscous}}(y) = \sum_{n \, \mathrm{odd}} v_n(\omega, \gamma_0) T_n(y),
\end{gather}
where \( x = \gamma / \gamma_0  \) and 
\( y = \dot{\gamma}/\dot{\gamma}_0 \) are the normalized variables, and \( T_n \) denotes the \( n \)-th order Chebyshev polynomial.

In the linear viscoelastic regime, at low strain values, \( e_1 \) and \( v_1 \) correspond to the storage modulus \( G' \) and the viscosity \( \eta \), respectively. 
The coefficients \( e_n \) and \( v_n \) provide insights into the elastic and viscous nonlinearities, respectively. 
According to these coefficients, the following intra-cycle nonlinear interpretations were proposed:
Strain-stiffening: \( e_3 > 0 \),
Strain-softening: \( e_3 < 0 \),
Shear-thickening: \( v_3 > 0 \),
and Shear-thinning: \( v_3 < 0 \).

\begin{figure}[htbp]
   \centering
    \includegraphics[width=0.95\textwidth]{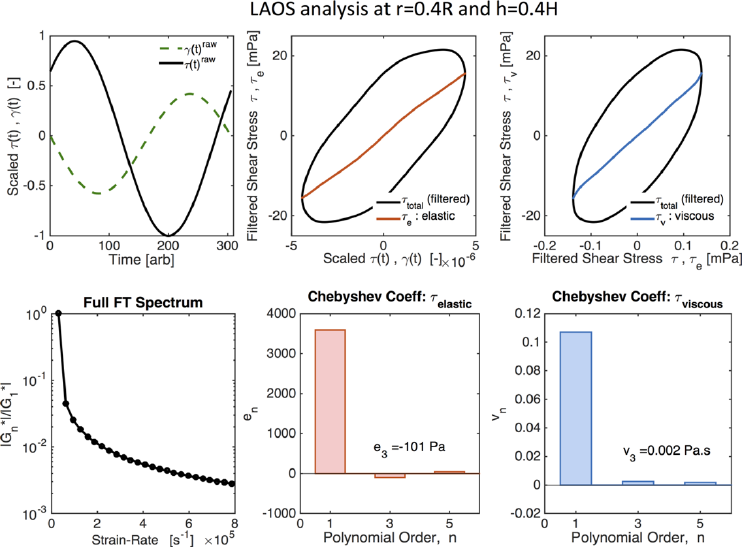}
  \caption{Stress decomposition (SD) analysis of airway mucus under LAOS, highlighting nonlinear coefficients (\(e_3, v_3\)) at different driving velocities and aspect ratios, adapted from \citep{Vasquez_2016}.}
  \label{fig:SD}
\end{figure}

Moreover, the \( n \)-th order Chebyshev coefficients and Fourier coefficients can be related to each other via the following equations:
\begin{equation}
e_n = G_n' (-1)^{\frac{n-1}{2}}, 
\quad 
v_n = \frac{G_n''}{\omega} = \eta_n'.
\end{equation}
As previously discussed, in the nonlinear regime, the first-order moduli alone are insufficient to fully capture the transient material response, despite their continued relevance for understanding energy storage and dissipation. \citet{Vasquez_2016} characterized the nonlinearity of mucus flow using a 5-mode Giesekus model, demonstrating that the third harmonic Chebyshev coefficients depend on aspect ratio and mean driving velocity (\cref{fig:SD}). Their findings indicate that elastic stress ($e_3$) dominates over viscous stress ($v_3$), suggesting that nonlinearities primarily result from strain softening induced by cilia-driven forces. However, mucus behavior in LAOS can vary across organs and species. For instance,  \citet{Vasquez_2014} reported strain-stiffening in horse mucus.

\citet{Ewoldt_2008} introduced a set of geometrically-motivated elastic moduli, which are then related to both conventional Fourier Transform (FT) rheology and the newly proposed Chebyshev stress decomposition. 
The following definitions are provided:
\begin{gather}
G'_M = \left.\frac{d\sigma}{d\gamma}\right|_{\gamma=0} = \sum_{n \, \mathrm{odd}} n G'_n (-1)^{(n-1)/2} = e_1 - 3e_3 + \cdots \\
G'_L = \left.\frac{\sigma}{\gamma}\right|_{\gamma = \gamma_0} = \sum_{n \, \mathrm{odd}} G'_n (-1)^{(n-1)/2} = e_1 + e_3 + \cdots
\end{gather}
Here, \( G'_M \) represents the minimum-strain modulus or tangent modulus at zero strain, while \( G'_L \) denotes the large-strain modulus or secant modulus evaluated at the maximum strain. These measures can be used to calculate the elastic modulus from raw data, and for a linear-viscoelastic response, both \( G'_M \) and \( G'_L \) will be equivalent to the conventional \( G' \).

\citet{Vasquez_2014} found that the minimum-strain modulus \( G'_M \) and the large-strain modulus \( G'_L \) provide crucial information about the nonlinear behavior of horse lung mucus. 
At low strain values, they observed that \( G'_M = G'_L = G'_1 = G' \), which indicates a linear viscoelastic regime. 
At larger strain values, slight variations in the strain moduli were obtained, suggesting the onset of a nonlinear viscoelastic regime.
Recently, \citet{Jory_2022} investigated these local moduli under different strain amplitudes with mucus samples obtained from a healthy human subject (control) or a patient with COPD.
Their results showed that these moduli in the mucus of the COPD patient are higher than those in the control mucus at the same strain amplitude. 
Moreover, both \( G'_M \) and \( G'_L \) for the control and COPD mucus showed significant deviations from the fundamental moduli.


These new alternative measures of elastic modulus and dynamic viscosity can be compared to quantify intracycle nonlinearities that distort the linear viscoelastic ellipse. 
For instance, if the large-strain modulus \( G'_L \) is greater than the minimum-strain modulus \( G'_M \), then the response is strain stiffening within that particular cycle (i.e., intracycle strain stiffening).
\citet{Ewoldt_2008} defined the following dimensionless index of nonlinearity:
\begin{equation}
S \equiv \frac{G'_L - G'_M}{G'_L} = \frac{4e_3 + \dots}{e_1 + e_3 + \dots}.
\end{equation}
Note that \( S = 0 \) for a linear elastic response, \( S > 0 \) indicates intracycle strain stiffening, and \( S < 0 \) corresponds to intracycle strain softening.

\citet{Jory_2022} recently conducted a LAOS flow protocol on mucus collected from human bronchial samples corresponding to healthy controls, smokers, and patients with COPD, using a strain-imposed rheometer over a \SIrange{10}{1000}{\percent} strain amplitude range. 
For all samples tested, increases in the nonlinear coefficients \( e_3 / G'_1 \) and \( S \) were observed.
They found that \( S \) value ranged from 0.3 to 0.64, indicating pathological strain-stiffening \citep{Jory_2022}.
This behavior correlates with impaired mucociliary clearance and aligns with findings in horse lung mucus and synthetic simulants by \citet{Vasquez_2014}, \citet{Ewoldt_2008}, \citet{Puchelle_1983} and \citet{Button_2018}. 
In contrast, the onset of a nonlinear response in the HBE-released mucus studied was recorded at much smaller deformations \citep{Jory_2022}. 
More experiments are needed to determine whether this approach has a relevant diagnostic value.

Further research is needed to explore the full potential of LAOS in studying airway mucus. 
This includes investigating the effects of mucolytic agents on the nonlinear rheology of mucus and developing more accurate models that represent mucus behavior under physiological conditions. 
Studies on the interactions between mucus and other airway components, such as cilia and epithelial cells, could provide a more comprehensive understanding of the factors influencing mucus clearance.
High-resolution imaging techniques combined with LAOS could offer valuable insights into the microstructural changes in mucus during large deformations.



\subsection{Extensional rheology of airway mucus}

Conventional rheometry, utilizing rotational or oscillatory rheometers, provides valuable insights into the shear properties of airway mucus. However, the flow in the airways of the lung is inherently complex, involving both shear and extensional components \citep{Quraishi_1998,Lai_2009,Wagner_2018}. 
Extensional flows, in particular, can significantly stretch mucus, amplifying elastic forces and extensional viscosity \citep{Puchelle_1983,Puchelle_1985,Radtke_2018}. 
Therefore, quantifying these extensional rheological properties is essential to understand mucus behavior under physiologically relevant conditions \citep{Girod_1992,Quraishi_1998}. Such properties play a critical role in physiological processes such as mucociliary clearance and cough clearance, making them vital for developing novel diagnostics and treatments for respiratory diseases \citep{Quraishi_1998,Lai_2009,Wagner_2018,Kavishvar_2023}.

Experimentally, extensional flows are generated by stretching mucus samples between two vertically aligned plates. Several techniques have been developed to measure extensional properties:
\begin{itemize}
    \item The \textit{Filament Stretching Extensional Rheometer (FiSER)} measures the force required to extend a filament until it breaks \citep{McKinley_2002}.
    \item The \textit{Capillary Breakup Extensional Rheometer (CaBER)} monitors the thinning filament's radius to determine extensional viscosity and relaxation time \citep{Zahm_1986,Anna_2001,Ahmad_2018}.
    \item The \textit{Filancemeter}, an older technique, measures the final height at breakup to assess spinnability, a key indicator of mucus extensibility.
\end{itemize}
The organization of the mucin network is a critical factor in determining the mucus's extensional rheological response \citep{Puchelle_1983}. However, \citet{Girod_1992} caution against applying linear properties like elasticity and viscosity to mucus, a highly nonlinear material. Instead, they suggest that spinnability may be a more appropriate measure. Studies simulating cough conditions and airflow interactions with mucus have shown that reduced viscosity or spinnability enhances mucus transport. For instance, \citet{Wagner_2017} demonstrated that mucin's extensional properties are highly sensitive to sample age, as evidenced by changes in relaxation time and filament breakup time.


In COPD, mucus becomes abnormally thick, obstructing airflow and exacerbating symptoms. 
Techniques such as CaBER have shown promise in identifying biomarkers for infection stages and disease severity, offering potential diagnostic and therapeutic applications \citep{King_1989,Tabatabaei_2015}.
A pioneering study by \citet{Tabatabaei_2015} investigated the extensional rheological behavior of sputum from COPD patients using both experimental and computational approaches. 
By employing CaBER, they characterized sputum in uninfected and infected states under extensional flow conditions mimicking lung-airway deformations.
Furthermore, the study employed the Single extended pom-pom (SXPP) model and a time-dependent thixotropic modified Bautista-Manero (MBM) model to describe the network structure and extensional properties. 
The strong correlation between the measured and predicted mid-filament diameters during necking and filament breakup highlights CaBER's potential as a sensitive biomarker for infection stages and a diagnostic tool for COPD \citep{Tabatabaei_2015} .

Moreover, \citet{King_1997} evaluated the effects of recombinant human deoxyribonuclease (rhDNase) and hypertonic saline (HS) on the spinnability of CF sputum using a Filancemeter. Sputum samples from CF patients were treated with normal saline (NS, \SI{0.9}{\percent} NaCl), HS (\SI{3}{\percent} NaCl), and combinations of rhDNase with NS and HS. 
The study showed that HS alone reduced spinnability by \SI{26}{\percent}, while NS alone resulted in a \SI{16}{\percent} reduction. 
Combined treatments of rhDNase with NS and HS further reduced spinnability by \SI{37}{\percent} and \SI{40}{\percent}, respectively.
Notably, both NS and rhDNase improved cough clearability, with the combination of rhDNase and HS demonstrating the most significant effect. 
These findings suggest that the combined treatment could enhance the mucociliary and cough clearability, offering a promising strategy for managing CF sputum \citep{King_1997}. 
Future research should standardize the measurement techniques and explore biochemical mechanisms regulating the mucus extensibility to improve clinical outcomes in respiratory diseases.%

\subsection{Rheological models of airway mucus}

Understanding the rheological behavior of airway mucus is essential for comprehending its complex viscoelastic properties, which play a critical role in both healthy and diseased states. 
These properties are typically characterized using experimental data from steady state shear, SAOS and LAOS tests \citep{Vanaki_2020,Nawroth_2020}. 
These tests capture the shear thinning, linear and non-linear viscoelastic regions of mucus, respectively, and have been foundational in studies by \citet{Dawson_2003,Lai_2009,Vanaki_2020}.
%
Shear-thinning, where viscosity decreases with increasing shear rate, is a fundamental characteristic of mucus rheology, particularly under pathological conditions. 
This behavior is crucial for effective mucus transport, allowing it to flow more easily under the shear forces generated by ciliary motion and coughing. 
For example, \citet{Craster_2000} explored the effects of yield stress and shear-thinning on mucus and surfactant spreading using a bilayer system of immiscible Herschel–Bulkley materials.
Similarly, \citet{Chatelin_2017} conducted experiments on CF and bronchiectasis mucus, measuring shear-thinning rheological indices and applying these findings within the Carreau model to simulate non-Newtonian behavior.
Further studies by \citet{Dawson_2003}, \citet{Serisier_2009}, and \citet{Tomaiuolo_2014} provided detailed rheological parameters for CF mucus, while \citet{Jeanneret-Grosjean_1988}, \citet{Zayas_1990}, and \citet{Serisier_2009} offered insights into the properties of healthy mucus.
 

The viscoelastic properties of airway mucus have been modeled using both linear and non-linear approaches, each with specific applications and limitations. 
Linear models, such as the Maxwell and Jeffrey models, simplify mucus behavior by representing it through combinations of mechanical elements such as Hookean springs and Newtonian dashpots \citep{Vanaki_2020}. 
These elements correspond to the elastic and viscous components of the material, respectively.
The Maxwell model, for instance, has been employed by \citet{Smith_2007} to simulate mucus transport, revealing that Newtonian mucus was transported more efficiently than viscoelastic mucus.
The Jeffrey model extends this approach by incorporating additional elements to better capture the viscoelastic behavior over a wider range of conditions, as demonstrated in studies by \citet{Lukens_2010} and \citet{Dillon_2007}. 
However, linear models are limited in their ability to capture the full complexity of mucus behavior, particularly under conditions involving large deformations or high stress. 


To address these limitations, non-linear models such as the Upper Convected Maxwell (UCM) model and the Oldroyd-B model have been developed. 
These models are better suited to describe the real-world behavior of mucus, especially under physiological conditions that involve significant deformation.
The UCM model, as used by \citet{Mitran_2007}, accounts for upper convective effects that are crucial in capturing the behavior of polymeric fluids like mucus.
Their findings revealed that increased stiffness in mucus significantly decreases the mucociliary clearance. Similarly, the Oldroyd-B model, which decomposes stress into contributions from a Newtonian solvent and a polymeric elastic solute, has been employed by \citet{Sedaghat_2016} and \citet{Guo_2017} to study mucus transport properties under various flow conditions.
 

Further advancing the understanding of mucus rheology, the Giesekus model incorporates both shear-thinning and strain-hardening effects, making it particularly effective for capturing the complex behavior of mucus under large deformations. \citet{Vasquez_2016} employed a five-mode Giesekus model using micro- and macro-rheology experimental data to approximate the relaxation spectrum of cell culture mucus. Their work provided valuable insights into the viscoelastic behavior of mucus, though it did not account for the pathological conditions that can alter mucus rheology, highlighting a significant gap in current modeling approaches.
More recently, \citet{Sedaghat_2021} integrated a nonlinear viscoelastic 5-mode Giesekus constitutive law with the dynamics of active ciliary motion, allowing for a more accurate understanding of the interaction between mucus viscoelasticity and ciliary forces.
Additionally, \citet{Liu_2024} employed a multi-nonlinear Giesekus model to simulate airway mucus in both healthy and asthmatic conditions, finding that asthmatic mucus exhibits greater nonlinearity compared to normal mucus.


Despite these advancements, no validated constitutive model that comprehensively captures the rheological behavior of airway mucus across all physiological and pathological conditions. 
Future research should focus on developing more sophisticated models that integrate both linear and non-linear behaviors and account for the shear-thinning and yield stress characteristics observed in pathological mucus. 
Such efforts will require interdisciplinary collaboration, combining expertise in rheology, respiratory physiology, and computational modeling. These models are crucial for improving our understanding of mucociliary clearance and guiding the development of effective therapeutic strategies to manage chronic respiratory diseases.

\section{MICRORHEOLOGY OF AIRWAY MUCUS}

Macrorheology has significantly advanced our understanding of the bulk properties of airway mucus, providing insights into its overall viscoelastic behavior.
However, this approach often averages out local variations within the sample, potentially missing important details. 
To address these limitations, microrheology offers a complementary perspective to understand mucus behavior by examining the heterogeneity, local mechanical properties, and diffusion characteristics of mucus at smaller scales \citep{Lai_2009,Squires_2010,MacKintosh_1999}.


Microrheology measures the viscoelastic properties of small volumes of airway mucus using embedded colloidal probes at nano- or microscale levels. 
These probes can be driven passively by thermal fluctuations or actively using methods such as optical or magnetic tweezers \citep{Squires_2010,Waigh_2016,Hill_2022}. 
Unlike bulk rheology, microrheology provides high spatial resolution and detects heterogeneities within the mucus.
This detailed analysis includes contributions from both the fluid within the mucus network and the network mesh itself, making it essential for understanding local rheological properties and diffusion behaviors often overlooked by bulk techniques \citep{Lai_2009,Squires_2010,Waigh_2016}


Microrheology is particularly useful for studying the transport and diffusion of small entities, such as mucolytic drugs or viruses, through airway mucus at scales comparable to or smaller than the pores of the mucus gel network. 
Techniques such as particle tracking microrheometry (PTM) \citep{MacKintosh_1999,Lai_2009,Squires_2010} and fluorescence recovery after photobleaching (FRAP) \citep{Waigh_2016,Schuster_2017} measure rheological parameters, estimate the diffusivity of viruses and drugs, and determine local viscoelastic properties. 
PTM typically demonstrates the linear elastic response of mucus at the nano- or microscale, while optical tweezers or magnetic fields can probe both linear and non-linear responses beyond its yielding point \citep{Squires_2010,Waigh_2016,Hill_2022,Kavishvar_2023,Pangeni_2023}.

\begin{table}[htbp]
\centering
\linespread{0.9}\selectfont 

\caption{Advantages and disadvantages of microrheology techniques for airway mucus}
\label{tab:microrheology_techniques}

\begin{tabular}{
|>{\raggedright\arraybackslash}p{5cm}|
>{\raggedright\arraybackslash}p{5cm}|
>{\raggedright\arraybackslash}p{5cm}|}
\hline
\textbf{Technique} & \textbf{Advantages} & \textbf{Disadvantages} \\ \hline
Particle Tracking Microrheometry (PTM) & 
Provides high spatial resolution and reveals local heterogeneities. Requires minimal sample volume. & 
May face challenges with probe size needing to be larger than the characteristic length scale. Complex gels like mucus might not meet all assumptions, leading to discrepancies between microscopic and macroscopic measurements. \\ \hline

Fluorescence Recovery After Photobleaching (FRAP) & 
Highly sensitive and can measure diffusion properties accurately. Requires minimal sample volume. & 
Necessitates fluorescent labeling, which may not always reflect viscoelastic properties accurately in heterogeneous samples. \\ \hline

Optical Tweezers & 
Offers precise control with high-resolution measurements. Captures both linear and non-linear responses. & 
Requires complex instrumentation, which may not be feasible for large-scale studies. \\ \hline

Magnetic Tweezers & 
Capable of applying larger forces and measuring nonlinear viscoelastic behavior. Ideal for larger probes. & 
Utilizes magnetic particles and requires complex instrumentation. May have limitations in resolution compared to optical tweezers. \\ \hline
\end{tabular}

\end{table}

\subsection{Passive microrheology}

Passive microrheology measures the thermal motion of colloidal beads within mucus to infer its viscoelastic properties. This involves two primary methods: particle tracking and fluorescence recovery after photobleaching (FRAP) \citep{Lai_2009,Duncan_2016a,Wagner_2018,Hill_2022}.

\subsubsection{Particle tracking microrheology}

Particle tracking microrheology (PTMR), a subset of passive microrheology, specifically focuses on the diffusion behavior of spherical colloidal particles to provide insights into the microrheological properties of mucus \citep{Lai_2009,Waigh_2016,Hill_2022}.
In PTMR, particles ranging from 
\SI{100}{\nano\metre} to 
\SI{1}{\micro\metre} are monitored to analyze their Brownian motion within the mucus. 
The mean squared displacement (MSD) of these particles is defined as:
\begin{equation}
\text{MSD}(\tau) = \left\langle \Delta r^2(\tau) 
\right\rangle
= 
\left\langle [r(t + \tau) - r(t)]^2 
\right\rangle,    
\end{equation}
where \( r(t) \) denotes the particle's position at time \( t \), and \( \tau \) is the time lag. The angle brackets indicate averaging over multiple particles and time \citep{MacKintosh_1999,Lai_2010,Waigh_2016}. In simple viscous fluids, the MSD increases linearly with time:
\begin{equation}
\text{MSD}(\tau) = 6D\tau,
\end{equation}
where \( D \) represents the diffusion coefficient. In contrast, in airway mucus, the MSD often follows a power-law dependence:
\begin{equation}
\text{MSD}(\tau) \propto \tau^\alpha,
\end{equation}
where \( \alpha \) is the diffusive exponent. For normal diffusion, \( \alpha = 1 \); subdiffusive behavior (\( \alpha < 1 \)) indicates that the motion of the particles is constrained by the microstructure of the mucus.
In contrast, superdiffusive behavior (\( \alpha > 1 \)) suggests enhanced particle movement beyond typical diffusive dynamics
\citep{Lai_2009,Lafforgue_2017,Waigh_2016}.

Analyzing the MSD allows for the determination of the diffusion coefficient $D$, which reflects the viscosity and elasticity of the mucus. In a purely viscous fluid, MSD increases linearly, and the shear viscosity can be computed using the Stokes-Einstein relation:
\begin{equation}
D = \frac{k_{\mathrm{B}} T}{6 \pi \eta a},    
\end{equation}
where \( k_{\mathrm{B}} T \) represents thermal energy, \( a \) is the particle radius, and \( \eta \) is the viscosity. 
In highly elastic materials, particles experience a restoring force, resulting in a constant MSD. 
For mucus, which exhibits both viscous and elastic properties, the MSD profile may be time-dependent. 
To accurately characterize these properties, the generalized Stokes-Einstein relation (GSER) is used to relate particle motion to the material's viscoelastic moduli.
By applying the Laplace transform to the time-dependent MSD \( \langle \Delta r^2(s) \rangle \), researchers derive the viscoelastic spectrum of mucus:
\begin{equation}
G(s) = \frac{k_{\mathrm{B}} T}{\pi a \langle \Delta r^2(s) \rangle},
\end{equation}
where \( G(s) \) represents the viscoelastic spectrum. 
The elastic and viscous moduli, \( G'(\omega) \) and \( G''(\omega) \), respectively, are components of the complex modulus \( G^*(\omega) \):
\begin{equation}
G^*(\omega) = G'(\omega) + i G''(\omega).    
\end{equation}

For a purely viscous liquid, \( G' = 0 \) and \( G'' = \eta \omega \); for an elastic solid, \( G'' = 0 \) and \( G' = G_0 \). 
Mucus exhibits non-zero values for both \( G' \) and \( G'' \), varying with frequency \( \omega \), indicating its viscoelastic nature \citep{Lai_2009,Lai_2007,Lai_2010,Schuster_2013}.


PTMR provides critical insights into the microrheological properties of airway mucus by analyzing the thermal motion of colloidal particles suspended within the mucus matrix \citep{Schuster_2013,Hill_2022}.
PTMR includes methodologies such as single-particle tracking (SPT) and multiple particle tracking (MPT), which enable researchers to monitor particle dynamics, thereby providing a comprehensive understanding of mucus rheology \citep{Metzler_2000,Metzler_2014,Suk_2009,Birket_2014,Hancock_2018,Morgan_2021}.
SPT involves recording the spatial trajectories of individual particles over time. The ensemble average MSD of tracked particles provides insights into their diffusive behavior and correlates with mucus solids concentration in various health conditions
\citep{Hill_2014,Georgiades_2014,Wagner_2017}.
Variations in particle diffusion due to factors such as pH and mucin concentration highlight the complex and heterogeneous nature of airway mucus
\citep{Bansil_2006,Bhaskar_1991,Maleki_2008}.
%
SPT has been used to assess the impact of surfactants, reducing agents, chaotropic agents, and other compounds on natural and synthetic airway mucus rheology
(\citep{Wagner_2017,Wang_2013,Georgiades_2014}.
%
High-resolution multiple particle tracking (MPT) has further advanced our understanding of mucus microrheology. 
MPT quantitatively describes the motion of numerous nanoparticles within mucus, leveraging high-speed cameras and fluorescence microscopy.
This technique has been pivotal in studying the barrier properties of airway mucus, demonstrating how factors like nanoparticle size and surface chemistry influence mucus penetration
\citep{Lai_2009,Lai_2007,Schuster_2013,
Schuster_2014,Cone_2009}.

\begin{figure}[htbp]
   \centering
    \includegraphics[width=0.99\textwidth]{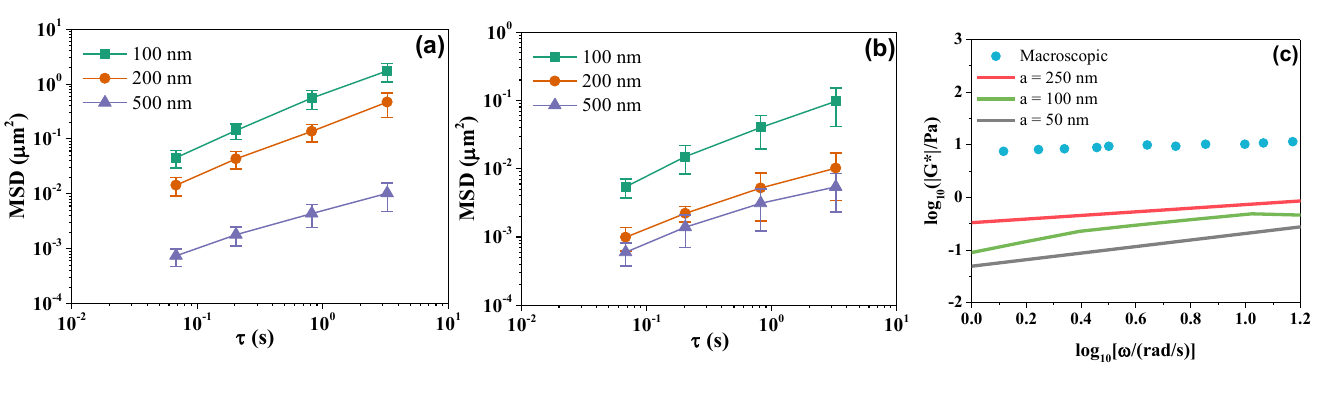}
  \caption{Effect of nanoparticle size on the ensemble-averaged geometric mean squared displacement (\(\langle \text{MSD} \rangle\)) as a function of time and the complex modulus magnitude (\(\log_{10}(|G^{*}|\)/Pa)) as a function of angular frequency (\(\log_{10}[\omega\)/(rad/s)]) for nanoparticles in airway mucus. (a/b) PS-PEG/PS-COOH  nanoparticles, average size: 100, 200, and \SI{500}{\nano \metre}, respectively, data represent means from measurements of five mucus samples, with at least 100 particles of each type tracked per sample, error bars represent the standard error of the mean, adapted from \citep{Schuster_2013}.  
    (c) CF sputum, bronchial mucus, and polymer analogs, measured by macroscopic technique, and PTMR with nanoparticle size of 50, 100, \SI{250} {\nano \metre}
    \citep{Dawson_2003, Tan_2020}.
    }
  \label{fig:PTM}
\end{figure}

To characterize the barrier properties of airway mucus, MPT has been used to quantify the diffusion of polymeric nanoparticles with various physicochemical properties in porcine or human airway mucus from healthy subjects \citep{Hill_2014,Schuster_2013,Lin_2020}, and those with disease~\citep{Suk_2009,Birket_2014,Duncan_2016,Fernandez-Petty_2019}. 
\citet{Schuster_2013} explored nanoparticle diffusion in airway mucus from healthy individuals, aiming to optimize drug delivery across this biological barrier (\cref{fig:PTM}). 
The researchers synthesized polymeric nanoparticles coated with low molecular weight polyethylene glycol (PEG) to reduce muco-adhesion and compared their transport efficiency to uncoated particles in human respiratory mucus collected from surgical patients without respiratory comorbidities by MPT. Their results demonstrated a striking size-dependent behavior: PEG-coated nanoparticles with diameters of (\SIrange{100}{200}{\nano\metre})  exhibited rapid mucus penetration, outperforming uncoated particles by approximately 15-35 fold, respectively. 
In contrast, larger PEG-coated nanoparticles (\SI{500}{\nano\metre}) were sterically immobilized by the mucus mesh. 
These studies have shown that nanoparticle size and surface chemistry significantly affect mucus penetration. 
Nanoparticles must be smaller than the characteristic mucus mesh size (\SIrange{100}{200}{\nano\metre}) and nonadhesive to efficiently penetrate the mucus barrier. 

\begin{figure}[htbp]
   \centering
    \includegraphics[width=0.99\textwidth]{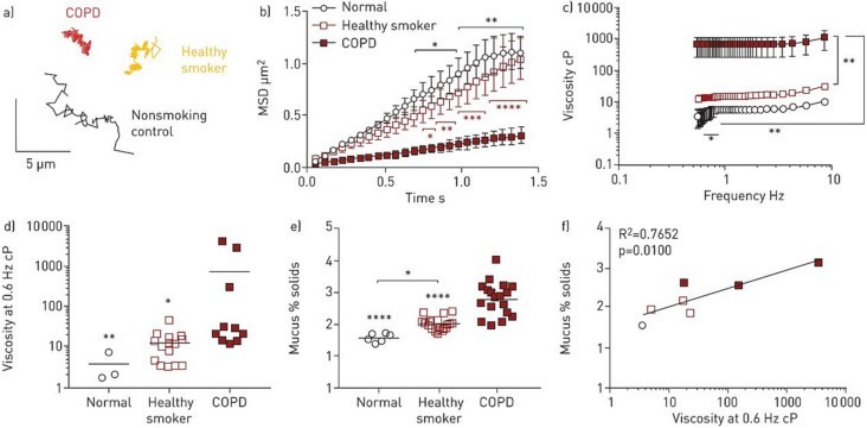}
  \caption{Persistent abnormal microrheological behaviors of airway mucus generated by human bronchial epithelial (HBE) cells derived from normal individuals, healthy smokers and COPD patients.  
(a) Representative tracings show distinct patterns of Brownian motion of \SI{1}{\micro\metre} particles in mucus between the three groups. 
(b) MSD of the particles over time in mucus between the groups.  
(c) Effective viscosity of the mucus between the groups as a funtion of frequency, calculated based on the MSD data.  
(d) Effective viscosity of the mucus at the frequency of \SI{0.6}{\hertz}  from the three groups.
(e) The percentage of solids by weight in the mucus from each group.  
(f) The relationship between mucus solid content and effective viscosity at \SI{0.6}{\hertz} as evaluated for individual donors. Data adapted from \citet{Lin_2020}.}
  \label{fig:PTMR}
\end{figure}

\citet{Chisholm_2019} utilized MPT to study the microstructure of sputum from smokers with and without airway obstruction. 
They observed that \SI{100}{\nano\metre} muco-inert particles (MIPs) diffused readily through sputum from smokers without obstruction, whereas muco-adhesive particles (MAPs) were significantly hindered. 
In COPD patients, MIPs exhibited higher MSD values compared to MAPs, indicating a tighter sputum mesh (smaller mesh size) in COPD. 
This study highlighted significantly higher solid content and mucin concentration in COPD sputum compared to non-COPD samples, although DNA content did not differ significantly between the two groups. 
\citet{Lin_2020} also used MPT to study mucus derived from primary human bronchial epithelial (HBE) cells exposed to exogenous cigarette smoke or cholinergic stimulant. 
They demonstrated that mucus from COPD patients had higher microviscosity and solid content than mucus from healthy smokers and non-smoking donors. 
The reduced transport of \SI{1}{\micro\metre} MIPs in COPD mucus indicated increased sputum viscosity and solid content in response to chronic exposure of cigarette smoke, suggesting changes in mucin structure, mucus mesh size, and mucociliary transport as shown in \cref{fig:PTMR}.

\citet{Morgan_2021} examined the biophysical properties of airway mucus from asthma patients using MPT.
They treated mucus samples with tris(2-carboxyethyl) phosphine (TCEP), a reducing agent, and observed increased MSD values, decreased viscoelasticity, and normalized homogeneity and MSD levels similar to controls. 
This study suggests that disrupting mucus disulfides can improve mucus gel structure and mucociliary clearance in asthma patients.


Research has shown that MPT can elucidate the microrheology and microstructure of mucus in respiratory diseases by tracking MIP probes \citep{Birket_2014,Duncan_2016,Duncan_2016a}. 
PEG-coated MIPs, as large as \SI{500}{\nano\metre}, can diffuse through normal airway mucus from healthy individuals, whereas uncoated polymeric particles of the same size are hindered due to adhesive interactions with mucus components~\citep{Suk_2009,Schuster_2013,Hill_2014}. Studies have demonstrated that PEG-coated MIPs up to \SI{200}{\nano\metre} in diameter diffuse significantly faster through CF sputum compared to uncoated particles~\citep{Suk_2009}. 
However, there is considerable variability in particle diffusion across sputum samples from different CF patients, indicating patient-specific differences in mucus microstructure~\citep{Lai_2009,Duncan_2016,Patarin_2020,Bansil_2013}. 
\citet{Suk_2009} estimated the mesh spacing of fresh CF sputum to be approximately $140 \pm \SI{50}{\nano\metre}$ using a fitted transport rate of various-sized MIPs. 
Furthermore, \citet{Suk_2009,Suk_2011,Suk_2011a} found that N-acetyl l-cysteine (NAC), a mucolytic drug that cleaves disulfide bonds, can enhance MIP penetration through airway mucus. 
The diffusion exponent, \( \alpha \), was 0.91 for \SI{200}{\nano\metre} MIPs in NAC-treated CF sputum, compared to 0.70 in untreated sputum.
These studies indicated significant inter-patient variability, with decreases in sputum mesh size correlating with increases in mucin, DNA, and total solid concentration~\citep{ Duncan_2016,Lai_2010,Suk_2009,Markovetz_2019,Schuster_2013,hill_2018}.
%
\citet{Dawson_2003} found that recombinant human deoxyribonuclease (rhDNase) reduced CF sputum viscosity by breaking down DNA aggregates, thereby increasing particle diffusion. 
MPT has also revealed that CF sputum exhibits lower microviscosity than macroviscosity, suggesting greater microheterogeneity \citep{Dawson_2003}.
These findings underscore the potential for therapeutic interventions to modify mucus properties and improve patient outcomes.


MPT has also been used to study the properties of artificial mucus. 
\citet{Song_2021,Song_2022} examined the effect of MUC5B:MUC5AC mucin ratios on the macro- and microrheology of mucin-based hydrogels. 
They found that higher MUC5AC content in synthetic mucus resulted in greater elastic and viscous moduli, reduced particle diffusion, and decreased mesh network size. 
They also evaluated the gel network mesh size via MPT analysis, reporting that MUC5B-rich hydrogels had larger mesh sizes, whereas MUC5AC-rich gels had reduced network mesh sizes. 
These studies emphasize the role of mucin composition in determining mucus rheological properties \citep{Thornton_2008,Voynow_2009,Atanasova_2019,Carpenter_2021,Song_2022,Caughman_2024}.

Despite its advantages, including accessibility and minimal sample volume requirements, PTMR has limitations. 
Particle--mucus interactions can introduce artifacts, such as stickiness, adhesiveness, which may affect rheological measurements  \citep{Lai_2009,Waigh_2016,Hill_2022}. 
Ensuring that probe-mucus interactions are minimal or representative of the biological object of interest is crucial, as discrepancies between microbead and macroscopic rheology may arise due to these interactions 
\citep{Hill_2022}. 
Studies have shown that the viscosity and moduli reported by microbead rheology often differ from those obtained through macroscopic rheology, typically being lower 
\citep{MacKintosh_1999,Crocker_2000,Lai_2009,Bansil_2013,Bokkasam_2016,Wagner_2017,Tan_2020}. 
This discrepancy likely arises from the differences between surface and bulk rheology of mucus 
\citep{Hill_2022,Schuster_2014}.


PTMR, particularly through MPT, provides valuable insights into the microrheological properties of airway mucus. 
These techniques enhance our understanding of mucus structure and function in both health and disease, offering critical information for developing better management strategies for chronic respiratory diseases. 
The practical applications and findings from MPT studies underscore its significance in characterizing the complex rheological behavior of mucus, ultimately contributing to more effective therapeutic interventions
\citep{Duncan_2016,Hancock_2018,Esther_2019,Murgia_2020,Rouillard_2020,Song_2021,Song_2022}.

\subsubsection{Fluorescence recovery after photobleaching (FRAP)}

Fluorescence recovery after photobleaching (FRAP) is a key microrheological technique used to study the transport behavior of nanoparticles (NPs) and solutes in airway mucus. 
By measuring a small area corresponding to the signal recovery of fluorescence after photobleaching, FRAP quantifies the effective diffusion coefficient and the fraction of immobile particles in fluid sample. 
This technique has been widely applied to investigate the diffusion of therapeutic agents, environmental toxins, and other small molecules in mucus, providing insights into their interactions with the mucus mesh ~\citep{Lai_2009,Duncan_2016,Waigh_2016,Schuster_2017,Wagner_2018,Hill_2022,Huck_2022}. 
The widespread use of FRAP to study mobility characteristics of molecules and particles in airway mucus has led to an increasing number of different FRAP models and analytical frameworks 
\citep{Periasamy_1998,Sengupta_2003,Loren_2015}. 
%

FRAP offers significant advantages; NPs can be added directly to mucus with minimal dilution, preserving the native mucus structure.
Additionally, FRAP experiments are completed within minutes, minimizing proteolytic degradation of mucus. 
This technique is particularly suitable for use in ciliated epithelial cultures, tracheal explants, or freshly collected mucus samples
\citep{Derichs_2011,Birket_2014,Shah_2016}.
When using FRAP to characterize airway mucus microrheology, key factors including size, adhesiveness, pH sensitibity, photostability, and brightness of the fluorescent probes should be considered \citep{Hill_2022,Duncan_2016}.
Accurate interpretation of results also depends on the mathematical models used to calculate rheological properties \citep{Suh_2005,Mastorakos_2015,Duncan_2016,Hill_2022}.

FRAP has been extensively used to study the diffusion of molecules in CF mucus. 
For example, \citet{Braeckmans_2003} found that FITC-labeled dextran molecules with hydrodynamic radii of 9, 15, and \SI{33}{\nano\metre} exhibited size-independent diffusion in CF mucus, with nearly complete fluorescence recovery (\SI{90}{\percent}). 
This suggests that these molecules diffuse relatively unhindered through the mucus mesh. 
The viscosity of the interstitial fluid in CF mucus was estimated to be only 4--6 times higher than that of water based on dextran diffusion rates~\citep{Braeckmans_2003}.
Similar studies using \SI{70}{\kilo\dalton} fluorescent dextran in HBE cultures and pig CF models reported the viscosity of interstitial fluid was 4--8 times higher than that of water~\citep{Tang_2016,Derichs_2011}.


FRAP studies on polystyrene nanoparticles (PS NPs) in CF mucus revealed significant fractions of immobile particles, with \SI{62}{\percent} and \SI{44}{\percent} immobility for 37 and \SI{89}{\nano\metre} PS NPs, respectively. 
This immobility is likely due to the adhesion to hydrophobic protein domains of mucins within the mucus mesh~\citep{Braeckmans_2003}. 
Similarly, \SI{76}{\percent} of \SI{100}{\nano\metre}, \SI{46}{\percent} of \SI{200}{\nano\metre}, and \SI{86}{\percent} of \SI{500}{\nano\metre} PS NPs were immobilized \textit{in ex vivo} porcine airway mucus collected from excised tracheas~\citep{Murgia_2016}.


A recent study combining FRAP and MPT found differing trends in particle mobility based on size. 
\citet{Murgia_2016} showed that 100 and \SI{200}{\nano\metre} PS NPs exhibited similar mobile fractions (\SI{43}{\percent} and \SI{51}{\percent}, respectively), but FRAP analysis indicated that the mobile fraction of \SI{100}{\nano\metre} PS NPs was half that of \SI{200}{\nano\metre} PS NPs.
These results suggest variations in short-versus long-time diffusion behavior of mucoadhesive PS NPs in porcine airway mucus. 
Conversely, previous studies demonstrated that non-mucoadhesive gene vectors, which exhibited higher diffusion rates in human CF mucus as assessed by MPT, also showed broader \textit{in vivo} distribution throughout mouse airways following inhalation~\citep{Mastorakos_2015,Suk_2014}. 
The contrasting findings may be attributed to differences in mucoadhesive versus non-mucoadhesive NPs and the sources of airway mucus samples collected~\citep{Birket_2014,Murgia_2016}. 
%

A key limitation of FRAP is the need for a high concentration of NPs to produce a bright and uniform fluorescent background. 
Highly concentrated NPs, especially mucoadhesive ones, may cause mucin fiber aggregation, altering the mucus microstructure and increasing mesh size~\citep{Vasconcellos_1994,Olmsted_2001,Wang_2011,Hill_2022,Duncan_2016}.
No single rheological technique can fully capture the complex behavior of airway mucus across different physiological and pathological conditions.
By combining FRAP, PTMR, and other techniques, researchers can gain a comprehensive understanding of mucus microrheology, advancing the development of effective therapies for respiratory diseases.

\subsection{Active microrheology}

Passive microrheological techniques are limited to probing the linear viscoelastic response. 
To investigate the nonlinear microscopic response of mucus and mucin gels, active microrheological techniques such as optical tweezers (OT) and magnetic tweezers (MT) must be used, wherein particles are driven within the material by an external force~\citep{Squires_2010,Waigh_2016,Lai_2009,MacKintosh_1999,Hill_2022}.

\subsubsection{Optical tweezers}

\begin{figure}[htbp]
   \centering
    \includegraphics[width=0.80\textwidth]{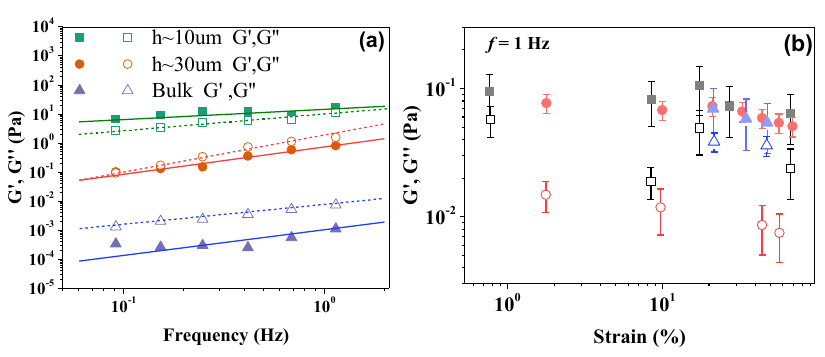}
\caption{Frequency or strain dependence of elastic and viscous moduli (\(G', G''\)) in airway mucus from healthy human bronchial epithelial (HBE) cultures measured by optical tweezers. (a) Bulk: measured with mucus extracted from BHE culture, h-10/30um: measured directly on HBE cultures with nanoparticle size of 10 or \SI{30}{\nano\metre}, adapted from \citet{Jory_2019}. (b) Colors and symbols indicate mucus samples taken from three distinct locations within the HUE culture, shear strain sweep at 1 Hz, adapted from \citet{Jory_2022}.}
  \label{fig:OT}
\end{figure}

OT coupled with an inverted microscope enable the isolation, trapping, and precise manipulation of microbeads within mucus, with their motion captured to infer local rheological properties from linear to nonlinear ~\citep{Squires_2010,Waigh_2016,Gross_2017} as shown in \cref{fig:OT}.
\citet{Kirch_2012} leveraged OT to probe native airway mucus, finding that most particles were immobilized by the stiff polymer scaffold, while those in larger pores or void regions could be displaced. 
Consistent with the heterogeneity previously investigated in native mucus using FRAP and PTMR~\citep{Murgia_2016}, underscoring the structural complexity of mucus. 
To further dissect this heterogeneity, 
\citet{Jory_2022} developed an OT-based method to quantify mucus adhesion at the microscale. 
Their measurements of adhesion forces, combined with rheological comparisons to mucus simulants and imaging data, support a model of mucus as a network of elastic adhesive filaments with a large mesh size, embedded within a soft gel matrix.
The local rheological landscape of mucus is further shaped by its proximity to the epithelium.
\citet{Jory_2019} showed that mucus near human bronchial epithelial surfaces (within tens of microns) exhibits higher viscoelasticity, forming a gel-like layer due to ciliary action preventing mucin chain relaxation. 
In contrast, mucus farther from the epithelium transitions to a liquid-like state with lower viscoelasticity, creating a viscoelastic gradient \cref{fig:OT}. 
These findings highlight a critical divergence between local OT-derived rheology and bulk rheometer measurements \citep{Jory_2022, Gross_2017}, emphasizing the need for multiscale approaches to fully characterize mucus behavior.

\subsubsection{Magnetic tweezers}

In addition to OT, magnetic tweezers (MT) have also been widely used to measure the viscoelastic properties of mucus and mucus simulants~\citep{Cribb_2013, Litt_1970, Khan_1976, King_1977, Rubin_1990, Sanctis_1994, King_1997, Braunreuther_2023, Hill_2022}. 
Unlike traditional methods, MT systems allow precise control of input stress and oscillation frequencies without introducing an air-mucus interface, which can alter mucus properties. 
However, the use of large foreign objects, such as magnetic beads, may disrupt the mucus microstructure, and interactions between the beads and the mucus layer can influence the measurement results~\citep{Squires_2010, Waigh_2016}. 
To mitigate these issues, one approach is to use beads significantly larger than the mucus correlation length and ensure their neutral interactions with the mucus network. 
This minimizes the bead-mucus adhesion and thus reduces artifacts in the measured rheological properties~\citep{Hill_2022}. 
Furthermore, to validate the accuracy of MT assays, it is critical to compare the micromagnetic data with macroscopic rheometer, which controls for air-mucus interfaces. 
Such comparisons should utilize both native mucus samples and well-characterized polymeric reagents to ensure consistency and reliability~\citep{Hill_2022}.


\subsubsection{Magnetic microwire microrheology} 

\citet{Radiom_2021} explored the viscoelastic properties of airway mucus obtained from healthy human bronchial tubes and cultured bronchial resections using a novel technique called magnetic rotational spectroscopy (MRS).
This method measures the behavior of magnetic wires (\SIrange{5}{80}{\micro\metre}) in response to a rotating magnetic field, allowing the study of mucus properties at a microscopic scale. 
By applying models such as Maxwell and Kelvin-Voigt, the researchers calculated key properties, including static shear viscosity and elastic modulus. 
Their results indicated that mucus behaves as a viscoelastic liquid, with an elastic modulus of $2.5 \pm \SI{0.5}{\pascal}$ and a viscosity of $100 \pm \SI{40}{\pascal \second}$. 
In addition to spatial variations due to microcavities, the study identified secondary inhomogeneities in the relaxation time of the mucin network, which may play a significant role in mucus behavior.
Furthermore, \citet{Braunreuther_2023} introduced the magnetic microwire rheometer (MMWR), a device designed to monitor the mechanical properties of hydrogels over time, during both gelation and degradation.
This device enables non-destructive measurements of small volumes and can be combined with live-cell imaging, making it especially useful for studying biological materials in real time. 
The ability to measure rheological properties without disturbing the material opens up new possibilities for researching dynamic in situ systems~\citep{Cai_2024}. 

Very recently, \citet{Liegeois_2024} further applied this technology to investigate the viscoelastic properties of mucus secreted by human airway epithelial cells (HAECs) under various conditions. 
The movement of the microwire within the mucus gel allows for precise, in situ measurement of mucus rheology without disturbing the surrounding epithelial cells. 
Their findings revealed that normal HAEC mucus behaves like a viscoelastic liquid, whereas mucus secreted by IL-13–activated HAECs exhibited solid-like characteristics due to mucin cross-linking. 
This solid behavior was further enhanced when a thiolated polymer (thiomer) solution was applied but could be reversed by inhibiting peroxidase activity—an enzyme involved in cross-linking. 
The study also found that IL-13–activated HAECs displayed increased expression of thyroid peroxidase (TPO), while lactoperoxidase (LPO) levels remained constant. 
Both enzymes catalyze reactions that lead to mucin cross-linking, contributing to the solid-like behavior of the mucus. 
In patients with asthma, higher levels of TPO gene expression were observed in airway epithelial cells, correlating with excessive mucus production and airway mucus plugs.
These findings suggest that IL-13–activated HAECs produce pathological mucus through enzyme-mediated mucin cross-linking, which may underlie mucus overproduction in asthma.
Notably, the successful application of MMWR to measure in situ airway mucus rheology is likely to receive increasing attention in future studies of other respiratory diseases~\citep{Cai_2024,Liegeois_2024}.

\subsubsection{Low field nuclear magnetic resonance (LF-NMR)}

\citet{Abrami_2020} employed low-field nuclear magnetic resonance (LF-NMR) to investigate mucus samples obtained through expectoration from CF patients and healthy controls. 
LF-NMR quantitatively assesses the response of hydrogen atom dipoles (\( \mu \)) to a constant external magnetic field, designated 
as \( B_0 \). 
Upon the application of a perpendicular radiofrequency pulse (\( B_1 \)), the dipoles realign with \( B_0 \), a process referred to as relaxation. 
The relaxation rate is characterized by the inverse of the spin–spin or transverse relaxation time, $T_2$ \citep{Abrami_2018}. 
Importantly, $T_2$ exhibits an inverse correlation with the solid content present in sputum. 
Prior investigations by \citet{Abrami_2018} established a pathological increase in proteins, biological polymers, and mucin within the sputum of CF and COPD patients. 
Their findings revealed a significant correlation between $T_2$ and the concentrations of these components in CF sputum. 
Furthermore, they demonstrated an inverse relationship between $T_2$ and systemic inflammatory markers such as C-reactive protein (CRP), 
as well as local inflammatory markers including interleukin-1{\textbeta} 
(IL-1\textbeta) 
 and tumor necrosis factor-alpha (TNF-\textalpha) \citep{Abrami_2020}.
In subsequent research, \citet{Abrami_2021}) integrated LF-NMR with rheological analysis to elucidate the characteristics of CF sputum. 
This study investigated the relationship between $T_2$ and (1) sputum viscoelasticity, (2) the mucociliary clearance index (MCI) or cough clearance index (CCI), and (3) the average mesh size of sputum. 
The authors identified an inverse correlation between $T_2$ and both the elastic and viscous properties of the mucus, suggesting that increased mucus viscoelasticity is associated with reduced lung function, as indicated by diminished forced expiratory volume in the first second (FEV$_1$). 
This change may contribute to mucus stasis and inflammation. Additionally, $T_2$ demonstrated a direct correlation with MCI and CCI indices, underscoring its relevance in evaluating airway mucus clearance. 
Moreover, $T_2$ was positively correlated with the average sputum mesh size \citep{Abrami_2020,Abrami_2021,Abrami_2022}. 
The estimated average mesh size of CF sputum was reported as 
$93 \pm \SI{38}{\nano \metre}$, which aligns with findings from \citet{Suh_2005} utilizing MPT techniques. 
\citet{Suh_2005} employed MIPs of various sizes in conjunction with an obstruction-scaling model to determine the average three-dimensional mesh spacing in CF sputum, which was measured with an average of $140 \pm \SI{50}{\nano \metre}$, with a range of 
$60$--\SI{300}{\nano\metre}. 
This study demonstrated that the combined approach of LF-NMR and rheological analysis provides a comprehensive understanding of the three-dimensional architecture and mesh size distribution of the airway mucus network \citep{Abrami_2021,Abrami_2022}.
In summary, both passive and active microrheological measurements of airway mucus can be performed with minimal sample volumes, allowing for detailed spatial characterization of the airway mucus's microstructure. Due to sample heterogeneity and probe-mucin biochemical interactions, micro- and macrorheological measurements often do not align. Nevertheless, characterizing rheological properties at the microscopic scale is crucial for understanding the environment encountered by small molecules and bioengineered nanoparticles in airway mucus~\citep{Abrami_2024}.

\subsection{Microfluidic devices} 

Microfluidic ``mucus-on-chip'' systems offer a powerful and innovative approach for studying the yield stress, nanoparticle diffusion, and interactions within airway mucus~\citep{Abdula_2017,Carpenter_2018,
Marczynski_2018,Elberskirch_2019,
Jia_2021,Huck_2022,Pednekar_2022,Kavishvar_2023,Wagner_2023,Suh_2024}. 
These systems typically consist of a chip coated with a layer of native mucus or a mucin solution, which is then exposed to a constant liquid flow. 
This configuration allows nanoparticles to be added and examined under dynamic flow conditions that closely mimic physiological environments. 
A recent study by \citet{Abdula_2017} evaluated the mechanical clearance of porcine gastric mucin (PGM) in a Y-junction microfluidic channel using bubble and foam-driven mechanisms such as vortex ear, bubble scouring, and foam scouring. 
The shear stress at the interface of mucus and the clearing agent (foam or bubbles) in the Y-junction was assessed, predicting the yield stress of mucus to be 35 Pa. 
Flow rates explored in \SI{1}{\milli \metre} channels are typically experienced by bronchioles in generations 8 and 9 of the human lung. 
These findings have significant implications for the treatment of CF and other lung diseases. 
\citet{Jia_2021} demonstrated that microfluidic systems can effectively distinguish the mucopenetration and mucoadhesion of PEGylated and pectin-functionalized nanoparticles of \SI{50}{\nano\metre} and \SI{200}{\nano\metre} sizes. 
The physiological mucus environment can be modified by adding mucolytic agents of N-acetylcysteine (NAC), which reduce barrier function and significantly accelerate nanoparticle penetration, regardless of size and bio interfacial properties. 
This ``mucus-on-chip'' methodology provides valuable insights into nanoparticle-mucus interactions and can enhance the design of particulate formulations for more efficient mucosal drug delivery. 
In a custom-made microfluidic chip, \citet{Marczynski_2018} explored the penetration and binding behavior of fluorescently labeled dextrans with neutral, negative, and positive charges to MUC5C and MUC2. 
Their findings indicated that only positively charged molecules accumulated at the buffer/gel interface. 
These microfluidic models not only provide mechanistic insights but also serve as a foundation for mathematical modeling that account for both the diffusive transport of molecules through the mucus hydrogel and its renewal processes.
Microfluidic devices have also gained considerable attention for point-of-care (POT) testing in various diseases. 
They enable precise control over experimental conditions and facilitate high-throughput analysis, making them an invaluable tool for investigating the rheological properties of mucus. 
Consequently, further research is necessary to explore the applications of microfluidic devices in disease detection and management based on mucus rheology~\citep{Kavishvar_2023,Suh_2024}.

\begin{figure}[htbp]
   \centering
    \includegraphics[width=0.6\textwidth]{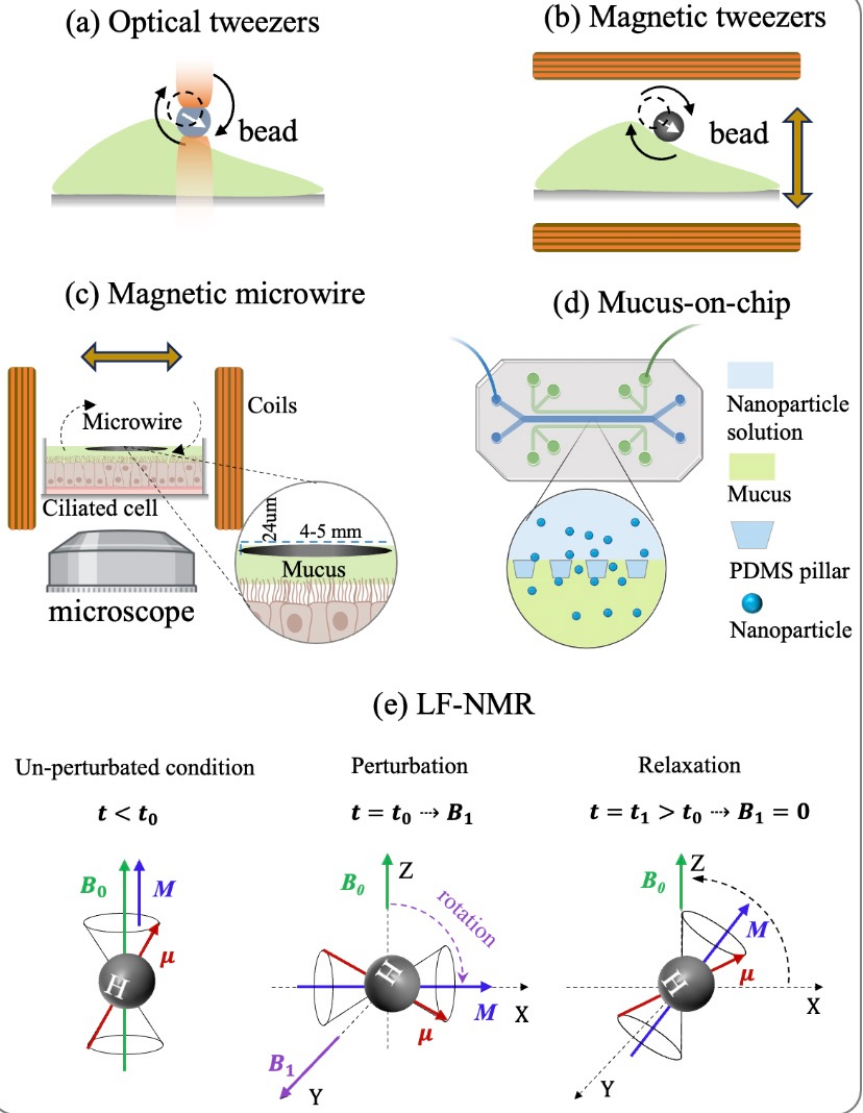}
  \caption{Schematic of active microrheology experimental methods.
  (a) Optical tweezers,
  (b) Magnetic tweezers,
  (c) Magnetic microwire rheometer,
  (d) Microfluidic ''mucus-on-chip''
  (e) Low field nuclear magnetc resonance (LF-NMR).}
  \label{fig:active_microrheology}
\end{figure}
\section{Airway Mucus Rheology in Clinical Applications}

The rheological properties of airway mucus play a critical role in the pathophysiology of chronic respiratory diseases such as CF, COPD, asthma, and bronchitis. In these conditions, the mucus undergoes significant changes in composition and concentration \citep{Lai_2009,Cohen_2012,Kesimer_2017,Thornton_2008,Voynow_2009,Atanasova_2019,Carpenter_2021,Song_2022,Caughman_2024}. 
Increased levels of mucins, along with the presence of non-mucin components like DNA, actin, and cellular debris, contribute to elevated solid concentrations, which can range from \SI{3}{\percent} to as much as \SI{15}{\percent} in pathological states~\citep{Kesimer_2017,Hill_2022,Wagner_2018}. 
These changes lead to higher viscosity and elasticity, impairing mucus clearance and leading to notable differences in rheological properties between healthy and diseased mucus~\citep{McShane_2021,Kavishvar_2023,Pangeni_2023,Wang_2024}.
\begin{figure}[htbp]
   \centering
    \includegraphics[width=0.9\textwidth]{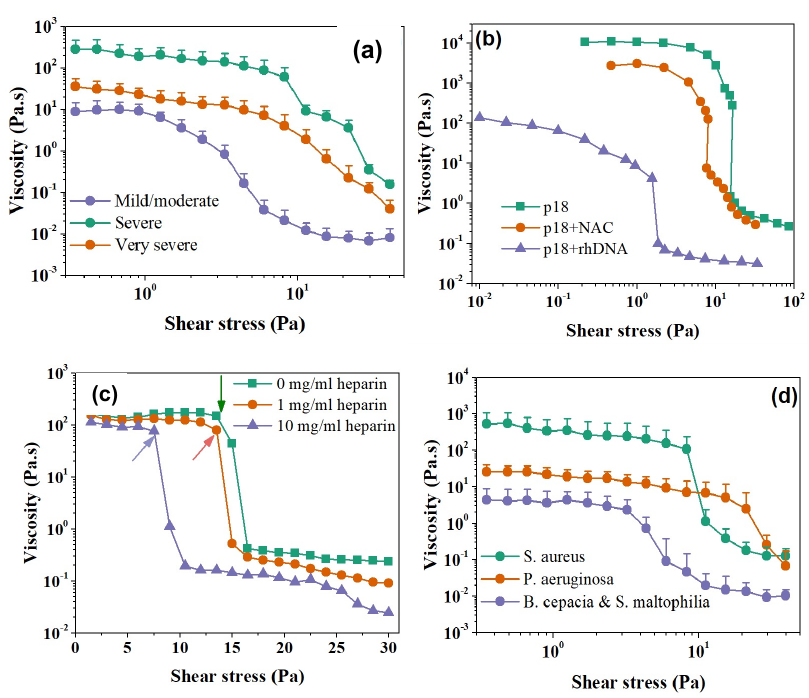}
  \caption{Flow curves of homogenized airway mucus in various disease conditions. (a) CF sputum samples from 33 donors with very severe, severe, and mild/moderate FEV1 levels, data present average viscosity values, adapted from~\citep{Tomaiuolo_2014}.
  (b) Sputum from patient 18 (p18) with or without N-acetylcysteine (NAC)/rhDNA treatment, indicating apparent yield stress defined as the stress at the point of discontinuity in the flow curve, adapted from~\citep{Ghanem_2021}. 
  (c) CF sputum measured in the absence or presence of unfractionated heparin (UFH: 0, 1, or 10 mg/ml), indicating different yield stress, adapted from \citep{Broughton-Head_2007}.
  (d) Sputum from donors with different bacterial colonization profiles (B. cepacia and S. maltophila, P. aeruginosa, and S. aureus), data present average viscosity values, adapted from~\citep{Tomaiuolo_2014}.} 
  \label{fig:ClinicApp}
\end{figure}
\subsection{Diagnostic tool} 

Rheological measurements of airway mucus serve as valuable diagnostic tools for chronic respiratory diseases. 
Typically, healthy mucus has a yield stress ranging from 0.05 to \SI{0.7}{\pascal}, remaining below \SI{1}{\pascal}~\citep{Kavishvar_2023}. 
In contrast, diseased mucus exhibits much higher yield stress: CF mucus ranges from 0.1 to \SI{100}{\pascal}, COPD mucus from 1 to \SI{40}{\pascal}, and asthmatic mucus from 0.2 to \SI{3}{\pascal}~\citep{Patarin_2020,Kavishvar_2023,Liu_2024}. 
Elevated yield stress in mucus correlates with disease severity, making rheological assessments useful for diagnosing and differentiating between conditions~\citep{Broughton-Head_2007,Lai_2007,
Tomaiuolo_2014,Patarin_2020,Ghanem_2021}. 

\subsection{Monitoring disease progression}

Monitoring disease progression through mucus rheology provides insights into the severity and course of chronic respiratory diseases~\citep{Duncan_2016a,Hancock_2018,Esther_2019,Murgia_2020,Rouillard_2020,Song_2022}. 
Studies have shown that the yield stress of CF mucus from severe patients is 4--5 times higher than that from mild patients, highlighting the relationship between mucus rheology and disease severity~\citep{Tomaiuolo_2014,Kavishvar_2023} as shown in \cref{fig:ClinicApp}(a). Changes in rheological properties can reflect the progression of the disease, offering a quantitative measure of how the condition is evolving over time~\citep{Ma_2018}.

\subsection{Evaluating treatment efficacy}

Rheological analysis is also crucial for evaluating the efficacy of treatments. 
Therapeutic interventions such as N-acetyl-cysteine (NAC) and rhDNAse have demonstrated effectiveness in modifying mucus rheology~\citep{Khan_1976,Shak_1990,Shak_1995,King_1997,Frederiksen_2006,Suk_2011,Schuster_2014,Duncan_2016,Duncan_2016a,Esther_2019,Morgan_2021} as shown in \cref{fig:ClinicApp}(b). 
NAC disrupts the mucin network, while rhDNAse breaks down actin-DNA bundles, leading to a significant reduction in yield stress. 
Similarly, the use of unfractionated heparin (UFH) has been shown to reduce yield stress by dissolving DNA-actin bundles as shown in \cref{fig:ClinicApp}(c)~\citep{Broughton-Head_2007}. 
These treatments highlight the potential of rheological measurements to assess and compare the effectiveness of different therapeutic strategies.

\subsection{Bacterial colonization}

Rheological properties of mucus can provide insights into bacterial colonization, which is particularly relevant in CF patients~\citep{Matsui_2006,Yang_2011,Tomaiuolo_2014,Abrami_2018,Ma_2018,Wheeler_2019,Murgia_2020,Batson_2022}.
Specific yield stress values are associated with different bacterial strains, such as S. aureus, P. aeruginosa, and B. cepacia as shown in \cref{fig:ClinicApp}(d) \citep{Tomaiuolo_2014}. 
Monitoring these properties can help identify and manage bacterial infections, as changes in mucus rheology can reflect shifts in bacterial presence and activity.


In conclusion, the rheology of airway mucus offers a comprehensive framework for understanding, diagnosing, and managing chronic respiratory diseases. 
By leveraging rheological measurements, clinicians can gain valuable insights into disease mechanisms, monitor progression, evaluate treatment efficacy, and assess bacterial colonization, ultimately improving patient care and outcomes~\citep{Chen_2010,Gustafsson_2012,
Hill_2014,Yuan_2015,Button_2016,Wagner_2018,Witten_2018}.

\section{CHALLENGES AND LIMITATIONS IN AIRWAY MUCUS RHEOLOGY}

\begin{figure}[htbp]
   \centering
    \includegraphics[width=0.85\textwidth]{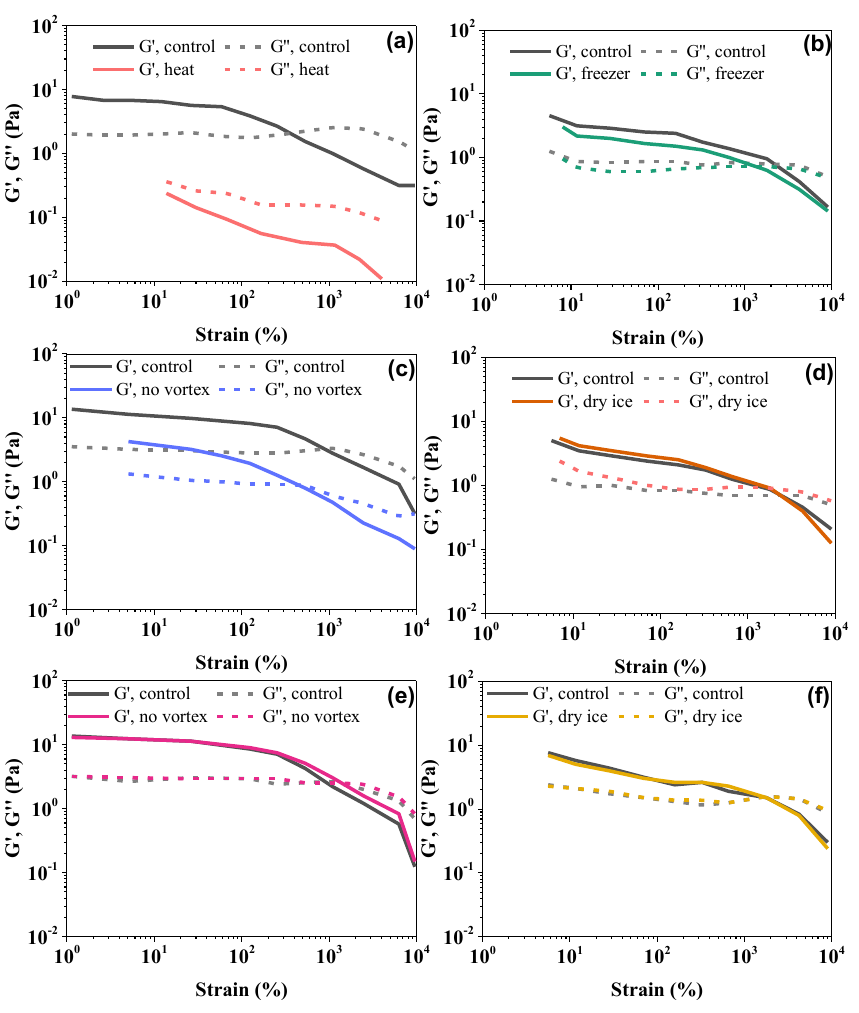}
  \caption{The effects of various factors on the evolution of elastic and viscous moduli (\(G',G''\)) of cystic fibrosis (CF) and non-cystic fibrosis bronchiectasis (NCFB) sputa during strain sweep tests. (a) Heat treatment, (b) freezer storage, (c, e) no vortex treatment, and (d, f) dry ice treatment~\citep{Esteban-Enjuto_2023}.} 
  \label{fig:Challenge}
\end{figure}

\subsection{Variability in native mucus samples}

Airway mucus rheology faces significant challenges due to the inherent variability of native mucus samples.
Mucus composition can differ widely between individuals and even within the same individual over time, influenced by multiple factors including but not limited to disease state, hydration level, and medication~\citep{Mellnik_2014,Mellnik_2016,Esther_2019,
Markovetz_2019,Hsu_1994,Hsu_1996}. 
This variability complicates the standardization of rheological measurements and makes comparisons across different studies difficult. 
Additionally, the method of native mucus collection---whether through spontaneous expectoration, induced sputum, or bronchoscopy---affects its composition and rheological properties~\citep{Lai_2009,Hill_2022,
Huck_2019,Huck_2022,Pangeni_2023}. 
Establishing and adhering to consistent protocols for native mucus collection and handling is crucial for obtaining reliable and comparable results.

\subsection{Storage, processing, and environmental factors} 

The proper storage and processing of mucus samples are also essential to preserving their native rheological properties. 
Freshly collected, undiluted samples should be used to maintain these properties accurately. 
Any storage or processing steps can significantly alter the physicochemical characteristics of mucus, affecting its viscosity and elasticity~\citep{Esteban-Enjuto_2023,Hill_2022}. 
For example, temperature fluctuations and freeze-thaw cycles during storage can break down mucins, cause dehydration, or alter ionic composition, thereby changing the rheological behavior 
\cref{fig:Challenge} 
\citep{Taylor_2005,Taylor_2005a,Innes_2009,Lafforgue_2017,Lafforgue_2018,Jory_2022}.
Therefore, it is crucial to store samples at consistent, controlled temperatures that mimic physiological conditions.


Processing techniques like dilution, centrifugation, or mechanical agitation can disrupt the natural gel structure, introducing artifacts and compromising the sample's native state~\citep{Esteban-Enjuto_2023,Hill_2022,Sharma_2021}. 
Gentle handling and minimal processing are recommended to avoid these issues and ensure reliable rheological measurements. 
Maintaining the physiological barrier function of mucus is crucial for evaluating its interactions with therapeutic agents or pathogens~\citep{Yuan_2015,Morgan_2021,Hsu_1994,
Lafforgue_2017,Jory_2019}. 
Any alterations during storage or processing can compromise this function, emphasizing the importance of meticulous handling.


Environmental factors such as temperature, humidity, and exposure to pollutants can also significantly influence mucus rheology \cref{fig:Challenge}.
Variations in environmental temperature can alter mucus viscosity, potentially affecting the accuracy of rheological measurements. 
Similarly, changes in humidity can impact the water content and rheological properties of mucus.
Additionally, exposure to pathogens or pollutants like cigarette smoke can modify mucus composition and its rheological behavior~\citep{Radtke_2018,Ahonen_2019,Rouillard_2020}.
Controlling and accounting for these environmental factors during experiments is essential to achieve accurate and reproducible results.

\subsection{Limitations of rheological techniques}

Current rheological techniques, including macrorheology and microrheology, have limitations when applied to airway mucus~\citep{Lai_2009,Kavishvar_2023,Hill_2022}.
Macrorheology, which assesses bulk properties, is more effective for understanding the overall behavior of mucus, but it may not detect fine-scale changes in highly viscous or elastic samples. 
Microrheology, on the other hand, can provide detailed spatial characterization of the material microstructure with very limited sample volume, making it a valuable tool for examining mucus at a microscopic level \citep{Lai_2009,Hill_2022,Waigh_2016,Wagner_2018}. 
Both active and passive microrheological measurements can be performed to achieve this. 
However, micro- and macrorheological measurements of mucus often do not align due to sample heterogeneity and probe-mucin biochemical interactions
\citep{MacKintosh_1999,Crocker_2000,Lai_2009,
Bansil_2013,Bokkasam_2016,
Wang_2017,Tan_2020,Jory_2022}. 
Despite these discrepancies, characterizing rheological properties at the microscopic scale is critical for understanding the airway mucus environment faced by small molecules and bioengineered nanoparticles for drug delivery~\citep{Araujo_2018,Bansil_2018,
Garcia-Diaz_2018,Huckaby_2018,Lock_2018,
Murgia_2018,Nordgard_2018,Taherali_2018,Wu_2018}.
No single rheological technique can fully capture the diverse behaviors of mucus in various biological and physiological contexts in the lung. 
Additionally, these techniques require sophisticated equipment and specialized expertise, which can limit their accessibility in clinical settings~\citep{Huck_2022,Pangeni_2023}. 
Moreover, obtaining high-quality native physiological mucus in sufficient volumes for repeated or even single measurements is challenging~\citep{Lai_2009,Duncan_2016,Atanasova_2019,
McCarron_2021,Bej_2022,Huck_2022,
Kavishvar_2023,Abrami_2024}.
Normal mucus from the lungs is typically collected via bronchoscopy brush or endotracheal tube techniques, which are limited by the small volume collected and potential mechanical stimulation that can alter rheological properties. 
\textit{In vitro} studies of airway mucus may not accurately represent in vivo conditions due to the small sample size and uncertain composition of the `gel' layer or periciliary layer.
To improve the study of mucus rheology, it is essential to establish standardized assays that ensure reproducibility across different laboratories~\citep{Esteban-Enjuto_2023,Hill_2022,
Kavishvar_2023,Pednekar_2022}. 
Additionally, it is vital that studies on the physical and physiological properties of mucus include detailed characterizations of mucus concentration, as well as measurements of mucin concentration and molecular weight~\citep{Hill_2022,Wagner_2018}. 
These efforts will help provide more accurate and comprehensive descriptions of mucus behavior.

\subsection{Translational challenges}

Bridging the gap between rheological research and clinical practice involves several challenges. 
While rheological measurements provide valuable insights into disease mechanisms and treatment efficacy, integrating these findings into routine clinical protocols is complex~\citep{Hill_2022,Kavishvar_2023,Radtke_2018}. 
Standardized methods and clear guidelines are necessary for effectively utilizing rheological data in clinical decision-making. 
Further research is required to establish how changes in mucus rheology correlate with clinical outcomes and patient symptoms~\citep{Esteban-Enjuto_2023,Abrami_2024,Pangeni_2023}.
The rheological study of airway mucus offers significant insights into chronic respiratory diseases, but several challenges must be addressed. 
These include variability in native mucus samples, the complexity of mucus composition, limitations of current rheological techniques, environmental influences, and translational challenges. Proper handling and processing of mucus samples are critical for maintaining their integrity~\citep{Lai_2009,Hill_2022,Abrami_2024,
Kamkar_2022,Pednekar_2022}. 
Addressing these issues will enhance the reliability and applicability of rheological research in understanding and managing chronic respiratory conditions.

\section{Conclusion}

The rheological properties of airway mucus are essential to its role as a dynamic and selective barrier and lubricant within the respiratory system. These properties vary significantly with changes in stress amplitude, shear rate, environmental conditions, and length scale. Comprehensive macrorheological characterization of airway mucus using established methods has greatly enhanced our understanding of physiological processes such as mucociliary and cough clearance, as well as the pathology of respiratory diseases. Recent advancements in microrheological techniques and the development of non-mucoadhesive probes have enabled detailed characterization of mucus at micro- and nano-length scales. These techniques have provided deeper insights into the rheological behavior of mucus, revealing its functional properties, pathogen transmission, and potential for nanoparticle-based therapies and preventive strategies. By integrating rheological characterizations across different length scales, this review offers a more comprehensive understanding of the complex behavior of airway mucus. This integration improves the study of disease mechanisms and supports the advancement of clinical applications, ultimately contributing to better diagnosis, treatment, and management of chronic respiratory diseases. Furthermore, this review highlights the potential of using mucus rheology as an indicator of disease for diagnosis in clinical settings. However, obtaining sufficient and uncontaminated mucus samples remains a significant challenge in clinical research. Additionally, variations in mucus rheology between patients and over time, dilution during collection, and the accurate determination of rheological properties present obstacles to the practical use of mucus rheology in clinical applications. In conclusion, while significant progress has been made in understanding and characterizing the rheological properties of airway mucus, further research is necessary to address these challenges and fully realize the potential of mucus rheology in clinical practice.

\section{Acknowlegdement}

This work was partially supported by NSFC Grant Nos. 12272063, 11532003, and 31670950; Jiangsu Provincial Natural Science Foundation Grant No. BK20151186; and Changzhou Municipal Natural Science Foundation Grant No. CJ20159033. We wish to acknowledge the support of the Wenzhou Institute, University of Chinese Academy of Sciences (No. WIUCASQD2020002). We also thank Professor G.G. Fuller, Professor A. J. Giacomin, Professor M. Doi, and Dr. Z. Xiong for very helpful and stimulating discussions.

\section{AUTHOR DECLARATIONS}
The authors have no conflicts to disclose.

\section{DATA AVAILABILITY}
Data sharing is not applicable to this article as no new data were created or analyzed in this study.

\appendix
\newpage 

\section{List of abbreviations and variables}

\begin{table}[htp!]
\centering
\begin{tabular}{|l|p{12cm}|}
\hline

\textbf{Abbreviation} & \textbf{Definition} \\ \hline
CCI & Cough clearance index \\ \hline
CF & Cystic fibrosis \\ \hline
COPD & Chronic obstructive pulmonary disease \\ \hline
ET & Endotracheal tube \\ \hline
FRAP & Fluorescence recovery after photobleaching \\ \hline
MCC & Mucociliary clearance \\ \hline
MCI &Mucociliary clearance index\\ \hline
MPT & Multiple particle tracking \\ \hline
MSD & Mean squared displacement \\ \hline
NAC & N-acetylcysteine \\ \hline
PGM & Porcine gastric mucin \\ \hline
PTM & Particle tracking microrheometry \\ \hline
SAOS & Small amplitude oscillatory shear \\ \hline
LAOS & Large amplitude oscillatory shear \\ \hline
rhDNase & Recombinant human deoxyribonuclease \\ \hline
HS & Hypertonic saline \\ \hline

\end{tabular}
\caption{List of abbreviations and their definitions.}
\label{tab:abbreviations}
\end{table}

\section{Rheological models}

\subsection{Carreau Model}

The Carreau model is a widely used constitutive model that describes shear-thinning behavior, where viscosity decreases with increasing shear rate \citep{Meldrum_2018}. 
This model is particularly relevant for airway mucus, which often exhibits shear-thinning characteristics under normal and pathological conditions.
The Carreau model is expressed as:
\begin{equation}
\eta(\dot{\gamma}) = \eta_\infty + (\eta_0 - \eta_\infty) \left[1 + (\lambda \dot{\gamma})^2\right]^{(n-1)/2},
\end{equation}
where \( \eta(\dot{\gamma}) \) is the viscosity as a function of the shear rate \( \dot{\gamma} \), \( \eta_0 \) is the zero-shear viscosity, \( \eta_\infty \) is the infinite-shear viscosity, \( \lambda \) is the relaxation time, and \( n \) is the power-law index, which determines the degree of shear-thinning.
This model is particularly effective in capturing the non-Newtonian behavior of mucus, especially under conditions where the mucus needs to flow more easily, such as during ciliary movement or coughing.

\subsection{Herschel-Bulkley Model}

The Herschel--Bulkley model extends the concept of shear-thinning by incorporating yield stress, which is the stress required to initiate flow in a material that behaves as a solid under low-stress conditions~\citep{Lafforgue_2017,Lafforgue_2017a,
Lafforgue_2018,Meldrum_2018}. 
This model is particularly useful for describing mucus in pathological conditions where increased viscosity and yield stress hinder mucociliary clearance. The Herschel-Bulkley model is given by:
\begin{equation}
\sigma = \sigma_y + K \dot{\gamma}^n,
\end{equation}
where \( \sigma \) is the shear stress, \( \sigma_y \) is the yield stress, \( K \) is the consistency index, \( \dot{\gamma} \) is the shear rate, and \( n \) is the flow index (shear-thinning if \( n < 1 \)). 
This model is applicable to mucus in conditions such as cystic fibrosis and chronic obstructive pulmonary disease (COPD), where the mucus exhibits significant yield stress and shear-thinning behavior.

\subsection{Linear Rheological Models}

\paragraph{Maxwell Model}

The Maxwell model, which is commonly used for viscoelastic fluids, can be expressed as \citep{Smith_2007,Vasquez_2016}:
\begin{equation}
\frac{D\sigma}{Dt} + \frac{\sigma}{\lambda} = 2G D,
\end{equation}
where \( \sigma \) is the stress tensor, \( \lambda \) is the relaxation time, \( G \) is the modulus, and \( D \) is the rate of deformation tensor. 
This model is suitable for capturing the stress relaxation behavior of mucus under small deformations but falls short in describing its response under large deformations.

\paragraph{Jeffrey Model}

The Jeffrey model extends the Maxwell model by incorporating an additional term to account for the Newtonian solvent contribution, thus providing a more comprehensive description of viscoelastic behavior over a broader range of conditions \citep{Dillon_2007,Lukens_2010}:
\begin{equation}
\frac{D\sigma}{Dt} + \frac{\sigma}{\lambda} + \frac{\eta_{\mathrm{s}}}{G\lambda} \sigma = 2G D,
\end{equation}
where \( \eta_{\mathrm{s}} \) is the solvent viscosity. 
This model has been used to simulate mucus transport and provides a better approximation of the material's behavior under steady-state shear.

\subsection{Nonlinear Rheological Models}

\paragraph{Upper-Convected Maxwell (UCM) Model}

The UCM model accounts for the upper convected effects in viscoelastic fluids, which are critical for capturing the deformation history in flowing materials \citep{Mitran_2007,Vasquez_2016}:
\begin{equation}
\frac{D\sigma}{Dt} - L \cdot \sigma - \sigma \cdot L^T + \frac{\sigma}{\lambda} = 2G D.
\end{equation}
This model is particularly useful for describing the behavior of mucus under physiological conditions that involve significant deformation.

\paragraph{Oldroyd-B Model}
The Oldroyd-B model builds on the UCM model by adding a solvent viscosity term, providing a more complete description of both elastic and viscous contributions \citep{Brinkmann_2004,Sedaghat_2016,Sedaghat_2016a,Sedaghat_2016b,Guo_2017}:
\begin{equation}
\frac{D\sigma}{Dt} - L \cdot \sigma - \sigma \cdot L^T + \frac{\sigma}{\lambda} = 2(G + \eta_{\mathrm{s}}) D.
\end{equation}
This model is particularly effective in capturing the flow behavior of mucus under various physiological conditions, including those encountered in disease states.

\paragraph{Giesekus Model}

The Giesekus model incorporates nonlinear effects such as strain hardening and anisotropic drag, making it suitable for describing complex fluids like mucus under large deformations~\citep{Vasquez_2016,Liu_2020,
Sedaghat_2021,Pednekar_2022,Liu_2024}. 
The model is expressed as:
\begin{equation}
\frac{D\sigma}{Dt} - L \cdot \sigma - \sigma \cdot L^T + \frac{\sigma}{\lambda} + \frac{\alpha}{\lambda G} \sigma \cdot \sigma = 2G D,
\end{equation}
where \( \alpha \) is the nonlinearity parameter, which allows the model to capture complex behaviors such as shear thinning and strain hardening. 
The Giesekus model has been successfully applied to simulate the rheological behavior of mucus under various physiological conditions.

\subsection{Pom-Pom Model Under Extensional Flow}

The establishment of the Pom-Pom model enables us to simultaneously understand the strong strain hardening phenomenon and shear thinning behavior of airway mucus under tensile conditions. 
The application of its correction in experiments has achieved great success. 
For example, using a multi-mode Pom-Pom model based on Maxwell linear spectra can fit the nonlinear rheology data of various branching polymers. 
Specifically, the eXtended Pom-Pom (XPP) model was introduced in 
\citet{Tabatabaei_2015}
%
The Single eXtended Pom-Pom (SXPP) model, in terms of the degree of system anisotropy \( \alpha_\mathrm{pom} \) and the relaxation moduli \( G \), may be given by:
\begin{equation}
\frac{D\tau}{Dt} - L \cdot \tau - \tau \cdot L^T + \frac{f(\tau)}{\DeNum} \tau + \frac{G}{\DeNum} \left(f(\tau) - 1\right) I + \frac{\alpha_\mathrm{pom}}{
\DeNum \, G} \tau \cdot \tau = 2G D,
\end{equation}
where \( D \equiv (L + L^T)/2 \), and the function \( f(\tau) \) is given by:
\begin{equation}
f(\tau) = \frac{2}{\epsilon_\mathrm{pom}} \left(1 - \frac{1}{\lambda_\mathrm{pom}}\right) 
e^{2(\lambda_\mathrm{pom}-1)/q} + \frac{1}{\lambda_\mathrm{pom}^2} \left(1 - \frac{\alpha_\mathrm{pom}}{3G^2} \operatorname{tr}(\tau \cdot \tau)\right).
\end{equation}

The physical meaning of \( q \) is the number of side-branch arms attached to the backbone segment. The parameter \( \epsilon_\mathrm{pom} \) represents the ratio of the polymer-chain backbone stretch to the orientation relaxation times. The stretch of the backbone segment is denoted as:
\begin{equation}
\lambda_\mathrm{pom} = 
\sqrt{1 + \frac{\left|\operatorname{tr}(\tau)\right|}{3G}}.
\end{equation}

\bibliography{ref}

\begin{thebibliography}{270}%
\makeatletter
\providecommand \@ifxundefined [1]{%
 \@ifx{#1\undefined}
}%
\providecommand \@ifnum [1]{%
 \ifnum #1\expandafter \@firstoftwo
 \else \expandafter \@secondoftwo
 \fi
}%
\providecommand \@ifx [1]{%
 \ifx #1\expandafter \@firstoftwo
 \else \expandafter \@secondoftwo
 \fi
}%
\providecommand \natexlab [1]{#1}%
\providecommand \enquote  [1]{``#1''}%
\providecommand \bibnamefont  [1]{#1}%
\providecommand \bibfnamefont [1]{#1}%
\providecommand \citenamefont [1]{#1}%
\providecommand \href@noop [0]{\@secondoftwo}%
\providecommand \href [0]{\begingroup \@sanitize@url \@href}%
\providecommand \@href[1]{\@@startlink{#1}\@@href}%
\providecommand \@@href[1]{\endgroup#1\@@endlink}%
\providecommand \@sanitize@url [0]{\catcode `\\12\catcode `\$12\catcode `\&12\catcode `\#12\catcode `\^12\catcode `\_12\catcode `\%12\relax}%
\providecommand \@@startlink[1]{}%
\providecommand \@@endlink[0]{}%
\providecommand \url  [0]{\begingroup\@sanitize@url \@url }%
\providecommand \@url [1]{\endgroup\@href {#1}{\urlprefix }}%
\providecommand \urlprefix  [0]{URL }%
\providecommand \Eprint [0]{\href }%
\providecommand \doibase [0]{https://doi.org/}%
\providecommand \selectlanguage [0]{\@gobble}%
\providecommand \bibinfo  [0]{\@secondoftwo}%
\providecommand \bibfield  [0]{\@secondoftwo}%
\providecommand \translation [1]{[#1]}%
\providecommand \BibitemOpen [0]{}%
\providecommand \bibitemStop [0]{}%
\providecommand \bibitemNoStop [0]{.\EOS\space}%
\providecommand \EOS [0]{\spacefactor3000\relax}%
\providecommand \BibitemShut  [1]{\csname bibitem#1\endcsname}%
\let\auto@bib@innerbib\@empty
\bibitem [{\citenamefont {Fischer}\ and\ \citenamefont {Widdicombe}(2006)}]{Fischer_2006}%
  \BibitemOpen
  \bibfield  {author} {\bibinfo {author} {\bibfnamefont {H.}~\bibnamefont {Fischer}}\ and\ \bibinfo {author} {\bibfnamefont {J.~H.}\ \bibnamefont {Widdicombe}},\ }\bibfield  {title} {\bibinfo {title} {Mechanisms of acid and base secretion by the airway epithelium},\ }\href@noop {} {\bibfield  {journal} {\bibinfo  {journal} {The Journal of Membrane Biology}\ }\textbf {\bibinfo {volume} {211}},\ \bibinfo {pages} {139} (\bibinfo {year} {2006})}\BibitemShut {NoStop}%
\bibitem [{\citenamefont {Cone}(2009)}]{Cone_2009}%
  \BibitemOpen
  \bibfield  {author} {\bibinfo {author} {\bibfnamefont {R.~A.}\ \bibnamefont {Cone}},\ }\bibfield  {title} {\bibinfo {title} {Barrier properties of mucus},\ }\href@noop {} {\bibfield  {journal} {\bibinfo  {journal} {Advanced drug delivery reviews}\ }\textbf {\bibinfo {volume} {61}},\ \bibinfo {pages} {75} (\bibinfo {year} {2009})}\BibitemShut {NoStop}%
\bibitem [{\citenamefont {King}(2009)}]{King_2009}%
  \BibitemOpen
  \bibfield  {author} {\bibinfo {author} {\bibfnamefont {P.~T.}\ \bibnamefont {King}},\ }\bibfield  {title} {\bibinfo {title} {The pathophysiology of bronchiectasis},\ }\href@noop {} {\bibfield  {journal} {\bibinfo  {journal} {International Journal of Chronic Obstructive Pulmonary Disease}\ }\textbf {\bibinfo {volume} {4}},\ \bibinfo {pages} {411} (\bibinfo {year} {2009})}\BibitemShut {NoStop}%
\bibitem [{\citenamefont {Lai}\ \emph {et~al.}(2009)\citenamefont {Lai}, \citenamefont {Wang}, \citenamefont {Wirtz},\ and\ \citenamefont {Hanes}}]{Lai_2009}%
  \BibitemOpen
  \bibfield  {author} {\bibinfo {author} {\bibfnamefont {S.~K.}\ \bibnamefont {Lai}}, \bibinfo {author} {\bibfnamefont {Y.-Y.}\ \bibnamefont {Wang}}, \bibinfo {author} {\bibfnamefont {D.}~\bibnamefont {Wirtz}},\ and\ \bibinfo {author} {\bibfnamefont {J.}~\bibnamefont {Hanes}},\ }\bibfield  {title} {\bibinfo {title} {Micro- and macrorheology of mucus},\ }\href@noop {} {\bibfield  {journal} {\bibinfo  {journal} {Advanced Drug Delivery Reviews}\ }\textbf {\bibinfo {volume} {61}},\ \bibinfo {pages} {86} (\bibinfo {year} {2009})}\BibitemShut {NoStop}%
\bibitem [{\citenamefont {Voynow}\ and\ \citenamefont {Rubin}(2009)}]{Voynow_2009}%
  \BibitemOpen
  \bibfield  {author} {\bibinfo {author} {\bibfnamefont {J.~A.}\ \bibnamefont {Voynow}}\ and\ \bibinfo {author} {\bibfnamefont {B.~K.}\ \bibnamefont {Rubin}},\ }\bibfield  {title} {\bibinfo {title} {Mucins, mucus, and sputum},\ }\href@noop {} {\bibfield  {journal} {\bibinfo  {journal} {Chest}\ }\textbf {\bibinfo {volume} {135}},\ \bibinfo {pages} {505} (\bibinfo {year} {2009})}\BibitemShut {NoStop}%
\bibitem [{\citenamefont {Fahy}\ and\ \citenamefont {Dickey}(2010)}]{Fahy_2010}%
  \BibitemOpen
  \bibfield  {author} {\bibinfo {author} {\bibfnamefont {J.~V.}\ \bibnamefont {Fahy}}\ and\ \bibinfo {author} {\bibfnamefont {B.~F.}\ \bibnamefont {Dickey}},\ }\bibfield  {title} {\bibinfo {title} {Airway mucus function and dysfunction},\ }\href@noop {} {\bibfield  {journal} {\bibinfo  {journal} {New England Journal of Medicine}\ }\textbf {\bibinfo {volume} {363}},\ \bibinfo {pages} {2233} (\bibinfo {year} {2010})}\BibitemShut {NoStop}%
\bibitem [{\citenamefont {Rubin}(2010)}]{Rubin_2010}%
  \BibitemOpen
  \bibfield  {author} {\bibinfo {author} {\bibfnamefont {B.~K.}\ \bibnamefont {Rubin}},\ }\bibfield  {title} {\bibinfo {title} {The role of mucus in cough research},\ }\href@noop {} {\bibfield  {journal} {\bibinfo  {journal} {Lung}\ }\textbf {\bibinfo {volume} {188}},\ \bibinfo {pages} {69} (\bibinfo {year} {2010})}\BibitemShut {NoStop}%
\bibitem [{\citenamefont {Cohen}\ and\ \citenamefont {Prince}(2012)}]{Cohen_2012}%
  \BibitemOpen
  \bibfield  {author} {\bibinfo {author} {\bibfnamefont {T.~S.}\ \bibnamefont {Cohen}}\ and\ \bibinfo {author} {\bibfnamefont {A.}~\bibnamefont {Prince}},\ }\bibfield  {title} {\bibinfo {title} {Cystic fibrosis: a mucosal immunodeficiency syndrome},\ }\href@noop {} {\bibfield  {journal} {\bibinfo  {journal} {Nature Medicine}\ }\textbf {\bibinfo {volume} {18}},\ \bibinfo {pages} {509} (\bibinfo {year} {2012})}\BibitemShut {NoStop}%
\bibitem [{\citenamefont {Hartl}\ \emph {et~al.}(2012)\citenamefont {Hartl}, \citenamefont {Gaggar}, \citenamefont {Bruscia}, \citenamefont {Hector}, \citenamefont {Marcos}, \citenamefont {Jung}, \citenamefont {Greene}, \citenamefont {McElvaney}, \citenamefont {Mall},\ and\ \citenamefont {D{\"o}ring}}]{Hartl_2012}%
  \BibitemOpen
  \bibfield  {author} {\bibinfo {author} {\bibfnamefont {D.}~\bibnamefont {Hartl}}, \bibinfo {author} {\bibfnamefont {A.}~\bibnamefont {Gaggar}}, \bibinfo {author} {\bibfnamefont {E.}~\bibnamefont {Bruscia}}, \bibinfo {author} {\bibfnamefont {A.}~\bibnamefont {Hector}}, \bibinfo {author} {\bibfnamefont {V.}~\bibnamefont {Marcos}}, \bibinfo {author} {\bibfnamefont {A.}~\bibnamefont {Jung}}, \bibinfo {author} {\bibfnamefont {C.}~\bibnamefont {Greene}}, \bibinfo {author} {\bibfnamefont {G.}~\bibnamefont {McElvaney}}, \bibinfo {author} {\bibfnamefont {M.}~\bibnamefont {Mall}},\ and\ \bibinfo {author} {\bibfnamefont {G.}~\bibnamefont {D{\"o}ring}},\ }\bibfield  {title} {\bibinfo {title} {Innate immunity in cystic fibrosis lung disease},\ }\href@noop {} {\bibfield  {journal} {\bibinfo  {journal} {Journal of Cystic Fibrosis}\ }\textbf {\bibinfo {volume} {11}},\ \bibinfo {pages} {363} (\bibinfo {year} {2012})}\BibitemShut {NoStop}%
\bibitem [{\citenamefont {Tuder}\ and\ \citenamefont {Petrache}(2012)}]{Tuder_2012}%
  \BibitemOpen
  \bibfield  {author} {\bibinfo {author} {\bibfnamefont {R.~M.}\ \bibnamefont {Tuder}}\ and\ \bibinfo {author} {\bibfnamefont {I.}~\bibnamefont {Petrache}},\ }\bibfield  {title} {\bibinfo {title} {Pathogenesis of chronic obstructive pulmonary disease},\ }\href@noop {} {\bibfield  {journal} {\bibinfo  {journal} {The Journal of Clinical Investigation}\ }\textbf {\bibinfo {volume} {122}},\ \bibinfo {pages} {2749} (\bibinfo {year} {2012})}\BibitemShut {NoStop}%
\bibitem [{\citenamefont {Broedersz}\ and\ \citenamefont {MacKintosh}(2014)}]{Broedersz_2014}%
  \BibitemOpen
  \bibfield  {author} {\bibinfo {author} {\bibfnamefont {C.~P.}\ \bibnamefont {Broedersz}}\ and\ \bibinfo {author} {\bibfnamefont {F.~C.}\ \bibnamefont {MacKintosh}},\ }\bibfield  {title} {\bibinfo {title} {Modeling semiflexible polymer networks},\ }\href@noop {} {\bibfield  {journal} {\bibinfo  {journal} {Reviews of Modern Physics}\ }\textbf {\bibinfo {volume} {86}},\ \bibinfo {pages} {995} (\bibinfo {year} {2014})}\BibitemShut {NoStop}%
\bibitem [{\citenamefont {Spagnolie}(2015)}]{Spagnolie_2015}%
  \BibitemOpen
  \bibfield  {author} {\bibinfo {author} {\bibfnamefont {S.~E.}\ \bibnamefont {Spagnolie}},\ }\href@noop {} {\emph {\bibinfo {title} {Complex Fluids in Biological Systems}}}\ (\bibinfo  {publisher} {Springer New York, NY},\ \bibinfo {year} {2015})\BibitemShut {NoStop}%
\bibitem [{\citenamefont {Duncan}\ \emph {et~al.}(2016{\natexlab{a}})\citenamefont {Duncan}, \citenamefont {Jung}, \citenamefont {Hanes},\ and\ \citenamefont {Suk}}]{Duncan_2016a}%
  \BibitemOpen
  \bibfield  {author} {\bibinfo {author} {\bibfnamefont {G.~A.}\ \bibnamefont {Duncan}}, \bibinfo {author} {\bibfnamefont {J.}~\bibnamefont {Jung}}, \bibinfo {author} {\bibfnamefont {J.}~\bibnamefont {Hanes}},\ and\ \bibinfo {author} {\bibfnamefont {J.~S.}\ \bibnamefont {Suk}},\ }\bibfield  {title} {\bibinfo {title} {The mucus barrier to inhaled gene therapy},\ }\href@noop {} {\bibfield  {journal} {\bibinfo  {journal} {Molecular Therapy}\ }\textbf {\bibinfo {volume} {24}},\ \bibinfo {pages} {2043} (\bibinfo {year} {2016}{\natexlab{a}})}\BibitemShut {NoStop}%
\bibitem [{\citenamefont {Kesimer}\ \emph {et~al.}(2017)\citenamefont {Kesimer}, \citenamefont {Ford}, \citenamefont {Ceppe}, \citenamefont {Radicioni}, \citenamefont {Cao}, \citenamefont {Davis}, \citenamefont {Doerschuk}, \citenamefont {Alexis}, \citenamefont {Anderson}, \citenamefont {Henderson}, \citenamefont {Barr}, \citenamefont {Bleecker}, \citenamefont {Christenson}, \citenamefont {Cooper}, \citenamefont {Han}, \citenamefont {Hansel}, \citenamefont {Hastie}, \citenamefont {Hoffman}, \citenamefont {Kanner}, \citenamefont {Martinez}, \citenamefont {Paine}, \citenamefont {Woodruff}, \citenamefont {O'Neal},\ and\ \citenamefont {Boucher}}]{Kesimer_2017}%
  \BibitemOpen
  \bibfield  {author} {\bibinfo {author} {\bibfnamefont {M.}~\bibnamefont {Kesimer}}, \bibinfo {author} {\bibfnamefont {A.~A.}\ \bibnamefont {Ford}}, \bibinfo {author} {\bibfnamefont {A.}~\bibnamefont {Ceppe}}, \bibinfo {author} {\bibfnamefont {G.}~\bibnamefont {Radicioni}}, \bibinfo {author} {\bibfnamefont {R.}~\bibnamefont {Cao}}, \bibinfo {author} {\bibfnamefont {C.~W.}\ \bibnamefont {Davis}}, \bibinfo {author} {\bibfnamefont {C.~M.}\ \bibnamefont {Doerschuk}}, \bibinfo {author} {\bibfnamefont {N.~E.}\ \bibnamefont {Alexis}}, \bibinfo {author} {\bibfnamefont {W.~H.}\ \bibnamefont {Anderson}}, \bibinfo {author} {\bibfnamefont {A.~G.}\ \bibnamefont {Henderson}}, \bibinfo {author} {\bibfnamefont {R.~G.}\ \bibnamefont {Barr}}, \bibinfo {author} {\bibfnamefont {E.~R.}\ \bibnamefont {Bleecker}}, \bibinfo {author} {\bibfnamefont {S.~A.}\ \bibnamefont {Christenson}}, \bibinfo {author} {\bibfnamefont {C.~B.}\ \bibnamefont {Cooper}}, \bibinfo {author} {\bibfnamefont {M.~K.}\ \bibnamefont {Han}}, \bibinfo {author}
  {\bibfnamefont {N.~N.}\ \bibnamefont {Hansel}}, \bibinfo {author} {\bibfnamefont {A.~T.}\ \bibnamefont {Hastie}}, \bibinfo {author} {\bibfnamefont {E.~A.}\ \bibnamefont {Hoffman}}, \bibinfo {author} {\bibfnamefont {R.~E.}\ \bibnamefont {Kanner}}, \bibinfo {author} {\bibfnamefont {F.}~\bibnamefont {Martinez}}, \bibinfo {author} {\bibfnamefont {R.}~\bibnamefont {Paine}}, \bibinfo {author} {\bibfnamefont {P.~G.}\ \bibnamefont {Woodruff}}, \bibinfo {author} {\bibfnamefont {W.~K.}\ \bibnamefont {O'Neal}},\ and\ \bibinfo {author} {\bibfnamefont {R.~C.}\ \bibnamefont {Boucher}},\ }\bibfield  {title} {\bibinfo {title} {Airway mucin concentration as a marker of chronic bronchitis},\ }\href@noop {} {\bibfield  {journal} {\bibinfo  {journal} {New England Journal of Medicine}\ }\textbf {\bibinfo {volume} {377}},\ \bibinfo {pages} {911} (\bibinfo {year} {2017})}\BibitemShut {NoStop}%
\bibitem [{\citenamefont {Witten}\ \emph {et~al.}(2018)\citenamefont {Witten}, \citenamefont {Samad},\ and\ \citenamefont {Ribbeck}}]{Witten_2018}%
  \BibitemOpen
  \bibfield  {author} {\bibinfo {author} {\bibfnamefont {J.}~\bibnamefont {Witten}}, \bibinfo {author} {\bibfnamefont {T.}~\bibnamefont {Samad}},\ and\ \bibinfo {author} {\bibfnamefont {K.}~\bibnamefont {Ribbeck}},\ }\bibfield  {title} {\bibinfo {title} {Selective permeability of mucus barriers},\ }\href@noop {} {\bibfield  {journal} {\bibinfo  {journal} {Current Opinion in Biotechnology}\ }\textbf {\bibinfo {volume} {52}},\ \bibinfo {pages} {124} (\bibinfo {year} {2018})}\BibitemShut {NoStop}%
\bibitem [{\citenamefont {Boucher}(2019)}]{Boucher_2019}%
  \BibitemOpen
  \bibfield  {author} {\bibinfo {author} {\bibfnamefont {R.~C.}\ \bibnamefont {Boucher}},\ }\bibfield  {title} {\bibinfo {title} {Muco-obstructive lung diseases},\ }\href@noop {} {\bibfield  {journal} {\bibinfo  {journal} {New England Journal of Medicine}\ }\textbf {\bibinfo {volume} {380}},\ \bibinfo {pages} {1941} (\bibinfo {year} {2019})}\BibitemShut {NoStop}%
\bibitem [{\citenamefont {Nawroth}\ \emph {et~al.}(2020)\citenamefont {Nawroth}, \citenamefont {van~der Does}, \citenamefont {Ryan},\ and\ \citenamefont {Kanso}}]{Nawroth_2020}%
  \BibitemOpen
  \bibfield  {author} {\bibinfo {author} {\bibfnamefont {J.~C.}\ \bibnamefont {Nawroth}}, \bibinfo {author} {\bibfnamefont {A.~M.}\ \bibnamefont {van~der Does}}, \bibinfo {author} {\bibfnamefont {A.}~\bibnamefont {Ryan}},\ and\ \bibinfo {author} {\bibfnamefont {E.}~\bibnamefont {Kanso}},\ }\bibfield  {title} {\bibinfo {title} {Multiscale mechanics of mucociliary clearance in the lung},\ }\href@noop {} {\bibfield  {journal} {\bibinfo  {journal} {Philosophical Transactions of the Royal Society B: Biological Sciences}\ }\textbf {\bibinfo {volume} {375}},\ \bibinfo {pages} {20190160} (\bibinfo {year} {2020})}\BibitemShut {NoStop}%
\bibitem [{\citenamefont {McShane}\ \emph {et~al.}(2021)\citenamefont {McShane}, \citenamefont {Bath}, \citenamefont {Jaramillo}, \citenamefont {Ridley}, \citenamefont {Walsh}, \citenamefont {Evans}, \citenamefont {Thornton},\ and\ \citenamefont {Ribbeck}}]{McShane_2021}%
  \BibitemOpen
  \bibfield  {author} {\bibinfo {author} {\bibfnamefont {A.}~\bibnamefont {McShane}}, \bibinfo {author} {\bibfnamefont {J.}~\bibnamefont {Bath}}, \bibinfo {author} {\bibfnamefont {A.~M.}\ \bibnamefont {Jaramillo}}, \bibinfo {author} {\bibfnamefont {C.}~\bibnamefont {Ridley}}, \bibinfo {author} {\bibfnamefont {A.~A.}\ \bibnamefont {Walsh}}, \bibinfo {author} {\bibfnamefont {C.~M.}\ \bibnamefont {Evans}}, \bibinfo {author} {\bibfnamefont {D.~J.}\ \bibnamefont {Thornton}},\ and\ \bibinfo {author} {\bibfnamefont {K.}~\bibnamefont {Ribbeck}},\ }\bibfield  {title} {\bibinfo {title} {Mucus},\ }\href@noop {} {\bibfield  {journal} {\bibinfo  {journal} {Current Biology}\ }\textbf {\bibinfo {volume} {31}},\ \bibinfo {pages} {R938} (\bibinfo {year} {2021})}\BibitemShut {NoStop}%
\bibitem [{\citenamefont {Meldrum}\ and\ \citenamefont {Chotirmall}(2021)}]{Meldrum_2021}%
  \BibitemOpen
  \bibfield  {author} {\bibinfo {author} {\bibfnamefont {O.~W.}\ \bibnamefont {Meldrum}}\ and\ \bibinfo {author} {\bibfnamefont {S.~H.}\ \bibnamefont {Chotirmall}},\ }\bibfield  {title} {\bibinfo {title} {Mucus, microbiomes and pulmonary disease},\ }\href@noop {} {\bibfield  {journal} {\bibinfo  {journal} {Biomedicines}\ }\textbf {\bibinfo {volume} {9}},\ \bibinfo {pages} {675} (\bibinfo {year} {2021})}\BibitemShut {NoStop}%
\bibitem [{\citenamefont {Hill}\ \emph {et~al.}(2022)\citenamefont {Hill}, \citenamefont {Button}, \citenamefont {Rubinstein},\ and\ \citenamefont {Boucher}}]{Hill_2022}%
  \BibitemOpen
  \bibfield  {author} {\bibinfo {author} {\bibfnamefont {D.~B.}\ \bibnamefont {Hill}}, \bibinfo {author} {\bibfnamefont {B.}~\bibnamefont {Button}}, \bibinfo {author} {\bibfnamefont {M.}~\bibnamefont {Rubinstein}},\ and\ \bibinfo {author} {\bibfnamefont {R.~C.}\ \bibnamefont {Boucher}},\ }\bibfield  {title} {\bibinfo {title} {Physiology and pathophysiology of human airway mucus},\ }\href@noop {} {\bibfield  {journal} {\bibinfo  {journal} {Physiological Reviews}\ }\textbf {\bibinfo {volume} {102}},\ \bibinfo {pages} {1757} (\bibinfo {year} {2022})}\BibitemShut {NoStop}%
\bibitem [{\citenamefont {Kavishvar}\ and\ \citenamefont {Ramachandran}(2023)}]{Kavishvar_2023}%
  \BibitemOpen
  \bibfield  {author} {\bibinfo {author} {\bibfnamefont {D.}~\bibnamefont {Kavishvar}}\ and\ \bibinfo {author} {\bibfnamefont {A.}~\bibnamefont {Ramachandran}},\ }\bibfield  {title} {\bibinfo {title} {The yielding behaviour of human mucus},\ }\href@noop {} {\bibfield  {journal} {\bibinfo  {journal} {Advances in Colloid and Interface Science}\ }\textbf {\bibinfo {volume} {322}},\ \bibinfo {pages} {103049} (\bibinfo {year} {2023})}\BibitemShut {NoStop}%
\bibitem [{\citenamefont {Pangeni}\ \emph {et~al.}(2023)\citenamefont {Pangeni}, \citenamefont {Meng}, \citenamefont {Poudel}, \citenamefont {Sharma}, \citenamefont {Hutsell}, \citenamefont {Ma}, \citenamefont {Rubin}, \citenamefont {Longest}, \citenamefont {Hindle},\ and\ \citenamefont {Xu}}]{Pangeni_2023}%
  \BibitemOpen
  \bibfield  {author} {\bibinfo {author} {\bibfnamefont {R.}~\bibnamefont {Pangeni}}, \bibinfo {author} {\bibfnamefont {T.}~\bibnamefont {Meng}}, \bibinfo {author} {\bibfnamefont {S.}~\bibnamefont {Poudel}}, \bibinfo {author} {\bibfnamefont {D.}~\bibnamefont {Sharma}}, \bibinfo {author} {\bibfnamefont {H.}~\bibnamefont {Hutsell}}, \bibinfo {author} {\bibfnamefont {J.}~\bibnamefont {Ma}}, \bibinfo {author} {\bibfnamefont {B.~K.}\ \bibnamefont {Rubin}}, \bibinfo {author} {\bibfnamefont {W.}~\bibnamefont {Longest}}, \bibinfo {author} {\bibfnamefont {M.}~\bibnamefont {Hindle}},\ and\ \bibinfo {author} {\bibfnamefont {Q.}~\bibnamefont {Xu}},\ }\bibfield  {title} {\bibinfo {title} {Airway mucus in pulmonary diseases: Muco-adhesive and muco-penetrating particles to overcome the airway mucus barriers},\ }\href@noop {} {\bibfield  {journal} {\bibinfo  {journal} {International Journal of Pharmaceutics}\ }\textbf {\bibinfo {volume} {634}},\ \bibinfo {pages} {122661} (\bibinfo {year} {2023})}\BibitemShut {NoStop}%
\bibitem [{\citenamefont {Abrami}\ \emph {et~al.}(2024)\citenamefont {Abrami}, \citenamefont {Biasin}, \citenamefont {Tescione}, \citenamefont {Tierno}, \citenamefont {Dapas}, \citenamefont {Carbone}, \citenamefont {Grassi}, \citenamefont {Conese}, \citenamefont {Di~Gioia}, \citenamefont {Larobina},\ and\ \citenamefont {Grassi}}]{Abrami_2024}%
  \BibitemOpen
  \bibfield  {author} {\bibinfo {author} {\bibfnamefont {M.}~\bibnamefont {Abrami}}, \bibinfo {author} {\bibfnamefont {A.}~\bibnamefont {Biasin}}, \bibinfo {author} {\bibfnamefont {F.}~\bibnamefont {Tescione}}, \bibinfo {author} {\bibfnamefont {D.}~\bibnamefont {Tierno}}, \bibinfo {author} {\bibfnamefont {B.}~\bibnamefont {Dapas}}, \bibinfo {author} {\bibfnamefont {A.}~\bibnamefont {Carbone}}, \bibinfo {author} {\bibfnamefont {G.}~\bibnamefont {Grassi}}, \bibinfo {author} {\bibfnamefont {M.}~\bibnamefont {Conese}}, \bibinfo {author} {\bibfnamefont {S.}~\bibnamefont {Di~Gioia}}, \bibinfo {author} {\bibfnamefont {D.}~\bibnamefont {Larobina}},\ and\ \bibinfo {author} {\bibfnamefont {M.}~\bibnamefont {Grassi}},\ }\bibfield  {title} {\bibinfo {title} {Mucus structure, viscoelastic properties, and composition in chronic respiratory diseases},\ }\href@noop {} {\bibfield  {journal} {\bibinfo  {journal} {International Journal of Molecular Sciences}\ }\textbf {\bibinfo {volume} {25}},\ \bibinfo {pages} {1933} (\bibinfo
  {year} {2024})}\BibitemShut {NoStop}%
\bibitem [{\citenamefont {Wagner}\ \emph {et~al.}(2017)\citenamefont {Wagner}, \citenamefont {Turner}, \citenamefont {Rubinstein}, \citenamefont {McKinley},\ and\ \citenamefont {Ribbeck}}]{Wagner_2017}%
  \BibitemOpen
  \bibfield  {author} {\bibinfo {author} {\bibfnamefont {C.~E.}\ \bibnamefont {Wagner}}, \bibinfo {author} {\bibfnamefont {B.~S.}\ \bibnamefont {Turner}}, \bibinfo {author} {\bibfnamefont {M.}~\bibnamefont {Rubinstein}}, \bibinfo {author} {\bibfnamefont {G.~H.}\ \bibnamefont {McKinley}},\ and\ \bibinfo {author} {\bibfnamefont {K.}~\bibnamefont {Ribbeck}},\ }\bibfield  {title} {\bibinfo {title} {A rheological study of the association and dynamics of muc5ac gels},\ }\href@noop {} {\bibfield  {journal} {\bibinfo  {journal} {Biomacromolecules}\ }\textbf {\bibinfo {volume} {18}},\ \bibinfo {pages} {3654} (\bibinfo {year} {2017})}\BibitemShut {NoStop}%
\bibitem [{\citenamefont {Vanaki}\ \emph {et~al.}(2020)\citenamefont {Vanaki}, \citenamefont {Holmes}, \citenamefont {Saha}, \citenamefont {Chen}, \citenamefont {Brown},\ and\ \citenamefont {Jayathilake}}]{Vanaki_2020}%
  \BibitemOpen
  \bibfield  {author} {\bibinfo {author} {\bibfnamefont {S.~M.}\ \bibnamefont {Vanaki}}, \bibinfo {author} {\bibfnamefont {D.}~\bibnamefont {Holmes}}, \bibinfo {author} {\bibfnamefont {S.~C.}\ \bibnamefont {Saha}}, \bibinfo {author} {\bibfnamefont {J.~J.}\ \bibnamefont {Chen}}, \bibinfo {author} {\bibfnamefont {R.~J.}\ \bibnamefont {Brown}},\ and\ \bibinfo {author} {\bibfnamefont {P.~G.}\ \bibnamefont {Jayathilake}},\ }\bibfield  {title} {\bibinfo {title} {Muco-ciliary clearance: A review of modelling techniques},\ }\href@noop {} {\bibfield  {journal} {\bibinfo  {journal} {Journal of Biomechanics}\ }\textbf {\bibinfo {volume} {99}} (\bibinfo {year} {2020})}\BibitemShut {NoStop}%
\bibitem [{\citenamefont {Huck}\ \emph {et~al.}(2022)\citenamefont {Huck}, \citenamefont {Murgia}, \citenamefont {Frisch}, \citenamefont {Hittinger}, \citenamefont {Hidalgo}, \citenamefont {Loretz},\ and\ \citenamefont {Lehr}}]{Huck_2022}%
  \BibitemOpen
  \bibfield  {author} {\bibinfo {author} {\bibfnamefont {B.~C.}\ \bibnamefont {Huck}}, \bibinfo {author} {\bibfnamefont {X.}~\bibnamefont {Murgia}}, \bibinfo {author} {\bibfnamefont {S.}~\bibnamefont {Frisch}}, \bibinfo {author} {\bibfnamefont {M.}~\bibnamefont {Hittinger}}, \bibinfo {author} {\bibfnamefont {A.}~\bibnamefont {Hidalgo}}, \bibinfo {author} {\bibfnamefont {B.}~\bibnamefont {Loretz}},\ and\ \bibinfo {author} {\bibfnamefont {C.-M.}\ \bibnamefont {Lehr}},\ }\bibfield  {title} {\bibinfo {title} {Models using native tracheobronchial mucus in the context of pulmonary drug delivery research: Composition, structure and barrier properties},\ }\href@noop {} {\bibfield  {journal} {\bibinfo  {journal} {Advanced Drug Delivery Reviews}\ }\textbf {\bibinfo {volume} {183}},\ \bibinfo {pages} {114141} (\bibinfo {year} {2022})}\BibitemShut {NoStop}%
\bibitem [{\citenamefont {Grotberg}(2011{\natexlab{a}})}]{Grotberg_2011}%
  \BibitemOpen
  \bibfield  {author} {\bibinfo {author} {\bibfnamefont {J.~B.}\ \bibnamefont {Grotberg}},\ }\bibfield  {title} {\bibinfo {title} {{Respiratory fluid mechanics}},\ }\href@noop {} {\bibfield  {journal} {\bibinfo  {journal} {Phys. Fluids}\ }\textbf {\bibinfo {volume} {23}},\ \bibinfo {pages} {021301} (\bibinfo {year} {2011}{\natexlab{a}})}\BibitemShut {NoStop}%
\bibitem [{\citenamefont {Wang}\ and\ \citenamefont {Shi}(2017)}]{Wang_2017}%
  \BibitemOpen
  \bibfield  {author} {\bibinfo {author} {\bibfnamefont {J.}~\bibnamefont {Wang}}\ and\ \bibinfo {author} {\bibfnamefont {X.}~\bibnamefont {Shi}},\ }\bibfield  {title} {\bibinfo {title} {Molecular dynamics simulation of diffusion of nanoparticles in mucus},\ }\href@noop {} {\bibfield  {journal} {\bibinfo  {journal} {Acta Mechanica Solida Sinica}\ }\textbf {\bibinfo {volume} {30}},\ \bibinfo {pages} {241} (\bibinfo {year} {2017})}\BibitemShut {NoStop}%
\bibitem [{\citenamefont {Wang}\ \emph {et~al.}(2020)\citenamefont {Wang}, \citenamefont {Zhang}, \citenamefont {Yu}, \citenamefont {Han}, \citenamefont {Zhou},\ and\ \citenamefont {Bi}}]{Wang_2020}%
  \BibitemOpen
  \bibfield  {author} {\bibinfo {author} {\bibfnamefont {Y.}~\bibnamefont {Wang}}, \bibinfo {author} {\bibfnamefont {M.}~\bibnamefont {Zhang}}, \bibinfo {author} {\bibfnamefont {Y.}~\bibnamefont {Yu}}, \bibinfo {author} {\bibfnamefont {T.}~\bibnamefont {Han}}, \bibinfo {author} {\bibfnamefont {J.}~\bibnamefont {Zhou}},\ and\ \bibinfo {author} {\bibfnamefont {L.}~\bibnamefont {Bi}},\ }\bibfield  {title} {\bibinfo {title} {Sputum characteristics and airway clearance methods in patients with severe covid-19},\ }\href@noop {} {\bibfield  {journal} {\bibinfo  {journal} {Medicine}\ }\textbf {\bibinfo {volume} {99}},\ \bibinfo {pages} {e23257} (\bibinfo {year} {2020})}\BibitemShut {NoStop}%
\bibitem [{\citenamefont {Kratochvil}\ \emph {et~al.}(2022)\citenamefont {Kratochvil}, \citenamefont {Kaber}, \citenamefont {Demirdjian}, \citenamefont {Cai}, \citenamefont {Burgener}, \citenamefont {Nagy}, \citenamefont {Barlow}, \citenamefont {Popescu}, \citenamefont {Nicolls}, \citenamefont {Ozawa}, \citenamefont {Regula}, \citenamefont {Pacheco-Navarro}, \citenamefont {Yang}, \citenamefont {de~Jesus~Perez}, \citenamefont {Karmouty-Quintana}, \citenamefont {Peters}, \citenamefont {Zhao}, \citenamefont {Buja}, \citenamefont {Johnson}, \citenamefont {Vernon}, \citenamefont {Wight}, \citenamefont {Milla}, \citenamefont {Rogers}, \citenamefont {Spakowitz}, \citenamefont {Heilshorn},\ and\ \citenamefont {Bollyky}}]{Kratochvil_2022}%
  \BibitemOpen
  \bibfield  {author} {\bibinfo {author} {\bibfnamefont {M.~J.}\ \bibnamefont {Kratochvil}}, \bibinfo {author} {\bibfnamefont {G.}~\bibnamefont {Kaber}}, \bibinfo {author} {\bibfnamefont {S.}~\bibnamefont {Demirdjian}}, \bibinfo {author} {\bibfnamefont {P.~C.}\ \bibnamefont {Cai}}, \bibinfo {author} {\bibfnamefont {E.~B.}\ \bibnamefont {Burgener}}, \bibinfo {author} {\bibfnamefont {N.}~\bibnamefont {Nagy}}, \bibinfo {author} {\bibfnamefont {G.~L.}\ \bibnamefont {Barlow}}, \bibinfo {author} {\bibfnamefont {M.}~\bibnamefont {Popescu}}, \bibinfo {author} {\bibfnamefont {M.~R.}\ \bibnamefont {Nicolls}}, \bibinfo {author} {\bibfnamefont {M.~G.}\ \bibnamefont {Ozawa}}, \bibinfo {author} {\bibfnamefont {D.~P.}\ \bibnamefont {Regula}}, \bibinfo {author} {\bibfnamefont {A.~E.}\ \bibnamefont {Pacheco-Navarro}}, \bibinfo {author} {\bibfnamefont {S.}~\bibnamefont {Yang}}, \bibinfo {author} {\bibfnamefont {V.~A.}\ \bibnamefont {de~Jesus~Perez}}, \bibinfo {author} {\bibfnamefont {H.}~\bibnamefont {Karmouty-Quintana}},
  \bibinfo {author} {\bibfnamefont {A.~M.}\ \bibnamefont {Peters}}, \bibinfo {author} {\bibfnamefont {B.}~\bibnamefont {Zhao}}, \bibinfo {author} {\bibfnamefont {M.~L.}\ \bibnamefont {Buja}}, \bibinfo {author} {\bibfnamefont {P.~Y.}\ \bibnamefont {Johnson}}, \bibinfo {author} {\bibfnamefont {R.~B.}\ \bibnamefont {Vernon}}, \bibinfo {author} {\bibfnamefont {T.~N.}\ \bibnamefont {Wight}}, \bibinfo {author} {\bibfnamefont {C.~E.}\ \bibnamefont {Milla}}, \bibinfo {author} {\bibfnamefont {A.~J.}\ \bibnamefont {Rogers}}, \bibinfo {author} {\bibfnamefont {A.~J.}\ \bibnamefont {Spakowitz}}, \bibinfo {author} {\bibfnamefont {S.~C.}\ \bibnamefont {Heilshorn}},\ and\ \bibinfo {author} {\bibfnamefont {P.~L.}\ \bibnamefont {Bollyky}},\ }\bibfield  {title} {\bibinfo {title} {Biochemical, biophysical, and immunological characterization of respiratory secretions in severe sars-cov-2 infections},\ }\href@noop {} {\bibfield  {journal} {\bibinfo  {journal} {JCI Insight}\ }\textbf {\bibinfo {volume} {7}} (\bibinfo {year}
  {2022})}\BibitemShut {NoStop}%
\bibitem [{\citenamefont {Bessonov}\ and\ \citenamefont {Volpert}(2023)}]{Bessonov_2023}%
  \BibitemOpen
  \bibfield  {author} {\bibinfo {author} {\bibfnamefont {N.}~\bibnamefont {Bessonov}}\ and\ \bibinfo {author} {\bibfnamefont {V.}~\bibnamefont {Volpert}},\ }\bibfield  {title} {\bibinfo {title} {Airway obstruction in respiratory viral infections due to impaired mucociliary clearance},\ }\href@noop {} {\bibfield  {journal} {\bibinfo  {journal} {International Journal for Numerical Methods in Biomedical Engineering}\ }\textbf {\bibinfo {volume} {39}},\ \bibinfo {pages} {e3707} (\bibinfo {year} {2023})}\BibitemShut {NoStop}%
\bibitem [{\citenamefont {Atanasova}\ and\ \citenamefont {Reznikov}(2019)}]{Atanasova_2019}%
  \BibitemOpen
  \bibfield  {author} {\bibinfo {author} {\bibfnamefont {K.~R.}\ \bibnamefont {Atanasova}}\ and\ \bibinfo {author} {\bibfnamefont {L.~R.}\ \bibnamefont {Reznikov}},\ }\bibfield  {title} {\bibinfo {title} {Strategies for measuring airway mucus and mucins},\ }\href@noop {} {\bibfield  {journal} {\bibinfo  {journal} {Respiratory Research}\ }\textbf {\bibinfo {volume} {20}},\ \bibinfo {pages} {261} (\bibinfo {year} {2019})}\BibitemShut {NoStop}%
\bibitem [{\citenamefont {McCarron}\ \emph {et~al.}(2021)\citenamefont {McCarron}, \citenamefont {Parsons},\ and\ \citenamefont {Donnelley}}]{McCarron_2021}%
  \BibitemOpen
  \bibfield  {author} {\bibinfo {author} {\bibfnamefont {A.}~\bibnamefont {McCarron}}, \bibinfo {author} {\bibfnamefont {D.}~\bibnamefont {Parsons}},\ and\ \bibinfo {author} {\bibfnamefont {M.}~\bibnamefont {Donnelley}},\ }\bibfield  {title} {\bibinfo {title} {Animal and cell culture models for cystic fibrosis: Which model is right for your application?},\ }\href@noop {} {\bibfield  {journal} {\bibinfo  {journal} {The American Journal of Pathology}\ }\textbf {\bibinfo {volume} {191}},\ \bibinfo {pages} {228} (\bibinfo {year} {2021})}\BibitemShut {NoStop}%
\bibitem [{\citenamefont {Abrami}\ \emph {et~al.}(2018)\citenamefont {Abrami}, \citenamefont {Ascenzioni}, \citenamefont {Di~Domenico}, \citenamefont {Maschio}, \citenamefont {Ventura}, \citenamefont {Confalonieri}, \citenamefont {Di~Gioia}, \citenamefont {Conese}, \citenamefont {Dapas}, \citenamefont {Grassi},\ and\ \citenamefont {Grassi}}]{Abrami_2018}%
  \BibitemOpen
  \bibfield  {author} {\bibinfo {author} {\bibfnamefont {M.}~\bibnamefont {Abrami}}, \bibinfo {author} {\bibfnamefont {F.}~\bibnamefont {Ascenzioni}}, \bibinfo {author} {\bibfnamefont {E.~G.}\ \bibnamefont {Di~Domenico}}, \bibinfo {author} {\bibfnamefont {M.}~\bibnamefont {Maschio}}, \bibinfo {author} {\bibfnamefont {A.}~\bibnamefont {Ventura}}, \bibinfo {author} {\bibfnamefont {M.}~\bibnamefont {Confalonieri}}, \bibinfo {author} {\bibfnamefont {S.}~\bibnamefont {Di~Gioia}}, \bibinfo {author} {\bibfnamefont {M.}~\bibnamefont {Conese}}, \bibinfo {author} {\bibfnamefont {B.}~\bibnamefont {Dapas}}, \bibinfo {author} {\bibfnamefont {G.}~\bibnamefont {Grassi}},\ and\ \bibinfo {author} {\bibfnamefont {M.}~\bibnamefont {Grassi}},\ }\bibfield  {title} {\bibinfo {title} {A novel approach based on low-field nmr for the detection of the pathological components of sputum in cystic fibrosis patients},\ }\href@noop {} {\bibfield  {journal} {\bibinfo  {journal} {Magnetic Resonance in Medicine}\ }\textbf {\bibinfo {volume}
  {79}},\ \bibinfo {pages} {2323} (\bibinfo {year} {2018})}\BibitemShut {NoStop}%
\bibitem [{\citenamefont {KARNAD}\ \emph {et~al.}(1990)\citenamefont {KARNAD}, \citenamefont {MHAISEKAR},\ and\ \citenamefont {MORALWAR}}]{KARNAD_1990}%
  \BibitemOpen
  \bibfield  {author} {\bibinfo {author} {\bibfnamefont {D.~R.}\ \bibnamefont {KARNAD}}, \bibinfo {author} {\bibfnamefont {D.~G.}\ \bibnamefont {MHAISEKAR}},\ and\ \bibinfo {author} {\bibfnamefont {K.~V.}\ \bibnamefont {MORALWAR}},\ }\bibfield  {title} {\bibinfo {title} {Respiratory mucus ph in tracheostomized intensive care unit patients: Effects of colonization and pneumonia},\ }\href@noop {} {\bibfield  {journal} {\bibinfo  {journal} {Critical Care Medicine}\ }\textbf {\bibinfo {volume} {18}},\ \bibinfo {pages} {699} (\bibinfo {year} {1990})}\BibitemShut {NoStop}%
\bibitem [{\citenamefont {Shah}\ \emph {et~al.}(2016)\citenamefont {Shah}, \citenamefont {Meyerholz}, \citenamefont {Tang}, \citenamefont {Reznikov}, \citenamefont {Abou~Alaiwa}, \citenamefont {Ernst}, \citenamefont {Karp}, \citenamefont {Wohlford-Lenane}, \citenamefont {Heilmann}, \citenamefont {Leidinger}, \citenamefont {Allen}, \citenamefont {Zabner}, \citenamefont {McCray}, \citenamefont {Ostedgaard}, \citenamefont {Stoltz}, \citenamefont {Randak},\ and\ \citenamefont {Welsh}}]{Shah_2016}%
  \BibitemOpen
  \bibfield  {author} {\bibinfo {author} {\bibfnamefont {V.~S.}\ \bibnamefont {Shah}}, \bibinfo {author} {\bibfnamefont {D.~K.}\ \bibnamefont {Meyerholz}}, \bibinfo {author} {\bibfnamefont {X.~X.}\ \bibnamefont {Tang}}, \bibinfo {author} {\bibfnamefont {L.}~\bibnamefont {Reznikov}}, \bibinfo {author} {\bibfnamefont {M.}~\bibnamefont {Abou~Alaiwa}}, \bibinfo {author} {\bibfnamefont {S.~E.}\ \bibnamefont {Ernst}}, \bibinfo {author} {\bibfnamefont {P.~H.}\ \bibnamefont {Karp}}, \bibinfo {author} {\bibfnamefont {C.~L.}\ \bibnamefont {Wohlford-Lenane}}, \bibinfo {author} {\bibfnamefont {K.~P.}\ \bibnamefont {Heilmann}}, \bibinfo {author} {\bibfnamefont {M.~R.}\ \bibnamefont {Leidinger}}, \bibinfo {author} {\bibfnamefont {P.~D.}\ \bibnamefont {Allen}}, \bibinfo {author} {\bibfnamefont {J.}~\bibnamefont {Zabner}}, \bibinfo {author} {\bibfnamefont {P.~B.}\ \bibnamefont {McCray}}, \bibinfo {author} {\bibfnamefont {L.~S.}\ \bibnamefont {Ostedgaard}}, \bibinfo {author} {\bibfnamefont {D.~A.}\ \bibnamefont {Stoltz}},
  \bibinfo {author} {\bibfnamefont {C.~O.}\ \bibnamefont {Randak}},\ and\ \bibinfo {author} {\bibfnamefont {M.~J.}\ \bibnamefont {Welsh}},\ }\bibfield  {title} {\bibinfo {title} {Airway acidification initiates host defense abnormalities in cystic fibrosis mice},\ }\href@noop {} {\bibfield  {journal} {\bibinfo  {journal} {Science}\ }\textbf {\bibinfo {volume} {351}},\ \bibinfo {pages} {503} (\bibinfo {year} {2016})}\BibitemShut {NoStop}%
\bibitem [{\citenamefont {Meldrum}\ \emph {et~al.}(2018)\citenamefont {Meldrum}, \citenamefont {Yakubov}, \citenamefont {Bonilla}, \citenamefont {Deshmukh}, \citenamefont {McGuckin},\ and\ \citenamefont {Gidley}}]{Meldrum_2018}%
  \BibitemOpen
  \bibfield  {author} {\bibinfo {author} {\bibfnamefont {O.~W.}\ \bibnamefont {Meldrum}}, \bibinfo {author} {\bibfnamefont {G.~E.}\ \bibnamefont {Yakubov}}, \bibinfo {author} {\bibfnamefont {M.~R.}\ \bibnamefont {Bonilla}}, \bibinfo {author} {\bibfnamefont {O.}~\bibnamefont {Deshmukh}}, \bibinfo {author} {\bibfnamefont {M.~A.}\ \bibnamefont {McGuckin}},\ and\ \bibinfo {author} {\bibfnamefont {M.~J.}\ \bibnamefont {Gidley}},\ }\bibfield  {title} {\bibinfo {title} {Mucin gel assembly is controlled by a collective action of non-mucin proteins, disulfide bridges, {Ca}$^{2+}$-mediated links, and hydrogen bonding},\ }\href@noop {} {\bibfield  {journal} {\bibinfo  {journal} {Scientific Reports}\ }\textbf {\bibinfo {volume} {8}},\ \bibinfo {pages} {5802} (\bibinfo {year} {2018})}\BibitemShut {NoStop}%
\bibitem [{\citenamefont {Ara{\'u}jo}\ \emph {et~al.}(2018)\citenamefont {Ara{\'u}jo}, \citenamefont {Martins}, \citenamefont {Azevedo},\ and\ \citenamefont {Sarmento}}]{Araujo_2018}%
  \BibitemOpen
  \bibfield  {author} {\bibinfo {author} {\bibfnamefont {F.}~\bibnamefont {Ara{\'u}jo}}, \bibinfo {author} {\bibfnamefont {C.}~\bibnamefont {Martins}}, \bibinfo {author} {\bibfnamefont {C.}~\bibnamefont {Azevedo}},\ and\ \bibinfo {author} {\bibfnamefont {B.}~\bibnamefont {Sarmento}},\ }\bibfield  {title} {\bibinfo {title} {Chemical modification of drug molecules as strategy to reduce interactions with mucus},\ }\href@noop {} {\bibfield  {journal} {\bibinfo  {journal} {Advanced Drug Delivery Reviews}\ }\textbf {\bibinfo {volume} {124}},\ \bibinfo {pages} {98} (\bibinfo {year} {2018})}\BibitemShut {NoStop}%
\bibitem [{\citenamefont {Serisier}\ \emph {et~al.}(2009)\citenamefont {Serisier}, \citenamefont {Carroll}, \citenamefont {Shute},\ and\ \citenamefont {Young}}]{Serisier_2009}%
  \BibitemOpen
  \bibfield  {author} {\bibinfo {author} {\bibfnamefont {D.~J.}\ \bibnamefont {Serisier}}, \bibinfo {author} {\bibfnamefont {M.~P.}\ \bibnamefont {Carroll}}, \bibinfo {author} {\bibfnamefont {J.~K.}\ \bibnamefont {Shute}},\ and\ \bibinfo {author} {\bibfnamefont {S.~A.}\ \bibnamefont {Young}},\ }\bibfield  {title} {\bibinfo {title} {Macrorheology of cystic fibrosis, chronic obstructive pulmonary disease {\&} normal sputum},\ }\href@noop {} {\bibfield  {journal} {\bibinfo  {journal} {Respiratory Research}\ }\textbf {\bibinfo {volume} {10}},\ \bibinfo {pages} {63} (\bibinfo {year} {2009})}\BibitemShut {NoStop}%
\bibitem [{\citenamefont {Weynants}\ \emph {et~al.}(1985)\citenamefont {Weynants}, \citenamefont {Cordier}, \citenamefont {Cellier}, \citenamefont {Pages}, \citenamefont {Loire},\ and\ \citenamefont {Brune}}]{Weynants_1985}%
  \BibitemOpen
  \bibfield  {author} {\bibinfo {author} {\bibfnamefont {P.}~\bibnamefont {Weynants}}, \bibinfo {author} {\bibfnamefont {J.~F.}\ \bibnamefont {Cordier}}, \bibinfo {author} {\bibfnamefont {C.~C.}\ \bibnamefont {Cellier}}, \bibinfo {author} {\bibfnamefont {J.}~\bibnamefont {Pages}}, \bibinfo {author} {\bibfnamefont {R.}~\bibnamefont {Loire}},\ and\ \bibinfo {author} {\bibfnamefont {J.}~\bibnamefont {Brune}},\ }\bibfield  {title} {\bibinfo {title} {Primary immunocytoma of the lung: the diagnostic value of bronchoalveolar lavage},\ }\href@noop {} {\bibfield  {journal} {\bibinfo  {journal} {Thorax}\ }\textbf {\bibinfo {volume} {40}},\ \bibinfo {pages} {542} (\bibinfo {year} {1985})}\BibitemShut {NoStop}%
\bibitem [{\citenamefont {Zayas}\ \emph {et~al.}(1990)\citenamefont {Zayas}, \citenamefont {Man},\ and\ \citenamefont {King}}]{Zayas_1990}%
  \BibitemOpen
  \bibfield  {author} {\bibinfo {author} {\bibfnamefont {J.~G.}\ \bibnamefont {Zayas}}, \bibinfo {author} {\bibfnamefont {G.~C.~W.}\ \bibnamefont {Man}},\ and\ \bibinfo {author} {\bibfnamefont {M.}~\bibnamefont {King}},\ }\bibfield  {title} {\bibinfo {title} {Tracheal mucus rheology in patients undergoing diagnostic bronchoscopy. interrelations with smoking and cancer},\ }\href@noop {} {\bibfield  {journal} {\bibinfo  {journal} {American Review of Respiratory Disease}\ }\textbf {\bibinfo {volume} {141}},\ \bibinfo {pages} {1107} (\bibinfo {year} {1990})}\BibitemShut {NoStop}%
\bibitem [{\citenamefont {Pizzichini}\ \emph {et~al.}(1996)\citenamefont {Pizzichini}, \citenamefont {Popov}, \citenamefont {Efthimiadis}, \citenamefont {Hussack}, \citenamefont {Evans}, \citenamefont {Pizzichini}, \citenamefont {Dolovich},\ and\ \citenamefont {Hargreave}}]{Pizzichini_1996}%
  \BibitemOpen
  \bibfield  {author} {\bibinfo {author} {\bibfnamefont {M.~M.}\ \bibnamefont {Pizzichini}}, \bibinfo {author} {\bibfnamefont {T.~A.}\ \bibnamefont {Popov}}, \bibinfo {author} {\bibfnamefont {A.}~\bibnamefont {Efthimiadis}}, \bibinfo {author} {\bibfnamefont {P.}~\bibnamefont {Hussack}}, \bibinfo {author} {\bibfnamefont {S.}~\bibnamefont {Evans}}, \bibinfo {author} {\bibfnamefont {E.}~\bibnamefont {Pizzichini}}, \bibinfo {author} {\bibfnamefont {J.}~\bibnamefont {Dolovich}},\ and\ \bibinfo {author} {\bibfnamefont {F.~E.}\ \bibnamefont {Hargreave}},\ }\bibfield  {title} {\bibinfo {title} {Spontaneous and induced sputum to measure indices of airway inflammation in asthma},\ }\href@noop {} {\bibfield  {journal} {\bibinfo  {journal} {American Journal of Respiratory and Critical Care Medicine}\ }\textbf {\bibinfo {volume} {154}},\ \bibinfo {pages} {866} (\bibinfo {year} {1996})}\BibitemShut {NoStop}%
\bibitem [{\citenamefont {Kamin}\ \emph {et~al.}(2006)\citenamefont {Kamin}, \citenamefont {Kl{\"a}r-Hlawatsch},\ and\ \citenamefont {Truebel}}]{Kamin_2006}%
  \BibitemOpen
  \bibfield  {author} {\bibinfo {author} {\bibfnamefont {W.}~\bibnamefont {Kamin}}, \bibinfo {author} {\bibfnamefont {B.}~\bibnamefont {Kl{\"a}r-Hlawatsch}},\ and\ \bibinfo {author} {\bibfnamefont {H.}~\bibnamefont {Truebel}},\ }\bibfield  {title} {\bibinfo {title} {Easy removal of a large mucus plug with a flexible paediatric bronchoscope after administration of rhdnase (pulmozyme)},\ }\href@noop {} {\bibfield  {journal} {\bibinfo  {journal} {Klin Padiatr}\ }\textbf {\bibinfo {volume} {218}},\ \bibinfo {pages} {88} (\bibinfo {year} {2006})}\BibitemShut {NoStop}%
\bibitem [{\citenamefont {Innes}\ \emph {et~al.}(2009)\citenamefont {Innes}, \citenamefont {Carrington}, \citenamefont {Thornton}, \citenamefont {Kirkham}, \citenamefont {Rousseau}, \citenamefont {Dougherty}, \citenamefont {Raymond}, \citenamefont {Caughey}, \citenamefont {Muller},\ and\ \citenamefont {Fahy}}]{Innes_2009}%
  \BibitemOpen
  \bibfield  {author} {\bibinfo {author} {\bibfnamefont {A.~L.}\ \bibnamefont {Innes}}, \bibinfo {author} {\bibfnamefont {S.~D.}\ \bibnamefont {Carrington}}, \bibinfo {author} {\bibfnamefont {D.~J.}\ \bibnamefont {Thornton}}, \bibinfo {author} {\bibfnamefont {S.}~\bibnamefont {Kirkham}}, \bibinfo {author} {\bibfnamefont {K.}~\bibnamefont {Rousseau}}, \bibinfo {author} {\bibfnamefont {R.~H.}\ \bibnamefont {Dougherty}}, \bibinfo {author} {\bibfnamefont {W.~W.}\ \bibnamefont {Raymond}}, \bibinfo {author} {\bibfnamefont {G.~H.}\ \bibnamefont {Caughey}}, \bibinfo {author} {\bibfnamefont {S.~J.}\ \bibnamefont {Muller}},\ and\ \bibinfo {author} {\bibfnamefont {J.~V.}\ \bibnamefont {Fahy}},\ }\bibfield  {title} {\bibinfo {title} {Ex vivo sputum analysis reveals impairment of protease-dependent mucus degradation by plasma proteins in acute asthma},\ }\href@noop {} {\bibfield  {journal} {\bibinfo  {journal} {American Journal of Respiratory and Critical Care Medicine}\ }\textbf {\bibinfo {volume} {180}},\ \bibinfo
  {pages} {203} (\bibinfo {year} {2009})}\BibitemShut {NoStop}%
\bibitem [{\citenamefont {Andreasson}\ \emph {et~al.}(2016)\citenamefont {Andreasson}, \citenamefont {Karamanou}, \citenamefont {Gillespie}, \citenamefont {{\"O}zalp}, \citenamefont {Butt}, \citenamefont {Hill}, \citenamefont {Jiwa}, \citenamefont {Walden}, \citenamefont {Green}, \citenamefont {Borthwick}, \citenamefont {Clark}, \citenamefont {Pauli}, \citenamefont {Gould}, \citenamefont {Corris}, \citenamefont {Ali}, \citenamefont {Dark},\ and\ \citenamefont {Fisher}}]{Andreasson_2016}%
  \BibitemOpen
  \bibfield  {author} {\bibinfo {author} {\bibfnamefont {A.~S.}\ \bibnamefont {Andreasson}}, \bibinfo {author} {\bibfnamefont {D.~M.}\ \bibnamefont {Karamanou}}, \bibinfo {author} {\bibfnamefont {C.~S.}\ \bibnamefont {Gillespie}}, \bibinfo {author} {\bibfnamefont {F.}~\bibnamefont {{\"O}zalp}}, \bibinfo {author} {\bibfnamefont {T.}~\bibnamefont {Butt}}, \bibinfo {author} {\bibfnamefont {P.}~\bibnamefont {Hill}}, \bibinfo {author} {\bibfnamefont {K.}~\bibnamefont {Jiwa}}, \bibinfo {author} {\bibfnamefont {H.~R.}\ \bibnamefont {Walden}}, \bibinfo {author} {\bibfnamefont {N.~J.}\ \bibnamefont {Green}}, \bibinfo {author} {\bibfnamefont {L.~A.}\ \bibnamefont {Borthwick}}, \bibinfo {author} {\bibfnamefont {S.~C.}\ \bibnamefont {Clark}}, \bibinfo {author} {\bibfnamefont {H.}~\bibnamefont {Pauli}}, \bibinfo {author} {\bibfnamefont {K.~F.}\ \bibnamefont {Gould}}, \bibinfo {author} {\bibfnamefont {P.~A.}\ \bibnamefont {Corris}}, \bibinfo {author} {\bibfnamefont {S.}~\bibnamefont {Ali}}, \bibinfo {author} {\bibfnamefont
  {J.~H.}\ \bibnamefont {Dark}},\ and\ \bibinfo {author} {\bibfnamefont {A.~J.}\ \bibnamefont {Fisher}},\ }\bibfield  {title} {\bibinfo {title} {Profiling inflammation and tissue injury markers in perfusate and bronchoalveolar lavage fluid during human ex vivo lung perfusion},\ }\href@noop {} {\bibfield  {journal} {\bibinfo  {journal} {European Journal of Cardio-Thoracic Surgery}\ }\textbf {\bibinfo {volume} {51}},\ \bibinfo {pages} {577} (\bibinfo {year} {2016})}\BibitemShut {NoStop}%
\bibitem [{\citenamefont {Fernandez-Petty}\ \emph {et~al.}(2019)\citenamefont {Fernandez-Petty}, \citenamefont {Hughes}, \citenamefont {Bowers}, \citenamefont {Watson}, \citenamefont {Rosen}, \citenamefont {Townsend}, \citenamefont {Santos}, \citenamefont {Ridley}, \citenamefont {Chu}, \citenamefont {Birket}, \citenamefont {Li}, \citenamefont {Leung}, \citenamefont {Mazur}, \citenamefont {Garcia}, \citenamefont {Evans}, \citenamefont {Libby}, \citenamefont {Hathorne}, \citenamefont {Hanes}, \citenamefont {Tearney}, \citenamefont {Clancy}, \citenamefont {Engelhardt}, \citenamefont {Swords}, \citenamefont {Thornton}, \citenamefont {Wiesmann}, \citenamefont {Baker},\ and\ \citenamefont {Rowe}}]{Fernandez-Petty_2019}%
  \BibitemOpen
  \bibfield  {author} {\bibinfo {author} {\bibfnamefont {C.~M.}\ \bibnamefont {Fernandez-Petty}}, \bibinfo {author} {\bibfnamefont {G.~W.}\ \bibnamefont {Hughes}}, \bibinfo {author} {\bibfnamefont {H.~L.}\ \bibnamefont {Bowers}}, \bibinfo {author} {\bibfnamefont {J.~D.}\ \bibnamefont {Watson}}, \bibinfo {author} {\bibfnamefont {B.~H.}\ \bibnamefont {Rosen}}, \bibinfo {author} {\bibfnamefont {S.~M.}\ \bibnamefont {Townsend}}, \bibinfo {author} {\bibfnamefont {C.}~\bibnamefont {Santos}}, \bibinfo {author} {\bibfnamefont {C.~E.}\ \bibnamefont {Ridley}}, \bibinfo {author} {\bibfnamefont {K.~K.}\ \bibnamefont {Chu}}, \bibinfo {author} {\bibfnamefont {S.~E.}\ \bibnamefont {Birket}}, \bibinfo {author} {\bibfnamefont {Y.}~\bibnamefont {Li}}, \bibinfo {author} {\bibfnamefont {H.~M.}\ \bibnamefont {Leung}}, \bibinfo {author} {\bibfnamefont {M.}~\bibnamefont {Mazur}}, \bibinfo {author} {\bibfnamefont {B.~A.}\ \bibnamefont {Garcia}}, \bibinfo {author} {\bibfnamefont {T.~I.~A.}\ \bibnamefont {Evans}}, \bibinfo {author}
  {\bibfnamefont {E.~F.}\ \bibnamefont {Libby}}, \bibinfo {author} {\bibfnamefont {H.}~\bibnamefont {Hathorne}}, \bibinfo {author} {\bibfnamefont {J.}~\bibnamefont {Hanes}}, \bibinfo {author} {\bibfnamefont {G.~J.}\ \bibnamefont {Tearney}}, \bibinfo {author} {\bibfnamefont {J.~P.}\ \bibnamefont {Clancy}}, \bibinfo {author} {\bibfnamefont {J.~F.}\ \bibnamefont {Engelhardt}}, \bibinfo {author} {\bibfnamefont {W.~E.}\ \bibnamefont {Swords}}, \bibinfo {author} {\bibfnamefont {D.~J.}\ \bibnamefont {Thornton}}, \bibinfo {author} {\bibfnamefont {W.~P.}\ \bibnamefont {Wiesmann}}, \bibinfo {author} {\bibfnamefont {S.~M.}\ \bibnamefont {Baker}},\ and\ \bibinfo {author} {\bibfnamefont {S.~M.}\ \bibnamefont {Rowe}},\ }\bibfield  {title} {\bibinfo {title} {A glycopolymer improves vascoelasticity and mucociliary transport of abnormal cystic fibrosis mucus},\ }\href@noop {} {\bibfield  {journal} {\bibinfo  {journal} {JCI Insight}\ }\textbf {\bibinfo {volume} {4}} (\bibinfo {year} {2019})}\BibitemShut {NoStop}%
\bibitem [{\citenamefont {Jeanneret-Grosjean}\ \emph {et~al.}(1988)\citenamefont {Jeanneret-Grosjean}, \citenamefont {King}, \citenamefont {Michoud}, \citenamefont {Liote},\ and\ \citenamefont {Amyot}}]{Jeanneret-Grosjean_1988}%
  \BibitemOpen
  \bibfield  {author} {\bibinfo {author} {\bibfnamefont {A.}~\bibnamefont {Jeanneret-Grosjean}}, \bibinfo {author} {\bibfnamefont {M.}~\bibnamefont {King}}, \bibinfo {author} {\bibfnamefont {M.~C.}\ \bibnamefont {Michoud}}, \bibinfo {author} {\bibfnamefont {H.}~\bibnamefont {Liote}},\ and\ \bibinfo {author} {\bibfnamefont {R.}~\bibnamefont {Amyot}},\ }\bibfield  {title} {\bibinfo {title} {Sampling technique and rheology of human tracheobronchial mucus},\ }\href@noop {} {\bibfield  {journal} {\bibinfo  {journal} {American Review of Respiratory Disease}\ }\textbf {\bibinfo {volume} {137}},\ \bibinfo {pages} {707} (\bibinfo {year} {1988})}\BibitemShut {NoStop}%
\bibitem [{\citenamefont {Bej}\ and\ \citenamefont {Haag}(2022)}]{Bej_2022}%
  \BibitemOpen
  \bibfield  {author} {\bibinfo {author} {\bibfnamefont {R.}~\bibnamefont {Bej}}\ and\ \bibinfo {author} {\bibfnamefont {R.}~\bibnamefont {Haag}},\ }\bibfield  {title} {\bibinfo {title} {Mucus-inspired dynamic hydrogels: Synthesis and future perspectives},\ }\href@noop {} {\bibfield  {journal} {\bibinfo  {journal} {Journal of the American Chemical Society}\ }\textbf {\bibinfo {volume} {144}},\ \bibinfo {pages} {20137} (\bibinfo {year} {2022})}\BibitemShut {NoStop}%
\bibitem [{\citenamefont {Rubin}\ \emph {et~al.}(1990)\citenamefont {Rubin}, \citenamefont {Ramirez}, \citenamefont {Zayas}, \citenamefont {Finegan},\ and\ \citenamefont {King}}]{Rubin_1990}%
  \BibitemOpen
  \bibfield  {author} {\bibinfo {author} {\bibfnamefont {B.~K.}\ \bibnamefont {Rubin}}, \bibinfo {author} {\bibfnamefont {O.}~\bibnamefont {Ramirez}}, \bibinfo {author} {\bibfnamefont {J.~G.}\ \bibnamefont {Zayas}}, \bibinfo {author} {\bibfnamefont {B.}~\bibnamefont {Finegan}},\ and\ \bibinfo {author} {\bibfnamefont {M.}~\bibnamefont {King}},\ }\bibfield  {title} {\bibinfo {title} {Collection and analysis of respiratory mucus from subjects without lung disease},\ }\href@noop {} {\bibfield  {journal} {\bibinfo  {journal} {American Review of Respiratory Disease}\ }\textbf {\bibinfo {volume} {141}},\ \bibinfo {pages} {1040} (\bibinfo {year} {1990})}\BibitemShut {NoStop}%
\bibitem [{\citenamefont {Lapinsky}(2015)}]{Lapinsky_2015}%
  \BibitemOpen
  \bibfield  {author} {\bibinfo {author} {\bibfnamefont {S.~E.}\ \bibnamefont {Lapinsky}},\ }\bibfield  {title} {\bibinfo {title} {Endotracheal intubation in the icu},\ }\href@noop {} {\bibfield  {journal} {\bibinfo  {journal} {Critical Care}\ }\textbf {\bibinfo {volume} {19}},\ \bibinfo {pages} {258} (\bibinfo {year} {2015})}\BibitemShut {NoStop}%
\bibitem [{\citenamefont {Mietto}\ \emph {et~al.}(2014)\citenamefont {Mietto}, \citenamefont {Foley}, \citenamefont {Salerno}, \citenamefont {Oleksak}, \citenamefont {Pinciroli}, \citenamefont {Goverman},\ and\ \citenamefont {Berra}}]{Mietto_2014}%
  \BibitemOpen
  \bibfield  {author} {\bibinfo {author} {\bibfnamefont {C.}~\bibnamefont {Mietto}}, \bibinfo {author} {\bibfnamefont {K.}~\bibnamefont {Foley}}, \bibinfo {author} {\bibfnamefont {L.}~\bibnamefont {Salerno}}, \bibinfo {author} {\bibfnamefont {J.}~\bibnamefont {Oleksak}}, \bibinfo {author} {\bibfnamefont {R.}~\bibnamefont {Pinciroli}}, \bibinfo {author} {\bibfnamefont {J.}~\bibnamefont {Goverman}},\ and\ \bibinfo {author} {\bibfnamefont {L.}~\bibnamefont {Berra}},\ }\bibfield  {title} {\bibinfo {title} {Removal of endotracheal tube obstruction with a secretion clearance device},\ }\href@noop {} {\bibfield  {journal} {\bibinfo  {journal} {Respiratory Care}\ }\textbf {\bibinfo {volume} {59}},\ \bibinfo {pages} {e122} (\bibinfo {year} {2014})}\BibitemShut {NoStop}%
\bibitem [{\citenamefont {Berra}\ \emph {et~al.}(2012)\citenamefont {Berra}, \citenamefont {Coppadoro}, \citenamefont {Bittner}, \citenamefont {Kolobow}, \citenamefont {Laquerriere}, \citenamefont {Pohlmann}, \citenamefont {Bramati}, \citenamefont {Moss},\ and\ \citenamefont {Pesenti}}]{Berra_2012}%
  \BibitemOpen
  \bibfield  {author} {\bibinfo {author} {\bibfnamefont {L.}~\bibnamefont {Berra}}, \bibinfo {author} {\bibfnamefont {A.}~\bibnamefont {Coppadoro}}, \bibinfo {author} {\bibfnamefont {E.~A.}\ \bibnamefont {Bittner}}, \bibinfo {author} {\bibfnamefont {T.}~\bibnamefont {Kolobow}}, \bibinfo {author} {\bibfnamefont {P.}~\bibnamefont {Laquerriere}}, \bibinfo {author} {\bibfnamefont {J.~R.}\ \bibnamefont {Pohlmann}}, \bibinfo {author} {\bibfnamefont {S.}~\bibnamefont {Bramati}}, \bibinfo {author} {\bibfnamefont {J.}~\bibnamefont {Moss}},\ and\ \bibinfo {author} {\bibfnamefont {A.}~\bibnamefont {Pesenti}},\ }\bibfield  {title} {\bibinfo {title} {A clinical assessment of the mucus shaver: A device to keep the endotracheal tube free from secretions*},\ }\href@noop {} {\bibfield  {journal} {\bibinfo  {journal} {Critical Care Medicine}\ }\textbf {\bibinfo {volume} {40}},\ \bibinfo {pages} {119} (\bibinfo {year} {2012})}\BibitemShut {NoStop}%
\bibitem [{\citenamefont {Markovetz}\ \emph {et~al.}(2022)\citenamefont {Markovetz}, \citenamefont {Garbarine}, \citenamefont {Morrison}, \citenamefont {Kissner}, \citenamefont {Seim}, \citenamefont {Forest}, \citenamefont {Papanikolas}, \citenamefont {Freeman}, \citenamefont {Ceppe}, \citenamefont {Ghio}, \citenamefont {Alexis}, \citenamefont {Stick}, \citenamefont {Ehre}, \citenamefont {Boucher}, \citenamefont {Esther}, \citenamefont {Muhlebach},\ and\ \citenamefont {Hill}}]{Markovetz_2022}%
  \BibitemOpen
  \bibfield  {author} {\bibinfo {author} {\bibfnamefont {M.~R.}\ \bibnamefont {Markovetz}}, \bibinfo {author} {\bibfnamefont {I.~C.}\ \bibnamefont {Garbarine}}, \bibinfo {author} {\bibfnamefont {C.~B.}\ \bibnamefont {Morrison}}, \bibinfo {author} {\bibfnamefont {W.~J.}\ \bibnamefont {Kissner}}, \bibinfo {author} {\bibfnamefont {I.}~\bibnamefont {Seim}}, \bibinfo {author} {\bibfnamefont {M.~G.}\ \bibnamefont {Forest}}, \bibinfo {author} {\bibfnamefont {M.~J.}\ \bibnamefont {Papanikolas}}, \bibinfo {author} {\bibfnamefont {R.}~\bibnamefont {Freeman}}, \bibinfo {author} {\bibfnamefont {A.}~\bibnamefont {Ceppe}}, \bibinfo {author} {\bibfnamefont {A.}~\bibnamefont {Ghio}}, \bibinfo {author} {\bibfnamefont {N.~E.}\ \bibnamefont {Alexis}}, \bibinfo {author} {\bibfnamefont {S.~M.}\ \bibnamefont {Stick}}, \bibinfo {author} {\bibfnamefont {C.}~\bibnamefont {Ehre}}, \bibinfo {author} {\bibfnamefont {R.~C.}\ \bibnamefont {Boucher}}, \bibinfo {author} {\bibfnamefont {C.~R.}\ \bibnamefont {Esther}}, \bibinfo {author}
  {\bibfnamefont {M.~S.}\ \bibnamefont {Muhlebach}},\ and\ \bibinfo {author} {\bibfnamefont {D.~B.}\ \bibnamefont {Hill}},\ }\bibfield  {title} {\bibinfo {title} {Mucus and mucus flake composition and abundance reflect inflammatory and infection status in cystic fibrosis},\ }\href@noop {} {\bibfield  {journal} {\bibinfo  {journal} {Journal of Cystic Fibrosis}\ }\textbf {\bibinfo {volume} {21}},\ \bibinfo {pages} {959} (\bibinfo {year} {2022})}\BibitemShut {NoStop}%
\bibitem [{\citenamefont {Radiom}\ \emph {et~al.}(2021)\citenamefont {Radiom}, \citenamefont {H{\'e}nault}, \citenamefont {Mani}, \citenamefont {Iankovski}, \citenamefont {Norel},\ and\ \citenamefont {Berret}}]{Radiom_2021}%
  \BibitemOpen
  \bibfield  {author} {\bibinfo {author} {\bibfnamefont {M.}~\bibnamefont {Radiom}}, \bibinfo {author} {\bibfnamefont {R.}~\bibnamefont {H{\'e}nault}}, \bibinfo {author} {\bibfnamefont {S.}~\bibnamefont {Mani}}, \bibinfo {author} {\bibfnamefont {A.~G.}\ \bibnamefont {Iankovski}}, \bibinfo {author} {\bibfnamefont {X.}~\bibnamefont {Norel}},\ and\ \bibinfo {author} {\bibfnamefont {J.-F.}\ \bibnamefont {Berret}},\ }\bibfield  {title} {\bibinfo {title} {Magnetic wire active microrheology of human respiratory mucus},\ }\href@noop {} {\bibfield  {journal} {\bibinfo  {journal} {Soft Matter}\ }\textbf {\bibinfo {volume} {17}},\ \bibinfo {pages} {7585} (\bibinfo {year} {2021})}\BibitemShut {NoStop}%
\bibitem [{\citenamefont {Song}\ \emph {et~al.}(2021)\citenamefont {Song}, \citenamefont {Iverson}, \citenamefont {Kaler}, \citenamefont {Bader}, \citenamefont {Scull},\ and\ \citenamefont {Duncan}}]{Song_2021}%
  \BibitemOpen
  \bibfield  {author} {\bibinfo {author} {\bibfnamefont {D.}~\bibnamefont {Song}}, \bibinfo {author} {\bibfnamefont {E.}~\bibnamefont {Iverson}}, \bibinfo {author} {\bibfnamefont {L.}~\bibnamefont {Kaler}}, \bibinfo {author} {\bibfnamefont {S.}~\bibnamefont {Bader}}, \bibinfo {author} {\bibfnamefont {M.~A.}\ \bibnamefont {Scull}},\ and\ \bibinfo {author} {\bibfnamefont {G.~A.}\ \bibnamefont {Duncan}},\ }\bibfield  {title} {\bibinfo {title} {Modeling airway dysfunction in asthma using synthetic mucus biomaterials},\ }\href@noop {} {\bibfield  {journal} {\bibinfo  {journal} {ACS Biomaterials Science {\&} Engineering}\ }\textbf {\bibinfo {volume} {7}},\ \bibinfo {pages} {2723} (\bibinfo {year} {2021})}\BibitemShut {NoStop}%
\bibitem [{\citenamefont {Markovetz}\ \emph {et~al.}(2019)\citenamefont {Markovetz}, \citenamefont {Subramani}, \citenamefont {Kissner}, \citenamefont {Morrison}, \citenamefont {Garbarine}, \citenamefont {Ghio}, \citenamefont {Ramsey}, \citenamefont {Arora}, \citenamefont {Kumar}, \citenamefont {Nix}, \citenamefont {Kumagai}, \citenamefont {Krunkosky}, \citenamefont {Krause}, \citenamefont {Radicioni}, \citenamefont {Alexis}, \citenamefont {Kesimer}, \citenamefont {Tiemeyer}, \citenamefont {Boucher}, \citenamefont {Ehre},\ and\ \citenamefont {Hill}}]{Markovetz_2019}%
  \BibitemOpen
  \bibfield  {author} {\bibinfo {author} {\bibfnamefont {M.~R.}\ \bibnamefont {Markovetz}}, \bibinfo {author} {\bibfnamefont {D.~B.}\ \bibnamefont {Subramani}}, \bibinfo {author} {\bibfnamefont {W.~J.}\ \bibnamefont {Kissner}}, \bibinfo {author} {\bibfnamefont {C.~B.}\ \bibnamefont {Morrison}}, \bibinfo {author} {\bibfnamefont {I.~C.}\ \bibnamefont {Garbarine}}, \bibinfo {author} {\bibfnamefont {A.}~\bibnamefont {Ghio}}, \bibinfo {author} {\bibfnamefont {K.~A.}\ \bibnamefont {Ramsey}}, \bibinfo {author} {\bibfnamefont {H.}~\bibnamefont {Arora}}, \bibinfo {author} {\bibfnamefont {P.}~\bibnamefont {Kumar}}, \bibinfo {author} {\bibfnamefont {D.~B.}\ \bibnamefont {Nix}}, \bibinfo {author} {\bibfnamefont {T.}~\bibnamefont {Kumagai}}, \bibinfo {author} {\bibfnamefont {T.~M.}\ \bibnamefont {Krunkosky}}, \bibinfo {author} {\bibfnamefont {D.~C.}\ \bibnamefont {Krause}}, \bibinfo {author} {\bibfnamefont {G.}~\bibnamefont {Radicioni}}, \bibinfo {author} {\bibfnamefont {N.~E.}\ \bibnamefont {Alexis}}, \bibinfo {author}
  {\bibfnamefont {M.}~\bibnamefont {Kesimer}}, \bibinfo {author} {\bibfnamefont {M.}~\bibnamefont {Tiemeyer}}, \bibinfo {author} {\bibfnamefont {R.~C.}\ \bibnamefont {Boucher}}, \bibinfo {author} {\bibfnamefont {C.}~\bibnamefont {Ehre}},\ and\ \bibinfo {author} {\bibfnamefont {D.~B.}\ \bibnamefont {Hill}},\ }\bibfield  {title} {\bibinfo {title} {Endotracheal tube mucus as a source of airway mucus for rheological study},\ }\href@noop {} {\bibfield  {journal} {\bibinfo  {journal} {American Journal of Physiology-Lung Cellular and Molecular Physiology}\ }\textbf {\bibinfo {volume} {317}},\ \bibinfo {pages} {L498} (\bibinfo {year} {2019})}\BibitemShut {NoStop}%
\bibitem [{\citenamefont {Duncan}\ \emph {et~al.}(2016{\natexlab{b}})\citenamefont {Duncan}, \citenamefont {Jung}, \citenamefont {Joseph}, \citenamefont {Thaxton}, \citenamefont {West}, \citenamefont {Boyle}, \citenamefont {Hanes},\ and\ \citenamefont {Suk}}]{Duncan_2016}%
  \BibitemOpen
  \bibfield  {author} {\bibinfo {author} {\bibfnamefont {G.~A.}\ \bibnamefont {Duncan}}, \bibinfo {author} {\bibfnamefont {J.}~\bibnamefont {Jung}}, \bibinfo {author} {\bibfnamefont {A.}~\bibnamefont {Joseph}}, \bibinfo {author} {\bibfnamefont {A.~L.}\ \bibnamefont {Thaxton}}, \bibinfo {author} {\bibfnamefont {N.~E.}\ \bibnamefont {West}}, \bibinfo {author} {\bibfnamefont {M.~P.}\ \bibnamefont {Boyle}}, \bibinfo {author} {\bibfnamefont {J.}~\bibnamefont {Hanes}},\ and\ \bibinfo {author} {\bibfnamefont {J.~S.}\ \bibnamefont {Suk}},\ }\bibfield  {title} {\bibinfo {title} {Microstructural alterations of sputum in cystic fibrosis lung disease},\ }\href@noop {} {\bibfield  {journal} {\bibinfo  {journal} {JCI Insight}\ }\textbf {\bibinfo {volume} {1}},\ \bibinfo {pages} {e88198} (\bibinfo {year} {2016}{\natexlab{b}})}\BibitemShut {NoStop}%
\bibitem [{\citenamefont {Hall}\ and\ \citenamefont {Gandevia}(1971)}]{Hall_1971}%
  \BibitemOpen
  \bibfield  {author} {\bibinfo {author} {\bibfnamefont {G.~J.}\ \bibnamefont {Hall}}\ and\ \bibinfo {author} {\bibfnamefont {B.}~\bibnamefont {Gandevia}},\ }\bibfield  {title} {\bibinfo {title} {Relationship of the loose cough sign to daily sputum volume. observer variation in its detection},\ }\href@noop {} {\bibfield  {journal} {\bibinfo  {journal} {Br J Prev Soc Med}\ }\textbf {\bibinfo {volume} {25}},\ \bibinfo {pages} {109} (\bibinfo {year} {1971})}\BibitemShut {NoStop}%
\bibitem [{\citenamefont {Bartoli}\ \emph {et~al.}(2002)\citenamefont {Bartoli}, \citenamefont {Bacci}, \citenamefont {Carnevali}, \citenamefont {Cianchetti}, \citenamefont {Dente}, \citenamefont {Di~Franco}, \citenamefont {Giannini}, \citenamefont {Taccola}, \citenamefont {Vagaggini},\ and\ \citenamefont {Paggiaro}}]{Bartoli_2002}%
  \BibitemOpen
  \bibfield  {author} {\bibinfo {author} {\bibfnamefont {M.~L.}\ \bibnamefont {Bartoli}}, \bibinfo {author} {\bibfnamefont {E.}~\bibnamefont {Bacci}}, \bibinfo {author} {\bibfnamefont {S.}~\bibnamefont {Carnevali}}, \bibinfo {author} {\bibfnamefont {S.}~\bibnamefont {Cianchetti}}, \bibinfo {author} {\bibfnamefont {F.~L.}\ \bibnamefont {Dente}}, \bibinfo {author} {\bibfnamefont {A.}~\bibnamefont {Di~Franco}}, \bibinfo {author} {\bibfnamefont {D.}~\bibnamefont {Giannini}}, \bibinfo {author} {\bibfnamefont {M.}~\bibnamefont {Taccola}}, \bibinfo {author} {\bibfnamefont {B.}~\bibnamefont {Vagaggini}},\ and\ \bibinfo {author} {\bibfnamefont {P.~L.}\ \bibnamefont {Paggiaro}},\ }\bibfield  {title} {\bibinfo {title} {Quality evaluation of samples obtained by spontaneous or induced sputum: Comparison between two methods of processing and relationship with clinical and functional findings},\ }\href@noop {} {\bibfield  {journal} {\bibinfo  {journal} {Journal of Asthma}\ }\textbf {\bibinfo {volume} {39}},\ \bibinfo
  {pages} {479} (\bibinfo {year} {2002})}\BibitemShut {NoStop}%
\bibitem [{\citenamefont {Sagel}\ \emph {et~al.}(2007)\citenamefont {Sagel}, \citenamefont {Chmiel},\ and\ \citenamefont {Konstan}}]{Sagel_2007}%
  \BibitemOpen
  \bibfield  {author} {\bibinfo {author} {\bibfnamefont {S.~D.}\ \bibnamefont {Sagel}}, \bibinfo {author} {\bibfnamefont {J.~F.}\ \bibnamefont {Chmiel}},\ and\ \bibinfo {author} {\bibfnamefont {M.~W.}\ \bibnamefont {Konstan}},\ }\bibfield  {title} {\bibinfo {title} {Sputum biomarkers of inflammation in cystic fibrosis lung disease},\ }\href@noop {} {\bibfield  {journal} {\bibinfo  {journal} {Proc Am Thorac Soc}\ }\textbf {\bibinfo {volume} {4}},\ \bibinfo {pages} {406} (\bibinfo {year} {2007})}\BibitemShut {NoStop}%
\bibitem [{\citenamefont {Suk}\ \emph {et~al.}(2009)\citenamefont {Suk}, \citenamefont {Lai}, \citenamefont {Wang}, \citenamefont {Ensign}, \citenamefont {Zeitlin}, \citenamefont {Boyle},\ and\ \citenamefont {Hanes}}]{Suk_2009}%
  \BibitemOpen
  \bibfield  {author} {\bibinfo {author} {\bibfnamefont {J.~S.}\ \bibnamefont {Suk}}, \bibinfo {author} {\bibfnamefont {S.~K.}\ \bibnamefont {Lai}}, \bibinfo {author} {\bibfnamefont {Y.-Y.}\ \bibnamefont {Wang}}, \bibinfo {author} {\bibfnamefont {L.~M.}\ \bibnamefont {Ensign}}, \bibinfo {author} {\bibfnamefont {P.~L.}\ \bibnamefont {Zeitlin}}, \bibinfo {author} {\bibfnamefont {M.~P.}\ \bibnamefont {Boyle}},\ and\ \bibinfo {author} {\bibfnamefont {J.}~\bibnamefont {Hanes}},\ }\bibfield  {title} {\bibinfo {title} {The penetration of fresh undiluted sputum expectorated by cystic fibrosis patients by non-adhesive polymer nanoparticles},\ }\href@noop {} {\bibfield  {journal} {\bibinfo  {journal} {Biomaterials}\ }\textbf {\bibinfo {volume} {30}},\ \bibinfo {pages} {2591} (\bibinfo {year} {2009})}\BibitemShut {NoStop}%
\bibitem [{\citenamefont {Vieira}\ \emph {et~al.}(2011)\citenamefont {Vieira}, \citenamefont {Pizzichini}, \citenamefont {Steidle}, \citenamefont {da~Silva},\ and\ \citenamefont {Pizzichini}}]{Vieira_2011}%
  \BibitemOpen
  \bibfield  {author} {\bibinfo {author} {\bibfnamefont {M.~O.}\ \bibnamefont {Vieira}}, \bibinfo {author} {\bibfnamefont {E.}~\bibnamefont {Pizzichini}}, \bibinfo {author} {\bibfnamefont {L.~J.}\ \bibnamefont {Steidle}}, \bibinfo {author} {\bibfnamefont {J.~K.}\ \bibnamefont {da~Silva}},\ and\ \bibinfo {author} {\bibfnamefont {M.~M.}\ \bibnamefont {Pizzichini}},\ }\bibfield  {title} {\bibinfo {title} {Sputum induction in severe exacerbations of asthma: safety of a modified method},\ }\href@noop {} {\bibfield  {journal} {\bibinfo  {journal} {Eur Respir J}\ }\textbf {\bibinfo {volume} {38}},\ \bibinfo {pages} {979} (\bibinfo {year} {2011})}\BibitemShut {NoStop}%
\bibitem [{\citenamefont {Weiszhar}\ and\ \citenamefont {Horvath}(2013)}]{Weiszhar_2013}%
  \BibitemOpen
  \bibfield  {author} {\bibinfo {author} {\bibfnamefont {Z.}~\bibnamefont {Weiszhar}}\ and\ \bibinfo {author} {\bibfnamefont {I.}~\bibnamefont {Horvath}},\ }\bibfield  {title} {\bibinfo {title} {Induced sputum analysis: step by step},\ }\href@noop {} {\bibfield  {journal} {\bibinfo  {journal} {Breathe}\ }\textbf {\bibinfo {volume} {9}},\ \bibinfo {pages} {300} (\bibinfo {year} {2013})}\BibitemShut {NoStop}%
\bibitem [{\citenamefont {Patarin}\ \emph {et~al.}(2020)\citenamefont {Patarin}, \citenamefont {Ghiringhelli}, \citenamefont {Darsy}, \citenamefont {Obamba}, \citenamefont {Bochu}, \citenamefont {Camara}, \citenamefont {Qu{\'e}tant}, \citenamefont {Cracowski}, \citenamefont {Cracowski},\ and\ \citenamefont {Robert~de Saint~Vincent}}]{Patarin_2020}%
  \BibitemOpen
  \bibfield  {author} {\bibinfo {author} {\bibfnamefont {J.}~\bibnamefont {Patarin}}, \bibinfo {author} {\bibfnamefont {{\'E}.}~\bibnamefont {Ghiringhelli}}, \bibinfo {author} {\bibfnamefont {G.}~\bibnamefont {Darsy}}, \bibinfo {author} {\bibfnamefont {M.}~\bibnamefont {Obamba}}, \bibinfo {author} {\bibfnamefont {P.}~\bibnamefont {Bochu}}, \bibinfo {author} {\bibfnamefont {B.}~\bibnamefont {Camara}}, \bibinfo {author} {\bibfnamefont {S.}~\bibnamefont {Qu{\'e}tant}}, \bibinfo {author} {\bibfnamefont {J.-L.}\ \bibnamefont {Cracowski}}, \bibinfo {author} {\bibfnamefont {C.}~\bibnamefont {Cracowski}},\ and\ \bibinfo {author} {\bibfnamefont {M.}~\bibnamefont {Robert~de Saint~Vincent}},\ }\bibfield  {title} {\bibinfo {title} {Rheological analysis of sputum from patients with chronic bronchial diseases},\ }\href@noop {} {\bibfield  {journal} {\bibinfo  {journal} {Scientific Reports}\ }\textbf {\bibinfo {volume} {10}},\ \bibinfo {pages} {15685} (\bibinfo {year} {2020})}\BibitemShut {NoStop}%
\bibitem [{\citenamefont {Davis}(2002)}]{Davis_2002}%
  \BibitemOpen
  \bibfield  {author} {\bibinfo {author} {\bibfnamefont {C.~W.}\ \bibnamefont {Davis}},\ }\bibfield  {title} {\bibinfo {title} {Regulation of mucin secretion from in vitro cellular models},\ }\href@noop {} {\bibfield  {journal} {\bibinfo  {journal} {Novartis Found Symp}\ }\textbf {\bibinfo {volume} {248}},\ \bibinfo {pages} {113} (\bibinfo {year} {2002})}\BibitemShut {NoStop}%
\bibitem [{\citenamefont {Kemp}\ \emph {et~al.}(2004)\citenamefont {Kemp}, \citenamefont {Sugar},\ and\ \citenamefont {Jackson}}]{Kemp_2004}%
  \BibitemOpen
  \bibfield  {author} {\bibinfo {author} {\bibfnamefont {P.~A.}\ \bibnamefont {Kemp}}, \bibinfo {author} {\bibfnamefont {R.~A.}\ \bibnamefont {Sugar}},\ and\ \bibinfo {author} {\bibfnamefont {A.~D.}\ \bibnamefont {Jackson}},\ }\bibfield  {title} {\bibinfo {title} {Nucleotide-mediated mucin secretion from differentiated human bronchial epithelial cells},\ }\href@noop {} {\bibfield  {journal} {\bibinfo  {journal} {Am J Respir Cell Mol Biol}\ }\textbf {\bibinfo {volume} {31}},\ \bibinfo {pages} {446} (\bibinfo {year} {2004})}\BibitemShut {NoStop}%
\bibitem [{\citenamefont {Matsui}\ \emph {et~al.}(2006)\citenamefont {Matsui}, \citenamefont {Wagner}, \citenamefont {Hill}, \citenamefont {Schwab}, \citenamefont {Rogers}, \citenamefont {Button}, \citenamefont {Taylor}, \citenamefont {Superfine}, \citenamefont {Rubinstein}, \citenamefont {Iglewski},\ and\ \citenamefont {Boucher}}]{Matsui_2006}%
  \BibitemOpen
  \bibfield  {author} {\bibinfo {author} {\bibfnamefont {H.}~\bibnamefont {Matsui}}, \bibinfo {author} {\bibfnamefont {V.~E.}\ \bibnamefont {Wagner}}, \bibinfo {author} {\bibfnamefont {D.~B.}\ \bibnamefont {Hill}}, \bibinfo {author} {\bibfnamefont {U.~E.}\ \bibnamefont {Schwab}}, \bibinfo {author} {\bibfnamefont {T.~D.}\ \bibnamefont {Rogers}}, \bibinfo {author} {\bibfnamefont {B.}~\bibnamefont {Button}}, \bibinfo {author} {\bibfnamefont {R.~M.}\ \bibnamefont {Taylor}}, \bibinfo {author} {\bibfnamefont {R.}~\bibnamefont {Superfine}}, \bibinfo {author} {\bibfnamefont {M.}~\bibnamefont {Rubinstein}}, \bibinfo {author} {\bibfnamefont {B.~H.}\ \bibnamefont {Iglewski}},\ and\ \bibinfo {author} {\bibfnamefont {R.~C.}\ \bibnamefont {Boucher}},\ }\bibfield  {title} {\bibinfo {title} {A physical linkage between cystic fibrosis airway surface dehydration and pseudomonas aeruginosa biofilms},\ }\href@noop {} {\bibfield  {journal} {\bibinfo  {journal} {Proceedings of the National Academy of Sciences}\ }\textbf {\bibinfo
  {volume} {103}},\ \bibinfo {pages} {18131} (\bibinfo {year} {2006})}\BibitemShut {NoStop}%
\bibitem [{\citenamefont {Kesimer}\ \emph {et~al.}(2009)\citenamefont {Kesimer}, \citenamefont {Kirkham}, \citenamefont {Pickles}, \citenamefont {Henderson}, \citenamefont {Alexis}, \citenamefont {Demaria}, \citenamefont {Knight}, \citenamefont {Thornton},\ and\ \citenamefont {Sheehan}}]{Kesimer_2009}%
  \BibitemOpen
  \bibfield  {author} {\bibinfo {author} {\bibfnamefont {M.}~\bibnamefont {Kesimer}}, \bibinfo {author} {\bibfnamefont {S.}~\bibnamefont {Kirkham}}, \bibinfo {author} {\bibfnamefont {R.~J.}\ \bibnamefont {Pickles}}, \bibinfo {author} {\bibfnamefont {A.~G.}\ \bibnamefont {Henderson}}, \bibinfo {author} {\bibfnamefont {N.~E.}\ \bibnamefont {Alexis}}, \bibinfo {author} {\bibfnamefont {G.}~\bibnamefont {Demaria}}, \bibinfo {author} {\bibfnamefont {D.}~\bibnamefont {Knight}}, \bibinfo {author} {\bibfnamefont {D.~J.}\ \bibnamefont {Thornton}},\ and\ \bibinfo {author} {\bibfnamefont {J.~K.}\ \bibnamefont {Sheehan}},\ }\bibfield  {title} {\bibinfo {title} {Tracheobronchial air-liquid interface cell culture: a model for innate mucosal defense of the upper airways?},\ }\href@noop {} {\bibfield  {journal} {\bibinfo  {journal} {Am J Physiol Lung Cell Mol Physiol}\ }\textbf {\bibinfo {volume} {296}},\ \bibinfo {pages} {L92} (\bibinfo {year} {2009})}\BibitemShut {NoStop}%
\bibitem [{\citenamefont {Sears}\ \emph {et~al.}(2011)\citenamefont {Sears}, \citenamefont {Davis}, \citenamefont {Chua},\ and\ \citenamefont {Sheehan}}]{Sears_2011}%
  \BibitemOpen
  \bibfield  {author} {\bibinfo {author} {\bibfnamefont {P.~R.}\ \bibnamefont {Sears}}, \bibinfo {author} {\bibfnamefont {C.~W.}\ \bibnamefont {Davis}}, \bibinfo {author} {\bibfnamefont {M.}~\bibnamefont {Chua}},\ and\ \bibinfo {author} {\bibfnamefont {J.~K.}\ \bibnamefont {Sheehan}},\ }\bibfield  {title} {\bibinfo {title} {Mucociliary interactions and mucus dynamics in ciliated human bronchial epithelial cell cultures},\ }\href@noop {} {\bibfield  {journal} {\bibinfo  {journal} {American Journal of Physiology-Lung Cellular and Molecular Physiology}\ }\textbf {\bibinfo {volume} {301}},\ \bibinfo {pages} {L181} (\bibinfo {year} {2011})}\BibitemShut {NoStop}%
\bibitem [{\citenamefont {Abdullah}\ \emph {et~al.}(2012)\citenamefont {Abdullah}, \citenamefont {Wolber}, \citenamefont {Kesimer}, \citenamefont {Sheehan},\ and\ \citenamefont {Davis}}]{Abdullah_2012}%
  \BibitemOpen
  \bibfield  {author} {\bibinfo {author} {\bibfnamefont {L.~H.}\ \bibnamefont {Abdullah}}, \bibinfo {author} {\bibfnamefont {C.}~\bibnamefont {Wolber}}, \bibinfo {author} {\bibfnamefont {M.}~\bibnamefont {Kesimer}}, \bibinfo {author} {\bibfnamefont {J.~K.}\ \bibnamefont {Sheehan}},\ and\ \bibinfo {author} {\bibfnamefont {C.~W.}\ \bibnamefont {Davis}},\ }\bibfield  {title} {\bibinfo {title} {Studying mucin secretion from human bronchial epithelial cell primary cultures},\ }\href@noop {} {\bibfield  {journal} {\bibinfo  {journal} {Methods Mol Biol}\ }\textbf {\bibinfo {volume} {842}},\ \bibinfo {pages} {259} (\bibinfo {year} {2012})}\BibitemShut {NoStop}%
\bibitem [{\citenamefont {Hill}\ \emph {et~al.}(2014)\citenamefont {Hill}, \citenamefont {Vasquez}, \citenamefont {Mellnik}, \citenamefont {McKinley}, \citenamefont {Vose}, \citenamefont {Mu}, \citenamefont {Henderson}, \citenamefont {Donaldson}, \citenamefont {Alexis}, \citenamefont {Boucher},\ and\ \citenamefont {Forest}}]{Hill_2014}%
  \BibitemOpen
  \bibfield  {author} {\bibinfo {author} {\bibfnamefont {D.~B.}\ \bibnamefont {Hill}}, \bibinfo {author} {\bibfnamefont {P.~A.}\ \bibnamefont {Vasquez}}, \bibinfo {author} {\bibfnamefont {J.}~\bibnamefont {Mellnik}}, \bibinfo {author} {\bibfnamefont {S.~A.}\ \bibnamefont {McKinley}}, \bibinfo {author} {\bibfnamefont {A.}~\bibnamefont {Vose}}, \bibinfo {author} {\bibfnamefont {F.}~\bibnamefont {Mu}}, \bibinfo {author} {\bibfnamefont {A.~G.}\ \bibnamefont {Henderson}}, \bibinfo {author} {\bibfnamefont {S.~H.}\ \bibnamefont {Donaldson}}, \bibinfo {author} {\bibfnamefont {N.~E.}\ \bibnamefont {Alexis}}, \bibinfo {author} {\bibfnamefont {R.~C.}\ \bibnamefont {Boucher}},\ and\ \bibinfo {author} {\bibfnamefont {M.~G.}\ \bibnamefont {Forest}},\ }\bibfield  {title} {\bibinfo {title} {A biophysical basis for mucus solids concentration as a candidate biomarker for airways disease},\ }\href@noop {} {\bibfield  {journal} {\bibinfo  {journal} {PLOS ONE}\ }\textbf {\bibinfo {volume} {9}},\ \bibinfo {pages} {e87681}
  (\bibinfo {year} {2014})}\BibitemShut {NoStop}%
\bibitem [{\citenamefont {Vasquez}\ \emph {et~al.}(2016)\citenamefont {Vasquez}, \citenamefont {Jin}, \citenamefont {Palmer}, \citenamefont {Hill},\ and\ \citenamefont {Forest}}]{Vasquez_2016}%
  \BibitemOpen
  \bibfield  {author} {\bibinfo {author} {\bibfnamefont {P.~A.}\ \bibnamefont {Vasquez}}, \bibinfo {author} {\bibfnamefont {Y.}~\bibnamefont {Jin}}, \bibinfo {author} {\bibfnamefont {E.}~\bibnamefont {Palmer}}, \bibinfo {author} {\bibfnamefont {D.}~\bibnamefont {Hill}},\ and\ \bibinfo {author} {\bibfnamefont {M.~G.}\ \bibnamefont {Forest}},\ }\bibfield  {title} {\bibinfo {title} {Modeling and simulation of mucus flow in human bronchial epithelial cell cultures - part {I}: Idealized axisymmetric swirling flow},\ }\href@noop {} {\bibfield  {journal} {\bibinfo  {journal} {Plos Computational Biology}\ }\textbf {\bibinfo {volume} {12}} (\bibinfo {year} {2016})}\BibitemShut {NoStop}%
\bibitem [{\citenamefont {Jory}\ \emph {et~al.}(2019)\citenamefont {Jory}, \citenamefont {Bellouma}, \citenamefont {Blanc}, \citenamefont {Casanellas}, \citenamefont {Petit}, \citenamefont {Reynaud}, \citenamefont {Vernisse}, \citenamefont {Vachier}, \citenamefont {Bourdin},\ and\ \citenamefont {Massiera}}]{Jory_2019}%
  \BibitemOpen
  \bibfield  {author} {\bibinfo {author} {\bibfnamefont {M.}~\bibnamefont {Jory}}, \bibinfo {author} {\bibfnamefont {K.}~\bibnamefont {Bellouma}}, \bibinfo {author} {\bibfnamefont {C.}~\bibnamefont {Blanc}}, \bibinfo {author} {\bibfnamefont {L.}~\bibnamefont {Casanellas}}, \bibinfo {author} {\bibfnamefont {A.}~\bibnamefont {Petit}}, \bibinfo {author} {\bibfnamefont {P.}~\bibnamefont {Reynaud}}, \bibinfo {author} {\bibfnamefont {C.}~\bibnamefont {Vernisse}}, \bibinfo {author} {\bibfnamefont {I.}~\bibnamefont {Vachier}}, \bibinfo {author} {\bibfnamefont {A.}~\bibnamefont {Bourdin}},\ and\ \bibinfo {author} {\bibfnamefont {G.}~\bibnamefont {Massiera}},\ }\bibfield  {title} {\bibinfo {title} {Mucus microrheology measured on human bronchial epithelium culture},\ }\href@noop {} {\bibfield  {journal} {\bibinfo  {journal} {Frontiers in Physics}\ }\textbf {\bibinfo {volume} {7}} (\bibinfo {year} {2019})}\BibitemShut {NoStop}%
\bibitem [{\citenamefont {Jory}\ \emph {et~al.}(2022)\citenamefont {Jory}, \citenamefont {Donnarumma}, \citenamefont {Blanc}, \citenamefont {Bellouma}, \citenamefont {Fort}, \citenamefont {Vachier}, \citenamefont {Casanellas}, \citenamefont {Bourdin},\ and\ \citenamefont {Massiera}}]{Jory_2022}%
  \BibitemOpen
  \bibfield  {author} {\bibinfo {author} {\bibfnamefont {M.}~\bibnamefont {Jory}}, \bibinfo {author} {\bibfnamefont {D.}~\bibnamefont {Donnarumma}}, \bibinfo {author} {\bibfnamefont {C.}~\bibnamefont {Blanc}}, \bibinfo {author} {\bibfnamefont {K.}~\bibnamefont {Bellouma}}, \bibinfo {author} {\bibfnamefont {A.}~\bibnamefont {Fort}}, \bibinfo {author} {\bibfnamefont {I.}~\bibnamefont {Vachier}}, \bibinfo {author} {\bibfnamefont {L.}~\bibnamefont {Casanellas}}, \bibinfo {author} {\bibfnamefont {A.}~\bibnamefont {Bourdin}},\ and\ \bibinfo {author} {\bibfnamefont {G.}~\bibnamefont {Massiera}},\ }\bibfield  {title} {\bibinfo {title} {Mucus from human bronchial epithelial cultures: rheology and adhesion across length scales},\ }\href@noop {} {\bibfield  {journal} {\bibinfo  {journal} {Interface Focus}\ }\textbf {\bibinfo {volume} {12}} (\bibinfo {year} {2022})}\BibitemShut {NoStop}%
\bibitem [{\citenamefont {Lin}\ \emph {et~al.}(2020)\citenamefont {Lin}, \citenamefont {Kaza}, \citenamefont {Birket}, \citenamefont {Kim}, \citenamefont {Edwards}, \citenamefont {LaFontaine}, \citenamefont {Liu}, \citenamefont {Mazur}, \citenamefont {Byzek}, \citenamefont {Hanes}, \citenamefont {Tearney}, \citenamefont {Raju},\ and\ \citenamefont {Rowe}}]{Lin_2020}%
  \BibitemOpen
  \bibfield  {author} {\bibinfo {author} {\bibfnamefont {V.~Y.}\ \bibnamefont {Lin}}, \bibinfo {author} {\bibfnamefont {N.}~\bibnamefont {Kaza}}, \bibinfo {author} {\bibfnamefont {S.~E.}\ \bibnamefont {Birket}}, \bibinfo {author} {\bibfnamefont {H.}~\bibnamefont {Kim}}, \bibinfo {author} {\bibfnamefont {L.~J.}\ \bibnamefont {Edwards}}, \bibinfo {author} {\bibfnamefont {J.}~\bibnamefont {LaFontaine}}, \bibinfo {author} {\bibfnamefont {L.}~\bibnamefont {Liu}}, \bibinfo {author} {\bibfnamefont {M.}~\bibnamefont {Mazur}}, \bibinfo {author} {\bibfnamefont {S.~A.}\ \bibnamefont {Byzek}}, \bibinfo {author} {\bibfnamefont {J.}~\bibnamefont {Hanes}}, \bibinfo {author} {\bibfnamefont {G.~J.}\ \bibnamefont {Tearney}}, \bibinfo {author} {\bibfnamefont {S.~V.}\ \bibnamefont {Raju}},\ and\ \bibinfo {author} {\bibfnamefont {S.~M.}\ \bibnamefont {Rowe}},\ }\bibfield  {title} {\bibinfo {title} {Excess mucus viscosity and airway dehydration impact copd airway clearance},\ }\href@noop {} {\bibfield  {journal} {\bibinfo
  {journal} {European Respiratory Journal}\ }\textbf {\bibinfo {volume} {55}},\ \bibinfo {pages} {1900419} (\bibinfo {year} {2020})}\BibitemShut {NoStop}%
\bibitem [{\citenamefont {Liegeois}\ \emph {et~al.}(2024)\citenamefont {Liegeois}, \citenamefont {Braunreuther}, \citenamefont {Charbit}, \citenamefont {Raymond}, \citenamefont {Tang}, \citenamefont {Woodruff}, \citenamefont {Christenson}, \citenamefont {Castro}, \citenamefont {Erzurum}, \citenamefont {Israel}, \citenamefont {Jarjour}, \citenamefont {Levy}, \citenamefont {Moore}, \citenamefont {Wenzel}, \citenamefont {Fuller},\ and\ \citenamefont {Fahy}}]{Liegeois_2024}%
  \BibitemOpen
  \bibfield  {author} {\bibinfo {author} {\bibfnamefont {M.~A.}\ \bibnamefont {Liegeois}}, \bibinfo {author} {\bibfnamefont {M.}~\bibnamefont {Braunreuther}}, \bibinfo {author} {\bibfnamefont {A.~R.}\ \bibnamefont {Charbit}}, \bibinfo {author} {\bibfnamefont {W.~W.}\ \bibnamefont {Raymond}}, \bibinfo {author} {\bibfnamefont {M.}~\bibnamefont {Tang}}, \bibinfo {author} {\bibfnamefont {P.~G.}\ \bibnamefont {Woodruff}}, \bibinfo {author} {\bibfnamefont {S.~A.}\ \bibnamefont {Christenson}}, \bibinfo {author} {\bibfnamefont {M.}~\bibnamefont {Castro}}, \bibinfo {author} {\bibfnamefont {S.~C.}\ \bibnamefont {Erzurum}}, \bibinfo {author} {\bibfnamefont {E.}~\bibnamefont {Israel}}, \bibinfo {author} {\bibfnamefont {N.~N.}\ \bibnamefont {Jarjour}}, \bibinfo {author} {\bibfnamefont {B.~D.}\ \bibnamefont {Levy}}, \bibinfo {author} {\bibfnamefont {W.~C.}\ \bibnamefont {Moore}}, \bibinfo {author} {\bibfnamefont {S.~E.}\ \bibnamefont {Wenzel}}, \bibinfo {author} {\bibfnamefont {G.~G.}\ \bibnamefont {Fuller}},\ and\
  \bibinfo {author} {\bibfnamefont {J.~V.}\ \bibnamefont {Fahy}},\ }\bibfield  {title} {\bibinfo {title} {Peroxidase-mediated mucin cross-linking drives pathologic mucus gel formation in il-13--stimulated airway epithelial cells},\ }\href@noop {} {\bibfield  {journal} {\bibinfo  {journal} {JCI Insight}\ }\textbf {\bibinfo {volume} {9}} (\bibinfo {year} {2024})}\BibitemShut {NoStop}%
\bibitem [{\citenamefont {Cai}\ \emph {et~al.}(2024)\citenamefont {Cai}, \citenamefont {Braunreuther}, \citenamefont {Shih}, \citenamefont {Spakowitz}, \citenamefont {Fuller},\ and\ \citenamefont {Heilshorn}}]{Cai_2024}%
  \BibitemOpen
  \bibfield  {author} {\bibinfo {author} {\bibfnamefont {P.~C.}\ \bibnamefont {Cai}}, \bibinfo {author} {\bibfnamefont {M.}~\bibnamefont {Braunreuther}}, \bibinfo {author} {\bibfnamefont {A.}~\bibnamefont {Shih}}, \bibinfo {author} {\bibfnamefont {A.~J.}\ \bibnamefont {Spakowitz}}, \bibinfo {author} {\bibfnamefont {G.~G.}\ \bibnamefont {Fuller}},\ and\ \bibinfo {author} {\bibfnamefont {S.~C.}\ \bibnamefont {Heilshorn}},\ }\bibfield  {title} {\bibinfo {title} {Air--liquid intestinal cell culture allows in situ rheological characterization of intestinal mucus},\ }\href@noop {} {\bibfield  {journal} {\bibinfo  {journal} {APL Bioengineering}\ }\textbf {\bibinfo {volume} {8}} (\bibinfo {year} {2024})}\BibitemShut {NoStop}%
\bibitem [{\citenamefont {Vasquez}\ \emph {et~al.}(2014)\citenamefont {Vasquez}, \citenamefont {Bowser}, \citenamefont {Swiderski}, \citenamefont {Walters},\ and\ \citenamefont {Kundu}}]{Vasquez_2014}%
  \BibitemOpen
  \bibfield  {author} {\bibinfo {author} {\bibfnamefont {E.~S.}\ \bibnamefont {Vasquez}}, \bibinfo {author} {\bibfnamefont {J.}~\bibnamefont {Bowser}}, \bibinfo {author} {\bibfnamefont {C.}~\bibnamefont {Swiderski}}, \bibinfo {author} {\bibfnamefont {K.~B.}\ \bibnamefont {Walters}},\ and\ \bibinfo {author} {\bibfnamefont {S.}~\bibnamefont {Kundu}},\ }\bibfield  {title} {\bibinfo {title} {Rheological characterization of mammalian lung mucus},\ }\href@noop {} {\bibfield  {journal} {\bibinfo  {journal} {RSC Advances}\ }\textbf {\bibinfo {volume} {4}},\ \bibinfo {pages} {34780} (\bibinfo {year} {2014})}\BibitemShut {NoStop}%
\bibitem [{\citenamefont {Murgia}\ \emph {et~al.}(2016)\citenamefont {Murgia}, \citenamefont {Pawelzyk}, \citenamefont {Schaefer}, \citenamefont {Wagner}, \citenamefont {Willenbacher},\ and\ \citenamefont {Lehr}}]{Murgia_2016}%
  \BibitemOpen
  \bibfield  {author} {\bibinfo {author} {\bibfnamefont {X.}~\bibnamefont {Murgia}}, \bibinfo {author} {\bibfnamefont {P.}~\bibnamefont {Pawelzyk}}, \bibinfo {author} {\bibfnamefont {U.~F.}\ \bibnamefont {Schaefer}}, \bibinfo {author} {\bibfnamefont {C.}~\bibnamefont {Wagner}}, \bibinfo {author} {\bibfnamefont {N.}~\bibnamefont {Willenbacher}},\ and\ \bibinfo {author} {\bibfnamefont {C.-M.}\ \bibnamefont {Lehr}},\ }\bibfield  {title} {\bibinfo {title} {Size-limited penetration of nanoparticles into porcine respiratory mucus after aerosol deposition},\ }\href@noop {} {\bibfield  {journal} {\bibinfo  {journal} {Biomacromolecules}\ }\textbf {\bibinfo {volume} {17}},\ \bibinfo {pages} {1536} (\bibinfo {year} {2016})}\BibitemShut {NoStop}%
\bibitem [{\citenamefont {Gross}\ \emph {et~al.}(2017)\citenamefont {Gross}, \citenamefont {Torge}, \citenamefont {Schaefer}, \citenamefont {Schneider}, \citenamefont {Lehr},\ and\ \citenamefont {Wagner}}]{Gross_2017}%
  \BibitemOpen
  \bibfield  {author} {\bibinfo {author} {\bibfnamefont {A.}~\bibnamefont {Gross}}, \bibinfo {author} {\bibfnamefont {A.}~\bibnamefont {Torge}}, \bibinfo {author} {\bibfnamefont {U.~F.}\ \bibnamefont {Schaefer}}, \bibinfo {author} {\bibfnamefont {M.}~\bibnamefont {Schneider}}, \bibinfo {author} {\bibfnamefont {C.-M.}\ \bibnamefont {Lehr}},\ and\ \bibinfo {author} {\bibfnamefont {C.}~\bibnamefont {Wagner}},\ }\bibfield  {title} {\bibinfo {title} {A foam model highlights the differences of the macro- and microrheology of respiratory horse mucus},\ }\href@noop {} {\bibfield  {journal} {\bibinfo  {journal} {Journal of the Mechanical Behavior of Biomedical Materials}\ }\textbf {\bibinfo {volume} {71}},\ \bibinfo {pages} {216} (\bibinfo {year} {2017})}\BibitemShut {NoStop}%
\bibitem [{\citenamefont {Schneider}\ \emph {et~al.}(2017)\citenamefont {Schneider}, \citenamefont {Xu}, \citenamefont {Boylan}, \citenamefont {Chisholm}, \citenamefont {Tang}, \citenamefont {Schuster}, \citenamefont {Henning}, \citenamefont {Ensign}, \citenamefont {Lee}, \citenamefont {Adstamongkonkul}, \citenamefont {Simons}, \citenamefont {Wang}, \citenamefont {Gong}, \citenamefont {Yu}, \citenamefont {Boyle}, \citenamefont {Suk},\ and\ \citenamefont {Hanes}}]{Schneider_2017}%
  \BibitemOpen
  \bibfield  {author} {\bibinfo {author} {\bibfnamefont {C.~S.}\ \bibnamefont {Schneider}}, \bibinfo {author} {\bibfnamefont {Q.}~\bibnamefont {Xu}}, \bibinfo {author} {\bibfnamefont {N.~J.}\ \bibnamefont {Boylan}}, \bibinfo {author} {\bibfnamefont {J.}~\bibnamefont {Chisholm}}, \bibinfo {author} {\bibfnamefont {B.~C.}\ \bibnamefont {Tang}}, \bibinfo {author} {\bibfnamefont {B.~S.}\ \bibnamefont {Schuster}}, \bibinfo {author} {\bibfnamefont {A.}~\bibnamefont {Henning}}, \bibinfo {author} {\bibfnamefont {L.~M.}\ \bibnamefont {Ensign}}, \bibinfo {author} {\bibfnamefont {E.}~\bibnamefont {Lee}}, \bibinfo {author} {\bibfnamefont {P.}~\bibnamefont {Adstamongkonkul}}, \bibinfo {author} {\bibfnamefont {B.~W.}\ \bibnamefont {Simons}}, \bibinfo {author} {\bibfnamefont {S.-Y.~S.}\ \bibnamefont {Wang}}, \bibinfo {author} {\bibfnamefont {X.}~\bibnamefont {Gong}}, \bibinfo {author} {\bibfnamefont {T.}~\bibnamefont {Yu}}, \bibinfo {author} {\bibfnamefont {M.~P.}\ \bibnamefont {Boyle}}, \bibinfo {author} {\bibfnamefont
  {J.~S.}\ \bibnamefont {Suk}},\ and\ \bibinfo {author} {\bibfnamefont {J.}~\bibnamefont {Hanes}},\ }\bibfield  {title} {\bibinfo {title} {Nanoparticles that do not adhere to mucus provide uniform and long-lasting drug delivery to airways following inhalation},\ }\href@noop {} {\bibfield  {journal} {\bibinfo  {journal} {Science Advances}\ }\textbf {\bibinfo {volume} {3}},\ \bibinfo {pages} {e1601556} (\bibinfo {year} {2017})}\BibitemShut {NoStop}%
\bibitem [{\citenamefont {Morgan}\ \emph {et~al.}(2021)\citenamefont {Morgan}, \citenamefont {Jaramillo}, \citenamefont {Shenoy}, \citenamefont {Raclawska}, \citenamefont {Emezienna}, \citenamefont {Richardson}, \citenamefont {Hara}, \citenamefont {Harder}, \citenamefont {NeeDell}, \citenamefont {Hennessy}, \citenamefont {El-Batal}, \citenamefont {Magin}, \citenamefont {Grove~Villalon}, \citenamefont {Duncan}, \citenamefont {Hanes}, \citenamefont {Suk}, \citenamefont {Thornton}, \citenamefont {Holguin}, \citenamefont {Janssen}, \citenamefont {Thelin},\ and\ \citenamefont {Evans}}]{Morgan_2021}%
  \BibitemOpen
  \bibfield  {author} {\bibinfo {author} {\bibfnamefont {L.~E.}\ \bibnamefont {Morgan}}, \bibinfo {author} {\bibfnamefont {A.~M.}\ \bibnamefont {Jaramillo}}, \bibinfo {author} {\bibfnamefont {S.~K.}\ \bibnamefont {Shenoy}}, \bibinfo {author} {\bibfnamefont {D.}~\bibnamefont {Raclawska}}, \bibinfo {author} {\bibfnamefont {N.~A.}\ \bibnamefont {Emezienna}}, \bibinfo {author} {\bibfnamefont {V.~L.}\ \bibnamefont {Richardson}}, \bibinfo {author} {\bibfnamefont {N.}~\bibnamefont {Hara}}, \bibinfo {author} {\bibfnamefont {A.~Q.}\ \bibnamefont {Harder}}, \bibinfo {author} {\bibfnamefont {J.~C.}\ \bibnamefont {NeeDell}}, \bibinfo {author} {\bibfnamefont {C.~E.}\ \bibnamefont {Hennessy}}, \bibinfo {author} {\bibfnamefont {H.~M.}\ \bibnamefont {El-Batal}}, \bibinfo {author} {\bibfnamefont {C.~M.}\ \bibnamefont {Magin}}, \bibinfo {author} {\bibfnamefont {D.~E.}\ \bibnamefont {Grove~Villalon}}, \bibinfo {author} {\bibfnamefont {G.}~\bibnamefont {Duncan}}, \bibinfo {author} {\bibfnamefont {J.~S.}\ \bibnamefont {Hanes}},
  \bibinfo {author} {\bibfnamefont {J.~S.}\ \bibnamefont {Suk}}, \bibinfo {author} {\bibfnamefont {D.~J.}\ \bibnamefont {Thornton}}, \bibinfo {author} {\bibfnamefont {F.}~\bibnamefont {Holguin}}, \bibinfo {author} {\bibfnamefont {W.~J.}\ \bibnamefont {Janssen}}, \bibinfo {author} {\bibfnamefont {W.~R.}\ \bibnamefont {Thelin}},\ and\ \bibinfo {author} {\bibfnamefont {C.~M.}\ \bibnamefont {Evans}},\ }\bibfield  {title} {\bibinfo {title} {Disulfide disruption reverses mucus dysfunction in allergic airway disease},\ }\href@noop {} {\bibfield  {journal} {\bibinfo  {journal} {Nature Communications}\ }\textbf {\bibinfo {volume} {12}},\ \bibinfo {pages} {249} (\bibinfo {year} {2021})}\BibitemShut {NoStop}%
\bibitem [{\citenamefont {Tomaiuolo}\ \emph {et~al.}(2014)\citenamefont {Tomaiuolo}, \citenamefont {Rusciano}, \citenamefont {Caserta}, \citenamefont {Carciati}, \citenamefont {Carnovale}, \citenamefont {Abete}, \citenamefont {Sasso},\ and\ \citenamefont {Guido}}]{Tomaiuolo_2014}%
  \BibitemOpen
  \bibfield  {author} {\bibinfo {author} {\bibfnamefont {G.}~\bibnamefont {Tomaiuolo}}, \bibinfo {author} {\bibfnamefont {G.}~\bibnamefont {Rusciano}}, \bibinfo {author} {\bibfnamefont {S.}~\bibnamefont {Caserta}}, \bibinfo {author} {\bibfnamefont {A.}~\bibnamefont {Carciati}}, \bibinfo {author} {\bibfnamefont {V.}~\bibnamefont {Carnovale}}, \bibinfo {author} {\bibfnamefont {P.}~\bibnamefont {Abete}}, \bibinfo {author} {\bibfnamefont {A.}~\bibnamefont {Sasso}},\ and\ \bibinfo {author} {\bibfnamefont {S.}~\bibnamefont {Guido}},\ }\bibfield  {title} {\bibinfo {title} {A new method to improve the clinical evaluation of cystic fibrosis patients by mucus viscoelastic properties},\ }\href@noop {} {\bibfield  {journal} {\bibinfo  {journal} {PLOS ONE}\ }\textbf {\bibinfo {volume} {9}},\ \bibinfo {pages} {e82297} (\bibinfo {year} {2014})}\BibitemShut {NoStop}%
\bibitem [{\citenamefont {King}\ and\ \citenamefont {Macklem}(1977)}]{King_1977}%
  \BibitemOpen
  \bibfield  {author} {\bibinfo {author} {\bibfnamefont {M.}~\bibnamefont {King}}\ and\ \bibinfo {author} {\bibfnamefont {P.~T.}\ \bibnamefont {Macklem}},\ }\bibfield  {title} {\bibinfo {title} {Rheological properties of microliter quantities of normal mucus},\ }\href@noop {} {\bibfield  {journal} {\bibinfo  {journal} {Journal of Applied Physiology}\ }\textbf {\bibinfo {volume} {42}},\ \bibinfo {pages} {797} (\bibinfo {year} {1977})}\BibitemShut {NoStop}%
\bibitem [{\citenamefont {King}\ \emph {et~al.}(1989)\citenamefont {King}, \citenamefont {Zahm}, \citenamefont {Pierrot}, \citenamefont {Vaquez-Girod},\ and\ \citenamefont {Puchelle}}]{King_1989}%
  \BibitemOpen
  \bibfield  {author} {\bibinfo {author} {\bibfnamefont {M.}~\bibnamefont {King}}, \bibinfo {author} {\bibfnamefont {J.~M.}\ \bibnamefont {Zahm}}, \bibinfo {author} {\bibfnamefont {D.}~\bibnamefont {Pierrot}}, \bibinfo {author} {\bibfnamefont {S.}~\bibnamefont {Vaquez-Girod}},\ and\ \bibinfo {author} {\bibfnamefont {E.}~\bibnamefont {Puchelle}},\ }\bibfield  {title} {\bibinfo {title} {The role of mucus gel viscosity, spinnability, and adhesive properties in clearance by simulated cough},\ }\href@noop {} {\bibfield  {journal} {\bibinfo  {journal} {Biorheology}\ }\textbf {\bibinfo {volume} {26}},\ \bibinfo {pages} {737} (\bibinfo {year} {1989})}\BibitemShut {NoStop}%
\bibitem [{\citenamefont {Khan}\ \emph {et~al.}(1976)\citenamefont {Khan}, \citenamefont {Wolf},\ and\ \citenamefont {Litt}}]{Khan_1976}%
  \BibitemOpen
  \bibfield  {author} {\bibinfo {author} {\bibfnamefont {M.~A.}\ \bibnamefont {Khan}}, \bibinfo {author} {\bibfnamefont {D.~P.}\ \bibnamefont {Wolf}},\ and\ \bibinfo {author} {\bibfnamefont {M.}~\bibnamefont {Litt}},\ }\bibfield  {title} {\bibinfo {title} {Effect of mucolytic agents on the rheological properties of tracheal mucos},\ }\href@noop {} {\bibfield  {journal} {\bibinfo  {journal} {Biochimica et Biophysica Acta (BBA) - General Subjects}\ }\textbf {\bibinfo {volume} {444}},\ \bibinfo {pages} {369} (\bibinfo {year} {1976})}\BibitemShut {NoStop}%
\bibitem [{\citenamefont {King}\ \emph {et~al.}(1997)\citenamefont {King}, \citenamefont {DASGUPTA}, \citenamefont {TOMKIEWICZ},\ and\ \citenamefont {BROWN}}]{King_1997}%
  \BibitemOpen
  \bibfield  {author} {\bibinfo {author} {\bibfnamefont {M.}~\bibnamefont {King}}, \bibinfo {author} {\bibfnamefont {B.}~\bibnamefont {DASGUPTA}}, \bibinfo {author} {\bibfnamefont {R.~P.}\ \bibnamefont {TOMKIEWICZ}},\ and\ \bibinfo {author} {\bibfnamefont {N.~E.}\ \bibnamefont {BROWN}},\ }\bibfield  {title} {\bibinfo {title} {Rheology of cystic fibrosis sputum after in vitro treatment with hypertonic saline alone and in combination with recombinant human deoxyribonuclease i},\ }\href@noop {} {\bibfield  {journal} {\bibinfo  {journal} {American Journal of Respiratory and Critical Care Medicine}\ }\textbf {\bibinfo {volume} {156}},\ \bibinfo {pages} {173} (\bibinfo {year} {1997})}\BibitemShut {NoStop}%
\bibitem [{\citenamefont {Yang}\ \emph {et~al.}(2011)\citenamefont {Yang}, \citenamefont {Tsifansky}, \citenamefont {Shin}, \citenamefont {Lin},\ and\ \citenamefont {Yeo}}]{Yang_2011}%
  \BibitemOpen
  \bibfield  {author} {\bibinfo {author} {\bibfnamefont {Y.}~\bibnamefont {Yang}}, \bibinfo {author} {\bibfnamefont {M.~D.}\ \bibnamefont {Tsifansky}}, \bibinfo {author} {\bibfnamefont {S.}~\bibnamefont {Shin}}, \bibinfo {author} {\bibfnamefont {Q.}~\bibnamefont {Lin}},\ and\ \bibinfo {author} {\bibfnamefont {Y.}~\bibnamefont {Yeo}},\ }\bibfield  {title} {\bibinfo {title} {Mannitol-guided delivery of ciprofloxacin in artificial cystic fibrosis mucus model},\ }\href@noop {} {\bibfield  {journal} {\bibinfo  {journal} {Biotechnology and Bioengineering}\ }\textbf {\bibinfo {volume} {108}},\ \bibinfo {pages} {1441} (\bibinfo {year} {2011})}\BibitemShut {NoStop}%
\bibitem [{\citenamefont {Craparo}\ \emph {et~al.}(2016)\citenamefont {Craparo}, \citenamefont {Porsio}, \citenamefont {Sardo}, \citenamefont {Giammona},\ and\ \citenamefont {Cavallaro}}]{Craparo_2016}%
  \BibitemOpen
  \bibfield  {author} {\bibinfo {author} {\bibfnamefont {E.~F.}\ \bibnamefont {Craparo}}, \bibinfo {author} {\bibfnamefont {B.}~\bibnamefont {Porsio}}, \bibinfo {author} {\bibfnamefont {C.}~\bibnamefont {Sardo}}, \bibinfo {author} {\bibfnamefont {G.}~\bibnamefont {Giammona}},\ and\ \bibinfo {author} {\bibfnamefont {G.}~\bibnamefont {Cavallaro}},\ }\bibfield  {title} {\bibinfo {title} {Pegylated polyaspartamide--polylactide-based nanoparticles penetrating cystic fibrosis artificial mucus},\ }\href@noop {} {\bibfield  {journal} {\bibinfo  {journal} {Biomacromolecules}\ }\textbf {\bibinfo {volume} {17}},\ \bibinfo {pages} {767} (\bibinfo {year} {2016})}\BibitemShut {NoStop}%
\bibitem [{\citenamefont {Lafforgue}\ \emph {et~al.}(2017{\natexlab{a}})\citenamefont {Lafforgue}, \citenamefont {Poncet}, \citenamefont {Seyssiecq},\ and\ \citenamefont {Favier}}]{Lafforgue_2017}%
  \BibitemOpen
  \bibfield  {author} {\bibinfo {author} {\bibfnamefont {O.}~\bibnamefont {Lafforgue}}, \bibinfo {author} {\bibfnamefont {S.}~\bibnamefont {Poncet}}, \bibinfo {author} {\bibfnamefont {I.}~\bibnamefont {Seyssiecq}},\ and\ \bibinfo {author} {\bibfnamefont {J.}~\bibnamefont {Favier}},\ }\bibfield  {title} {\bibinfo {title} {Rheological characterization of macromolecular colloidal gels as simulant of bronchial mucus},\ }\href@noop {} {\bibfield  {journal} {\bibinfo  {journal} {AIP Conference Proceedings}\ }\textbf {\bibinfo {volume} {1914}} (\bibinfo {year} {2017}{\natexlab{a}})}\BibitemShut {NoStop}%
\bibitem [{\citenamefont {Huck}\ \emph {et~al.}(2019)\citenamefont {Huck}, \citenamefont {Hartwig}, \citenamefont {Biehl}, \citenamefont {Schwarzkopf}, \citenamefont {Wagner}, \citenamefont {Loretz}, \citenamefont {Murgia},\ and\ \citenamefont {Lehr}}]{Huck_2019}%
  \BibitemOpen
  \bibfield  {author} {\bibinfo {author} {\bibfnamefont {B.~C.}\ \bibnamefont {Huck}}, \bibinfo {author} {\bibfnamefont {O.}~\bibnamefont {Hartwig}}, \bibinfo {author} {\bibfnamefont {A.}~\bibnamefont {Biehl}}, \bibinfo {author} {\bibfnamefont {K.}~\bibnamefont {Schwarzkopf}}, \bibinfo {author} {\bibfnamefont {C.}~\bibnamefont {Wagner}}, \bibinfo {author} {\bibfnamefont {B.}~\bibnamefont {Loretz}}, \bibinfo {author} {\bibfnamefont {X.}~\bibnamefont {Murgia}},\ and\ \bibinfo {author} {\bibfnamefont {C.-M.}\ \bibnamefont {Lehr}},\ }\bibfield  {title} {\bibinfo {title} {Macro- and microrheological properties of mucus surrogates in comparison to native intestinal and pulmonary mucus},\ }\href@noop {} {\bibfield  {journal} {\bibinfo  {journal} {Biomacromolecules}\ }\textbf {\bibinfo {volume} {20}},\ \bibinfo {pages} {3504} (\bibinfo {year} {2019})}\BibitemShut {NoStop}%
\bibitem [{\citenamefont {Carpenter}\ \emph {et~al.}(2021)\citenamefont {Carpenter}, \citenamefont {Wang}, \citenamefont {Gupta}, \citenamefont {Li}, \citenamefont {Haridass}, \citenamefont {Subramani}, \citenamefont {Reidel}, \citenamefont {Morton}, \citenamefont {Ridley}, \citenamefont {O'Neal}, \citenamefont {Buisine}, \citenamefont {Ehre}, \citenamefont {Thornton},\ and\ \citenamefont {Kesimer}}]{Carpenter_2021}%
  \BibitemOpen
  \bibfield  {author} {\bibinfo {author} {\bibfnamefont {J.}~\bibnamefont {Carpenter}}, \bibinfo {author} {\bibfnamefont {Y.}~\bibnamefont {Wang}}, \bibinfo {author} {\bibfnamefont {R.}~\bibnamefont {Gupta}}, \bibinfo {author} {\bibfnamefont {Y.}~\bibnamefont {Li}}, \bibinfo {author} {\bibfnamefont {P.}~\bibnamefont {Haridass}}, \bibinfo {author} {\bibfnamefont {D.~B.}\ \bibnamefont {Subramani}}, \bibinfo {author} {\bibfnamefont {B.}~\bibnamefont {Reidel}}, \bibinfo {author} {\bibfnamefont {L.}~\bibnamefont {Morton}}, \bibinfo {author} {\bibfnamefont {C.}~\bibnamefont {Ridley}}, \bibinfo {author} {\bibfnamefont {W.~K.}\ \bibnamefont {O'Neal}}, \bibinfo {author} {\bibfnamefont {M.-P.}\ \bibnamefont {Buisine}}, \bibinfo {author} {\bibfnamefont {C.}~\bibnamefont {Ehre}}, \bibinfo {author} {\bibfnamefont {D.~J.}\ \bibnamefont {Thornton}},\ and\ \bibinfo {author} {\bibfnamefont {M.}~\bibnamefont {Kesimer}},\ }\bibfield  {title} {\bibinfo {title} {Assembly and organization of the n-terminal region of mucin muc5ac:
  Indications for structural and functional distinction from muc5b},\ }\href@noop {} {\bibfield  {journal} {\bibinfo  {journal} {Proceedings of the National Academy of Sciences}\ }\textbf {\bibinfo {volume} {118}},\ \bibinfo {pages} {e2104490118} (\bibinfo {year} {2021})}\BibitemShut {NoStop}%
\bibitem [{\citenamefont {Tan}\ \emph {et~al.}(2020)\citenamefont {Tan}, \citenamefont {Mao},\ and\ \citenamefont {Walker}}]{Tan_2020}%
  \BibitemOpen
  \bibfield  {author} {\bibinfo {author} {\bibfnamefont {M.}~\bibnamefont {Tan}}, \bibinfo {author} {\bibfnamefont {Y.}~\bibnamefont {Mao}},\ and\ \bibinfo {author} {\bibfnamefont {T.~W.}\ \bibnamefont {Walker}},\ }\bibfield  {title} {\bibinfo {title} {Rheological enhancement of artificial sputum medium},\ }\href@noop {} {\bibfield  {journal} {\bibinfo  {journal} {Applied Rheology}\ }\textbf {\bibinfo {volume} {30}},\ \bibinfo {pages} {27} (\bibinfo {year} {2020})}\BibitemShut {NoStop}%
\bibitem [{\citenamefont {Hamed}\ and\ \citenamefont {Fiegel}(2014)}]{Hamed_2014}%
  \BibitemOpen
  \bibfield  {author} {\bibinfo {author} {\bibfnamefont {R.}~\bibnamefont {Hamed}}\ and\ \bibinfo {author} {\bibfnamefont {J.}~\bibnamefont {Fiegel}},\ }\bibfield  {title} {\bibinfo {title} {Synthetic tracheal mucus with native rheological and surface tension properties},\ }\href {https://doi.org/https://doi.org/10.1002/jbm.a.34851} {\bibfield  {journal} {\bibinfo  {journal} {Journal of Biomedical Materials Research Part A}\ }\textbf {\bibinfo {volume} {102}},\ \bibinfo {pages} {1788} (\bibinfo {year} {2014})}\BibitemShut {NoStop}%
\bibitem [{\citenamefont {Sharma}\ \emph {et~al.}(2021)\citenamefont {Sharma}, \citenamefont {Thongrom}, \citenamefont {Bhatia}, \citenamefont {von Lospichl}, \citenamefont {Addante}, \citenamefont {Graeber}, \citenamefont {Lauster}, \citenamefont {Mall}, \citenamefont {Gradzielski},\ and\ \citenamefont {Haag}}]{Sharma_2021}%
  \BibitemOpen
  \bibfield  {author} {\bibinfo {author} {\bibfnamefont {A.}~\bibnamefont {Sharma}}, \bibinfo {author} {\bibfnamefont {B.}~\bibnamefont {Thongrom}}, \bibinfo {author} {\bibfnamefont {S.}~\bibnamefont {Bhatia}}, \bibinfo {author} {\bibfnamefont {B.}~\bibnamefont {von Lospichl}}, \bibinfo {author} {\bibfnamefont {A.}~\bibnamefont {Addante}}, \bibinfo {author} {\bibfnamefont {S.~Y.}\ \bibnamefont {Graeber}}, \bibinfo {author} {\bibfnamefont {D.}~\bibnamefont {Lauster}}, \bibinfo {author} {\bibfnamefont {M.~A.}\ \bibnamefont {Mall}}, \bibinfo {author} {\bibfnamefont {M.}~\bibnamefont {Gradzielski}},\ and\ \bibinfo {author} {\bibfnamefont {R.}~\bibnamefont {Haag}},\ }\bibfield  {title} {\bibinfo {title} {Polyglycerol-based mucus-inspired hydrogels},\ }\href@noop {} {\bibfield  {journal} {\bibinfo  {journal} {Macromolecular Rapid Communications}\ }\textbf {\bibinfo {volume} {42}},\ \bibinfo {pages} {2100303} (\bibinfo {year} {2021})}\BibitemShut {NoStop}%
\bibitem [{\citenamefont {Wagner}\ \emph {et~al.}(2023)\citenamefont {Wagner}, \citenamefont {Krupkin}, \citenamefont {Smith-Dupont}, \citenamefont {Wu}, \citenamefont {Bustos}, \citenamefont {Witten},\ and\ \citenamefont {Ribbeck}}]{Wagner_2023}%
  \BibitemOpen
  \bibfield  {author} {\bibinfo {author} {\bibfnamefont {C.~E.}\ \bibnamefont {Wagner}}, \bibinfo {author} {\bibfnamefont {M.}~\bibnamefont {Krupkin}}, \bibinfo {author} {\bibfnamefont {K.~B.}\ \bibnamefont {Smith-Dupont}}, \bibinfo {author} {\bibfnamefont {C.~M.}\ \bibnamefont {Wu}}, \bibinfo {author} {\bibfnamefont {N.~A.}\ \bibnamefont {Bustos}}, \bibinfo {author} {\bibfnamefont {J.}~\bibnamefont {Witten}},\ and\ \bibinfo {author} {\bibfnamefont {K.}~\bibnamefont {Ribbeck}},\ }\bibfield  {title} {\bibinfo {title} {Comparison of physicochemical properties of native mucus and reconstituted mucin gels},\ }\href@noop {} {\bibfield  {journal} {\bibinfo  {journal} {Biomacromolecules}\ }\textbf {\bibinfo {volume} {24}},\ \bibinfo {pages} {628} (\bibinfo {year} {2023})}\BibitemShut {NoStop}%
\bibitem [{\citenamefont {Liu}\ \emph {et~al.}(2024)\citenamefont {Liu}, \citenamefont {Seto}, \citenamefont {Zhang}, \citenamefont {Che}, \citenamefont {Liu},\ and\ \citenamefont {Deng}}]{Liu_2024}%
  \BibitemOpen
  \bibfield  {author} {\bibinfo {author} {\bibfnamefont {Z.}~\bibnamefont {Liu}}, \bibinfo {author} {\bibfnamefont {R.}~\bibnamefont {Seto}}, \bibinfo {author} {\bibfnamefont {H.}~\bibnamefont {Zhang}}, \bibinfo {author} {\bibfnamefont {B.}~\bibnamefont {Che}}, \bibinfo {author} {\bibfnamefont {L.}~\bibnamefont {Liu}},\ and\ \bibinfo {author} {\bibfnamefont {L.}~\bibnamefont {Deng}},\ }\bibfield  {title} {\bibinfo {title} {Highly distinctive linear and nonlinear rheological behaviors of mucin-based protein solutions as simulated normal and asthmatic human airway mucus},\ }\href@noop {} {\bibfield  {journal} {\bibinfo  {journal} {Phys. Fluids}\ }\textbf {\bibinfo {volume} {36}},\ \bibinfo {pages} {043108} (\bibinfo {year} {2024})}\BibitemShut {NoStop}%
\bibitem [{\citenamefont {Degen}\ \emph {et~al.}(2025)\citenamefont {Degen}, \citenamefont {Stevens}, \citenamefont {Cárcamo-Oyarce}, \citenamefont {Song}, \citenamefont {Bej}, \citenamefont {Tang}, \citenamefont {Ribbeck}, \citenamefont {Haag},\ and\ \citenamefont {McKinley}}]{Degen_2025}%
  \BibitemOpen
  \bibfield  {author} {\bibinfo {author} {\bibfnamefont {G.~D.}\ \bibnamefont {Degen}}, \bibinfo {author} {\bibfnamefont {C.~A.}\ \bibnamefont {Stevens}}, \bibinfo {author} {\bibfnamefont {G.}~\bibnamefont {Cárcamo-Oyarce}}, \bibinfo {author} {\bibfnamefont {J.}~\bibnamefont {Song}}, \bibinfo {author} {\bibfnamefont {R.}~\bibnamefont {Bej}}, \bibinfo {author} {\bibfnamefont {P.}~\bibnamefont {Tang}}, \bibinfo {author} {\bibfnamefont {K.}~\bibnamefont {Ribbeck}}, \bibinfo {author} {\bibfnamefont {R.}~\bibnamefont {Haag}},\ and\ \bibinfo {author} {\bibfnamefont {G.~H.}\ \bibnamefont {McKinley}},\ }\bibfield  {title} {\bibinfo {title} {Mussel-inspired cross-linking mechanisms enhance gelation and adhesion of multifunctional mucin-derived hydrogels},\ }\href {https://doi.org/10.1073/pnas.2415927122} {\bibfield  {journal} {\bibinfo  {journal} {Proceedings of the National Academy of Sciences}\ }\textbf {\bibinfo {volume} {122}},\ \bibinfo {pages} {e2415927122} (\bibinfo {year} {2025})},\ \Eprint
  {https://arxiv.org/abs/https://www.pnas.org/doi/pdf/10.1073/pnas.2415927122} {https://www.pnas.org/doi/pdf/10.1073/pnas.2415927122} \BibitemShut {NoStop}%
\bibitem [{\citenamefont {Schuster}\ \emph {et~al.}(2017)\citenamefont {Schuster}, \citenamefont {Allan}, \citenamefont {Kays}, \citenamefont {Hanes},\ and\ \citenamefont {Leheny}}]{Schuster_2017}%
  \BibitemOpen
  \bibfield  {author} {\bibinfo {author} {\bibfnamefont {B.~S.}\ \bibnamefont {Schuster}}, \bibinfo {author} {\bibfnamefont {D.~B.}\ \bibnamefont {Allan}}, \bibinfo {author} {\bibfnamefont {J.~C.}\ \bibnamefont {Kays}}, \bibinfo {author} {\bibfnamefont {J.}~\bibnamefont {Hanes}},\ and\ \bibinfo {author} {\bibfnamefont {R.~L.}\ \bibnamefont {Leheny}},\ }\bibfield  {title} {\bibinfo {title} {Photoactivatable fluorescent probes reveal heterogeneous nanoparticle permeation through biological gels at multiple scales},\ }\href@noop {} {\bibfield  {journal} {\bibinfo  {journal} {Journal of Controlled Release}\ }\textbf {\bibinfo {volume} {260}},\ \bibinfo {pages} {124} (\bibinfo {year} {2017})}\BibitemShut {NoStop}%
\bibitem [{\citenamefont {Kruger}\ \emph {et~al.}(2021)\citenamefont {Kruger}, \citenamefont {Brucks}, \citenamefont {Yan}, \citenamefont {C{\'a}rcarmo-Oyarce}, \citenamefont {Wei}, \citenamefont {Wen}, \citenamefont {Carvalho}, \citenamefont {Hore}, \citenamefont {Ribbeck}, \citenamefont {Schrock},\ and\ \citenamefont {Kiessling}}]{Kruger_2021}%
  \BibitemOpen
  \bibfield  {author} {\bibinfo {author} {\bibfnamefont {A.~G.}\ \bibnamefont {Kruger}}, \bibinfo {author} {\bibfnamefont {S.~D.}\ \bibnamefont {Brucks}}, \bibinfo {author} {\bibfnamefont {T.}~\bibnamefont {Yan}}, \bibinfo {author} {\bibfnamefont {G.}~\bibnamefont {C{\'a}rcarmo-Oyarce}}, \bibinfo {author} {\bibfnamefont {Y.}~\bibnamefont {Wei}}, \bibinfo {author} {\bibfnamefont {D.~H.}\ \bibnamefont {Wen}}, \bibinfo {author} {\bibfnamefont {D.~R.}\ \bibnamefont {Carvalho}}, \bibinfo {author} {\bibfnamefont {M.~J.~A.}\ \bibnamefont {Hore}}, \bibinfo {author} {\bibfnamefont {K.}~\bibnamefont {Ribbeck}}, \bibinfo {author} {\bibfnamefont {R.~R.}\ \bibnamefont {Schrock}},\ and\ \bibinfo {author} {\bibfnamefont {L.~L.}\ \bibnamefont {Kiessling}},\ }\bibfield  {title} {\bibinfo {title} {Stereochemical control yields mucin mimetic polymers},\ }\href@noop {} {\bibfield  {journal} {\bibinfo  {journal} {ACS Central Science}\ }\textbf {\bibinfo {volume} {7}},\ \bibinfo {pages} {624} (\bibinfo {year} {2021})}\BibitemShut
  {NoStop}%
\bibitem [{\citenamefont {Milian}\ \emph {et~al.}(2024)\citenamefont {Milian}, \citenamefont {Robert~de Saint~Vincent}, \citenamefont {Patarin},\ and\ \citenamefont {Bodiguel}}]{Milian_2024}%
  \BibitemOpen
  \bibfield  {author} {\bibinfo {author} {\bibfnamefont {D.}~\bibnamefont {Milian}}, \bibinfo {author} {\bibfnamefont {M.}~\bibnamefont {Robert~de Saint~Vincent}}, \bibinfo {author} {\bibfnamefont {J.}~\bibnamefont {Patarin}},\ and\ \bibinfo {author} {\bibfnamefont {H.}~\bibnamefont {Bodiguel}},\ }\bibfield  {title} {\bibinfo {title} {Gastropod slime-based gel as an adjustable synthetic model for human airway mucus},\ }\href@noop {} {\bibfield  {journal} {\bibinfo  {journal} {Biomacromolecules}\ }\textbf {\bibinfo {volume} {25}},\ \bibinfo {pages} {400} (\bibinfo {year} {2024})}\BibitemShut {NoStop}%
\bibitem [{\citenamefont {Wagner}\ \emph {et~al.}(2018)\citenamefont {Wagner}, \citenamefont {Wheeler},\ and\ \citenamefont {Ribbeck}}]{Wagner_2018}%
  \BibitemOpen
  \bibfield  {author} {\bibinfo {author} {\bibfnamefont {C.}~\bibnamefont {Wagner}}, \bibinfo {author} {\bibfnamefont {K.}~\bibnamefont {Wheeler}},\ and\ \bibinfo {author} {\bibfnamefont {K.}~\bibnamefont {Ribbeck}},\ }\bibfield  {title} {\bibinfo {title} {Mucins and their role in shaping the functions of mucus barriers},\ }\href@noop {} {\bibfield  {journal} {\bibinfo  {journal} {Annual Review of Cell and Developmental Biology}\ }\textbf {\bibinfo {volume} {34}},\ \bibinfo {pages} {189} (\bibinfo {year} {2018})}\BibitemShut {NoStop}%
\bibitem [{\citenamefont {Murgia}\ \emph {et~al.}(2018)\citenamefont {Murgia}, \citenamefont {Loretz}, \citenamefont {Hartwig}, \citenamefont {Hittinger},\ and\ \citenamefont {Lehr}}]{Murgia_2018}%
  \BibitemOpen
  \bibfield  {author} {\bibinfo {author} {\bibfnamefont {X.}~\bibnamefont {Murgia}}, \bibinfo {author} {\bibfnamefont {B.}~\bibnamefont {Loretz}}, \bibinfo {author} {\bibfnamefont {O.}~\bibnamefont {Hartwig}}, \bibinfo {author} {\bibfnamefont {M.}~\bibnamefont {Hittinger}},\ and\ \bibinfo {author} {\bibfnamefont {C.-M.}\ \bibnamefont {Lehr}},\ }\bibfield  {title} {\bibinfo {title} {The role of mucus on drug transport and its potential to affect therapeutic outcomes},\ }\href@noop {} {\bibfield  {journal} {\bibinfo  {journal} {Advanced Drug Delivery Reviews}\ }\textbf {\bibinfo {volume} {124}},\ \bibinfo {pages} {82} (\bibinfo {year} {2018})}\BibitemShut {NoStop}%
\bibitem [{\citenamefont {Huckaby}\ and\ \citenamefont {Lai}(2018)}]{Huckaby_2018}%
  \BibitemOpen
  \bibfield  {author} {\bibinfo {author} {\bibfnamefont {J.~T.}\ \bibnamefont {Huckaby}}\ and\ \bibinfo {author} {\bibfnamefont {S.~K.}\ \bibnamefont {Lai}},\ }\bibfield  {title} {\bibinfo {title} {Pegylation for enhancing nanoparticle diffusion in mucus},\ }\href@noop {} {\bibfield  {journal} {\bibinfo  {journal} {Advanced Drug Delivery Reviews}\ }\textbf {\bibinfo {volume} {124}},\ \bibinfo {pages} {125} (\bibinfo {year} {2018})}\BibitemShut {NoStop}%
\bibitem [{\citenamefont {Ma}\ \emph {et~al.}(2018)\citenamefont {Ma}, \citenamefont {Tang}, \citenamefont {Kang}, \citenamefont {Voynow},\ and\ \citenamefont {Rubin}}]{Ma_2018}%
  \BibitemOpen
  \bibfield  {author} {\bibinfo {author} {\bibfnamefont {J.~T.}\ \bibnamefont {Ma}}, \bibinfo {author} {\bibfnamefont {C.}~\bibnamefont {Tang}}, \bibinfo {author} {\bibfnamefont {L.}~\bibnamefont {Kang}}, \bibinfo {author} {\bibfnamefont {J.~A.}\ \bibnamefont {Voynow}},\ and\ \bibinfo {author} {\bibfnamefont {B.~K.}\ \bibnamefont {Rubin}},\ }\bibfield  {title} {\bibinfo {title} {Cystic fibrosis sputum rheology correlates with both acute and longitudinal changes in lung function},\ }\href@noop {} {\bibfield  {journal} {\bibinfo  {journal} {Chest}\ }\textbf {\bibinfo {volume} {154}},\ \bibinfo {pages} {370} (\bibinfo {year} {2018})}\BibitemShut {NoStop}%
\bibitem [{\citenamefont {Thornton}\ \emph {et~al.}(2008)\citenamefont {Thornton}, \citenamefont {Rousseau},\ and\ \citenamefont {McGuckin}}]{Thornton_2008}%
  \BibitemOpen
  \bibfield  {author} {\bibinfo {author} {\bibfnamefont {D.~J.}\ \bibnamefont {Thornton}}, \bibinfo {author} {\bibfnamefont {K.}~\bibnamefont {Rousseau}},\ and\ \bibinfo {author} {\bibfnamefont {M.~A.}\ \bibnamefont {McGuckin}},\ }\bibfield  {title} {\bibinfo {title} {Structure and function of the polymeric mucins in airways mucus},\ }\href@noop {} {\bibfield  {journal} {\bibinfo  {journal} {Annual Review of Physiology}\ }\textbf {\bibinfo {volume} {70}},\ \bibinfo {pages} {459} (\bibinfo {year} {2008})}\BibitemShut {NoStop}%
\bibitem [{\citenamefont {Song}\ \emph {et~al.}(2022)\citenamefont {Song}, \citenamefont {Iverson}, \citenamefont {Kaler}, \citenamefont {Boboltz}, \citenamefont {Scull},\ and\ \citenamefont {Duncan}}]{Song_2022}%
  \BibitemOpen
  \bibfield  {author} {\bibinfo {author} {\bibfnamefont {D.}~\bibnamefont {Song}}, \bibinfo {author} {\bibfnamefont {E.}~\bibnamefont {Iverson}}, \bibinfo {author} {\bibfnamefont {L.}~\bibnamefont {Kaler}}, \bibinfo {author} {\bibfnamefont {A.}~\bibnamefont {Boboltz}}, \bibinfo {author} {\bibfnamefont {M.~A.}\ \bibnamefont {Scull}},\ and\ \bibinfo {author} {\bibfnamefont {G.~A.}\ \bibnamefont {Duncan}},\ }\bibfield  {title} {\bibinfo {title} {Muc5b mobilizes and muc5ac spatially aligns mucociliary transport on human airway epithelium},\ }\href@noop {} {\bibfield  {journal} {\bibinfo  {journal} {Science Advances}\ }\textbf {\bibinfo {volume} {8}},\ \bibinfo {pages} {eabq5049} (\bibinfo {year} {2022})}\BibitemShut {NoStop}%
\bibitem [{\citenamefont {Caughman}\ \emph {et~al.}(2024)\citenamefont {Caughman}, \citenamefont {Papanikolas}, \citenamefont {Markovetz}, \citenamefont {Freeman}, \citenamefont {Hill}, \citenamefont {Forest},\ and\ \citenamefont {Lysy}}]{Caughman_2024}%
  \BibitemOpen
  \bibfield  {author} {\bibinfo {author} {\bibfnamefont {N.}~\bibnamefont {Caughman}}, \bibinfo {author} {\bibfnamefont {M.}~\bibnamefont {Papanikolas}}, \bibinfo {author} {\bibfnamefont {M.}~\bibnamefont {Markovetz}}, \bibinfo {author} {\bibfnamefont {R.}~\bibnamefont {Freeman}}, \bibinfo {author} {\bibfnamefont {D.~B.}\ \bibnamefont {Hill}}, \bibinfo {author} {\bibfnamefont {M.~G.}\ \bibnamefont {Forest}},\ and\ \bibinfo {author} {\bibfnamefont {M.}~\bibnamefont {Lysy}},\ }\bibfield  {title} {\bibinfo {title} {Experimental and statistical methods for microrheological characterization of heterogeneity in human respiratory mucus mimics of health and disease progression},\ }\href@noop {} {\bibfield  {journal} {\bibinfo  {journal} {Journal of Rheology}\ }\textbf {\bibinfo {volume} {68}},\ \bibinfo {pages} {995} (\bibinfo {year} {2024})}\BibitemShut {NoStop}%
\bibitem [{\citenamefont {Pednekar}\ \emph {et~al.}(2022)\citenamefont {Pednekar}, \citenamefont {Liguori}, \citenamefont {Marques}, \citenamefont {Zhang}, \citenamefont {Zhang}, \citenamefont {Zhou}, \citenamefont {Amoako},\ and\ \citenamefont {Gu}}]{Pednekar_2022}%
  \BibitemOpen
  \bibfield  {author} {\bibinfo {author} {\bibfnamefont {D.~D.}\ \bibnamefont {Pednekar}}, \bibinfo {author} {\bibfnamefont {M.~A.}\ \bibnamefont {Liguori}}, \bibinfo {author} {\bibfnamefont {C.~N.~H.}\ \bibnamefont {Marques}}, \bibinfo {author} {\bibfnamefont {T.}~\bibnamefont {Zhang}}, \bibinfo {author} {\bibfnamefont {N.}~\bibnamefont {Zhang}}, \bibinfo {author} {\bibfnamefont {Z.}~\bibnamefont {Zhou}}, \bibinfo {author} {\bibfnamefont {K.}~\bibnamefont {Amoako}},\ and\ \bibinfo {author} {\bibfnamefont {H.}~\bibnamefont {Gu}},\ }\bibfield  {title} {\bibinfo {title} {From static to dynamic: A review on the role of mucus heterogeneity in particle and microbial transport},\ }\href@noop {} {\bibfield  {journal} {\bibinfo  {journal} {ACS Biomaterials Science {\&} Engineering}\ }\textbf {\bibinfo {volume} {8}},\ \bibinfo {pages} {2825} (\bibinfo {year} {2022})}\BibitemShut {NoStop}%
\bibitem [{\citenamefont {Garc{\'i}a-D{\'i}az}\ \emph {et~al.}(2018)\citenamefont {Garc{\'i}a-D{\'i}az}, \citenamefont {Birch}, \citenamefont {Wan},\ and\ \citenamefont {Nielsen}}]{Garcia-Diaz_2018}%
  \BibitemOpen
  \bibfield  {author} {\bibinfo {author} {\bibfnamefont {M.}~\bibnamefont {Garc{\'i}a-D{\'i}az}}, \bibinfo {author} {\bibfnamefont {D.}~\bibnamefont {Birch}}, \bibinfo {author} {\bibfnamefont {F.}~\bibnamefont {Wan}},\ and\ \bibinfo {author} {\bibfnamefont {H.~M.}\ \bibnamefont {Nielsen}},\ }\bibfield  {title} {\bibinfo {title} {The role of mucus as an invisible cloak to transepithelial drug delivery by nanoparticles},\ }\href@noop {} {\bibfield  {journal} {\bibinfo  {journal} {Advanced Drug Delivery Reviews}\ }\textbf {\bibinfo {volume} {124}},\ \bibinfo {pages} {107} (\bibinfo {year} {2018})}\BibitemShut {NoStop}%
\bibitem [{\citenamefont {Lopez}\ \emph {et~al.}(2006)\citenamefont {Lopez}, \citenamefont {Shibuya}, \citenamefont {Rao}, \citenamefont {Mathers}, \citenamefont {Hansell}, \citenamefont {Held}, \citenamefont {Schmid},\ and\ \citenamefont {Buist}}]{Lopez_2006}%
  \BibitemOpen
  \bibfield  {author} {\bibinfo {author} {\bibfnamefont {A.~D.}\ \bibnamefont {Lopez}}, \bibinfo {author} {\bibfnamefont {K.}~\bibnamefont {Shibuya}}, \bibinfo {author} {\bibfnamefont {C.}~\bibnamefont {Rao}}, \bibinfo {author} {\bibfnamefont {C.~D.}\ \bibnamefont {Mathers}}, \bibinfo {author} {\bibfnamefont {A.~L.}\ \bibnamefont {Hansell}}, \bibinfo {author} {\bibfnamefont {L.~S.}\ \bibnamefont {Held}}, \bibinfo {author} {\bibfnamefont {V.}~\bibnamefont {Schmid}},\ and\ \bibinfo {author} {\bibfnamefont {S.}~\bibnamefont {Buist}},\ }\bibfield  {title} {\bibinfo {title} {Chronic obstructive pulmonary disease: current burden and future projections},\ }\href@noop {} {\bibfield  {journal} {\bibinfo  {journal} {European Respiratory Journal}\ }\textbf {\bibinfo {volume} {27}},\ \bibinfo {pages} {397} (\bibinfo {year} {2006})}\BibitemShut {NoStop}%
\bibitem [{\citenamefont {Decramer}\ \emph {et~al.}(2012)\citenamefont {Decramer}, \citenamefont {Janssens},\ and\ \citenamefont {Miravitlles}}]{Decramer_2012}%
  \BibitemOpen
  \bibfield  {author} {\bibinfo {author} {\bibfnamefont {M.}~\bibnamefont {Decramer}}, \bibinfo {author} {\bibfnamefont {W.}~\bibnamefont {Janssens}},\ and\ \bibinfo {author} {\bibfnamefont {M.}~\bibnamefont {Miravitlles}},\ }\bibfield  {title} {\bibinfo {title} {Chronic obstructive pulmonary disease},\ }\href@noop {} {\bibfield  {journal} {\bibinfo  {journal} {The Lancet}\ }\textbf {\bibinfo {volume} {379}},\ \bibinfo {pages} {1341} (\bibinfo {year} {2012})}\BibitemShut {NoStop}%
\bibitem [{\citenamefont {PILEWSKI}\ and\ \citenamefont {FRIZZELL}(1999)}]{PILEWSKI_1999}%
  \BibitemOpen
  \bibfield  {author} {\bibinfo {author} {\bibfnamefont {J.~M.}\ \bibnamefont {PILEWSKI}}\ and\ \bibinfo {author} {\bibfnamefont {R.~A.}\ \bibnamefont {FRIZZELL}},\ }\bibfield  {title} {\bibinfo {title} {Role of cftr in airway disease},\ }\href@noop {} {\bibfield  {journal} {\bibinfo  {journal} {Physiological Reviews}\ }\textbf {\bibinfo {volume} {79}},\ \bibinfo {pages} {S215} (\bibinfo {year} {1999})}\BibitemShut {NoStop}%
\bibitem [{\citenamefont {Collawn}\ and\ \citenamefont {Matalon}(2014)}]{Collawn_2014}%
  \BibitemOpen
  \bibfield  {author} {\bibinfo {author} {\bibfnamefont {J.~F.}\ \bibnamefont {Collawn}}\ and\ \bibinfo {author} {\bibfnamefont {S.}~\bibnamefont {Matalon}},\ }\bibfield  {title} {\bibinfo {title} {Cftr and lung homeostasis},\ }\href@noop {} {\bibfield  {journal} {\bibinfo  {journal} {American Journal of Physiology-Lung Cellular and Molecular Physiology}\ }\textbf {\bibinfo {volume} {307}},\ \bibinfo {pages} {L917} (\bibinfo {year} {2014})}\BibitemShut {NoStop}%
\bibitem [{\citenamefont {Rowe}\ \emph {et~al.}(2005)\citenamefont {Rowe}, \citenamefont {Miller},\ and\ \citenamefont {Sorscher}}]{Rowe_2005}%
  \BibitemOpen
  \bibfield  {author} {\bibinfo {author} {\bibfnamefont {S.~M.}\ \bibnamefont {Rowe}}, \bibinfo {author} {\bibfnamefont {S.}~\bibnamefont {Miller}},\ and\ \bibinfo {author} {\bibfnamefont {E.~J.}\ \bibnamefont {Sorscher}},\ }\bibfield  {title} {\bibinfo {title} {Cystic fibrosis},\ }\href@noop {} {\bibfield  {journal} {\bibinfo  {journal} {New England Journal of Medicine}\ }\textbf {\bibinfo {volume} {352}},\ \bibinfo {pages} {1992} (\bibinfo {year} {2005})}\BibitemShut {NoStop}%
\bibitem [{\citenamefont {Hauser}\ \emph {et~al.}(2011)\citenamefont {Hauser}, \citenamefont {Jain}, \citenamefont {Bar-Meir},\ and\ \citenamefont {McColley}}]{Hauser_2011}%
  \BibitemOpen
  \bibfield  {author} {\bibinfo {author} {\bibfnamefont {A.~R.}\ \bibnamefont {Hauser}}, \bibinfo {author} {\bibfnamefont {M.}~\bibnamefont {Jain}}, \bibinfo {author} {\bibfnamefont {M.}~\bibnamefont {Bar-Meir}},\ and\ \bibinfo {author} {\bibfnamefont {S.~A.}\ \bibnamefont {McColley}},\ }\bibfield  {title} {\bibinfo {title} {Clinical significance of microbial infection and adaptation in cystic fibrosis},\ }\href@noop {} {\bibfield  {journal} {\bibinfo  {journal} {Clinical Microbiology Reviews}\ }\textbf {\bibinfo {volume} {24}},\ \bibinfo {pages} {29} (\bibinfo {year} {2011})}\BibitemShut {NoStop}%
\bibitem [{\citenamefont {Morcillo}\ and\ \citenamefont {Cortijo}(2006)}]{Morcillo_2006}%
  \BibitemOpen
  \bibfield  {author} {\bibinfo {author} {\bibfnamefont {E.~J.}\ \bibnamefont {Morcillo}}\ and\ \bibinfo {author} {\bibfnamefont {J.}~\bibnamefont {Cortijo}},\ }\bibfield  {title} {\bibinfo {title} {Mucus and muc in asthma},\ }\href@noop {} {\bibfield  {journal} {\bibinfo  {journal} {Current Opinion in Pulmonary Medicine}\ }\textbf {\bibinfo {volume} {12}},\ \bibinfo {pages} {1} (\bibinfo {year} {2006})}\BibitemShut {NoStop}%
\bibitem [{\citenamefont {Evans}\ \emph {et~al.}(2009{\natexlab{a}})\citenamefont {Evans}, \citenamefont {Kim}, \citenamefont {Tuvim},\ and\ \citenamefont {Dickey}}]{Evans_2009a}%
  \BibitemOpen
  \bibfield  {author} {\bibinfo {author} {\bibfnamefont {C.~M.}\ \bibnamefont {Evans}}, \bibinfo {author} {\bibfnamefont {K.}~\bibnamefont {Kim}}, \bibinfo {author} {\bibfnamefont {M.~J.}\ \bibnamefont {Tuvim}},\ and\ \bibinfo {author} {\bibfnamefont {B.~F.}\ \bibnamefont {Dickey}},\ }\bibfield  {title} {\bibinfo {title} {Mucus hypersecretion in asthma: causes and effects},\ }\href@noop {} {\bibfield  {journal} {\bibinfo  {journal} {Current Opinion in Pulmonary Medicine}\ }\textbf {\bibinfo {volume} {15}},\ \bibinfo {pages} {4} (\bibinfo {year} {2009}{\natexlab{a}})}\BibitemShut {NoStop}%
\bibitem [{\citenamefont {Izuhara}\ \emph {et~al.}(2009)\citenamefont {Izuhara}, \citenamefont {Ohta}, \citenamefont {Shiraishi}, \citenamefont {Suzuki}, \citenamefont {Taniguchi}, \citenamefont {Toda}, \citenamefont {Tanabe}, \citenamefont {Yasuo}, \citenamefont {Kubo}, \citenamefont {Hoshino},\ and\ \citenamefont {Aizawa}}]{Izuhara_2009}%
  \BibitemOpen
  \bibfield  {author} {\bibinfo {author} {\bibfnamefont {K.}~\bibnamefont {Izuhara}}, \bibinfo {author} {\bibfnamefont {S.}~\bibnamefont {Ohta}}, \bibinfo {author} {\bibfnamefont {H.}~\bibnamefont {Shiraishi}}, \bibinfo {author} {\bibfnamefont {S.}~\bibnamefont {Suzuki}}, \bibinfo {author} {\bibfnamefont {K.}~\bibnamefont {Taniguchi}}, \bibinfo {author} {\bibfnamefont {S.}~\bibnamefont {Toda}}, \bibinfo {author} {\bibfnamefont {T.}~\bibnamefont {Tanabe}}, \bibinfo {author} {\bibfnamefont {M.}~\bibnamefont {Yasuo}}, \bibinfo {author} {\bibfnamefont {K.}~\bibnamefont {Kubo}}, \bibinfo {author} {\bibfnamefont {T.}~\bibnamefont {Hoshino}},\ and\ \bibinfo {author} {\bibfnamefont {H.}~\bibnamefont {Aizawa}},\ }\bibfield  {title} {\bibinfo {title} {The mechanism of mucus production in bronchial asthma},\ }\href@noop {} {\bibfield  {journal} {\bibinfo  {journal} {Current Medicinal Chemistry}\ }\textbf {\bibinfo {volume} {16}},\ \bibinfo {pages} {2867} (\bibinfo {year} {2009})}\BibitemShut {NoStop}%
\bibitem [{\citenamefont {Suh}\ \emph {et~al.}(2005)\citenamefont {Suh}, \citenamefont {Dawson},\ and\ \citenamefont {Hanes}}]{Suh_2005}%
  \BibitemOpen
  \bibfield  {author} {\bibinfo {author} {\bibfnamefont {J.}~\bibnamefont {Suh}}, \bibinfo {author} {\bibfnamefont {M.}~\bibnamefont {Dawson}},\ and\ \bibinfo {author} {\bibfnamefont {J.}~\bibnamefont {Hanes}},\ }\bibfield  {title} {\bibinfo {title} {Real-time multiple-particle tracking: applications to drug and gene delivery},\ }\href@noop {} {\bibfield  {journal} {\bibinfo  {journal} {Advanced Drug Delivery Reviews}\ }\textbf {\bibinfo {volume} {57}},\ \bibinfo {pages} {63} (\bibinfo {year} {2005})}\BibitemShut {NoStop}%
\bibitem [{\citenamefont {Nordg{\aa}rd}\ and\ \citenamefont {Draget}(2018)}]{Nordgard_2018}%
  \BibitemOpen
  \bibfield  {author} {\bibinfo {author} {\bibfnamefont {C.~T.}\ \bibnamefont {Nordg{\aa}rd}}\ and\ \bibinfo {author} {\bibfnamefont {K.~I.}\ \bibnamefont {Draget}},\ }\bibfield  {title} {\bibinfo {title} {Co association of mucus modulating agents and nanoparticles for mucosal drug delivery},\ }\href@noop {} {\bibfield  {journal} {\bibinfo  {journal} {Advanced Drug Delivery Reviews}\ }\textbf {\bibinfo {volume} {124}},\ \bibinfo {pages} {175} (\bibinfo {year} {2018})}\BibitemShut {NoStop}%
\bibitem [{\citenamefont {Woodruff}\ \emph {et~al.}(2015)\citenamefont {Woodruff}, \citenamefont {Agusti}, \citenamefont {Roche}, \citenamefont {Singh},\ and\ \citenamefont {Martinez}}]{Woodruff_2015}%
  \BibitemOpen
  \bibfield  {author} {\bibinfo {author} {\bibfnamefont {P.~G.}\ \bibnamefont {Woodruff}}, \bibinfo {author} {\bibfnamefont {A.}~\bibnamefont {Agusti}}, \bibinfo {author} {\bibfnamefont {N.}~\bibnamefont {Roche}}, \bibinfo {author} {\bibfnamefont {D.}~\bibnamefont {Singh}},\ and\ \bibinfo {author} {\bibfnamefont {F.~J.}\ \bibnamefont {Martinez}},\ }\bibfield  {title} {\bibinfo {title} {Current concepts in targeting chronic obstructive pulmonary disease pharmacotherapy: making progress towards personalised management},\ }\href@noop {} {\bibfield  {journal} {\bibinfo  {journal} {The Lancet}\ }\textbf {\bibinfo {volume} {385}},\ \bibinfo {pages} {1789} (\bibinfo {year} {2015})}\BibitemShut {NoStop}%
\bibitem [{\citenamefont {Raju}\ \emph {et~al.}(2016)\citenamefont {Raju}, \citenamefont {Solomon}, \citenamefont {Dransfield},\ and\ \citenamefont {Rowe}}]{Raju_2016}%
  \BibitemOpen
  \bibfield  {author} {\bibinfo {author} {\bibfnamefont {S.~V.}\ \bibnamefont {Raju}}, \bibinfo {author} {\bibfnamefont {G.~M.}\ \bibnamefont {Solomon}}, \bibinfo {author} {\bibfnamefont {M.~T.}\ \bibnamefont {Dransfield}},\ and\ \bibinfo {author} {\bibfnamefont {S.~M.}\ \bibnamefont {Rowe}},\ }\bibfield  {title} {\bibinfo {title} {Acquired cystic fibrosis transmembrane conductance regulator dysfunction in chronic bronchitis and other diseases of mucus clearance},\ }\href@noop {} {\bibfield  {journal} {\bibinfo  {journal} {Clinics in Chest Medicine}\ }\textbf {\bibinfo {volume} {37}},\ \bibinfo {pages} {147} (\bibinfo {year} {2016})}\BibitemShut {NoStop}%
\bibitem [{\citenamefont {Solomon}\ \emph {et~al.}(2017)\citenamefont {Solomon}, \citenamefont {Fu}, \citenamefont {Rowe},\ and\ \citenamefont {Collawn}}]{Solomon_2017}%
  \BibitemOpen
  \bibfield  {author} {\bibinfo {author} {\bibfnamefont {G.~M.}\ \bibnamefont {Solomon}}, \bibinfo {author} {\bibfnamefont {L.}~\bibnamefont {Fu}}, \bibinfo {author} {\bibfnamefont {S.~M.}\ \bibnamefont {Rowe}},\ and\ \bibinfo {author} {\bibfnamefont {J.~F.}\ \bibnamefont {Collawn}},\ }\bibfield  {title} {\bibinfo {title} {The therapeutic potential of cftr modulators for copd and other airway diseases},\ }\href@noop {} {\bibfield  {journal} {\bibinfo  {journal} {Current Opinion in Pharmacology}\ }\textbf {\bibinfo {volume} {34}},\ \bibinfo {pages} {132} (\bibinfo {year} {2017})}\BibitemShut {NoStop}%
\bibitem [{\citenamefont {Chisholm}\ \emph {et~al.}(2019)\citenamefont {Chisholm}, \citenamefont {Shenoy}, \citenamefont {Shade}, \citenamefont {Kim}, \citenamefont {Putcha}, \citenamefont {Carson}, \citenamefont {Wise}, \citenamefont {Hansel}, \citenamefont {Hanes}, \citenamefont {Suk},\ and\ \citenamefont {Neptune}}]{Chisholm_2019}%
  \BibitemOpen
  \bibfield  {author} {\bibinfo {author} {\bibfnamefont {J.~F.}\ \bibnamefont {Chisholm}}, \bibinfo {author} {\bibfnamefont {S.~K.}\ \bibnamefont {Shenoy}}, \bibinfo {author} {\bibfnamefont {J.~K.}\ \bibnamefont {Shade}}, \bibinfo {author} {\bibfnamefont {V.}~\bibnamefont {Kim}}, \bibinfo {author} {\bibfnamefont {N.}~\bibnamefont {Putcha}}, \bibinfo {author} {\bibfnamefont {K.~A.}\ \bibnamefont {Carson}}, \bibinfo {author} {\bibfnamefont {R.}~\bibnamefont {Wise}}, \bibinfo {author} {\bibfnamefont {N.~N.}\ \bibnamefont {Hansel}}, \bibinfo {author} {\bibfnamefont {J.~S.}\ \bibnamefont {Hanes}}, \bibinfo {author} {\bibfnamefont {J.~S.}\ \bibnamefont {Suk}},\ and\ \bibinfo {author} {\bibfnamefont {E.}~\bibnamefont {Neptune}},\ }\bibfield  {title} {\bibinfo {title} {Nanoparticle diffusion in spontaneously expectorated sputum as a biophysical tool to probe disease severity in copd},\ }\href@noop {} {\bibfield  {journal} {\bibinfo  {journal} {European Respiratory Journal}\ }\textbf {\bibinfo {volume} {54}},\
  \bibinfo {pages} {1900088} (\bibinfo {year} {2019})}\BibitemShut {NoStop}%
\bibitem [{\citenamefont {Thulborn}\ \emph {et~al.}(2019)\citenamefont {Thulborn}, \citenamefont {Mistry}, \citenamefont {Brightling}, \citenamefont {Moffitt}, \citenamefont {Ribeiro},\ and\ \citenamefont {Bafadhel}}]{Thulborn_2019}%
  \BibitemOpen
  \bibfield  {author} {\bibinfo {author} {\bibfnamefont {S.~J.}\ \bibnamefont {Thulborn}}, \bibinfo {author} {\bibfnamefont {V.}~\bibnamefont {Mistry}}, \bibinfo {author} {\bibfnamefont {C.~E.}\ \bibnamefont {Brightling}}, \bibinfo {author} {\bibfnamefont {K.~L.}\ \bibnamefont {Moffitt}}, \bibinfo {author} {\bibfnamefont {D.}~\bibnamefont {Ribeiro}},\ and\ \bibinfo {author} {\bibfnamefont {M.}~\bibnamefont {Bafadhel}},\ }\bibfield  {title} {\bibinfo {title} {Neutrophil elastase as a biomarker for bacterial infection in copd},\ }\href@noop {} {\bibfield  {journal} {\bibinfo  {journal} {Respiratory Research}\ }\textbf {\bibinfo {volume} {20}},\ \bibinfo {pages} {170} (\bibinfo {year} {2019})}\BibitemShut {NoStop}%
\bibitem [{\citenamefont {Linssen}\ \emph {et~al.}(2021)\citenamefont {Linssen}, \citenamefont {Chai}, \citenamefont {Ma}, \citenamefont {Kummarapurugu}, \citenamefont {Woensel}, \citenamefont {Bem}, \citenamefont {Kaler}, \citenamefont {Duncan}, \citenamefont {Zhou}, \citenamefont {Rubin},\ and\ \citenamefont {Xu}}]{Linssen_2021}%
  \BibitemOpen
  \bibfield  {author} {\bibinfo {author} {\bibfnamefont {R.~S.}\ \bibnamefont {Linssen}}, \bibinfo {author} {\bibfnamefont {G.}~\bibnamefont {Chai}}, \bibinfo {author} {\bibfnamefont {J.}~\bibnamefont {Ma}}, \bibinfo {author} {\bibfnamefont {A.~B.}\ \bibnamefont {Kummarapurugu}}, \bibinfo {author} {\bibfnamefont {J.~B. M.~v.}\ \bibnamefont {Woensel}}, \bibinfo {author} {\bibfnamefont {R.~A.}\ \bibnamefont {Bem}}, \bibinfo {author} {\bibfnamefont {L.}~\bibnamefont {Kaler}}, \bibinfo {author} {\bibfnamefont {G.~A.}\ \bibnamefont {Duncan}}, \bibinfo {author} {\bibfnamefont {L.}~\bibnamefont {Zhou}}, \bibinfo {author} {\bibfnamefont {B.~K.}\ \bibnamefont {Rubin}},\ and\ \bibinfo {author} {\bibfnamefont {Q.}~\bibnamefont {Xu}},\ }\bibfield  {title} {\bibinfo {title} {Neutrophil extracellular traps increase airway mucus viscoelasticity and slow mucus particle transit},\ }\href@noop {} {\bibfield  {journal} {\bibinfo  {journal} {American Journal of Respiratory Cell and Molecular Biology}\ }\textbf {\bibinfo {volume}
  {64}},\ \bibinfo {pages} {69} (\bibinfo {year} {2021})}\BibitemShut {NoStop}%
\bibitem [{\citenamefont {Volpato}\ \emph {et~al.}(2022)\citenamefont {Volpato}, \citenamefont {Vialaret}, \citenamefont {Hirtz}, \citenamefont {Petit}, \citenamefont {Suehs}, \citenamefont {Patarin}, \citenamefont {Matzner-Lober}, \citenamefont {Vachier}, \citenamefont {Molinari}, \citenamefont {Bourdin},\ and\ \citenamefont {Charriot}}]{Volpato_2022}%
  \BibitemOpen
  \bibfield  {author} {\bibinfo {author} {\bibfnamefont {M.}~\bibnamefont {Volpato}}, \bibinfo {author} {\bibfnamefont {J.}~\bibnamefont {Vialaret}}, \bibinfo {author} {\bibfnamefont {C.}~\bibnamefont {Hirtz}}, \bibinfo {author} {\bibfnamefont {A.}~\bibnamefont {Petit}}, \bibinfo {author} {\bibfnamefont {C.}~\bibnamefont {Suehs}}, \bibinfo {author} {\bibfnamefont {J.}~\bibnamefont {Patarin}}, \bibinfo {author} {\bibfnamefont {E.}~\bibnamefont {Matzner-Lober}}, \bibinfo {author} {\bibfnamefont {I.}~\bibnamefont {Vachier}}, \bibinfo {author} {\bibfnamefont {N.}~\bibnamefont {Molinari}}, \bibinfo {author} {\bibfnamefont {A.}~\bibnamefont {Bourdin}},\ and\ \bibinfo {author} {\bibfnamefont {J.}~\bibnamefont {Charriot}},\ }\bibfield  {title} {\bibinfo {title} {Rheology predicts sputum eosinophilia in patients with muco-obstructive lung diseases},\ }\href@noop {} {\bibfield  {journal} {\bibinfo  {journal} {Biochemical and Biophysical Research Communications}\ }\textbf {\bibinfo {volume} {622}},\ \bibinfo {pages}
  {64} (\bibinfo {year} {2022})}\BibitemShut {NoStop}%
\bibitem [{\citenamefont {Lambrecht}\ and\ \citenamefont {Hammad}(2015)}]{Lambrecht_2015}%
  \BibitemOpen
  \bibfield  {author} {\bibinfo {author} {\bibfnamefont {B.~N.}\ \bibnamefont {Lambrecht}}\ and\ \bibinfo {author} {\bibfnamefont {H.}~\bibnamefont {Hammad}},\ }\bibfield  {title} {\bibinfo {title} {The immunology of asthma},\ }\href@noop {} {\bibfield  {journal} {\bibinfo  {journal} {Nature Immunology}\ }\textbf {\bibinfo {volume} {16}},\ \bibinfo {pages} {45} (\bibinfo {year} {2015})}\BibitemShut {NoStop}%
\bibitem [{\citenamefont {Schulz}\ and\ \citenamefont {T{\"u}mmler}(2016)}]{Schulz_2016}%
  \BibitemOpen
  \bibfield  {author} {\bibinfo {author} {\bibfnamefont {A.}~\bibnamefont {Schulz}}\ and\ \bibinfo {author} {\bibfnamefont {B.}~\bibnamefont {T{\"u}mmler}},\ }\bibfield  {title} {\bibinfo {title} {Non-allergic asthma as a cftr-related disorder},\ }\href@noop {} {\bibfield  {journal} {\bibinfo  {journal} {Journal of Cystic Fibrosis}\ }\textbf {\bibinfo {volume} {15}},\ \bibinfo {pages} {641} (\bibinfo {year} {2016})}\BibitemShut {NoStop}%
\bibitem [{\citenamefont {Hammad}\ and\ \citenamefont {Lambrecht}(2021)}]{Hammad_2021}%
  \BibitemOpen
  \bibfield  {author} {\bibinfo {author} {\bibfnamefont {H.}~\bibnamefont {Hammad}}\ and\ \bibinfo {author} {\bibfnamefont {B.~N.}\ \bibnamefont {Lambrecht}},\ }\bibfield  {title} {\bibinfo {title} {The basic immunology of asthma},\ }\href@noop {} {\bibfield  {journal} {\bibinfo  {journal} {Cell}\ }\textbf {\bibinfo {volume} {184}},\ \bibinfo {pages} {1469} (\bibinfo {year} {2021})}\BibitemShut {NoStop}%
\bibitem [{\citenamefont {Livraghi}\ and\ \citenamefont {Randell}(2007)}]{Livraghi_2007}%
  \BibitemOpen
  \bibfield  {author} {\bibinfo {author} {\bibfnamefont {A.}~\bibnamefont {Livraghi}}\ and\ \bibinfo {author} {\bibfnamefont {S.~H.}\ \bibnamefont {Randell}},\ }\bibfield  {title} {\bibinfo {title} {Cystic fibrosis and other respiratory diseases of impaired mucus clearance},\ }\href@noop {} {\bibfield  {journal} {\bibinfo  {journal} {Toxicologic Pathology}\ }\textbf {\bibinfo {volume} {35}},\ \bibinfo {pages} {116} (\bibinfo {year} {2007})}\BibitemShut {NoStop}%
\bibitem [{\citenamefont {Button}\ \emph {et~al.}(2018)\citenamefont {Button}, \citenamefont {Goodell}, \citenamefont {Atieh}, \citenamefont {Chen}, \citenamefont {Williams}, \citenamefont {Shenoy}, \citenamefont {Lackey}, \citenamefont {Shenkute}, \citenamefont {Cai}, \citenamefont {Dennis}, \citenamefont {Boucher},\ and\ \citenamefont {Rubinstein}}]{Button_2018}%
  \BibitemOpen
  \bibfield  {author} {\bibinfo {author} {\bibfnamefont {B.}~\bibnamefont {Button}}, \bibinfo {author} {\bibfnamefont {H.~P.}\ \bibnamefont {Goodell}}, \bibinfo {author} {\bibfnamefont {E.}~\bibnamefont {Atieh}}, \bibinfo {author} {\bibfnamefont {Y.-C.}\ \bibnamefont {Chen}}, \bibinfo {author} {\bibfnamefont {R.}~\bibnamefont {Williams}}, \bibinfo {author} {\bibfnamefont {S.}~\bibnamefont {Shenoy}}, \bibinfo {author} {\bibfnamefont {E.}~\bibnamefont {Lackey}}, \bibinfo {author} {\bibfnamefont {N.~T.}\ \bibnamefont {Shenkute}}, \bibinfo {author} {\bibfnamefont {L.-H.}\ \bibnamefont {Cai}}, \bibinfo {author} {\bibfnamefont {R.~G.}\ \bibnamefont {Dennis}}, \bibinfo {author} {\bibfnamefont {R.~C.}\ \bibnamefont {Boucher}},\ and\ \bibinfo {author} {\bibfnamefont {M.}~\bibnamefont {Rubinstein}},\ }\bibfield  {title} {\bibinfo {title} {Roles of mucus adhesion and cohesion in cough clearance},\ }\href@noop {} {\bibfield  {journal} {\bibinfo  {journal} {Proceedings of the National Academy of Sciences}\ }\textbf
  {\bibinfo {volume} {115}},\ \bibinfo {pages} {12501} (\bibinfo {year} {2018})}\BibitemShut {NoStop}%
\bibitem [{\citenamefont {Abrami}\ \emph {et~al.}(2020)\citenamefont {Abrami}, \citenamefont {Maschio}, \citenamefont {Conese}, \citenamefont {Confalonieri}, \citenamefont {Di~Gioia}, \citenamefont {Gerin}, \citenamefont {Dapas}, \citenamefont {Tonon}, \citenamefont {Farra}, \citenamefont {Murano}, \citenamefont {Zanella}, \citenamefont {Salton}, \citenamefont {Torelli}, \citenamefont {Grassi},\ and\ \citenamefont {Grassi}}]{Abrami_2020}%
  \BibitemOpen
  \bibfield  {author} {\bibinfo {author} {\bibfnamefont {M.}~\bibnamefont {Abrami}}, \bibinfo {author} {\bibfnamefont {M.}~\bibnamefont {Maschio}}, \bibinfo {author} {\bibfnamefont {M.}~\bibnamefont {Conese}}, \bibinfo {author} {\bibfnamefont {M.}~\bibnamefont {Confalonieri}}, \bibinfo {author} {\bibfnamefont {S.}~\bibnamefont {Di~Gioia}}, \bibinfo {author} {\bibfnamefont {F.}~\bibnamefont {Gerin}}, \bibinfo {author} {\bibfnamefont {B.}~\bibnamefont {Dapas}}, \bibinfo {author} {\bibfnamefont {F.}~\bibnamefont {Tonon}}, \bibinfo {author} {\bibfnamefont {R.}~\bibnamefont {Farra}}, \bibinfo {author} {\bibfnamefont {E.}~\bibnamefont {Murano}}, \bibinfo {author} {\bibfnamefont {G.}~\bibnamefont {Zanella}}, \bibinfo {author} {\bibfnamefont {F.}~\bibnamefont {Salton}}, \bibinfo {author} {\bibfnamefont {L.}~\bibnamefont {Torelli}}, \bibinfo {author} {\bibfnamefont {G.}~\bibnamefont {Grassi}},\ and\ \bibinfo {author} {\bibfnamefont {M.}~\bibnamefont {Grassi}},\ }\bibfield  {title} {\bibinfo {title} {Use of low field
  nuclear magnetic resonance to monitor lung inflammation and the amount of pathological components in the sputum of cystic fibrosis patients},\ }\href@noop {} {\bibfield  {journal} {\bibinfo  {journal} {Magnetic Resonance in Medicine}\ }\textbf {\bibinfo {volume} {84}},\ \bibinfo {pages} {427} (\bibinfo {year} {2020})}\BibitemShut {NoStop}%
\bibitem [{\citenamefont {Abrami}\ \emph {et~al.}(2021)\citenamefont {Abrami}, \citenamefont {Maschio}, \citenamefont {Conese}, \citenamefont {Confalonieri}, \citenamefont {Gerin}, \citenamefont {Dapas}, \citenamefont {Farra}, \citenamefont {Adrover}, \citenamefont {Torelli}, \citenamefont {Ruaro}, \citenamefont {Grassi},\ and\ \citenamefont {Grassi}}]{Abrami_2021}%
  \BibitemOpen
  \bibfield  {author} {\bibinfo {author} {\bibfnamefont {M.}~\bibnamefont {Abrami}}, \bibinfo {author} {\bibfnamefont {M.}~\bibnamefont {Maschio}}, \bibinfo {author} {\bibfnamefont {M.}~\bibnamefont {Conese}}, \bibinfo {author} {\bibfnamefont {M.}~\bibnamefont {Confalonieri}}, \bibinfo {author} {\bibfnamefont {F.}~\bibnamefont {Gerin}}, \bibinfo {author} {\bibfnamefont {B.}~\bibnamefont {Dapas}}, \bibinfo {author} {\bibfnamefont {R.}~\bibnamefont {Farra}}, \bibinfo {author} {\bibfnamefont {A.}~\bibnamefont {Adrover}}, \bibinfo {author} {\bibfnamefont {L.}~\bibnamefont {Torelli}}, \bibinfo {author} {\bibfnamefont {B.}~\bibnamefont {Ruaro}}, \bibinfo {author} {\bibfnamefont {G.}~\bibnamefont {Grassi}},\ and\ \bibinfo {author} {\bibfnamefont {M.}~\bibnamefont {Grassi}},\ }\bibfield  {title} {\bibinfo {title} {Combined use of rheology and portable low-field nmr in cystic fibrosis patients},\ }\href@noop {} {\bibfield  {journal} {\bibinfo  {journal} {Respiratory Medicine}\ }\textbf {\bibinfo {volume} {189}},\
  \bibinfo {pages} {106623} (\bibinfo {year} {2021})}\BibitemShut {NoStop}%
\bibitem [{\citenamefont {Abrami}\ \emph {et~al.}(2022)\citenamefont {Abrami}, \citenamefont {Maschio}, \citenamefont {Conese}, \citenamefont {Confalonieri}, \citenamefont {Salton}, \citenamefont {Gerin}, \citenamefont {Dapas}, \citenamefont {Farra}, \citenamefont {Adrover}, \citenamefont {Milcovich}, \citenamefont {Fornasier}, \citenamefont {Biasin}, \citenamefont {Grassi},\ and\ \citenamefont {Grassi}}]{Abrami_2022}%
  \BibitemOpen
  \bibfield  {author} {\bibinfo {author} {\bibfnamefont {M.}~\bibnamefont {Abrami}}, \bibinfo {author} {\bibfnamefont {M.}~\bibnamefont {Maschio}}, \bibinfo {author} {\bibfnamefont {M.}~\bibnamefont {Conese}}, \bibinfo {author} {\bibfnamefont {M.}~\bibnamefont {Confalonieri}}, \bibinfo {author} {\bibfnamefont {F.}~\bibnamefont {Salton}}, \bibinfo {author} {\bibfnamefont {F.}~\bibnamefont {Gerin}}, \bibinfo {author} {\bibfnamefont {B.}~\bibnamefont {Dapas}}, \bibinfo {author} {\bibfnamefont {R.}~\bibnamefont {Farra}}, \bibinfo {author} {\bibfnamefont {A.}~\bibnamefont {Adrover}}, \bibinfo {author} {\bibfnamefont {G.}~\bibnamefont {Milcovich}}, \bibinfo {author} {\bibfnamefont {C.}~\bibnamefont {Fornasier}}, \bibinfo {author} {\bibfnamefont {A.}~\bibnamefont {Biasin}}, \bibinfo {author} {\bibfnamefont {M.}~\bibnamefont {Grassi}},\ and\ \bibinfo {author} {\bibfnamefont {G.}~\bibnamefont {Grassi}},\ }\bibfield  {title} {\bibinfo {title} {Effect of chest physiotherapy on cystic fibrosis sputum nanostructure: an
  experimental and theoretical approach},\ }\href@noop {} {\bibfield  {journal} {\bibinfo  {journal} {Drug Delivery and Translational Research}\ }\textbf {\bibinfo {volume} {12}},\ \bibinfo {pages} {1943} (\bibinfo {year} {2022})}\BibitemShut {NoStop}%
\bibitem [{\citenamefont {Batson}\ \emph {et~al.}(2022)\citenamefont {Batson}, \citenamefont {Zorn}, \citenamefont {Radicioni}, \citenamefont {Livengood}, \citenamefont {Kumagai}, \citenamefont {Dang}, \citenamefont {Ceppe}, \citenamefont {Clapp}, \citenamefont {Tunney}, \citenamefont {Elborn}, \citenamefont {McElvaney}, \citenamefont {Muhlebach}, \citenamefont {Boucher}, \citenamefont {Tiemeyer}, \citenamefont {Wolfgang},\ and\ \citenamefont {Kesimer}}]{Batson_2022}%
  \BibitemOpen
  \bibfield  {author} {\bibinfo {author} {\bibfnamefont {B.~D.}\ \bibnamefont {Batson}}, \bibinfo {author} {\bibfnamefont {B.~T.}\ \bibnamefont {Zorn}}, \bibinfo {author} {\bibfnamefont {G.}~\bibnamefont {Radicioni}}, \bibinfo {author} {\bibfnamefont {S.~S.}\ \bibnamefont {Livengood}}, \bibinfo {author} {\bibfnamefont {T.}~\bibnamefont {Kumagai}}, \bibinfo {author} {\bibfnamefont {H.}~\bibnamefont {Dang}}, \bibinfo {author} {\bibfnamefont {A.}~\bibnamefont {Ceppe}}, \bibinfo {author} {\bibfnamefont {P.~W.}\ \bibnamefont {Clapp}}, \bibinfo {author} {\bibfnamefont {M.}~\bibnamefont {Tunney}}, \bibinfo {author} {\bibfnamefont {J.~S.}\ \bibnamefont {Elborn}}, \bibinfo {author} {\bibfnamefont {N.~G.}\ \bibnamefont {McElvaney}}, \bibinfo {author} {\bibfnamefont {M.~S.}\ \bibnamefont {Muhlebach}}, \bibinfo {author} {\bibfnamefont {R.~C.}\ \bibnamefont {Boucher}}, \bibinfo {author} {\bibfnamefont {M.}~\bibnamefont {Tiemeyer}}, \bibinfo {author} {\bibfnamefont {M.~C.}\ \bibnamefont {Wolfgang}},\ and\ \bibinfo {author}
  {\bibfnamefont {M.}~\bibnamefont {Kesimer}},\ }\bibfield  {title} {\bibinfo {title} {Cystic fibrosis airway mucus hyperconcentration produces a vicious cycle of mucin, pathogen, and inflammatory interactions that promotes disease persistence},\ }\href@noop {} {\bibfield  {journal} {\bibinfo  {journal} {American Journal of Respiratory Cell and Molecular Biology}\ }\textbf {\bibinfo {volume} {67}},\ \bibinfo {pages} {253} (\bibinfo {year} {2022})}\BibitemShut {NoStop}%
\bibitem [{\citenamefont {Leith}(1968)}]{Leith_1968}%
  \BibitemOpen
  \bibfield  {author} {\bibinfo {author} {\bibfnamefont {D.~E.}\ \bibnamefont {Leith}},\ }\bibfield  {title} {\bibinfo {title} {Cough},\ }\href@noop {} {\bibfield  {journal} {\bibinfo  {journal} {Physical Therapy}\ }\textbf {\bibinfo {volume} {48}},\ \bibinfo {pages} {439} (\bibinfo {year} {1968})}\BibitemShut {NoStop}%
\bibitem [{\citenamefont {King}\ \emph {et~al.}(1985)\citenamefont {King}, \citenamefont {Brock},\ and\ \citenamefont {Lundell}}]{King_1985}%
  \BibitemOpen
  \bibfield  {author} {\bibinfo {author} {\bibfnamefont {M.}~\bibnamefont {King}}, \bibinfo {author} {\bibfnamefont {G.}~\bibnamefont {Brock}},\ and\ \bibinfo {author} {\bibfnamefont {C.}~\bibnamefont {Lundell}},\ }\bibfield  {title} {\bibinfo {title} {Clearance of mucus by simulated cough},\ }\href@noop {} {\bibfield  {journal} {\bibinfo  {journal} {Journal of Applied Physiology}\ }\textbf {\bibinfo {volume} {58}},\ \bibinfo {pages} {1776} (\bibinfo {year} {1985})}\BibitemShut {NoStop}%
\bibitem [{\citenamefont {Leith}(1985)}]{Leith_1985}%
  \BibitemOpen
  \bibfield  {author} {\bibinfo {author} {\bibfnamefont {D.~E.}\ \bibnamefont {Leith}},\ }\bibfield  {title} {\bibinfo {title} {The development of cough},\ }\href@noop {} {\bibfield  {journal} {\bibinfo  {journal} {American Review of Respiratory Disease}\ }\textbf {\bibinfo {volume} {131}},\ \bibinfo {pages} {S39} (\bibinfo {year} {1985})}\BibitemShut {NoStop}%
\bibitem [{\citenamefont {King}(1987)}]{King_1987}%
  \BibitemOpen
  \bibfield  {author} {\bibinfo {author} {\bibfnamefont {M.}~\bibnamefont {King}},\ }\bibfield  {title} {\bibinfo {title} {The role of mucus viscoelasticity in cough clearance},\ }\href@noop {} {\bibfield  {journal} {\bibinfo  {journal} {Biorheology}\ }\textbf {\bibinfo {volume} {24}},\ \bibinfo {pages} {589} (\bibinfo {year} {1987})}\BibitemShut {NoStop}%
\bibitem [{\citenamefont {Zahm}\ \emph {et~al.}(1986)\citenamefont {Zahm}, \citenamefont {Puchelle}, \citenamefont {Duvivier},\ and\ \citenamefont {Didelon}}]{Zahm_1986}%
  \BibitemOpen
  \bibfield  {author} {\bibinfo {author} {\bibfnamefont {J.~M.}\ \bibnamefont {Zahm}}, \bibinfo {author} {\bibfnamefont {E.}~\bibnamefont {Puchelle}}, \bibinfo {author} {\bibfnamefont {C.}~\bibnamefont {Duvivier}},\ and\ \bibinfo {author} {\bibfnamefont {J.}~\bibnamefont {Didelon}},\ }\bibfield  {title} {\bibinfo {title} {Spinability of respiratory mucous. validation of a new apparatus: the filancemeter},\ }\href@noop {} {\bibfield  {journal} {\bibinfo  {journal} {Bulletin europeen de physiopathologie respiratoire}\ }\textbf {\bibinfo {volume} {22}},\ \bibinfo {pages} {609} (\bibinfo {year} {1986})}\BibitemShut {NoStop}%
\bibitem [{\citenamefont {Agarwal}\ \emph {et~al.}(1994)\citenamefont {Agarwal}, \citenamefont {King},\ and\ \citenamefont {Shukla}}]{Agarwal_1994}%
  \BibitemOpen
  \bibfield  {author} {\bibinfo {author} {\bibfnamefont {M.}~\bibnamefont {Agarwal}}, \bibinfo {author} {\bibfnamefont {M.}~\bibnamefont {King}},\ and\ \bibinfo {author} {\bibfnamefont {J.~B.}\ \bibnamefont {Shukla}},\ }\bibfield  {title} {\bibinfo {title} {Mucous gel transport in a simulated cough machine: Effects of longitudinal grooves representing spacings between arrays of cilia},\ }\href@noop {} {\bibfield  {journal} {\bibinfo  {journal} {Biorheology}\ }\textbf {\bibinfo {volume} {31}},\ \bibinfo {pages} {11} (\bibinfo {year} {1994})}\BibitemShut {NoStop}%
\bibitem [{\citenamefont {Mahajan}\ \emph {et~al.}(1994)\citenamefont {Mahajan}, \citenamefont {Singh}, \citenamefont {Murty},\ and\ \citenamefont {Aitkenhead}}]{Mahajan_1994}%
  \BibitemOpen
  \bibfield  {author} {\bibinfo {author} {\bibfnamefont {R.~P.}\ \bibnamefont {Mahajan}}, \bibinfo {author} {\bibfnamefont {P.}~\bibnamefont {Singh}}, \bibinfo {author} {\bibfnamefont {G.~E.}\ \bibnamefont {Murty}},\ and\ \bibinfo {author} {\bibfnamefont {A.~R.}\ \bibnamefont {Aitkenhead}},\ }\bibfield  {title} {\bibinfo {title} {Relationship between expired lung volume, peak flow rate and peak velocity time during a voluntary cough manoeuvre},\ }\href {https://doi.org/10.1093/bja/72.3.298} {\bibfield  {journal} {\bibinfo  {journal} {British Journal of Anaesthesia}\ }\textbf {\bibinfo {volume} {72}},\ \bibinfo {pages} {298} (\bibinfo {year} {1994})}\BibitemShut {NoStop}%
\bibitem [{\citenamefont {Lauga}\ and\ \citenamefont {Hosoi}(2006)}]{Lauga_2006}%
  \BibitemOpen
  \bibfield  {author} {\bibinfo {author} {\bibfnamefont {E.}~\bibnamefont {Lauga}}\ and\ \bibinfo {author} {\bibfnamefont {A.~E.}\ \bibnamefont {Hosoi}},\ }\bibfield  {title} {\bibinfo {title} {Tuning gastropod locomotion: Modeling the influence of mucus rheology on the cost of crawling},\ }\href {https://doi.org/10.1063/1.2382591} {\bibfield  {journal} {\bibinfo  {journal} {Physics of Fluids}\ }\textbf {\bibinfo {volume} {18}},\ \bibinfo {pages} {113102} (\bibinfo {year} {2006})}\BibitemShut {NoStop}%
\bibitem [{\citenamefont {Gupta}\ \emph {et~al.}(2009)\citenamefont {Gupta}, \citenamefont {Lin},\ and\ \citenamefont {Chen}}]{Gupta_2009}%
  \BibitemOpen
  \bibfield  {author} {\bibinfo {author} {\bibfnamefont {J.~K.}\ \bibnamefont {Gupta}}, \bibinfo {author} {\bibfnamefont {C.~H.}\ \bibnamefont {Lin}},\ and\ \bibinfo {author} {\bibfnamefont {Q.}~\bibnamefont {Chen}},\ }\bibfield  {title} {\bibinfo {title} {Flow dynamics and characterization of a cough},\ }\href {https://doi.org/10.1111/j.1600-0668.2009.00619.x} {\bibfield  {journal} {\bibinfo  {journal} {Indoor Air}\ }\textbf {\bibinfo {volume} {19}},\ \bibinfo {pages} {517} (\bibinfo {year} {2009})}\BibitemShut {NoStop}%
\bibitem [{\citenamefont {Grotberg}(2011{\natexlab{b}})}]{Grotberg_2011a}%
  \BibitemOpen
  \bibfield  {author} {\bibinfo {author} {\bibfnamefont {J.~B.}\ \bibnamefont {Grotberg}},\ }\bibfield  {title} {\bibinfo {title} {Respiratory fluid mechanics},\ }\href {https://doi.org/10.1063/1.3517737} {\bibfield  {journal} {\bibinfo  {journal} {Physics of Fluids}\ }\textbf {\bibinfo {volume} {23}},\ \bibinfo {pages} {021301} (\bibinfo {year} {2011}{\natexlab{b}})}\BibitemShut {NoStop}%
\bibitem [{\citenamefont {Gupta}\ \emph {et~al.}(2011)\citenamefont {Gupta}, \citenamefont {Lin},\ and\ \citenamefont {Chen}}]{Gupta_2017}%
  \BibitemOpen
  \bibfield  {author} {\bibinfo {author} {\bibfnamefont {J.~K.}\ \bibnamefont {Gupta}}, \bibinfo {author} {\bibfnamefont {C.~H.}\ \bibnamefont {Lin}},\ and\ \bibinfo {author} {\bibfnamefont {Q.}~\bibnamefont {Chen}},\ }\bibfield  {title} {\bibinfo {title} {Transport of expiratory droplets in an aircraft cabin},\ }\href {https://doi.org/10.1111/j.1600-0668.2010.00676.x} {\bibfield  {journal} {\bibinfo  {journal} {Indoor Air}\ }\textbf {\bibinfo {volume} {21}},\ \bibinfo {pages} {3} (\bibinfo {year} {2011})}\BibitemShut {NoStop}%
\bibitem [{\citenamefont {Montenegro-Johnson}\ \emph {et~al.}(2013)\citenamefont {Montenegro-Johnson}, \citenamefont {Smith},\ and\ \citenamefont {Loghin}}]{Montenegro-Johnson_2013}%
  \BibitemOpen
  \bibfield  {author} {\bibinfo {author} {\bibfnamefont {T.~D.}\ \bibnamefont {Montenegro-Johnson}}, \bibinfo {author} {\bibfnamefont {D.~J.}\ \bibnamefont {Smith}},\ and\ \bibinfo {author} {\bibfnamefont {D.}~\bibnamefont {Loghin}},\ }\bibfield  {title} {\bibinfo {title} {Physics of rheologically enhanced propulsion: Different strokes in generalized stokes},\ }\href {https://doi.org/10.1063/1.4818640} {\bibfield  {journal} {\bibinfo  {journal} {Physics of Fluids}\ }\textbf {\bibinfo {volume} {25}},\ \bibinfo {pages} {081903} (\bibinfo {year} {2013})}\BibitemShut {NoStop}%
\bibitem [{\citenamefont {Yang}\ \emph {et~al.}(2017)\citenamefont {Yang}, \citenamefont {Li}, \citenamefont {Yan},\ and\ \citenamefont {Tu}}]{Yang_2017}%
  \BibitemOpen
  \bibfield  {author} {\bibinfo {author} {\bibfnamefont {L.}~\bibnamefont {Yang}}, \bibinfo {author} {\bibfnamefont {X.}~\bibnamefont {Li}}, \bibinfo {author} {\bibfnamefont {Y.}~\bibnamefont {Yan}},\ and\ \bibinfo {author} {\bibfnamefont {J.}~\bibnamefont {Tu}},\ }\bibfield  {title} {\bibinfo {title} {Effects of cough-jet on airflow and contaminant transport in an airliner cabin section},\ }\href {https://doi.org/10.1177/1757482X17746920} {\bibfield  {journal} {\bibinfo  {journal} {Journal of Computational Multiphase Flows}\ }\textbf {\bibinfo {volume} {10}},\ \bibinfo {pages} {72} (\bibinfo {year} {2017})}\BibitemShut {NoStop}%
\bibitem [{\citenamefont {Dudalski}\ \emph {et~al.}(2020)\citenamefont {Dudalski}, \citenamefont {Mohamed}, \citenamefont {Mubareka}, \citenamefont {Bi}, \citenamefont {Zhang},\ and\ \citenamefont {Savory}}]{Dudalski_2020}%
  \BibitemOpen
  \bibfield  {author} {\bibinfo {author} {\bibfnamefont {N.}~\bibnamefont {Dudalski}}, \bibinfo {author} {\bibfnamefont {A.}~\bibnamefont {Mohamed}}, \bibinfo {author} {\bibfnamefont {S.}~\bibnamefont {Mubareka}}, \bibinfo {author} {\bibfnamefont {R.}~\bibnamefont {Bi}}, \bibinfo {author} {\bibfnamefont {C.}~\bibnamefont {Zhang}},\ and\ \bibinfo {author} {\bibfnamefont {E.}~\bibnamefont {Savory}},\ }\bibfield  {title} {\bibinfo {title} {Experimental investigation of far-field human cough airflows from healthy and influenza-infected subjects},\ }\href {https://doi.org/10.1111/ina.12680} {\bibfield  {journal} {\bibinfo  {journal} {Indoor Air}\ }\textbf {\bibinfo {volume} {30}},\ \bibinfo {pages} {966} (\bibinfo {year} {2020})}\BibitemShut {NoStop}%
\bibitem [{\citenamefont {Feng}\ \emph {et~al.}(2020)\citenamefont {Feng}, \citenamefont {Marchal}, \citenamefont {Sperry},\ and\ \citenamefont {Yi}}]{Feng_2020}%
  \BibitemOpen
  \bibfield  {author} {\bibinfo {author} {\bibfnamefont {Y.}~\bibnamefont {Feng}}, \bibinfo {author} {\bibfnamefont {T.}~\bibnamefont {Marchal}}, \bibinfo {author} {\bibfnamefont {T.}~\bibnamefont {Sperry}},\ and\ \bibinfo {author} {\bibfnamefont {H.}~\bibnamefont {Yi}},\ }\bibfield  {title} {\bibinfo {title} {Influence of wind and relative humidity on the social distancing effectiveness to prevent covid-19 airborne transmission: A numerical study},\ }\bibfield  {journal} {\bibinfo  {journal} {Journal of Aerosol Science}\ }\textbf {\bibinfo {volume} {147}},\ \href {https://doi.org/10.1016/j.jaerosci.2020.105585} {10.1016/j.jaerosci.2020.105585} (\bibinfo {year} {2020})\BibitemShut {NoStop}%
\bibitem [{\citenamefont {Ren}\ \emph {et~al.}(2022)\citenamefont {Ren}, \citenamefont {Cai}, \citenamefont {Shi}, \citenamefont {Luo},\ and\ \citenamefont {Wang}}]{Ren_2022}%
  \BibitemOpen
  \bibfield  {author} {\bibinfo {author} {\bibfnamefont {S.}~\bibnamefont {Ren}}, \bibinfo {author} {\bibfnamefont {M.}~\bibnamefont {Cai}}, \bibinfo {author} {\bibfnamefont {Y.}~\bibnamefont {Shi}}, \bibinfo {author} {\bibfnamefont {Z.}~\bibnamefont {Luo}},\ and\ \bibinfo {author} {\bibfnamefont {T.}~\bibnamefont {Wang}},\ }\bibfield  {title} {\bibinfo {title} {Influence of cough airflow characteristics on respiratory mucus clearance},\ }\href {https://doi.org/10.1063/5.0088100} {\bibfield  {journal} {\bibinfo  {journal} {Physics of Fluids}\ }\textbf {\bibinfo {volume} {34}},\ \bibinfo {pages} {041911} (\bibinfo {year} {2022})}\BibitemShut {NoStop}%
\bibitem [{\citenamefont {Yi}\ \emph {et~al.}(2021)\citenamefont {Yi}, \citenamefont {Wang},\ and\ \citenamefont {Feng}}]{Yi_2021}%
  \BibitemOpen
  \bibfield  {author} {\bibinfo {author} {\bibfnamefont {H.}~\bibnamefont {Yi}}, \bibinfo {author} {\bibfnamefont {Q.}~\bibnamefont {Wang}},\ and\ \bibinfo {author} {\bibfnamefont {Y.}~\bibnamefont {Feng}},\ }\bibfield  {title} {\bibinfo {title} {Computational analysis of obstructive disease and cough intensity effects on the mucus transport and clearance in an idealized upper airway model using the volume of fluid method},\ }\href {https://doi.org/10.1063/5.0037764} {\bibfield  {journal} {\bibinfo  {journal} {Physics of Fluids}\ }\textbf {\bibinfo {volume} {33}},\ \bibinfo {pages} {021903} (\bibinfo {year} {2021})}\BibitemShut {NoStop}%
\bibitem [{\citenamefont {JoongKim}\ \emph {et~al.}(2020)\citenamefont {JoongKim}, \citenamefont {NyeokKim}, \citenamefont {JoonLee},\ and\ \citenamefont {JinSung}}]{Seung_2020}%
  \BibitemOpen
  \bibfield  {author} {\bibinfo {author} {\bibfnamefont {S.}~\bibnamefont {JoongKim}}, \bibinfo {author} {\bibfnamefont {H.}~\bibnamefont {NyeokKim}}, \bibinfo {author} {\bibfnamefont {S.}~\bibnamefont {JoonLee}},\ and\ \bibinfo {author} {\bibfnamefont {H.}~\bibnamefont {JinSung}},\ }\bibfield  {title} {\bibinfo {title} {A lubricant-infused slip surface for drag reduction},\ }\href {https://doi.org/10.1063/5.0018460} {\bibfield  {journal} {\bibinfo  {journal} {Physics of Fluids}\ }\textbf {\bibinfo {volume} {32}},\ \bibinfo {pages} {091901} (\bibinfo {year} {2020})}\BibitemShut {NoStop}%
\bibitem [{\citenamefont {Puchelle}\ \emph {et~al.}(1987)\citenamefont {Puchelle}, \citenamefont {Zahm},\ and\ \citenamefont {Quemada}}]{Puchelle_1987a}%
  \BibitemOpen
  \bibfield  {author} {\bibinfo {author} {\bibfnamefont {E.}~\bibnamefont {Puchelle}}, \bibinfo {author} {\bibfnamefont {J.}~\bibnamefont {Zahm}},\ and\ \bibinfo {author} {\bibfnamefont {D.}~\bibnamefont {Quemada}},\ }\bibfield  {title} {\bibinfo {title} {Rheological properties controlling mucociliary frequency and respiratory mucus transport},\ }\href@noop {} {\bibfield  {journal} {\bibinfo  {journal} {Biorheology}\ }\textbf {\bibinfo {volume} {24}},\ \bibinfo {pages} {557} (\bibinfo {year} {1987})}\BibitemShut {NoStop}%
\bibitem [{\citenamefont {App}\ \emph {et~al.}(1993)\citenamefont {App}, \citenamefont {Zayas},\ and\ \citenamefont {King}}]{App_1993}%
  \BibitemOpen
  \bibfield  {author} {\bibinfo {author} {\bibfnamefont {E.}~\bibnamefont {App}}, \bibinfo {author} {\bibfnamefont {J.}~\bibnamefont {Zayas}},\ and\ \bibinfo {author} {\bibfnamefont {M.}~\bibnamefont {King}},\ }\bibfield  {title} {\bibinfo {title} {Rheology of mucus and transepithelial potential difference: small airways versus trachea},\ }\href@noop {} {\bibfield  {journal} {\bibinfo  {journal} {European Respiratory Journal}\ }\textbf {\bibinfo {volume} {6}},\ \bibinfo {pages} {67} (\bibinfo {year} {1993})}\BibitemShut {NoStop}%
\bibitem [{\citenamefont {Dawson}\ \emph {et~al.}(2003)\citenamefont {Dawson}, \citenamefont {Wirtz},\ and\ \citenamefont {Hanes}}]{Dawson_2003}%
  \BibitemOpen
  \bibfield  {author} {\bibinfo {author} {\bibfnamefont {M.}~\bibnamefont {Dawson}}, \bibinfo {author} {\bibfnamefont {D.}~\bibnamefont {Wirtz}},\ and\ \bibinfo {author} {\bibfnamefont {J.}~\bibnamefont {Hanes}},\ }\bibfield  {title} {\bibinfo {title} {Enhanced viscoelasticity of human cystic fibrotic sputum correlates with increasing microheterogeneity in particle transport},\ }\href@noop {} {\bibfield  {journal} {\bibinfo  {journal} {Journal of Biological Chemistry}\ }\textbf {\bibinfo {volume} {278}},\ \bibinfo {pages} {50393} (\bibinfo {year} {2003})}\BibitemShut {NoStop}%
\bibitem [{\citenamefont {Puchelle}\ \emph {et~al.}(1985)\citenamefont {Puchelle}, \citenamefont {Zahm}, \citenamefont {Duvivier}, \citenamefont {Didelon}, \citenamefont {Jacquot},\ and\ \citenamefont {Quemada}}]{Puchelle_1985}%
  \BibitemOpen
  \bibfield  {author} {\bibinfo {author} {\bibfnamefont {E.}~\bibnamefont {Puchelle}}, \bibinfo {author} {\bibfnamefont {J.~M.}\ \bibnamefont {Zahm}}, \bibinfo {author} {\bibfnamefont {C.}~\bibnamefont {Duvivier}}, \bibinfo {author} {\bibfnamefont {J.}~\bibnamefont {Didelon}}, \bibinfo {author} {\bibfnamefont {J.}~\bibnamefont {Jacquot}},\ and\ \bibinfo {author} {\bibfnamefont {D.}~\bibnamefont {Quemada}},\ }\bibfield  {title} {\bibinfo {title} {Elasto-thixotropic properties of bronchial mucus and polymer analogs},\ }\href@noop {} {\bibfield  {journal} {\bibinfo  {journal} {Biorheology}\ }\textbf {\bibinfo {volume} {22}},\ \bibinfo {pages} {415} (\bibinfo {year} {1985})}\BibitemShut {NoStop}%
\bibitem [{\citenamefont {Girod}\ \emph {et~al.}(1992)\citenamefont {Girod}, \citenamefont {Zahm}, \citenamefont {Plotkowski}, \citenamefont {Beck},\ and\ \citenamefont {Puchelle}}]{Girod_1992}%
  \BibitemOpen
  \bibfield  {author} {\bibinfo {author} {\bibfnamefont {S.}~\bibnamefont {Girod}}, \bibinfo {author} {\bibfnamefont {J.}~\bibnamefont {Zahm}}, \bibinfo {author} {\bibfnamefont {C.}~\bibnamefont {Plotkowski}}, \bibinfo {author} {\bibfnamefont {G.}~\bibnamefont {Beck}},\ and\ \bibinfo {author} {\bibfnamefont {E.}~\bibnamefont {Puchelle}},\ }\bibfield  {title} {\bibinfo {title} {Role of the physiochemical properties of mucus in the protection of the respiratory epithelium},\ }\href@noop {} {\bibfield  {journal} {\bibinfo  {journal} {European Respiratory Journal}\ }\textbf {\bibinfo {volume} {5}},\ \bibinfo {pages} {477} (\bibinfo {year} {1992})}\BibitemShut {NoStop}%
\bibitem [{\citenamefont {Banerjee}\ \emph {et~al.}(2001)\citenamefont {Banerjee}, \citenamefont {Bellare},\ and\ \citenamefont {Puniyani}}]{Banerjee_2001}%
  \BibitemOpen
  \bibfield  {author} {\bibinfo {author} {\bibfnamefont {R.}~\bibnamefont {Banerjee}}, \bibinfo {author} {\bibfnamefont {J.~R.}\ \bibnamefont {Bellare}},\ and\ \bibinfo {author} {\bibfnamefont {R.~R.}\ \bibnamefont {Puniyani}},\ }\bibfield  {title} {\bibinfo {title} {Effect of phospholipid mixtures and surfactant formulations on rheology of polymeric gels, simulating mucus, at shear rates experienced in the tracheobronchial tree},\ }\href@noop {} {\bibfield  {journal} {\bibinfo  {journal} {Biochemical Engineering Journal}\ }\textbf {\bibinfo {volume} {7}},\ \bibinfo {pages} {195} (\bibinfo {year} {2001})}\BibitemShut {NoStop}%
\bibitem [{\citenamefont {Puchelle}\ \emph {et~al.}(1981)\citenamefont {Puchelle}, \citenamefont {Zahm},\ and\ \citenamefont {Aug}}]{Puchelle_1981}%
  \BibitemOpen
  \bibfield  {author} {\bibinfo {author} {\bibfnamefont {E.}~\bibnamefont {Puchelle}}, \bibinfo {author} {\bibfnamefont {J.-M.}\ \bibnamefont {Zahm}},\ and\ \bibinfo {author} {\bibfnamefont {F.}~\bibnamefont {Aug}},\ }\bibfield  {title} {\bibinfo {title} {Viscoelasticity, protein content and ciliary transport rate of sputum in patients with recurrent and chronic bronchitis},\ }\href@noop {} {\bibfield  {journal} {\bibinfo  {journal} {Biorheology}\ }\textbf {\bibinfo {volume} {18}},\ \bibinfo {pages} {659} (\bibinfo {year} {1981})}\BibitemShut {NoStop}%
\bibitem [{\citenamefont {Puchelle}\ \emph {et~al.}(1983)\citenamefont {Puchelle}, \citenamefont {Zahm},\ and\ \citenamefont {Duvivier}}]{Puchelle_1983}%
  \BibitemOpen
  \bibfield  {author} {\bibinfo {author} {\bibfnamefont {E.}~\bibnamefont {Puchelle}}, \bibinfo {author} {\bibfnamefont {J.~M.}\ \bibnamefont {Zahm}},\ and\ \bibinfo {author} {\bibfnamefont {C.}~\bibnamefont {Duvivier}},\ }\bibfield  {title} {\bibinfo {title} {Spinability of bronchial mucus. relationship with viscoelasticity and mucous transport properties},\ }\href@noop {} {\bibfield  {journal} {\bibinfo  {journal} {Biorheology}\ }\textbf {\bibinfo {volume} {20}},\ \bibinfo {pages} {239} (\bibinfo {year} {1983})}\BibitemShut {NoStop}%
\bibitem [{\citenamefont {Baconnais}\ \emph {et~al.}(1999)\citenamefont {Baconnais}, \citenamefont {Tirouvanziam}, \citenamefont {Zahm}, \citenamefont {Bentzmann}, \citenamefont {P{\'e}ault}, \citenamefont {Balossier},\ and\ \citenamefont {Puchelle}}]{Baconnais_1999}%
  \BibitemOpen
  \bibfield  {author} {\bibinfo {author} {\bibfnamefont {S.}~\bibnamefont {Baconnais}}, \bibinfo {author} {\bibfnamefont {R.}~\bibnamefont {Tirouvanziam}}, \bibinfo {author} {\bibfnamefont {J.-M.}\ \bibnamefont {Zahm}}, \bibinfo {author} {\bibfnamefont {S.~d.}\ \bibnamefont {Bentzmann}}, \bibinfo {author} {\bibfnamefont {B.}~\bibnamefont {P{\'e}ault}}, \bibinfo {author} {\bibfnamefont {G.}~\bibnamefont {Balossier}},\ and\ \bibinfo {author} {\bibfnamefont {E.}~\bibnamefont {Puchelle}},\ }\bibfield  {title} {\bibinfo {title} {Ion composition and rheology of airway liquid from cystic fibrosis fetal tracheal xenografts},\ }\href@noop {} {\bibfield  {journal} {\bibinfo  {journal} {American Journal of Respiratory Cell and Molecular Biology}\ }\textbf {\bibinfo {volume} {20}},\ \bibinfo {pages} {605} (\bibinfo {year} {1999})}\BibitemShut {NoStop}%
\bibitem [{\citenamefont {Tang}(2021)}]{Tang_2021}%
  \BibitemOpen
  \bibfield  {author} {\bibinfo {author} {\bibfnamefont {Q.}~\bibnamefont {Tang}},\ }\emph {\bibinfo {title} {Structural and Rheological Properties of Airway Mucus {\&} Counterion Condensation around Charged Nanoparticles}},\ \href@noop {} {\bibinfo {type} {Thesis}},\ \bibinfo  {school} {Chapel Hill, University of North Carolina} (\bibinfo {year} {2021})\BibitemShut {NoStop}%
\bibitem [{\citenamefont {Wong}\ \emph {et~al.}(1977)\citenamefont {Wong}, \citenamefont {Keens}, \citenamefont {Wannamaker}, \citenamefont {Crozier}, \citenamefont {Levison},\ and\ \citenamefont {Aspin}}]{Wong_1977}%
  \BibitemOpen
  \bibfield  {author} {\bibinfo {author} {\bibfnamefont {J.~W.}\ \bibnamefont {Wong}}, \bibinfo {author} {\bibfnamefont {T.~G.}\ \bibnamefont {Keens}}, \bibinfo {author} {\bibfnamefont {E.~M.}\ \bibnamefont {Wannamaker}}, \bibinfo {author} {\bibfnamefont {D.~N.}\ \bibnamefont {Crozier}}, \bibinfo {author} {\bibfnamefont {H.}~\bibnamefont {Levison}},\ and\ \bibinfo {author} {\bibfnamefont {N.}~\bibnamefont {Aspin}},\ }\bibfield  {title} {\bibinfo {title} {Effects of gravity on tracheal mucus transport rates in normal subjects and in patients with cystic fibrosis},\ }\href@noop {} {\bibfield  {journal} {\bibinfo  {journal} {Pediatrics}\ }\textbf {\bibinfo {volume} {60}},\ \bibinfo {pages} {146} (\bibinfo {year} {1977})}\BibitemShut {NoStop}%
\bibitem [{\citenamefont {Naylor}\ \emph {et~al.}(2006)\citenamefont {Naylor}, \citenamefont {McLean}, \citenamefont {Chow}, \citenamefont {Heard}, \citenamefont {Ting},\ and\ \citenamefont {Avolio}}]{Naylor_2006}%
  \BibitemOpen
  \bibfield  {author} {\bibinfo {author} {\bibfnamefont {J.~M.}\ \bibnamefont {Naylor}}, \bibinfo {author} {\bibfnamefont {A.}~\bibnamefont {McLean}}, \bibinfo {author} {\bibfnamefont {C.-M.}\ \bibnamefont {Chow}}, \bibinfo {author} {\bibfnamefont {R.}~\bibnamefont {Heard}}, \bibinfo {author} {\bibfnamefont {I.}~\bibnamefont {Ting}},\ and\ \bibinfo {author} {\bibfnamefont {A.}~\bibnamefont {Avolio}},\ }\bibfield  {title} {\bibinfo {title} {A modified postural drainage position produces less cardiovascular stress than a head-down position in patients with severe heart disease: A quasi-experimental study},\ }\href@noop {} {\bibfield  {journal} {\bibinfo  {journal} {Australian Journal of Physiotherapy}\ }\textbf {\bibinfo {volume} {52}},\ \bibinfo {pages} {201} (\bibinfo {year} {2006})}\BibitemShut {NoStop}%
\bibitem [{\citenamefont {Carpenter}\ \emph {et~al.}(2018)\citenamefont {Carpenter}, \citenamefont {Lynch}, \citenamefont {Cribb}, \citenamefont {Kylstra}, \citenamefont {Hill},\ and\ \citenamefont {Superfine}}]{Carpenter_2018}%
  \BibitemOpen
  \bibfield  {author} {\bibinfo {author} {\bibfnamefont {J.}~\bibnamefont {Carpenter}}, \bibinfo {author} {\bibfnamefont {S.~E.}\ \bibnamefont {Lynch}}, \bibinfo {author} {\bibfnamefont {J.~A.}\ \bibnamefont {Cribb}}, \bibinfo {author} {\bibfnamefont {S.}~\bibnamefont {Kylstra}}, \bibinfo {author} {\bibfnamefont {D.~B.}\ \bibnamefont {Hill}},\ and\ \bibinfo {author} {\bibfnamefont {R.}~\bibnamefont {Superfine}},\ }\bibfield  {title} {\bibinfo {title} {Buffer drains and mucus is transported upward in a tilted mucus clearance assay},\ }\href@noop {} {\bibfield  {journal} {\bibinfo  {journal} {American Journal of Physiology-Lung Cellular and Molecular Physiology}\ }\textbf {\bibinfo {volume} {315}},\ \bibinfo {pages} {L910} (\bibinfo {year} {2018})}\BibitemShut {NoStop}%
\bibitem [{\citenamefont {Ghanem}\ \emph {et~al.}(2021)\citenamefont {Ghanem}, \citenamefont {Roquefort}, \citenamefont {Ramel}, \citenamefont {Laurent}, \citenamefont {Haute}, \citenamefont {Le~Gall}, \citenamefont {Aubry},\ and\ \citenamefont {Montier}}]{Ghanem_2021}%
  \BibitemOpen
  \bibfield  {author} {\bibinfo {author} {\bibfnamefont {R.}~\bibnamefont {Ghanem}}, \bibinfo {author} {\bibfnamefont {P.}~\bibnamefont {Roquefort}}, \bibinfo {author} {\bibfnamefont {S.}~\bibnamefont {Ramel}}, \bibinfo {author} {\bibfnamefont {V.}~\bibnamefont {Laurent}}, \bibinfo {author} {\bibfnamefont {T.}~\bibnamefont {Haute}}, \bibinfo {author} {\bibfnamefont {T.}~\bibnamefont {Le~Gall}}, \bibinfo {author} {\bibfnamefont {T.}~\bibnamefont {Aubry}},\ and\ \bibinfo {author} {\bibfnamefont {T.}~\bibnamefont {Montier}},\ }\bibfield  {title} {\bibinfo {title} {Apparent yield stress of sputum as a relevant biomarker in cystic fibrosis},\ }\href@noop {} {\bibfield  {journal} {\bibinfo  {journal} {Cells}\ }\textbf {\bibinfo {volume} {10}},\ \bibinfo {pages} {3107} (\bibinfo {year} {2021})}\BibitemShut {NoStop}%
\bibitem [{\citenamefont {Broughton-Head}\ \emph {et~al.}(2007)\citenamefont {Broughton-Head}, \citenamefont {Shur}, \citenamefont {Carroll}, \citenamefont {Smith},\ and\ \citenamefont {Shute}}]{Broughton-Head_2007}%
  \BibitemOpen
  \bibfield  {author} {\bibinfo {author} {\bibfnamefont {V.~J.}\ \bibnamefont {Broughton-Head}}, \bibinfo {author} {\bibfnamefont {J.}~\bibnamefont {Shur}}, \bibinfo {author} {\bibfnamefont {M.~P.}\ \bibnamefont {Carroll}}, \bibinfo {author} {\bibfnamefont {J.~R.}\ \bibnamefont {Smith}},\ and\ \bibinfo {author} {\bibfnamefont {J.~K.}\ \bibnamefont {Shute}},\ }\bibfield  {title} {\bibinfo {title} {Unfractionated heparin reduces the elasticity of sputum from patients with cystic fibrosis},\ }\href@noop {} {\bibfield  {journal} {\bibinfo  {journal} {American Journal of Physiology-Lung Cellular and Molecular Physiology}\ }\textbf {\bibinfo {volume} {293}},\ \bibinfo {pages} {L1240} (\bibinfo {year} {2007})}\BibitemShut {NoStop}%
\bibitem [{\citenamefont {Lai}\ \emph {et~al.}(2007)\citenamefont {Lai}, \citenamefont {O'Hanlon}, \citenamefont {Harrold}, \citenamefont {Man}, \citenamefont {Wang}, \citenamefont {Cone},\ and\ \citenamefont {Hanes}}]{Lai_2007}%
  \BibitemOpen
  \bibfield  {author} {\bibinfo {author} {\bibfnamefont {S.~K.}\ \bibnamefont {Lai}}, \bibinfo {author} {\bibfnamefont {D.~E.}\ \bibnamefont {O'Hanlon}}, \bibinfo {author} {\bibfnamefont {S.}~\bibnamefont {Harrold}}, \bibinfo {author} {\bibfnamefont {S.~T.}\ \bibnamefont {Man}}, \bibinfo {author} {\bibfnamefont {Y.~Y.}\ \bibnamefont {Wang}}, \bibinfo {author} {\bibfnamefont {R.}~\bibnamefont {Cone}},\ and\ \bibinfo {author} {\bibfnamefont {J.}~\bibnamefont {Hanes}},\ }\bibfield  {title} {\bibinfo {title} {Rapid transport of large polymeric nanoparticles in fresh undiluted human mucus},\ }\href@noop {} {\bibfield  {journal} {\bibinfo  {journal} {Proc Natl Acad Sci U S A}\ }\textbf {\bibinfo {volume} {104}},\ \bibinfo {pages} {1482} (\bibinfo {year} {2007})}\BibitemShut {NoStop}%
\bibitem [{\citenamefont {Lafforgue}\ \emph {et~al.}(2017{\natexlab{b}})\citenamefont {Lafforgue}, \citenamefont {Bouguerra}, \citenamefont {Poncet}, \citenamefont {Seyssiecq}, \citenamefont {Favier},\ and\ \citenamefont {Elkoun}}]{Lafforgue_2017a}%
  \BibitemOpen
  \bibfield  {author} {\bibinfo {author} {\bibfnamefont {O.}~\bibnamefont {Lafforgue}}, \bibinfo {author} {\bibfnamefont {N.}~\bibnamefont {Bouguerra}}, \bibinfo {author} {\bibfnamefont {S.}~\bibnamefont {Poncet}}, \bibinfo {author} {\bibfnamefont {I.}~\bibnamefont {Seyssiecq}}, \bibinfo {author} {\bibfnamefont {J.}~\bibnamefont {Favier}},\ and\ \bibinfo {author} {\bibfnamefont {S.}~\bibnamefont {Elkoun}},\ }\bibfield  {title} {\bibinfo {title} {Thermo-physical properties of synthetic mucus for the study of airway clearance},\ }\href@noop {} {\bibfield  {journal} {\bibinfo  {journal} {Journal of Biomedical Materials Research Part A}\ }\textbf {\bibinfo {volume} {105}},\ \bibinfo {pages} {3025} (\bibinfo {year} {2017}{\natexlab{b}})}\BibitemShut {NoStop}%
\bibitem [{\citenamefont {Nielsen}\ \emph {et~al.}(2004)\citenamefont {Nielsen}, \citenamefont {Hvidt}, \citenamefont {Sheils},\ and\ \citenamefont {Janmey}}]{Nielsen_2004}%
  \BibitemOpen
  \bibfield  {author} {\bibinfo {author} {\bibfnamefont {H.}~\bibnamefont {Nielsen}}, \bibinfo {author} {\bibfnamefont {S.}~\bibnamefont {Hvidt}}, \bibinfo {author} {\bibfnamefont {C.~A.}\ \bibnamefont {Sheils}},\ and\ \bibinfo {author} {\bibfnamefont {P.~A.}\ \bibnamefont {Janmey}},\ }\bibfield  {title} {\bibinfo {title} {Elastic contributions dominate the viscoelastic properties of sputum from cystic fibrosis patients},\ }\href@noop {} {\bibfield  {journal} {\bibinfo  {journal} {Biophysical Chemistry}\ }\textbf {\bibinfo {volume} {112}},\ \bibinfo {pages} {193} (\bibinfo {year} {2004})}\BibitemShut {NoStop}%
\bibitem [{\citenamefont {Evans}\ \emph {et~al.}(2009{\natexlab{b}})\citenamefont {Evans}, \citenamefont {Tassieri}, \citenamefont {Auhl},\ and\ \citenamefont {Waigh}}]{Evans_2009}%
  \BibitemOpen
  \bibfield  {author} {\bibinfo {author} {\bibfnamefont {R.~M.~L.}\ \bibnamefont {Evans}}, \bibinfo {author} {\bibfnamefont {M.}~\bibnamefont {Tassieri}}, \bibinfo {author} {\bibfnamefont {D.}~\bibnamefont {Auhl}},\ and\ \bibinfo {author} {\bibfnamefont {T.~A.}\ \bibnamefont {Waigh}},\ }\bibfield  {title} {\bibinfo {title} {Direct conversion of rheological compliance measurements into storage and loss moduli},\ }\href@noop {} {\bibfield  {journal} {\bibinfo  {journal} {Physical Review E}\ }\textbf {\bibinfo {volume} {80}},\ \bibinfo {pages} {012501} (\bibinfo {year} {2009}{\natexlab{b}})}\BibitemShut {NoStop}%
\bibitem [{\citenamefont {Lafforgue}\ \emph {et~al.}(2018)\citenamefont {Lafforgue}, \citenamefont {Seyssiecq}, \citenamefont {Poncet},\ and\ \citenamefont {Favier}}]{Lafforgue_2018}%
  \BibitemOpen
  \bibfield  {author} {\bibinfo {author} {\bibfnamefont {O.}~\bibnamefont {Lafforgue}}, \bibinfo {author} {\bibfnamefont {I.}~\bibnamefont {Seyssiecq}}, \bibinfo {author} {\bibfnamefont {S.}~\bibnamefont {Poncet}},\ and\ \bibinfo {author} {\bibfnamefont {J.}~\bibnamefont {Favier}},\ }\bibfield  {title} {\bibinfo {title} {Rheological properties of synthetic mucus for airway clearance},\ }\href@noop {} {\bibfield  {journal} {\bibinfo  {journal} {Journal of Biomedical Materials Research Part A}\ }\textbf {\bibinfo {volume} {106}},\ \bibinfo {pages} {386} (\bibinfo {year} {2018})}\BibitemShut {NoStop}%
\bibitem [{\citenamefont {David}\ \emph {et~al.}(2018)\citenamefont {David}, \citenamefont {Robert}, \citenamefont {William}, \citenamefont {Eyad}, \citenamefont {Ian}, \citenamefont {Matthew}, \citenamefont {Nicholas}, \citenamefont {Matthew}, \citenamefont {Mehdi}, \citenamefont {Robert}, \citenamefont {Forest}, \citenamefont {Richard},\ and\ \citenamefont {Brian}}]{hill_2018}%
  \BibitemOpen
  \bibfield  {author} {\bibinfo {author} {\bibfnamefont {B.~H.}\ \bibnamefont {David}}, \bibinfo {author} {\bibfnamefont {F.~L.}\ \bibnamefont {Robert}}, \bibinfo {author} {\bibfnamefont {J.~K.}\ \bibnamefont {William}}, \bibinfo {author} {\bibfnamefont {A.}~\bibnamefont {Eyad}}, \bibinfo {author} {\bibfnamefont {C.~G.}\ \bibnamefont {Ian}}, \bibinfo {author} {\bibfnamefont {R.~M.}\ \bibnamefont {Matthew}}, \bibinfo {author} {\bibfnamefont {C.~F.}\ \bibnamefont {Nicholas}}, \bibinfo {author} {\bibfnamefont {C.}~\bibnamefont {Matthew}}, \bibinfo {author} {\bibfnamefont {H.}~\bibnamefont {Mehdi}}, \bibinfo {author} {\bibfnamefont {T.}~\bibnamefont {Robert}}, \bibinfo {author} {\bibfnamefont {M.~G.}\ \bibnamefont {Forest}}, \bibinfo {author} {\bibfnamefont {C.~B.}\ \bibnamefont {Richard}},\ and\ \bibinfo {author} {\bibfnamefont {B.}~\bibnamefont {Brian}},\ }\bibfield  {title} {\bibinfo {title} {Pathological mucus and impaired mucus clearance in cystic fibrosis patients result from increased concentration, not
  altered ph},\ }\href@noop {} {\bibfield  {journal} {\bibinfo  {journal} {European Respiratory Journal}\ }\textbf {\bibinfo {volume} {52}},\ \bibinfo {pages} {1801297} (\bibinfo {year} {2018})}\BibitemShut {NoStop}%
\bibitem [{\citenamefont {Ahonen}\ \emph {et~al.}(2019)\citenamefont {Ahonen}, \citenamefont {Hill},\ and\ \citenamefont {Schoenfisch}}]{Ahonen_2019}%
  \BibitemOpen
  \bibfield  {author} {\bibinfo {author} {\bibfnamefont {M.~J.~R.}\ \bibnamefont {Ahonen}}, \bibinfo {author} {\bibfnamefont {D.~B.}\ \bibnamefont {Hill}},\ and\ \bibinfo {author} {\bibfnamefont {M.~H.}\ \bibnamefont {Schoenfisch}},\ }\bibfield  {title} {\bibinfo {title} {Nitric oxide-releasing alginates as mucolytic agents},\ }\href@noop {} {\bibfield  {journal} {\bibinfo  {journal} {ACS Biomaterials Science {\&} Engineering}\ }\textbf {\bibinfo {volume} {5}},\ \bibinfo {pages} {3409} (\bibinfo {year} {2019})}\BibitemShut {NoStop}%
\bibitem [{\citenamefont {Rouillard}\ \emph {et~al.}(2020)\citenamefont {Rouillard}, \citenamefont {Hill},\ and\ \citenamefont {Schoenfisch}}]{Rouillard_2020}%
  \BibitemOpen
  \bibfield  {author} {\bibinfo {author} {\bibfnamefont {K.~R.}\ \bibnamefont {Rouillard}}, \bibinfo {author} {\bibfnamefont {D.~B.}\ \bibnamefont {Hill}},\ and\ \bibinfo {author} {\bibfnamefont {M.~H.}\ \bibnamefont {Schoenfisch}},\ }\bibfield  {title} {\bibinfo {title} {Antibiofilm and mucolytic action of nitric oxide delivered via gas or macromolecular donor using in vitro and ex vivo models},\ }\href@noop {} {\bibfield  {journal} {\bibinfo  {journal} {Journal of Cystic Fibrosis}\ }\textbf {\bibinfo {volume} {19}},\ \bibinfo {pages} {1004} (\bibinfo {year} {2020})}\BibitemShut {NoStop}%
\bibitem [{\citenamefont {Celli}\ \emph {et~al.}(2007)\citenamefont {Celli}, \citenamefont {Turner}, \citenamefont {Afdhal}, \citenamefont {Ewoldt}, \citenamefont {McKinley}, \citenamefont {Bansil},\ and\ \citenamefont {Erramilli}}]{Celli_2007}%
  \BibitemOpen
  \bibfield  {author} {\bibinfo {author} {\bibfnamefont {J.~P.}\ \bibnamefont {Celli}}, \bibinfo {author} {\bibfnamefont {B.~S.}\ \bibnamefont {Turner}}, \bibinfo {author} {\bibfnamefont {N.~H.}\ \bibnamefont {Afdhal}}, \bibinfo {author} {\bibfnamefont {R.~H.}\ \bibnamefont {Ewoldt}}, \bibinfo {author} {\bibfnamefont {G.~H.}\ \bibnamefont {McKinley}}, \bibinfo {author} {\bibfnamefont {R.}~\bibnamefont {Bansil}},\ and\ \bibinfo {author} {\bibfnamefont {S.}~\bibnamefont {Erramilli}},\ }\bibfield  {title} {\bibinfo {title} {Rheology of gastric mucin exhibits a ph-dependent sol-gel transition},\ }\href@noop {} {\bibfield  {journal} {\bibinfo  {journal} {Biomacromolecules}\ }\textbf {\bibinfo {volume} {8}},\ \bibinfo {pages} {1580} (\bibinfo {year} {2007})}\BibitemShut {NoStop}%
\bibitem [{\citenamefont {Mellnik}\ \emph {et~al.}(2014)\citenamefont {Mellnik}, \citenamefont {Vasquez}, \citenamefont {McKinley}, \citenamefont {Witten}, \citenamefont {Hill},\ and\ \citenamefont {Forest}}]{Mellnik_2014}%
  \BibitemOpen
  \bibfield  {author} {\bibinfo {author} {\bibfnamefont {J.}~\bibnamefont {Mellnik}}, \bibinfo {author} {\bibfnamefont {P.~A.}\ \bibnamefont {Vasquez}}, \bibinfo {author} {\bibfnamefont {S.~A.}\ \bibnamefont {McKinley}}, \bibinfo {author} {\bibfnamefont {J.}~\bibnamefont {Witten}}, \bibinfo {author} {\bibfnamefont {D.~B.}\ \bibnamefont {Hill}},\ and\ \bibinfo {author} {\bibfnamefont {M.~G.}\ \bibnamefont {Forest}},\ }\bibfield  {title} {\bibinfo {title} {Micro-heterogeneity metrics for diffusion in soft matter},\ }\href@noop {} {\bibfield  {journal} {\bibinfo  {journal} {Soft Matter}\ }\textbf {\bibinfo {volume} {10}},\ \bibinfo {pages} {7781} (\bibinfo {year} {2014})}\BibitemShut {NoStop}%
\bibitem [{\citenamefont {Caicedo}\ and\ \citenamefont {Perilla}(2015)}]{Caicedo_2015}%
  \BibitemOpen
  \bibfield  {author} {\bibinfo {author} {\bibfnamefont {J.}~\bibnamefont {Caicedo}}\ and\ \bibinfo {author} {\bibfnamefont {J.~E.}\ \bibnamefont {Perilla}},\ }\bibfield  {title} {\bibinfo {title} {Effect of ph on the rheological response of reconstituted gastric mucin},\ }\href@noop {} {\bibfield  {journal} {\bibinfo  {journal} {Ingenier{\'i}a e Investigaci{\'o}n}\ }\textbf {\bibinfo {volume} {35}},\ \bibinfo {pages} {43} (\bibinfo {year} {2015})}\BibitemShut {NoStop}%
\bibitem [{\citenamefont {Philippe}\ \emph {et~al.}(2017)\citenamefont {Philippe}, \citenamefont {Cipelletti},\ and\ \citenamefont {Larobina}}]{Philippe_2017}%
  \BibitemOpen
  \bibfield  {author} {\bibinfo {author} {\bibfnamefont {A.-M.}\ \bibnamefont {Philippe}}, \bibinfo {author} {\bibfnamefont {L.}~\bibnamefont {Cipelletti}},\ and\ \bibinfo {author} {\bibfnamefont {D.}~\bibnamefont {Larobina}},\ }\bibfield  {title} {\bibinfo {title} {Mucus as an arrested phase separation gel},\ }\href@noop {} {\bibfield  {journal} {\bibinfo  {journal} {Macromolecules}\ }\textbf {\bibinfo {volume} {50}},\ \bibinfo {pages} {8221} (\bibinfo {year} {2017})}\BibitemShut {NoStop}%
\bibitem [{\citenamefont {Brinkmann}\ \emph {et~al.}(2004)\citenamefont {Brinkmann}, \citenamefont {Reichard}, \citenamefont {Goosmann}, \citenamefont {Fauler}, \citenamefont {Uhlemann}, \citenamefont {Weiss}, \citenamefont {Weinrauch},\ and\ \citenamefont {Zychlinsky}}]{Brinkmann_2004}%
  \BibitemOpen
  \bibfield  {author} {\bibinfo {author} {\bibfnamefont {V.}~\bibnamefont {Brinkmann}}, \bibinfo {author} {\bibfnamefont {U.}~\bibnamefont {Reichard}}, \bibinfo {author} {\bibfnamefont {C.}~\bibnamefont {Goosmann}}, \bibinfo {author} {\bibfnamefont {B.}~\bibnamefont {Fauler}}, \bibinfo {author} {\bibfnamefont {Y.}~\bibnamefont {Uhlemann}}, \bibinfo {author} {\bibfnamefont {D.~S.}\ \bibnamefont {Weiss}}, \bibinfo {author} {\bibfnamefont {Y.}~\bibnamefont {Weinrauch}},\ and\ \bibinfo {author} {\bibfnamefont {A.}~\bibnamefont {Zychlinsky}},\ }\bibfield  {title} {\bibinfo {title} {Neutrophil extracellular traps kill bacteria},\ }\href@noop {} {\bibfield  {journal} {\bibinfo  {journal} {Science}\ }\textbf {\bibinfo {volume} {303}},\ \bibinfo {pages} {1532} (\bibinfo {year} {2004})}\BibitemShut {NoStop}%
\bibitem [{\citenamefont {Yuan}\ \emph {et~al.}(2015)\citenamefont {Yuan}, \citenamefont {Hollinger}, \citenamefont {Lachowicz-Scroggins}, \citenamefont {Kerr}, \citenamefont {Dunican}, \citenamefont {Daniel}, \citenamefont {Ghosh}, \citenamefont {Erzurum}, \citenamefont {Willard}, \citenamefont {Hazen}, \citenamefont {Huang}, \citenamefont {Carrington}, \citenamefont {Oscarson},\ and\ \citenamefont {Fahy}}]{Yuan_2015}%
  \BibitemOpen
  \bibfield  {author} {\bibinfo {author} {\bibfnamefont {S.}~\bibnamefont {Yuan}}, \bibinfo {author} {\bibfnamefont {M.}~\bibnamefont {Hollinger}}, \bibinfo {author} {\bibfnamefont {M.~E.}\ \bibnamefont {Lachowicz-Scroggins}}, \bibinfo {author} {\bibfnamefont {S.~C.}\ \bibnamefont {Kerr}}, \bibinfo {author} {\bibfnamefont {E.~M.}\ \bibnamefont {Dunican}}, \bibinfo {author} {\bibfnamefont {B.~M.}\ \bibnamefont {Daniel}}, \bibinfo {author} {\bibfnamefont {S.}~\bibnamefont {Ghosh}}, \bibinfo {author} {\bibfnamefont {S.~C.}\ \bibnamefont {Erzurum}}, \bibinfo {author} {\bibfnamefont {B.}~\bibnamefont {Willard}}, \bibinfo {author} {\bibfnamefont {S.~L.}\ \bibnamefont {Hazen}}, \bibinfo {author} {\bibfnamefont {X.}~\bibnamefont {Huang}}, \bibinfo {author} {\bibfnamefont {S.~D.}\ \bibnamefont {Carrington}}, \bibinfo {author} {\bibfnamefont {S.}~\bibnamefont {Oscarson}},\ and\ \bibinfo {author} {\bibfnamefont {J.~V.}\ \bibnamefont {Fahy}},\ }\bibfield  {title} {\bibinfo {title} {Oxidation increases mucin polymer
  cross-links to stiffen airway mucus gels},\ }\href@noop {} {\bibfield  {journal} {\bibinfo  {journal} {Science Translational Medicine}\ }\textbf {\bibinfo {volume} {7}},\ \bibinfo {pages} {276ra27} (\bibinfo {year} {2015})}\BibitemShut {NoStop}%
\bibitem [{\citenamefont {Esteban~Enjuto}\ \emph {et~al.}(2023)\citenamefont {Esteban~Enjuto}, \citenamefont {Robert~de Saint~Vincent}, \citenamefont {Maurin}, \citenamefont {Degano},\ and\ \citenamefont {Bodiguel}}]{Esteban-Enjuto_2023}%
  \BibitemOpen
  \bibfield  {author} {\bibinfo {author} {\bibfnamefont {L.}~\bibnamefont {Esteban~Enjuto}}, \bibinfo {author} {\bibfnamefont {M.}~\bibnamefont {Robert~de Saint~Vincent}}, \bibinfo {author} {\bibfnamefont {M.}~\bibnamefont {Maurin}}, \bibinfo {author} {\bibfnamefont {B.}~\bibnamefont {Degano}},\ and\ \bibinfo {author} {\bibfnamefont {H.}~\bibnamefont {Bodiguel}},\ }\bibfield  {title} {\bibinfo {title} {Sputum handling for rheology},\ }\href@noop {} {\bibfield  {journal} {\bibinfo  {journal} {Scientific Reports}\ }\textbf {\bibinfo {volume} {13}},\ \bibinfo {pages} {7695} (\bibinfo {year} {2023})}\BibitemShut {NoStop}%
\bibitem [{\citenamefont {Taylor}\ \emph {et~al.}(2005{\natexlab{a}})\citenamefont {Taylor}, \citenamefont {Draget}, \citenamefont {Pearson},\ and\ \citenamefont {Smidsr{\o}d}}]{Taylor_2005}%
  \BibitemOpen
  \bibfield  {author} {\bibinfo {author} {\bibfnamefont {C.}~\bibnamefont {Taylor}}, \bibinfo {author} {\bibfnamefont {K.~I.}\ \bibnamefont {Draget}}, \bibinfo {author} {\bibfnamefont {J.~P.}\ \bibnamefont {Pearson}},\ and\ \bibinfo {author} {\bibfnamefont {O.}~\bibnamefont {Smidsr{\o}d}},\ }\bibfield  {title} {\bibinfo {title} {Mucous systems show a novel mechanical response to applied deformation},\ }\href@noop {} {\bibfield  {journal} {\bibinfo  {journal} {Biomacromolecules}\ }\textbf {\bibinfo {volume} {6}},\ \bibinfo {pages} {1524} (\bibinfo {year} {2005}{\natexlab{a}})}\BibitemShut {NoStop}%
\bibitem [{\citenamefont {Taylor}\ \emph {et~al.}(2005{\natexlab{b}})\citenamefont {Taylor}, \citenamefont {Pearson}, \citenamefont {Draget}, \citenamefont {Dettmar},\ and\ \citenamefont {Smidsr{\o}d}}]{Taylor_2005a}%
  \BibitemOpen
  \bibfield  {author} {\bibinfo {author} {\bibfnamefont {C.}~\bibnamefont {Taylor}}, \bibinfo {author} {\bibfnamefont {J.~P.}\ \bibnamefont {Pearson}}, \bibinfo {author} {\bibfnamefont {K.~I.}\ \bibnamefont {Draget}}, \bibinfo {author} {\bibfnamefont {P.~W.}\ \bibnamefont {Dettmar}},\ and\ \bibinfo {author} {\bibfnamefont {O.}~\bibnamefont {Smidsr{\o}d}},\ }\bibfield  {title} {\bibinfo {title} {Rheological characterisation of mixed gels of mucin and alginate},\ }\href@noop {} {\bibfield  {journal} {\bibinfo  {journal} {Carbohydrate Polymers}\ }\textbf {\bibinfo {volume} {59}},\ \bibinfo {pages} {189} (\bibinfo {year} {2005}{\natexlab{b}})}\BibitemShut {NoStop}%
\bibitem [{\citenamefont {Kamkar}\ \emph {et~al.}(2022)\citenamefont {Kamkar}, \citenamefont {Salehiyan}, \citenamefont {Goudoulas}, \citenamefont {Abbasi}, \citenamefont {Saengow}, \citenamefont {Erfanian}, \citenamefont {Sadeghi}, \citenamefont {Natale}, \citenamefont {Rogers}, \citenamefont {Giacomin},\ and\ \citenamefont {Sundararaj}}]{Kamkar_2022}%
  \BibitemOpen
  \bibfield  {author} {\bibinfo {author} {\bibfnamefont {M.}~\bibnamefont {Kamkar}}, \bibinfo {author} {\bibfnamefont {R.}~\bibnamefont {Salehiyan}}, \bibinfo {author} {\bibfnamefont {T.~B.}\ \bibnamefont {Goudoulas}}, \bibinfo {author} {\bibfnamefont {M.}~\bibnamefont {Abbasi}}, \bibinfo {author} {\bibfnamefont {C.}~\bibnamefont {Saengow}}, \bibinfo {author} {\bibfnamefont {E.}~\bibnamefont {Erfanian}}, \bibinfo {author} {\bibfnamefont {S.}~\bibnamefont {Sadeghi}}, \bibinfo {author} {\bibfnamefont {G.}~\bibnamefont {Natale}}, \bibinfo {author} {\bibfnamefont {S.~A.}\ \bibnamefont {Rogers}}, \bibinfo {author} {\bibfnamefont {A.~J.}\ \bibnamefont {Giacomin}},\ and\ \bibinfo {author} {\bibfnamefont {U.}~\bibnamefont {Sundararaj}},\ }\bibfield  {title} {\bibinfo {title} {Large amplitude oscillatory shear flow: Microstructural assessment of polymeric systems},\ }\href@noop {} {\bibfield  {journal} {\bibinfo  {journal} {Progress in Polymer Science}\ }\textbf {\bibinfo {volume} {132}},\ \bibinfo {pages} {101580}
  (\bibinfo {year} {2022})}\BibitemShut {NoStop}%
\bibitem [{\citenamefont {Liu}\ \emph {et~al.}(2020)\citenamefont {Liu}, \citenamefont {Xiong}, \citenamefont {Nie},\ and\ \citenamefont {Yu}}]{Liu_2020}%
  \BibitemOpen
  \bibfield  {author} {\bibinfo {author} {\bibfnamefont {Z.}~\bibnamefont {Liu}}, \bibinfo {author} {\bibfnamefont {Z.}~\bibnamefont {Xiong}}, \bibinfo {author} {\bibfnamefont {Z.}~\bibnamefont {Nie}},\ and\ \bibinfo {author} {\bibfnamefont {W.}~\bibnamefont {Yu}},\ }\bibfield  {title} {\bibinfo {title} {Correlation between linear and nonlinear material functions under large amplitude oscillatory shear},\ }\href@noop {} {\bibfield  {journal} {\bibinfo  {journal} {Physics of Fluids}\ }\textbf {\bibinfo {volume} {32}} (\bibinfo {year} {2020})}\BibitemShut {NoStop}%
\bibitem [{\citenamefont {Schuster}\ \emph {et~al.}(2013)\citenamefont {Schuster}, \citenamefont {Suk}, \citenamefont {Woodworth},\ and\ \citenamefont {Hanes}}]{Schuster_2013}%
  \BibitemOpen
  \bibfield  {author} {\bibinfo {author} {\bibfnamefont {B.~S.}\ \bibnamefont {Schuster}}, \bibinfo {author} {\bibfnamefont {J.~S.}\ \bibnamefont {Suk}}, \bibinfo {author} {\bibfnamefont {G.~F.}\ \bibnamefont {Woodworth}},\ and\ \bibinfo {author} {\bibfnamefont {J.}~\bibnamefont {Hanes}},\ }\bibfield  {title} {\bibinfo {title} {Nanoparticle diffusion in respiratory mucus from humans without lung disease},\ }\href@noop {} {\bibfield  {journal} {\bibinfo  {journal} {Biomaterials}\ }\textbf {\bibinfo {volume} {34}},\ \bibinfo {pages} {3439} (\bibinfo {year} {2013})}\BibitemShut {NoStop}%
\bibitem [{\citenamefont {Duvivier}\ \emph {et~al.}(1984)\citenamefont {Duvivier}, \citenamefont {Didelon}, \citenamefont {Arnould}, \citenamefont {Zahm}, \citenamefont {Puchelle}, \citenamefont {Kopp},\ and\ \citenamefont {Obrecht}}]{Duvivier_1984}%
  \BibitemOpen
  \bibfield  {author} {\bibinfo {author} {\bibfnamefont {C.}~\bibnamefont {Duvivier}}, \bibinfo {author} {\bibfnamefont {J.}~\bibnamefont {Didelon}}, \bibinfo {author} {\bibfnamefont {J.~P.}\ \bibnamefont {Arnould}}, \bibinfo {author} {\bibfnamefont {J.~M.}\ \bibnamefont {Zahm}}, \bibinfo {author} {\bibfnamefont {E.}~\bibnamefont {Puchelle}}, \bibinfo {author} {\bibfnamefont {C.}~\bibnamefont {Kopp}},\ and\ \bibinfo {author} {\bibfnamefont {B.}~\bibnamefont {Obrecht}},\ }\bibfield  {title} {\bibinfo {title} {A new viscoelastometer for studying the rheological properties of bronchial mucus in clinical practice},\ }\href@noop {} {\bibfield  {journal} {\bibinfo  {journal} {Biorheology}\ }\textbf {\bibinfo {volume} {23}},\ \bibinfo {pages} {119} (\bibinfo {year} {1984})}\BibitemShut {NoStop}%
\bibitem [{\citenamefont {Radtke}\ \emph {et~al.}(2018)\citenamefont {Radtke}, \citenamefont {B{\"o}ni}, \citenamefont {Bohnacker}, \citenamefont {Maggi-Beba}, \citenamefont {Fischer}, \citenamefont {Kriemler}, \citenamefont {Benden},\ and\ \citenamefont {Dressel}}]{Radtke_2018}%
  \BibitemOpen
  \bibfield  {author} {\bibinfo {author} {\bibfnamefont {T.}~\bibnamefont {Radtke}}, \bibinfo {author} {\bibfnamefont {L.}~\bibnamefont {B{\"o}ni}}, \bibinfo {author} {\bibfnamefont {P.}~\bibnamefont {Bohnacker}}, \bibinfo {author} {\bibfnamefont {M.}~\bibnamefont {Maggi-Beba}}, \bibinfo {author} {\bibfnamefont {P.}~\bibnamefont {Fischer}}, \bibinfo {author} {\bibfnamefont {S.}~\bibnamefont {Kriemler}}, \bibinfo {author} {\bibfnamefont {C.}~\bibnamefont {Benden}},\ and\ \bibinfo {author} {\bibfnamefont {H.}~\bibnamefont {Dressel}},\ }\bibfield  {title} {\bibinfo {title} {Acute effects of combined exercise and oscillatory positive expiratory pressure therapy on sputum properties and lung diffusing capacity in cystic fibrosis: a randomized, controlled, crossover trial},\ }\href@noop {} {\bibfield  {journal} {\bibinfo  {journal} {BMC Pulmonary Medicine}\ }\textbf {\bibinfo {volume} {18}},\ \bibinfo {pages} {99} (\bibinfo {year} {2018})}\BibitemShut {NoStop}%
\bibitem [{\citenamefont {Taylor}\ \emph {et~al.}(2003)\citenamefont {Taylor}, \citenamefont {Allen}, \citenamefont {Dettmar},\ and\ \citenamefont {Pearson}}]{Taylor_2003}%
  \BibitemOpen
  \bibfield  {author} {\bibinfo {author} {\bibfnamefont {C.}~\bibnamefont {Taylor}}, \bibinfo {author} {\bibfnamefont {A.}~\bibnamefont {Allen}}, \bibinfo {author} {\bibfnamefont {P.~W.}\ \bibnamefont {Dettmar}},\ and\ \bibinfo {author} {\bibfnamefont {J.~P.}\ \bibnamefont {Pearson}},\ }\bibfield  {title} {\bibinfo {title} {The gel matrix of gastric mucus is maintained by a complex interplay of transient and nontransient associations},\ }\href@noop {} {\bibfield  {journal} {\bibinfo  {journal} {Biomacromolecules}\ }\textbf {\bibinfo {volume} {4}},\ \bibinfo {pages} {922} (\bibinfo {year} {2003})}\BibitemShut {NoStop}%
\bibitem [{\citenamefont {Larobina}\ \emph {et~al.}(2021)\citenamefont {Larobina}, \citenamefont {Pommella}, \citenamefont {Philippe}, \citenamefont {Nagazi},\ and\ \citenamefont {Cipelletti}}]{Larobina_2021}%
  \BibitemOpen
  \bibfield  {author} {\bibinfo {author} {\bibfnamefont {D.}~\bibnamefont {Larobina}}, \bibinfo {author} {\bibfnamefont {A.}~\bibnamefont {Pommella}}, \bibinfo {author} {\bibfnamefont {A.-M.}\ \bibnamefont {Philippe}}, \bibinfo {author} {\bibfnamefont {M.~Y.}\ \bibnamefont {Nagazi}},\ and\ \bibinfo {author} {\bibfnamefont {L.}~\bibnamefont {Cipelletti}},\ }\bibfield  {title} {\bibinfo {title} {Enhanced microscopic dynamics in mucus gels under a mechanical load in the linear viscoelastic regime},\ }\href@noop {} {\bibfield  {journal} {\bibinfo  {journal} {Proceedings of the National Academy of Sciences}\ }\textbf {\bibinfo {volume} {118}},\ \bibinfo {pages} {e2103995118} (\bibinfo {year} {2021})}\BibitemShut {NoStop}%
\bibitem [{\citenamefont {Hyun}\ \emph {et~al.}(2002)\citenamefont {Hyun}, \citenamefont {Kim}, \citenamefont {Ahn},\ and\ \citenamefont {Lee}}]{Hyun_2002}%
  \BibitemOpen
  \bibfield  {author} {\bibinfo {author} {\bibfnamefont {K.}~\bibnamefont {Hyun}}, \bibinfo {author} {\bibfnamefont {S.~H.}\ \bibnamefont {Kim}}, \bibinfo {author} {\bibfnamefont {K.~H.}\ \bibnamefont {Ahn}},\ and\ \bibinfo {author} {\bibfnamefont {S.~J.}\ \bibnamefont {Lee}},\ }\bibfield  {title} {\bibinfo {title} {Large amplitude oscillatory shear as a way to classify the complex fluids},\ }\href@noop {} {\bibfield  {journal} {\bibinfo  {journal} {Journal of Non-Newtonian Fluid Mechanics}\ }\textbf {\bibinfo {volume} {107}},\ \bibinfo {pages} {51} (\bibinfo {year} {2002})}\BibitemShut {NoStop}%
\bibitem [{\citenamefont {Ewoldt}\ \emph {et~al.}(2007)\citenamefont {Ewoldt}, \citenamefont {Clasen}, \citenamefont {Hosoi},\ and\ \citenamefont {McKinley}}]{Ewoldt_2007}%
  \BibitemOpen
  \bibfield  {author} {\bibinfo {author} {\bibfnamefont {R.~H.}\ \bibnamefont {Ewoldt}}, \bibinfo {author} {\bibfnamefont {C.}~\bibnamefont {Clasen}}, \bibinfo {author} {\bibfnamefont {A.~E.}\ \bibnamefont {Hosoi}},\ and\ \bibinfo {author} {\bibfnamefont {G.~H.}\ \bibnamefont {McKinley}},\ }\bibfield  {title} {\bibinfo {title} {Rheological fingerprinting of gastropod pedal mucus and synthetic complex fluids for biomimicking adhesive locomotion},\ }\href@noop {} {\bibfield  {journal} {\bibinfo  {journal} {Soft Matter}\ }\textbf {\bibinfo {volume} {3}},\ \bibinfo {pages} {634} (\bibinfo {year} {2007})}\BibitemShut {NoStop}%
\bibitem [{\citenamefont {Ewoldt}\ \emph {et~al.}(2008)\citenamefont {Ewoldt}, \citenamefont {Hosoi},\ and\ \citenamefont {McKinley}}]{Ewoldt_2008}%
  \BibitemOpen
  \bibfield  {author} {\bibinfo {author} {\bibfnamefont {R.~H.}\ \bibnamefont {Ewoldt}}, \bibinfo {author} {\bibfnamefont {A.}~\bibnamefont {Hosoi}},\ and\ \bibinfo {author} {\bibfnamefont {G.~H.}\ \bibnamefont {McKinley}},\ }\bibfield  {title} {\bibinfo {title} {New measures for characterizing nonlinear viscoelasticity in large amplitude oscillatory shear},\ }\href@noop {} {\bibfield  {journal} {\bibinfo  {journal} {Journal of Rheology (1978-present)}\ }\textbf {\bibinfo {volume} {52}},\ \bibinfo {pages} {1427} (\bibinfo {year} {2008})}\BibitemShut {NoStop}%
\bibitem [{\citenamefont {Quraishi}\ \emph {et~al.}(1998)\citenamefont {Quraishi}, \citenamefont {Jones},\ and\ \citenamefont {Mason}}]{Quraishi_1998}%
  \BibitemOpen
  \bibfield  {author} {\bibinfo {author} {\bibfnamefont {M.~S.}\ \bibnamefont {Quraishi}}, \bibinfo {author} {\bibfnamefont {N.~S.}\ \bibnamefont {Jones}},\ and\ \bibinfo {author} {\bibfnamefont {J.}~\bibnamefont {Mason}},\ }\bibfield  {title} {\bibinfo {title} {The rheology of nasal mucus: a review},\ }\href@noop {} {\bibfield  {journal} {\bibinfo  {journal} {Clinical Otolaryngology {\&} Allied Sciences}\ }\textbf {\bibinfo {volume} {23}},\ \bibinfo {pages} {403} (\bibinfo {year} {1998})}\BibitemShut {NoStop}%
\bibitem [{\citenamefont {McKinley}\ and\ \citenamefont {Sridhar}(2002)}]{McKinley_2002}%
  \BibitemOpen
  \bibfield  {author} {\bibinfo {author} {\bibfnamefont {G.~H.}\ \bibnamefont {McKinley}}\ and\ \bibinfo {author} {\bibfnamefont {T.}~\bibnamefont {Sridhar}},\ }\bibfield  {title} {\bibinfo {title} {Filament-stretching rheometry of complex fluids},\ }\href@noop {} {\bibfield  {journal} {\bibinfo  {journal} {Annual Review of Fluid Mechanics}\ }\textbf {\bibinfo {volume} {34}},\ \bibinfo {pages} {375} (\bibinfo {year} {2002})}\BibitemShut {NoStop}%
\bibitem [{\citenamefont {Anna}\ and\ \citenamefont {McKinley}(2001)}]{Anna_2001}%
  \BibitemOpen
  \bibfield  {author} {\bibinfo {author} {\bibfnamefont {S.~L.}\ \bibnamefont {Anna}}\ and\ \bibinfo {author} {\bibfnamefont {G.~H.}\ \bibnamefont {McKinley}},\ }\bibfield  {title} {\bibinfo {title} {Elasto-capillary thinning and breakup of model elastic liquids},\ }\href@noop {} {\bibfield  {journal} {\bibinfo  {journal} {Journal of Rheology}\ }\textbf {\bibinfo {volume} {45}},\ \bibinfo {pages} {115} (\bibinfo {year} {2001})}\BibitemShut {NoStop}%
\bibitem [{\citenamefont {Ahmad}\ \emph {et~al.}(2018)\citenamefont {Ahmad}, \citenamefont {Ritzoulis},\ and\ \citenamefont {Chen}}]{Ahmad_2018}%
  \BibitemOpen
  \bibfield  {author} {\bibinfo {author} {\bibfnamefont {M.}~\bibnamefont {Ahmad}}, \bibinfo {author} {\bibfnamefont {C.}~\bibnamefont {Ritzoulis}},\ and\ \bibinfo {author} {\bibfnamefont {J.}~\bibnamefont {Chen}},\ }\bibfield  {title} {\bibinfo {title} {Shear and extensional rheological characterisation of mucin solutions},\ }\href@noop {} {\bibfield  {journal} {\bibinfo  {journal} {Colloids and Surfaces B: Biointerfaces}\ }\textbf {\bibinfo {volume} {171}},\ \bibinfo {pages} {614} (\bibinfo {year} {2018})}\BibitemShut {NoStop}%
\bibitem [{\citenamefont {Tabatabaei}\ \emph {et~al.}(2015)\citenamefont {Tabatabaei}, \citenamefont {Jahromi}, \citenamefont {Webster}, \citenamefont {Williams}, \citenamefont {Holder}, \citenamefont {Lewis}, \citenamefont {Davies}, \citenamefont {Griffin}, \citenamefont {Ebden},\ and\ \citenamefont {Askill}}]{Tabatabaei_2015}%
  \BibitemOpen
  \bibfield  {author} {\bibinfo {author} {\bibfnamefont {S.}~\bibnamefont {Tabatabaei}}, \bibinfo {author} {\bibfnamefont {H.~T.}\ \bibnamefont {Jahromi}}, \bibinfo {author} {\bibfnamefont {M.~F.}\ \bibnamefont {Webster}}, \bibinfo {author} {\bibfnamefont {P.~R.}\ \bibnamefont {Williams}}, \bibinfo {author} {\bibfnamefont {A.~J.}\ \bibnamefont {Holder}}, \bibinfo {author} {\bibfnamefont {K.~E.}\ \bibnamefont {Lewis}}, \bibinfo {author} {\bibfnamefont {G.~A.}\ \bibnamefont {Davies}}, \bibinfo {author} {\bibfnamefont {L.}~\bibnamefont {Griffin}}, \bibinfo {author} {\bibfnamefont {P.}~\bibnamefont {Ebden}},\ and\ \bibinfo {author} {\bibfnamefont {C.}~\bibnamefont {Askill}},\ }\bibfield  {title} {\bibinfo {title} {A caber computational--experimental rheological study on human sputum},\ }\href@noop {} {\bibfield  {journal} {\bibinfo  {journal} {Journal of Non-Newtonian Fluid Mechanics}\ }\textbf {\bibinfo {volume} {222}},\ \bibinfo {pages} {272} (\bibinfo {year} {2015})}\BibitemShut {NoStop}%
\bibitem [{\citenamefont {Craster}\ and\ \citenamefont {Matar}(2000)}]{Craster_2000}%
  \BibitemOpen
  \bibfield  {author} {\bibinfo {author} {\bibfnamefont {R.~V.}\ \bibnamefont {Craster}}\ and\ \bibinfo {author} {\bibfnamefont {O.~K.}\ \bibnamefont {Matar}},\ }\bibfield  {title} {\bibinfo {title} {Surfactant transport on mucus films},\ }\href {https://doi.org/10.1017/S0022112000002317} {\bibfield  {journal} {\bibinfo  {journal} {Journal of Fluid Mechanics}\ }\textbf {\bibinfo {volume} {425}},\ \bibinfo {pages} {235} (\bibinfo {year} {2000})}\BibitemShut {NoStop}%
\bibitem [{\citenamefont {Chatelin}\ \emph {et~al.}(2017)\citenamefont {Chatelin}, \citenamefont {Anne-Archard}, \citenamefont {Murris-Espin}, \citenamefont {Thiriet},\ and\ \citenamefont {Poncet}}]{Chatelin_2017}%
  \BibitemOpen
  \bibfield  {author} {\bibinfo {author} {\bibfnamefont {R.}~\bibnamefont {Chatelin}}, \bibinfo {author} {\bibfnamefont {D.}~\bibnamefont {Anne-Archard}}, \bibinfo {author} {\bibfnamefont {M.}~\bibnamefont {Murris-Espin}}, \bibinfo {author} {\bibfnamefont {M.}~\bibnamefont {Thiriet}},\ and\ \bibinfo {author} {\bibfnamefont {P.}~\bibnamefont {Poncet}},\ }\bibfield  {title} {\bibinfo {title} {Numerical and experimental investigation of mucociliary clearance breakdown in cystic fibrosis},\ }\href@noop {} {\bibfield  {journal} {\bibinfo  {journal} {Journal of Biomechanics}\ }\textbf {\bibinfo {volume} {53}},\ \bibinfo {pages} {56} (\bibinfo {year} {2017})}\BibitemShut {NoStop}%
\bibitem [{\citenamefont {Smith}\ \emph {et~al.}(2007)\citenamefont {Smith}, \citenamefont {Gaffney},\ and\ \citenamefont {Blake}}]{Smith_2007}%
  \BibitemOpen
  \bibfield  {author} {\bibinfo {author} {\bibfnamefont {D.~J.}\ \bibnamefont {Smith}}, \bibinfo {author} {\bibfnamefont {E.~A.}\ \bibnamefont {Gaffney}},\ and\ \bibinfo {author} {\bibfnamefont {J.~R.}\ \bibnamefont {Blake}},\ }\bibfield  {title} {\bibinfo {title} {A viscoelastic traction layer model of muco-ciliary transport},\ }\href@noop {} {\bibfield  {journal} {\bibinfo  {journal} {Bulletin of Mathematical Biology}\ }\textbf {\bibinfo {volume} {69}},\ \bibinfo {pages} {289} (\bibinfo {year} {2007})}\BibitemShut {NoStop}%
\bibitem [{\citenamefont {Lukens}\ \emph {et~al.}(2010)\citenamefont {Lukens}, \citenamefont {Yang},\ and\ \citenamefont {Fauci}}]{Lukens_2010}%
  \BibitemOpen
  \bibfield  {author} {\bibinfo {author} {\bibfnamefont {S.}~\bibnamefont {Lukens}}, \bibinfo {author} {\bibfnamefont {X.}~\bibnamefont {Yang}},\ and\ \bibinfo {author} {\bibfnamefont {L.}~\bibnamefont {Fauci}},\ }\bibfield  {title} {\bibinfo {title} {Using lagrangian coherent structures to analyze fluid mixing by cilia},\ }\href@noop {} {\bibfield  {journal} {\bibinfo  {journal} {Chaos: An Interdisciplinary Journal of Nonlinear Science}\ }\textbf {\bibinfo {volume} {20}},\ \bibinfo {pages} {017511} (\bibinfo {year} {2010})}\BibitemShut {NoStop}%
\bibitem [{\citenamefont {Dillon}\ \emph {et~al.}(2007)\citenamefont {Dillon}, \citenamefont {Fauci}, \citenamefont {Omoto},\ and\ \citenamefont {Yang}}]{Dillon_2007}%
  \BibitemOpen
  \bibfield  {author} {\bibinfo {author} {\bibfnamefont {R.~H.}\ \bibnamefont {Dillon}}, \bibinfo {author} {\bibfnamefont {L.~J.}\ \bibnamefont {Fauci}}, \bibinfo {author} {\bibfnamefont {C.}~\bibnamefont {Omoto}},\ and\ \bibinfo {author} {\bibfnamefont {X.}~\bibnamefont {Yang}},\ }\bibfield  {title} {\bibinfo {title} {Fluid dynamic models of flagellar and ciliary beating},\ }\href {https://doi.org/https://doi.org/10.1196/annals.1389.016} {\bibfield  {journal} {\bibinfo  {journal} {Annals of the New York Academy of Sciences}\ }\textbf {\bibinfo {volume} {1101}},\ \bibinfo {pages} {494} (\bibinfo {year} {2007})}\BibitemShut {NoStop}%
\bibitem [{\citenamefont {Mitran}(2007)}]{Mitran_2007}%
  \BibitemOpen
  \bibfield  {author} {\bibinfo {author} {\bibfnamefont {S.~M.}\ \bibnamefont {Mitran}},\ }\bibfield  {title} {\bibinfo {title} {Metachronal wave formation in a model of pulmonary cilia},\ }\href@noop {} {\bibfield  {journal} {\bibinfo  {journal} {Computers {\&} Structures}\ }\textbf {\bibinfo {volume} {85}},\ \bibinfo {pages} {763} (\bibinfo {year} {2007})}\BibitemShut {NoStop}%
\bibitem [{\citenamefont {Sedaghat}\ \emph {et~al.}(2016{\natexlab{a}})\citenamefont {Sedaghat}, \citenamefont {Shahmardan}, \citenamefont {Norouzi},\ and\ \citenamefont {Heydari}}]{Sedaghat_2016}%
  \BibitemOpen
  \bibfield  {author} {\bibinfo {author} {\bibfnamefont {M.}~\bibnamefont {Sedaghat}}, \bibinfo {author} {\bibfnamefont {M.}~\bibnamefont {Shahmardan}}, \bibinfo {author} {\bibfnamefont {M.}~\bibnamefont {Norouzi}},\ and\ \bibinfo {author} {\bibfnamefont {M.}~\bibnamefont {Heydari}},\ }\bibfield  {title} {\bibinfo {title} {Effect of cilia beat frequency on muco-ciliary clearance},\ }\href@noop {} {\bibfield  {journal} {\bibinfo  {journal} {Journal of biomedical physics {\&} engineering}\ }\textbf {\bibinfo {volume} {6}},\ \bibinfo {pages} {265} (\bibinfo {year} {2016}{\natexlab{a}})}\BibitemShut {NoStop}%
\bibitem [{\citenamefont {Guo}\ and\ \citenamefont {Kanso}(2017)}]{Guo_2017}%
  \BibitemOpen
  \bibfield  {author} {\bibinfo {author} {\bibfnamefont {H.}~\bibnamefont {Guo}}\ and\ \bibinfo {author} {\bibfnamefont {E.}~\bibnamefont {Kanso}},\ }\bibfield  {title} {\bibinfo {title} {A computational study of mucociliary transport in healthy and diseased environments},\ }\href@noop {} {\bibfield  {journal} {\bibinfo  {journal} {European Journal of Computational Mechanics}\ }\textbf {\bibinfo {volume} {26}},\ \bibinfo {pages} {4} (\bibinfo {year} {2017})}\BibitemShut {NoStop}%
\bibitem [{\citenamefont {Sedaghat}\ \emph {et~al.}(2021)\citenamefont {Sedaghat}, \citenamefont {George},\ and\ \citenamefont {Abouali}}]{Sedaghat_2021}%
  \BibitemOpen
  \bibfield  {author} {\bibinfo {author} {\bibfnamefont {M.~H.}\ \bibnamefont {Sedaghat}}, \bibinfo {author} {\bibfnamefont {U.~Z.}\ \bibnamefont {George}},\ and\ \bibinfo {author} {\bibfnamefont {O.}~\bibnamefont {Abouali}},\ }\bibfield  {title} {\bibinfo {title} {A nonlinear viscoelastic model of mucociliary clearance},\ }\href@noop {} {\bibfield  {journal} {\bibinfo  {journal} {Rheologica Acta}\ }\textbf {\bibinfo {volume} {60}},\ \bibinfo {pages} {371} (\bibinfo {year} {2021})}\BibitemShut {NoStop}%
\bibitem [{\citenamefont {Squires}\ and\ \citenamefont {Mason}(2010)}]{Squires_2010}%
  \BibitemOpen
  \bibfield  {author} {\bibinfo {author} {\bibfnamefont {T.~M.}\ \bibnamefont {Squires}}\ and\ \bibinfo {author} {\bibfnamefont {T.~G.}\ \bibnamefont {Mason}},\ }\bibfield  {title} {\bibinfo {title} {Fluid mechanics of microrheology},\ }\href@noop {} {\bibfield  {journal} {\bibinfo  {journal} {Annual Review of Fluid Mechanics}\ }\textbf {\bibinfo {volume} {42}},\ \bibinfo {pages} {413} (\bibinfo {year} {2010})}\BibitemShut {NoStop}%
\bibitem [{\citenamefont {MacKintosh}\ and\ \citenamefont {Schmidt}(1999)}]{MacKintosh_1999}%
  \BibitemOpen
  \bibfield  {author} {\bibinfo {author} {\bibfnamefont {F.~C.}\ \bibnamefont {MacKintosh}}\ and\ \bibinfo {author} {\bibfnamefont {C.~F.}\ \bibnamefont {Schmidt}},\ }\bibfield  {title} {\bibinfo {title} {Microrheology},\ }\href@noop {} {\bibfield  {journal} {\bibinfo  {journal} {Current Opinion in Colloid {\&} Interface Science}\ }\textbf {\bibinfo {volume} {4}},\ \bibinfo {pages} {300} (\bibinfo {year} {1999})}\BibitemShut {NoStop}%
\bibitem [{\citenamefont {Waigh}(2016)}]{Waigh_2016}%
  \BibitemOpen
  \bibfield  {author} {\bibinfo {author} {\bibfnamefont {T.~A.}\ \bibnamefont {Waigh}},\ }\bibfield  {title} {\bibinfo {title} {Advances in the microrheology of complex fluids},\ }\href@noop {} {\bibfield  {journal} {\bibinfo  {journal} {Reports on Progress in Physics}\ }\textbf {\bibinfo {volume} {79}},\ \bibinfo {pages} {074601} (\bibinfo {year} {2016})}\BibitemShut {NoStop}%
\bibitem [{\citenamefont {Lai}\ \emph {et~al.}(2010)\citenamefont {Lai}, \citenamefont {Wang}, \citenamefont {Hida}, \citenamefont {Cone},\ and\ \citenamefont {Hanes}}]{Lai_2010}%
  \BibitemOpen
  \bibfield  {author} {\bibinfo {author} {\bibfnamefont {S.~K.}\ \bibnamefont {Lai}}, \bibinfo {author} {\bibfnamefont {Y.-Y.}\ \bibnamefont {Wang}}, \bibinfo {author} {\bibfnamefont {K.}~\bibnamefont {Hida}}, \bibinfo {author} {\bibfnamefont {R.}~\bibnamefont {Cone}},\ and\ \bibinfo {author} {\bibfnamefont {J.}~\bibnamefont {Hanes}},\ }\bibfield  {title} {\bibinfo {title} {Nanoparticles reveal that human cervicovaginal mucus is riddled with pores larger than viruses},\ }\href@noop {} {\bibfield  {journal} {\bibinfo  {journal} {Proceedings of the National Academy of Sciences}\ }\textbf {\bibinfo {volume} {107}},\ \bibinfo {pages} {598} (\bibinfo {year} {2010})}\BibitemShut {NoStop}%
\bibitem [{\citenamefont {Metzler}\ and\ \citenamefont {Klafter}(2000)}]{Metzler_2000}%
  \BibitemOpen
  \bibfield  {author} {\bibinfo {author} {\bibfnamefont {R.}~\bibnamefont {Metzler}}\ and\ \bibinfo {author} {\bibfnamefont {J.}~\bibnamefont {Klafter}},\ }\bibfield  {title} {\bibinfo {title} {The random walk's guide to anomalous diffusion: a fractional dynamics approach},\ }\href@noop {} {\bibfield  {journal} {\bibinfo  {journal} {Physics Reports}\ }\textbf {\bibinfo {volume} {339}},\ \bibinfo {pages} {1} (\bibinfo {year} {2000})}\BibitemShut {NoStop}%
\bibitem [{\citenamefont {Metzler}\ \emph {et~al.}(2014)\citenamefont {Metzler}, \citenamefont {Jeon}, \citenamefont {Cherstvy},\ and\ \citenamefont {Barkai}}]{Metzler_2014}%
  \BibitemOpen
  \bibfield  {author} {\bibinfo {author} {\bibfnamefont {R.}~\bibnamefont {Metzler}}, \bibinfo {author} {\bibfnamefont {J.-H.}\ \bibnamefont {Jeon}}, \bibinfo {author} {\bibfnamefont {A.~G.}\ \bibnamefont {Cherstvy}},\ and\ \bibinfo {author} {\bibfnamefont {E.}~\bibnamefont {Barkai}},\ }\bibfield  {title} {\bibinfo {title} {Anomalous diffusion models and their properties: non-stationarity, non-ergodicity, and ageing at the centenary of single particle tracking},\ }\href@noop {} {\bibfield  {journal} {\bibinfo  {journal} {Physical Chemistry Chemical Physics}\ }\textbf {\bibinfo {volume} {16}},\ \bibinfo {pages} {24128} (\bibinfo {year} {2014})}\BibitemShut {NoStop}%
\bibitem [{\citenamefont {Birket}\ \emph {et~al.}(2014)\citenamefont {Birket}, \citenamefont {Chu}, \citenamefont {Liu}, \citenamefont {Houser}, \citenamefont {Diephuis}, \citenamefont {Wilsterman}, \citenamefont {Dierksen}, \citenamefont {Mazur}, \citenamefont {Shastry}, \citenamefont {Li}, \citenamefont {Watson}, \citenamefont {Smith}, \citenamefont {Schuster}, \citenamefont {Hanes}, \citenamefont {Grizzle}, \citenamefont {Sorscher}, \citenamefont {Tearney},\ and\ \citenamefont {Rowe}}]{Birket_2014}%
  \BibitemOpen
  \bibfield  {author} {\bibinfo {author} {\bibfnamefont {S.~E.}\ \bibnamefont {Birket}}, \bibinfo {author} {\bibfnamefont {K.~K.}\ \bibnamefont {Chu}}, \bibinfo {author} {\bibfnamefont {L.}~\bibnamefont {Liu}}, \bibinfo {author} {\bibfnamefont {G.~H.}\ \bibnamefont {Houser}}, \bibinfo {author} {\bibfnamefont {B.~J.}\ \bibnamefont {Diephuis}}, \bibinfo {author} {\bibfnamefont {E.~J.}\ \bibnamefont {Wilsterman}}, \bibinfo {author} {\bibfnamefont {G.}~\bibnamefont {Dierksen}}, \bibinfo {author} {\bibfnamefont {M.}~\bibnamefont {Mazur}}, \bibinfo {author} {\bibfnamefont {S.}~\bibnamefont {Shastry}}, \bibinfo {author} {\bibfnamefont {Y.}~\bibnamefont {Li}}, \bibinfo {author} {\bibfnamefont {J.~D.}\ \bibnamefont {Watson}}, \bibinfo {author} {\bibfnamefont {A.~T.}\ \bibnamefont {Smith}}, \bibinfo {author} {\bibfnamefont {B.~S.}\ \bibnamefont {Schuster}}, \bibinfo {author} {\bibfnamefont {J.}~\bibnamefont {Hanes}}, \bibinfo {author} {\bibfnamefont {W.~E.}\ \bibnamefont {Grizzle}}, \bibinfo {author} {\bibfnamefont
  {E.~J.}\ \bibnamefont {Sorscher}}, \bibinfo {author} {\bibfnamefont {G.~J.}\ \bibnamefont {Tearney}},\ and\ \bibinfo {author} {\bibfnamefont {S.~M.}\ \bibnamefont {Rowe}},\ }\bibfield  {title} {\bibinfo {title} {A functional anatomic defect of the cystic fibrosis airway},\ }\href@noop {} {\bibfield  {journal} {\bibinfo  {journal} {American Journal of Respiratory and Critical Care Medicine}\ }\textbf {\bibinfo {volume} {190}},\ \bibinfo {pages} {421} (\bibinfo {year} {2014})}\BibitemShut {NoStop}%
\bibitem [{\citenamefont {Hancock}\ \emph {et~al.}(2018)\citenamefont {Hancock}, \citenamefont {Hennessy}, \citenamefont {Solomon}, \citenamefont {Dobrinskikh}, \citenamefont {Estrella}, \citenamefont {Hara}, \citenamefont {Hill}, \citenamefont {Kissner}, \citenamefont {Markovetz}, \citenamefont {Grove~Villalon}, \citenamefont {Voss}, \citenamefont {Tearney}, \citenamefont {Carroll}, \citenamefont {Shi}, \citenamefont {Schwarz}, \citenamefont {Thelin}, \citenamefont {Rowe}, \citenamefont {Yang}, \citenamefont {Evans},\ and\ \citenamefont {Schwartz}}]{Hancock_2018}%
  \BibitemOpen
  \bibfield  {author} {\bibinfo {author} {\bibfnamefont {L.~A.}\ \bibnamefont {Hancock}}, \bibinfo {author} {\bibfnamefont {C.~E.}\ \bibnamefont {Hennessy}}, \bibinfo {author} {\bibfnamefont {G.~M.}\ \bibnamefont {Solomon}}, \bibinfo {author} {\bibfnamefont {E.}~\bibnamefont {Dobrinskikh}}, \bibinfo {author} {\bibfnamefont {A.}~\bibnamefont {Estrella}}, \bibinfo {author} {\bibfnamefont {N.}~\bibnamefont {Hara}}, \bibinfo {author} {\bibfnamefont {D.~B.}\ \bibnamefont {Hill}}, \bibinfo {author} {\bibfnamefont {W.~J.}\ \bibnamefont {Kissner}}, \bibinfo {author} {\bibfnamefont {M.~R.}\ \bibnamefont {Markovetz}}, \bibinfo {author} {\bibfnamefont {D.~E.}\ \bibnamefont {Grove~Villalon}}, \bibinfo {author} {\bibfnamefont {M.~E.}\ \bibnamefont {Voss}}, \bibinfo {author} {\bibfnamefont {G.~J.}\ \bibnamefont {Tearney}}, \bibinfo {author} {\bibfnamefont {K.~S.}\ \bibnamefont {Carroll}}, \bibinfo {author} {\bibfnamefont {Y.}~\bibnamefont {Shi}}, \bibinfo {author} {\bibfnamefont {M.~I.}\ \bibnamefont {Schwarz}}, \bibinfo
  {author} {\bibfnamefont {W.~R.}\ \bibnamefont {Thelin}}, \bibinfo {author} {\bibfnamefont {S.~M.}\ \bibnamefont {Rowe}}, \bibinfo {author} {\bibfnamefont {I.~V.}\ \bibnamefont {Yang}}, \bibinfo {author} {\bibfnamefont {C.~M.}\ \bibnamefont {Evans}},\ and\ \bibinfo {author} {\bibfnamefont {D.~A.}\ \bibnamefont {Schwartz}},\ }\bibfield  {title} {\bibinfo {title} {Muc5b overexpression causes mucociliary dysfunction and enhances lung fibrosis in mice},\ }\href@noop {} {\bibfield  {journal} {\bibinfo  {journal} {Nature Communications}\ }\textbf {\bibinfo {volume} {9}},\ \bibinfo {pages} {5363} (\bibinfo {year} {2018})}\BibitemShut {NoStop}%
\bibitem [{\citenamefont {Georgiades}\ \emph {et~al.}(2014)\citenamefont {Georgiades}, \citenamefont {Pudney}, \citenamefont {Thornton},\ and\ \citenamefont {Waigh}}]{Georgiades_2014}%
  \BibitemOpen
  \bibfield  {author} {\bibinfo {author} {\bibfnamefont {P.}~\bibnamefont {Georgiades}}, \bibinfo {author} {\bibfnamefont {P.~D.~A.}\ \bibnamefont {Pudney}}, \bibinfo {author} {\bibfnamefont {D.~J.}\ \bibnamefont {Thornton}},\ and\ \bibinfo {author} {\bibfnamefont {T.~A.}\ \bibnamefont {Waigh}},\ }\bibfield  {title} {\bibinfo {title} {Particle tracking microrheology of purified gastrointestinal mucins},\ }\href@noop {} {\bibfield  {journal} {\bibinfo  {journal} {Biopolymers}\ }\textbf {\bibinfo {volume} {101}},\ \bibinfo {pages} {366} (\bibinfo {year} {2014})}\BibitemShut {NoStop}%
\bibitem [{\citenamefont {Bansil}\ and\ \citenamefont {Turner}(2006)}]{Bansil_2006}%
  \BibitemOpen
  \bibfield  {author} {\bibinfo {author} {\bibfnamefont {R.}~\bibnamefont {Bansil}}\ and\ \bibinfo {author} {\bibfnamefont {B.~S.}\ \bibnamefont {Turner}},\ }\bibfield  {title} {\bibinfo {title} {Mucin structure, aggregation, physiological functions and biomedical applications},\ }\href@noop {} {\bibfield  {journal} {\bibinfo  {journal} {Current Opinion in Colloid {\&} Interface Science}\ }\textbf {\bibinfo {volume} {11}},\ \bibinfo {pages} {164} (\bibinfo {year} {2006})}\BibitemShut {NoStop}%
\bibitem [{\citenamefont {Bhaskar}\ \emph {et~al.}(1991)\citenamefont {Bhaskar}, \citenamefont {Gong}, \citenamefont {Bansil}, \citenamefont {Pajevic}, \citenamefont {Hamilton}, \citenamefont {Turner},\ and\ \citenamefont {LaMont}}]{Bhaskar_1991}%
  \BibitemOpen
  \bibfield  {author} {\bibinfo {author} {\bibfnamefont {K.~R.}\ \bibnamefont {Bhaskar}}, \bibinfo {author} {\bibfnamefont {D.~H.}\ \bibnamefont {Gong}}, \bibinfo {author} {\bibfnamefont {R.}~\bibnamefont {Bansil}}, \bibinfo {author} {\bibfnamefont {S.}~\bibnamefont {Pajevic}}, \bibinfo {author} {\bibfnamefont {J.~A.}\ \bibnamefont {Hamilton}}, \bibinfo {author} {\bibfnamefont {B.~S.}\ \bibnamefont {Turner}},\ and\ \bibinfo {author} {\bibfnamefont {J.~T.}\ \bibnamefont {LaMont}},\ }\bibfield  {title} {\bibinfo {title} {Profound increase in viscosity and aggregation of pig gastric mucin at low ph},\ }\href@noop {} {\bibfield  {journal} {\bibinfo  {journal} {American Journal of Physiology-Gastrointestinal and Liver Physiology}\ }\textbf {\bibinfo {volume} {261}},\ \bibinfo {pages} {G827} (\bibinfo {year} {1991})}\BibitemShut {NoStop}%
\bibitem [{\citenamefont {Maleki}\ \emph {et~al.}(2008)\citenamefont {Maleki}, \citenamefont {Lafitte}, \citenamefont {Kj{\o}niksen}, \citenamefont {Thuresson},\ and\ \citenamefont {Nystr{\"o}m}}]{Maleki_2008}%
  \BibitemOpen
  \bibfield  {author} {\bibinfo {author} {\bibfnamefont {A.}~\bibnamefont {Maleki}}, \bibinfo {author} {\bibfnamefont {G.}~\bibnamefont {Lafitte}}, \bibinfo {author} {\bibfnamefont {A.-L.}\ \bibnamefont {Kj{\o}niksen}}, \bibinfo {author} {\bibfnamefont {K.}~\bibnamefont {Thuresson}},\ and\ \bibinfo {author} {\bibfnamefont {B.}~\bibnamefont {Nystr{\"o}m}},\ }\bibfield  {title} {\bibinfo {title} {Effect of ph on the association behavior in aqueous solutions of pig gastric mucin},\ }\href@noop {} {\bibfield  {journal} {\bibinfo  {journal} {Carbohydrate Research}\ }\textbf {\bibinfo {volume} {343}},\ \bibinfo {pages} {328} (\bibinfo {year} {2008})}\BibitemShut {NoStop}%
\bibitem [{\citenamefont {Wang}\ \emph {et~al.}(2013)\citenamefont {Wang}, \citenamefont {Lai}, \citenamefont {Ensign}, \citenamefont {Zhong}, \citenamefont {Cone},\ and\ \citenamefont {Hanes}}]{Wang_2013}%
  \BibitemOpen
  \bibfield  {author} {\bibinfo {author} {\bibfnamefont {Y.-Y.}\ \bibnamefont {Wang}}, \bibinfo {author} {\bibfnamefont {S.~K.}\ \bibnamefont {Lai}}, \bibinfo {author} {\bibfnamefont {L.~M.}\ \bibnamefont {Ensign}}, \bibinfo {author} {\bibfnamefont {W.}~\bibnamefont {Zhong}}, \bibinfo {author} {\bibfnamefont {R.}~\bibnamefont {Cone}},\ and\ \bibinfo {author} {\bibfnamefont {J.}~\bibnamefont {Hanes}},\ }\bibfield  {title} {\bibinfo {title} {The microstructure and bulk rheology of human cervicovaginal mucus are remarkably resistant to changes in ph},\ }\href@noop {} {\bibfield  {journal} {\bibinfo  {journal} {Biomacromolecules}\ }\textbf {\bibinfo {volume} {14}},\ \bibinfo {pages} {4429} (\bibinfo {year} {2013})}\BibitemShut {NoStop}%
\bibitem [{\citenamefont {Schuster}\ \emph {et~al.}(2014)\citenamefont {Schuster}, \citenamefont {Kim}, \citenamefont {Kays}, \citenamefont {Kanzawa}, \citenamefont {Guggino}, \citenamefont {Boyle}, \citenamefont {Rowe}, \citenamefont {Muzyczka}, \citenamefont {Suk},\ and\ \citenamefont {Hanes}}]{Schuster_2014}%
  \BibitemOpen
  \bibfield  {author} {\bibinfo {author} {\bibfnamefont {B.~S.}\ \bibnamefont {Schuster}}, \bibinfo {author} {\bibfnamefont {A.~J.}\ \bibnamefont {Kim}}, \bibinfo {author} {\bibfnamefont {J.~C.}\ \bibnamefont {Kays}}, \bibinfo {author} {\bibfnamefont {M.~M.}\ \bibnamefont {Kanzawa}}, \bibinfo {author} {\bibfnamefont {W.~B.}\ \bibnamefont {Guggino}}, \bibinfo {author} {\bibfnamefont {M.~P.}\ \bibnamefont {Boyle}}, \bibinfo {author} {\bibfnamefont {S.~M.}\ \bibnamefont {Rowe}}, \bibinfo {author} {\bibfnamefont {N.}~\bibnamefont {Muzyczka}}, \bibinfo {author} {\bibfnamefont {J.~S.}\ \bibnamefont {Suk}},\ and\ \bibinfo {author} {\bibfnamefont {J.}~\bibnamefont {Hanes}},\ }\bibfield  {title} {\bibinfo {title} {Overcoming the cystic fibrosis sputum barrier to leading adeno-associated virus gene therapy vectors},\ }\href@noop {} {\bibfield  {journal} {\bibinfo  {journal} {Molecular Therapy}\ }\textbf {\bibinfo {volume} {22}},\ \bibinfo {pages} {1484} (\bibinfo {year} {2014})}\BibitemShut {NoStop}%
\bibitem [{\citenamefont {Bansil}\ \emph {et~al.}(2013)\citenamefont {Bansil}, \citenamefont {Celli}, \citenamefont {Hardcastle},\ and\ \citenamefont {Turner}}]{Bansil_2013}%
  \BibitemOpen
  \bibfield  {author} {\bibinfo {author} {\bibfnamefont {R.}~\bibnamefont {Bansil}}, \bibinfo {author} {\bibfnamefont {J.}~\bibnamefont {Celli}}, \bibinfo {author} {\bibfnamefont {J.}~\bibnamefont {Hardcastle}},\ and\ \bibinfo {author} {\bibfnamefont {B.}~\bibnamefont {Turner}},\ }\bibfield  {title} {\bibinfo {title} {The influence of mucus microstructure and rheology in helicobacter pylori infection},\ }\href@noop {} {\bibfield  {journal} {\bibinfo  {journal} {Frontiers in Immunology}\ }\textbf {\bibinfo {volume} {4}} (\bibinfo {year} {2013})}\BibitemShut {NoStop}%
\bibitem [{\citenamefont {Suk}\ \emph {et~al.}(2011{\natexlab{a}})\citenamefont {Suk}, \citenamefont {Boylan}, \citenamefont {Trehan}, \citenamefont {Tang}, \citenamefont {Schneider}, \citenamefont {Lin}, \citenamefont {Boyle}, \citenamefont {Zeitlin}, \citenamefont {Lai}, \citenamefont {Cooper},\ and\ \citenamefont {Hanes}}]{Suk_2011}%
  \BibitemOpen
  \bibfield  {author} {\bibinfo {author} {\bibfnamefont {J.~S.}\ \bibnamefont {Suk}}, \bibinfo {author} {\bibfnamefont {N.~J.}\ \bibnamefont {Boylan}}, \bibinfo {author} {\bibfnamefont {K.}~\bibnamefont {Trehan}}, \bibinfo {author} {\bibfnamefont {B.~C.}\ \bibnamefont {Tang}}, \bibinfo {author} {\bibfnamefont {C.~S.}\ \bibnamefont {Schneider}}, \bibinfo {author} {\bibfnamefont {J.-M.~G.}\ \bibnamefont {Lin}}, \bibinfo {author} {\bibfnamefont {M.~P.}\ \bibnamefont {Boyle}}, \bibinfo {author} {\bibfnamefont {P.~L.}\ \bibnamefont {Zeitlin}}, \bibinfo {author} {\bibfnamefont {S.~K.}\ \bibnamefont {Lai}}, \bibinfo {author} {\bibfnamefont {M.~J.}\ \bibnamefont {Cooper}},\ and\ \bibinfo {author} {\bibfnamefont {J.}~\bibnamefont {Hanes}},\ }\bibfield  {title} {\bibinfo {title} {N-acetylcysteine enhances cystic fibrosis sputum penetration and airway gene transfer by highly compacted dna nanoparticles},\ }\href@noop {} {\bibfield  {journal} {\bibinfo  {journal} {Molecular Therapy}\ }\textbf {\bibinfo {volume} {19}},\
  \bibinfo {pages} {1981} (\bibinfo {year} {2011}{\natexlab{a}})}\BibitemShut {NoStop}%
\bibitem [{\citenamefont {Suk}\ \emph {et~al.}(2011{\natexlab{b}})\citenamefont {Suk}, \citenamefont {Lai}, \citenamefont {Boylan}, \citenamefont {Dawson}, \citenamefont {Boyle},\ and\ \citenamefont {Hanes}}]{Suk_2011a}%
  \BibitemOpen
  \bibfield  {author} {\bibinfo {author} {\bibfnamefont {J.~S.}\ \bibnamefont {Suk}}, \bibinfo {author} {\bibfnamefont {S.~K.}\ \bibnamefont {Lai}}, \bibinfo {author} {\bibfnamefont {N.~J.}\ \bibnamefont {Boylan}}, \bibinfo {author} {\bibfnamefont {M.~R.}\ \bibnamefont {Dawson}}, \bibinfo {author} {\bibfnamefont {M.~P.}\ \bibnamefont {Boyle}},\ and\ \bibinfo {author} {\bibfnamefont {J.}~\bibnamefont {Hanes}},\ }\bibfield  {title} {\bibinfo {title} {Rapid transport of muco-inert nanoparticles in cystic fibrosis sputum treated with n-acetyl cysteine},\ }\href@noop {} {\bibfield  {journal} {\bibinfo  {journal} {Nanomedicine}\ }\textbf {\bibinfo {volume} {6}},\ \bibinfo {pages} {365} (\bibinfo {year} {2011}{\natexlab{b}})}\BibitemShut {NoStop}%
\bibitem [{\citenamefont {Crocker}\ \emph {et~al.}(2000)\citenamefont {Crocker}, \citenamefont {Valentine}, \citenamefont {Weeks}, \citenamefont {Gisler}, \citenamefont {Kaplan}, \citenamefont {Yodh},\ and\ \citenamefont {Weitz}}]{Crocker_2000}%
  \BibitemOpen
  \bibfield  {author} {\bibinfo {author} {\bibfnamefont {J.~C.}\ \bibnamefont {Crocker}}, \bibinfo {author} {\bibfnamefont {M.~T.}\ \bibnamefont {Valentine}}, \bibinfo {author} {\bibfnamefont {E.~R.}\ \bibnamefont {Weeks}}, \bibinfo {author} {\bibfnamefont {T.}~\bibnamefont {Gisler}}, \bibinfo {author} {\bibfnamefont {P.~D.}\ \bibnamefont {Kaplan}}, \bibinfo {author} {\bibfnamefont {A.~G.}\ \bibnamefont {Yodh}},\ and\ \bibinfo {author} {\bibfnamefont {D.~A.}\ \bibnamefont {Weitz}},\ }\bibfield  {title} {\bibinfo {title} {Two-point microrheology of inhomogeneous soft materials},\ }\href@noop {} {\bibfield  {journal} {\bibinfo  {journal} {Physical Review Letters}\ }\textbf {\bibinfo {volume} {85}},\ \bibinfo {pages} {888} (\bibinfo {year} {2000})}\BibitemShut {NoStop}%
\bibitem [{\citenamefont {Bokkasam}\ \emph {et~al.}(2016)\citenamefont {Bokkasam}, \citenamefont {Ernst}, \citenamefont {Guenther}, \citenamefont {Wagner}, \citenamefont {Schaefer},\ and\ \citenamefont {Lehr}}]{Bokkasam_2016}%
  \BibitemOpen
  \bibfield  {author} {\bibinfo {author} {\bibfnamefont {H.}~\bibnamefont {Bokkasam}}, \bibinfo {author} {\bibfnamefont {M.}~\bibnamefont {Ernst}}, \bibinfo {author} {\bibfnamefont {M.}~\bibnamefont {Guenther}}, \bibinfo {author} {\bibfnamefont {C.}~\bibnamefont {Wagner}}, \bibinfo {author} {\bibfnamefont {U.~F.}\ \bibnamefont {Schaefer}},\ and\ \bibinfo {author} {\bibfnamefont {C.-M.}\ \bibnamefont {Lehr}},\ }\bibfield  {title} {\bibinfo {title} {Different macro- and micro-rheological properties of native porcine respiratory and intestinal mucus},\ }\href@noop {} {\bibfield  {journal} {\bibinfo  {journal} {International Journal of Pharmaceutics}\ }\textbf {\bibinfo {volume} {510}},\ \bibinfo {pages} {164} (\bibinfo {year} {2016})}\BibitemShut {NoStop}%
\bibitem [{\citenamefont {Esther}\ \emph {et~al.}(2019)\citenamefont {Esther}, \citenamefont {Muhlebach}, \citenamefont {Ehre}, \citenamefont {Hill}, \citenamefont {Wolfgang}, \citenamefont {Kesimer}, \citenamefont {Ramsey}, \citenamefont {Markovetz}, \citenamefont {Garbarine}, \citenamefont {Forest}, \citenamefont {Seim}, \citenamefont {Zorn}, \citenamefont {Morrison}, \citenamefont {Delion}, \citenamefont {Thelin}, \citenamefont {Villalon}, \citenamefont {Sabater}, \citenamefont {Turkovic}, \citenamefont {Ranganathan}, \citenamefont {Stick},\ and\ \citenamefont {Boucher}}]{Esther_2019}%
  \BibitemOpen
  \bibfield  {author} {\bibinfo {author} {\bibfnamefont {C.~R.}\ \bibnamefont {Esther}}, \bibinfo {author} {\bibfnamefont {M.~S.}\ \bibnamefont {Muhlebach}}, \bibinfo {author} {\bibfnamefont {C.}~\bibnamefont {Ehre}}, \bibinfo {author} {\bibfnamefont {D.~B.}\ \bibnamefont {Hill}}, \bibinfo {author} {\bibfnamefont {M.~C.}\ \bibnamefont {Wolfgang}}, \bibinfo {author} {\bibfnamefont {M.}~\bibnamefont {Kesimer}}, \bibinfo {author} {\bibfnamefont {K.~A.}\ \bibnamefont {Ramsey}}, \bibinfo {author} {\bibfnamefont {M.~R.}\ \bibnamefont {Markovetz}}, \bibinfo {author} {\bibfnamefont {I.~C.}\ \bibnamefont {Garbarine}}, \bibinfo {author} {\bibfnamefont {M.~G.}\ \bibnamefont {Forest}}, \bibinfo {author} {\bibfnamefont {I.}~\bibnamefont {Seim}}, \bibinfo {author} {\bibfnamefont {B.}~\bibnamefont {Zorn}}, \bibinfo {author} {\bibfnamefont {C.~B.}\ \bibnamefont {Morrison}}, \bibinfo {author} {\bibfnamefont {M.~F.}\ \bibnamefont {Delion}}, \bibinfo {author} {\bibfnamefont {W.~R.}\ \bibnamefont {Thelin}}, \bibinfo {author}
  {\bibfnamefont {D.}~\bibnamefont {Villalon}}, \bibinfo {author} {\bibfnamefont {J.~R.}\ \bibnamefont {Sabater}}, \bibinfo {author} {\bibfnamefont {L.}~\bibnamefont {Turkovic}}, \bibinfo {author} {\bibfnamefont {S.}~\bibnamefont {Ranganathan}}, \bibinfo {author} {\bibfnamefont {S.~M.}\ \bibnamefont {Stick}},\ and\ \bibinfo {author} {\bibfnamefont {R.~C.}\ \bibnamefont {Boucher}},\ }\bibfield  {title} {\bibinfo {title} {Mucus accumulation in the lungs precedes structural changes and infection in children with cystic fibrosis},\ }\href@noop {} {\bibfield  {journal} {\bibinfo  {journal} {Science Translational Medicine}\ }\textbf {\bibinfo {volume} {11}},\ \bibinfo {pages} {eaav3488} (\bibinfo {year} {2019})}\BibitemShut {NoStop}%
\bibitem [{\citenamefont {Murgia}\ \emph {et~al.}(2020)\citenamefont {Murgia}, \citenamefont {Kany}, \citenamefont {Herr}, \citenamefont {Ho}, \citenamefont {De~Rossi}, \citenamefont {Bals}, \citenamefont {Lehr}, \citenamefont {Hirsch}, \citenamefont {Hartmann}, \citenamefont {Empting},\ and\ \citenamefont {R{\"o}hrig}}]{Murgia_2020}%
  \BibitemOpen
  \bibfield  {author} {\bibinfo {author} {\bibfnamefont {X.}~\bibnamefont {Murgia}}, \bibinfo {author} {\bibfnamefont {A.~M.}\ \bibnamefont {Kany}}, \bibinfo {author} {\bibfnamefont {C.}~\bibnamefont {Herr}}, \bibinfo {author} {\bibfnamefont {D.-K.}\ \bibnamefont {Ho}}, \bibinfo {author} {\bibfnamefont {C.}~\bibnamefont {De~Rossi}}, \bibinfo {author} {\bibfnamefont {R.}~\bibnamefont {Bals}}, \bibinfo {author} {\bibfnamefont {C.-M.}\ \bibnamefont {Lehr}}, \bibinfo {author} {\bibfnamefont {A.~K.~H.}\ \bibnamefont {Hirsch}}, \bibinfo {author} {\bibfnamefont {R.~W.}\ \bibnamefont {Hartmann}}, \bibinfo {author} {\bibfnamefont {M.}~\bibnamefont {Empting}},\ and\ \bibinfo {author} {\bibfnamefont {T.}~\bibnamefont {R{\"o}hrig}},\ }\bibfield  {title} {\bibinfo {title} {Micro-rheological properties of lung homogenates correlate with infection severity in a mouse model of pseudomonas aeruginosa lung infection},\ }\href@noop {} {\bibfield  {journal} {\bibinfo  {journal} {Scientific Reports}\ }\textbf {\bibinfo {volume}
  {10}},\ \bibinfo {pages} {16502} (\bibinfo {year} {2020})}\BibitemShut {NoStop}%
\bibitem [{\citenamefont {Periasamy}\ and\ \citenamefont {Verkman}(1998)}]{Periasamy_1998}%
  \BibitemOpen
  \bibfield  {author} {\bibinfo {author} {\bibfnamefont {N.}~\bibnamefont {Periasamy}}\ and\ \bibinfo {author} {\bibfnamefont {A.~S.}\ \bibnamefont {Verkman}},\ }\bibfield  {title} {\bibinfo {title} {Analysis of fluorophore diffusion by continuous distributions of diffusion coefficients: Application to photobleaching measurements of multicomponent and anomalous diffusion},\ }\href@noop {} {\bibfield  {journal} {\bibinfo  {journal} {Biophysical Journal}\ }\textbf {\bibinfo {volume} {75}},\ \bibinfo {pages} {557} (\bibinfo {year} {1998})}\BibitemShut {NoStop}%
\bibitem [{\citenamefont {Sengupta}\ \emph {et~al.}(2003)\citenamefont {Sengupta}, \citenamefont {Garai}, \citenamefont {Balaji}, \citenamefont {Periasamy},\ and\ \citenamefont {Maiti}}]{Sengupta_2003}%
  \BibitemOpen
  \bibfield  {author} {\bibinfo {author} {\bibfnamefont {P.}~\bibnamefont {Sengupta}}, \bibinfo {author} {\bibfnamefont {K.}~\bibnamefont {Garai}}, \bibinfo {author} {\bibfnamefont {J.}~\bibnamefont {Balaji}}, \bibinfo {author} {\bibfnamefont {N.}~\bibnamefont {Periasamy}},\ and\ \bibinfo {author} {\bibfnamefont {S.}~\bibnamefont {Maiti}},\ }\bibfield  {title} {\bibinfo {title} {Measuring size distribution in highly heterogeneous systems with fluorescence correlation spectroscopy},\ }\href@noop {} {\bibfield  {journal} {\bibinfo  {journal} {Biophysical Journal}\ }\textbf {\bibinfo {volume} {84}},\ \bibinfo {pages} {1977} (\bibinfo {year} {2003})}\BibitemShut {NoStop}%
\bibitem [{\citenamefont {Lor{\'e}n}\ \emph {et~al.}(2015)\citenamefont {Lor{\'e}n}, \citenamefont {Hagman}, \citenamefont {Jonasson}, \citenamefont {Deschout}, \citenamefont {Bernin}, \citenamefont {Cella-Zanacchi}, \citenamefont {Diaspro}, \citenamefont {McNally}, \citenamefont {Ameloot}, \citenamefont {Smisdom}, \citenamefont {Nyd{\'e}n}, \citenamefont {Hermansson}, \citenamefont {Rudemo},\ and\ \citenamefont {Braeckmans}}]{Loren_2015}%
  \BibitemOpen
  \bibfield  {author} {\bibinfo {author} {\bibfnamefont {N.}~\bibnamefont {Lor{\'e}n}}, \bibinfo {author} {\bibfnamefont {J.}~\bibnamefont {Hagman}}, \bibinfo {author} {\bibfnamefont {J.~K.}\ \bibnamefont {Jonasson}}, \bibinfo {author} {\bibfnamefont {H.}~\bibnamefont {Deschout}}, \bibinfo {author} {\bibfnamefont {D.}~\bibnamefont {Bernin}}, \bibinfo {author} {\bibfnamefont {F.}~\bibnamefont {Cella-Zanacchi}}, \bibinfo {author} {\bibfnamefont {A.}~\bibnamefont {Diaspro}}, \bibinfo {author} {\bibfnamefont {J.~G.}\ \bibnamefont {McNally}}, \bibinfo {author} {\bibfnamefont {M.}~\bibnamefont {Ameloot}}, \bibinfo {author} {\bibfnamefont {N.}~\bibnamefont {Smisdom}}, \bibinfo {author} {\bibfnamefont {M.}~\bibnamefont {Nyd{\'e}n}}, \bibinfo {author} {\bibfnamefont {A.-M.}\ \bibnamefont {Hermansson}}, \bibinfo {author} {\bibfnamefont {M.}~\bibnamefont {Rudemo}},\ and\ \bibinfo {author} {\bibfnamefont {K.}~\bibnamefont {Braeckmans}},\ }\bibfield  {title} {\bibinfo {title} {Fluorescence recovery after photobleaching in
  material and life sciences: putting theory into practice},\ }\href@noop {} {\bibfield  {journal} {\bibinfo  {journal} {Quarterly Reviews of Biophysics}\ }\textbf {\bibinfo {volume} {48}},\ \bibinfo {pages} {323} (\bibinfo {year} {2015})}\BibitemShut {NoStop}%
\bibitem [{\citenamefont {Derichs}\ \emph {et~al.}(2011)\citenamefont {Derichs}, \citenamefont {Jin}, \citenamefont {Song}, \citenamefont {Finkbeiner},\ and\ \citenamefont {Verkman}}]{Derichs_2011}%
  \BibitemOpen
  \bibfield  {author} {\bibinfo {author} {\bibfnamefont {N.}~\bibnamefont {Derichs}}, \bibinfo {author} {\bibfnamefont {B.-J.}\ \bibnamefont {Jin}}, \bibinfo {author} {\bibfnamefont {Y.}~\bibnamefont {Song}}, \bibinfo {author} {\bibfnamefont {W.~E.}\ \bibnamefont {Finkbeiner}},\ and\ \bibinfo {author} {\bibfnamefont {A.~S.}\ \bibnamefont {Verkman}},\ }\bibfield  {title} {\bibinfo {title} {Hyperviscous airway periciliary and mucous liquid layers in cystic fibrosis measured by confocal fluorescence photobleaching},\ }\href@noop {} {\bibfield  {journal} {\bibinfo  {journal} {The FASEB Journal}\ }\textbf {\bibinfo {volume} {25}},\ \bibinfo {pages} {2325} (\bibinfo {year} {2011})}\BibitemShut {NoStop}%
\bibitem [{\citenamefont {Mastorakos}\ \emph {et~al.}(2015)\citenamefont {Mastorakos}, \citenamefont {da~Silva}, \citenamefont {Chisholm}, \citenamefont {Song}, \citenamefont {Choi}, \citenamefont {Boyle}, \citenamefont {Morales}, \citenamefont {Hanes},\ and\ \citenamefont {Suk}}]{Mastorakos_2015}%
  \BibitemOpen
  \bibfield  {author} {\bibinfo {author} {\bibfnamefont {P.}~\bibnamefont {Mastorakos}}, \bibinfo {author} {\bibfnamefont {A.~L.}\ \bibnamefont {da~Silva}}, \bibinfo {author} {\bibfnamefont {J.}~\bibnamefont {Chisholm}}, \bibinfo {author} {\bibfnamefont {E.}~\bibnamefont {Song}}, \bibinfo {author} {\bibfnamefont {W.~K.}\ \bibnamefont {Choi}}, \bibinfo {author} {\bibfnamefont {M.~P.}\ \bibnamefont {Boyle}}, \bibinfo {author} {\bibfnamefont {M.~M.}\ \bibnamefont {Morales}}, \bibinfo {author} {\bibfnamefont {J.}~\bibnamefont {Hanes}},\ and\ \bibinfo {author} {\bibfnamefont {J.~S.}\ \bibnamefont {Suk}},\ }\bibfield  {title} {\bibinfo {title} {Highly compacted biodegradable dna nanoparticles capable of overcoming the mucus barrier for inhaled lung gene therapy},\ }\href@noop {} {\bibfield  {journal} {\bibinfo  {journal} {Proceedings of the National Academy of Sciences}\ }\textbf {\bibinfo {volume} {112}},\ \bibinfo {pages} {8720} (\bibinfo {year} {2015})}\BibitemShut {NoStop}%
\bibitem [{\citenamefont {Braeckmans}\ \emph {et~al.}(2003)\citenamefont {Braeckmans}, \citenamefont {Peeters}, \citenamefont {Sanders}, \citenamefont {De~Smedt},\ and\ \citenamefont {Demeester}}]{Braeckmans_2003}%
  \BibitemOpen
  \bibfield  {author} {\bibinfo {author} {\bibfnamefont {K.}~\bibnamefont {Braeckmans}}, \bibinfo {author} {\bibfnamefont {L.}~\bibnamefont {Peeters}}, \bibinfo {author} {\bibfnamefont {N.~N.}\ \bibnamefont {Sanders}}, \bibinfo {author} {\bibfnamefont {S.~C.}\ \bibnamefont {De~Smedt}},\ and\ \bibinfo {author} {\bibfnamefont {J.}~\bibnamefont {Demeester}},\ }\bibfield  {title} {\bibinfo {title} {Three-dimensional fluorescence recovery after photobleaching with the confocal scanning laser microscope},\ }\href@noop {} {\bibfield  {journal} {\bibinfo  {journal} {Biophysical Journal}\ }\textbf {\bibinfo {volume} {85}},\ \bibinfo {pages} {2240} (\bibinfo {year} {2003})}\BibitemShut {NoStop}%
\bibitem [{\citenamefont {Tang}\ \emph {et~al.}(2016)\citenamefont {Tang}, \citenamefont {Ostedgaard}, \citenamefont {Hoegger}, \citenamefont {Moninger}, \citenamefont {Karp}, \citenamefont {McMenimen}, \citenamefont {Choudhury}, \citenamefont {Varki}, \citenamefont {Stoltz},\ and\ \citenamefont {Welsh}}]{Tang_2016}%
  \BibitemOpen
  \bibfield  {author} {\bibinfo {author} {\bibfnamefont {X.~X.}\ \bibnamefont {Tang}}, \bibinfo {author} {\bibfnamefont {L.~S.}\ \bibnamefont {Ostedgaard}}, \bibinfo {author} {\bibfnamefont {M.~J.}\ \bibnamefont {Hoegger}}, \bibinfo {author} {\bibfnamefont {T.~O.}\ \bibnamefont {Moninger}}, \bibinfo {author} {\bibfnamefont {P.~H.}\ \bibnamefont {Karp}}, \bibinfo {author} {\bibfnamefont {J.~D.}\ \bibnamefont {McMenimen}}, \bibinfo {author} {\bibfnamefont {B.}~\bibnamefont {Choudhury}}, \bibinfo {author} {\bibfnamefont {A.}~\bibnamefont {Varki}}, \bibinfo {author} {\bibfnamefont {D.~A.}\ \bibnamefont {Stoltz}},\ and\ \bibinfo {author} {\bibfnamefont {M.~J.}\ \bibnamefont {Welsh}},\ }\bibfield  {title} {\bibinfo {title} {Acidic ph increases airway surface liquid viscosity in cystic fibrosis},\ }\href@noop {} {\bibfield  {journal} {\bibinfo  {journal} {J Clin Invest}\ }\textbf {\bibinfo {volume} {126}},\ \bibinfo {pages} {879} (\bibinfo {year} {2016})}\BibitemShut {NoStop}%
\bibitem [{\citenamefont {Suk}\ \emph {et~al.}(2014)\citenamefont {Suk}, \citenamefont {Kim}, \citenamefont {Trehan}, \citenamefont {Schneider}, \citenamefont {Cebotaru}, \citenamefont {Woodward}, \citenamefont {Boylan}, \citenamefont {Boyle}, \citenamefont {Lai}, \citenamefont {Guggino},\ and\ \citenamefont {Hanes}}]{Suk_2014}%
  \BibitemOpen
  \bibfield  {author} {\bibinfo {author} {\bibfnamefont {J.~S.}\ \bibnamefont {Suk}}, \bibinfo {author} {\bibfnamefont {A.~J.}\ \bibnamefont {Kim}}, \bibinfo {author} {\bibfnamefont {K.}~\bibnamefont {Trehan}}, \bibinfo {author} {\bibfnamefont {C.~S.}\ \bibnamefont {Schneider}}, \bibinfo {author} {\bibfnamefont {L.}~\bibnamefont {Cebotaru}}, \bibinfo {author} {\bibfnamefont {O.~M.}\ \bibnamefont {Woodward}}, \bibinfo {author} {\bibfnamefont {N.~J.}\ \bibnamefont {Boylan}}, \bibinfo {author} {\bibfnamefont {M.~P.}\ \bibnamefont {Boyle}}, \bibinfo {author} {\bibfnamefont {S.~K.}\ \bibnamefont {Lai}}, \bibinfo {author} {\bibfnamefont {W.~B.}\ \bibnamefont {Guggino}},\ and\ \bibinfo {author} {\bibfnamefont {J.}~\bibnamefont {Hanes}},\ }\bibfield  {title} {\bibinfo {title} {Lung gene therapy with highly compacted dna nanoparticles that overcome the mucus barrier},\ }\href {https://doi.org/10.1016/j.jconrel.2014.01.007} {\bibfield  {journal} {\bibinfo  {journal} {Journal of Controlled Release}\ }\textbf {\bibinfo
  {volume} {178}},\ \bibinfo {pages} {8} (\bibinfo {year} {2014})}\BibitemShut {NoStop}%
\bibitem [{\citenamefont {Vasconcellos}\ \emph {et~al.}(1994)\citenamefont {Vasconcellos}, \citenamefont {Allen}, \citenamefont {Wohl}, \citenamefont {Drazen}, \citenamefont {Janmey},\ and\ \citenamefont {Stossel}}]{Vasconcellos_1994}%
  \BibitemOpen
  \bibfield  {author} {\bibinfo {author} {\bibfnamefont {C.~A.}\ \bibnamefont {Vasconcellos}}, \bibinfo {author} {\bibfnamefont {P.~G.}\ \bibnamefont {Allen}}, \bibinfo {author} {\bibfnamefont {M.~E.}\ \bibnamefont {Wohl}}, \bibinfo {author} {\bibfnamefont {J.~M.}\ \bibnamefont {Drazen}}, \bibinfo {author} {\bibfnamefont {P.~A.}\ \bibnamefont {Janmey}},\ and\ \bibinfo {author} {\bibfnamefont {T.~P.}\ \bibnamefont {Stossel}},\ }\bibfield  {title} {\bibinfo {title} {Reduction in viscosity of cystic fibrosis sputum in vitro by gelsolin},\ }\href@noop {} {\bibfield  {journal} {\bibinfo  {journal} {Science}\ }\textbf {\bibinfo {volume} {263}},\ \bibinfo {pages} {969} (\bibinfo {year} {1994})}\BibitemShut {NoStop}%
\bibitem [{\citenamefont {Olmsted}\ \emph {et~al.}(2001)\citenamefont {Olmsted}, \citenamefont {Padgett}, \citenamefont {Yudin}, \citenamefont {Whaley}, \citenamefont {Moench},\ and\ \citenamefont {Cone}}]{Olmsted_2001}%
  \BibitemOpen
  \bibfield  {author} {\bibinfo {author} {\bibfnamefont {S.~S.}\ \bibnamefont {Olmsted}}, \bibinfo {author} {\bibfnamefont {J.~L.}\ \bibnamefont {Padgett}}, \bibinfo {author} {\bibfnamefont {A.~I.}\ \bibnamefont {Yudin}}, \bibinfo {author} {\bibfnamefont {K.~J.}\ \bibnamefont {Whaley}}, \bibinfo {author} {\bibfnamefont {T.~R.}\ \bibnamefont {Moench}},\ and\ \bibinfo {author} {\bibfnamefont {R.~A.}\ \bibnamefont {Cone}},\ }\bibfield  {title} {\bibinfo {title} {Diffusion of macromolecules and virus-like particles in human cervical mucus},\ }\href@noop {} {\bibfield  {journal} {\bibinfo  {journal} {Biophysical Journal}\ }\textbf {\bibinfo {volume} {81}},\ \bibinfo {pages} {1930} (\bibinfo {year} {2001})}\BibitemShut {NoStop}%
\bibitem [{\citenamefont {Wang}\ \emph {et~al.}(2011)\citenamefont {Wang}, \citenamefont {Lai}, \citenamefont {So}, \citenamefont {Schneider}, \citenamefont {Cone},\ and\ \citenamefont {Hanes}}]{Wang_2011}%
  \BibitemOpen
  \bibfield  {author} {\bibinfo {author} {\bibfnamefont {Y.-Y.}\ \bibnamefont {Wang}}, \bibinfo {author} {\bibfnamefont {S.~K.}\ \bibnamefont {Lai}}, \bibinfo {author} {\bibfnamefont {C.}~\bibnamefont {So}}, \bibinfo {author} {\bibfnamefont {C.}~\bibnamefont {Schneider}}, \bibinfo {author} {\bibfnamefont {R.}~\bibnamefont {Cone}},\ and\ \bibinfo {author} {\bibfnamefont {J.}~\bibnamefont {Hanes}},\ }\bibfield  {title} {\bibinfo {title} {Mucoadhesive nanoparticles may disrupt the protective human mucus barrier by altering its microstructure},\ }\href@noop {} {\bibfield  {journal} {\bibinfo  {journal} {PLOS ONE}\ }\textbf {\bibinfo {volume} {6}},\ \bibinfo {pages} {e21547} (\bibinfo {year} {2011})}\BibitemShut {NoStop}%
\bibitem [{\citenamefont {Kirch}\ \emph {et~al.}(2012)\citenamefont {Kirch}, \citenamefont {Schneider}, \citenamefont {Abou}, \citenamefont {Hopf}, \citenamefont {Schaefer}, \citenamefont {Schneider}, \citenamefont {Schall}, \citenamefont {Wagner},\ and\ \citenamefont {Lehr}}]{Kirch_2012}%
  \BibitemOpen
  \bibfield  {author} {\bibinfo {author} {\bibfnamefont {J.}~\bibnamefont {Kirch}}, \bibinfo {author} {\bibfnamefont {A.}~\bibnamefont {Schneider}}, \bibinfo {author} {\bibfnamefont {B.}~\bibnamefont {Abou}}, \bibinfo {author} {\bibfnamefont {A.}~\bibnamefont {Hopf}}, \bibinfo {author} {\bibfnamefont {U.~F.}\ \bibnamefont {Schaefer}}, \bibinfo {author} {\bibfnamefont {M.}~\bibnamefont {Schneider}}, \bibinfo {author} {\bibfnamefont {C.}~\bibnamefont {Schall}}, \bibinfo {author} {\bibfnamefont {C.}~\bibnamefont {Wagner}},\ and\ \bibinfo {author} {\bibfnamefont {C.-M.}\ \bibnamefont {Lehr}},\ }\bibfield  {title} {\bibinfo {title} {Optical tweezers reveal relationship between microstructure and nanoparticle penetration of pulmonary mucus},\ }\href@noop {} {\bibfield  {journal} {\bibinfo  {journal} {Proceedings of the National Academy of Sciences}\ }\textbf {\bibinfo {volume} {109}},\ \bibinfo {pages} {18355} (\bibinfo {year} {2012})}\BibitemShut {NoStop}%
\bibitem [{\citenamefont {Cribb}\ \emph {et~al.}(2013)\citenamefont {Cribb}, \citenamefont {Vasquez}, \citenamefont {Moore}, \citenamefont {Norris}, \citenamefont {Shah}, \citenamefont {Forest},\ and\ \citenamefont {Superfine}}]{Cribb_2013}%
  \BibitemOpen
  \bibfield  {author} {\bibinfo {author} {\bibfnamefont {J.~A.}\ \bibnamefont {Cribb}}, \bibinfo {author} {\bibfnamefont {P.~A.}\ \bibnamefont {Vasquez}}, \bibinfo {author} {\bibfnamefont {P.}~\bibnamefont {Moore}}, \bibinfo {author} {\bibfnamefont {S.}~\bibnamefont {Norris}}, \bibinfo {author} {\bibfnamefont {S.}~\bibnamefont {Shah}}, \bibinfo {author} {\bibfnamefont {M.~G.}\ \bibnamefont {Forest}},\ and\ \bibinfo {author} {\bibfnamefont {R.}~\bibnamefont {Superfine}},\ }\bibfield  {title} {\bibinfo {title} {Nonlinear signatures in active microbead rheology of entangled polymer solutions},\ }\href@noop {} {\bibfield  {journal} {\bibinfo  {journal} {Journal of Rheology}\ }\textbf {\bibinfo {volume} {57}},\ \bibinfo {pages} {1247} (\bibinfo {year} {2013})}\BibitemShut {NoStop}%
\bibitem [{\citenamefont {Litt}(1970)}]{Litt_1970}%
  \BibitemOpen
  \bibfield  {author} {\bibinfo {author} {\bibfnamefont {M.}~\bibnamefont {Litt}},\ }\bibfield  {title} {\bibinfo {title} {Mucus rheology: Relevance to mucociliary clearance},\ }\href@noop {} {\bibfield  {journal} {\bibinfo  {journal} {Archives of Internal Medicine}\ }\textbf {\bibinfo {volume} {126}},\ \bibinfo {pages} {417} (\bibinfo {year} {1970})}\BibitemShut {NoStop}%
\bibitem [{\citenamefont {Sanctis}\ \emph {et~al.}(1994)\citenamefont {Sanctis}, \citenamefont {Tomkiewicz}, \citenamefont {Rubin}, \citenamefont {Schurch},\ and\ \citenamefont {King}}]{Sanctis_1994}%
  \BibitemOpen
  \bibfield  {author} {\bibinfo {author} {\bibfnamefont {G.~D.}\ \bibnamefont {Sanctis}}, \bibinfo {author} {\bibfnamefont {R.}~\bibnamefont {Tomkiewicz}}, \bibinfo {author} {\bibfnamefont {B.}~\bibnamefont {Rubin}}, \bibinfo {author} {\bibfnamefont {S.}~\bibnamefont {Schurch}},\ and\ \bibinfo {author} {\bibfnamefont {M.}~\bibnamefont {King}},\ }\bibfield  {title} {\bibinfo {title} {Exogenous surfactant enhances mucociliary clearance in the anaesthetized dog},\ }\href@noop {} {\bibfield  {journal} {\bibinfo  {journal} {European Respiratory Journal}\ }\textbf {\bibinfo {volume} {7}},\ \bibinfo {pages} {1616} (\bibinfo {year} {1994})}\BibitemShut {NoStop}%
\bibitem [{\citenamefont {Braunreuther}\ \emph {et~al.}(2023)\citenamefont {Braunreuther}, \citenamefont {Liegeois}, \citenamefont {Fahy},\ and\ \citenamefont {Fuller}}]{Braunreuther_2023}%
  \BibitemOpen
  \bibfield  {author} {\bibinfo {author} {\bibfnamefont {M.}~\bibnamefont {Braunreuther}}, \bibinfo {author} {\bibfnamefont {M.}~\bibnamefont {Liegeois}}, \bibinfo {author} {\bibfnamefont {J.~V.}\ \bibnamefont {Fahy}},\ and\ \bibinfo {author} {\bibfnamefont {G.~G.}\ \bibnamefont {Fuller}},\ }\bibfield  {title} {\bibinfo {title} {Nondestructive rheological measurements of biomaterials with a magnetic microwire rheometer},\ }\href@noop {} {\bibfield  {journal} {\bibinfo  {journal} {Journal of Rheology}\ }\textbf {\bibinfo {volume} {67}},\ \bibinfo {pages} {579} (\bibinfo {year} {2023})}\BibitemShut {NoStop}%
\bibitem [{\citenamefont {Abdula}\ \emph {et~al.}(2017)\citenamefont {Abdula}, \citenamefont {Lerud},\ and\ \citenamefont {Rananavare}}]{Abdula_2017}%
  \BibitemOpen
  \bibfield  {author} {\bibinfo {author} {\bibfnamefont {D.}~\bibnamefont {Abdula}}, \bibinfo {author} {\bibfnamefont {R.}~\bibnamefont {Lerud}},\ and\ \bibinfo {author} {\bibfnamefont {S.}~\bibnamefont {Rananavare}},\ }\bibfield  {title} {\bibinfo {title} {Bubbling and foaming assisted clearing of mucin plugs in microfluidic y-junctions},\ }\href@noop {} {\bibfield  {journal} {\bibinfo  {journal} {Journal of Biomechanics}\ }\textbf {\bibinfo {volume} {64}},\ \bibinfo {pages} {1} (\bibinfo {year} {2017})}\BibitemShut {NoStop}%
\bibitem [{\citenamefont {Marczynski}\ \emph {et~al.}(2018)\citenamefont {Marczynski}, \citenamefont {K{\"a}sdorf}, \citenamefont {Altaner}, \citenamefont {Wenzler}, \citenamefont {Gerland},\ and\ \citenamefont {Lieleg}}]{Marczynski_2018}%
  \BibitemOpen
  \bibfield  {author} {\bibinfo {author} {\bibfnamefont {M.}~\bibnamefont {Marczynski}}, \bibinfo {author} {\bibfnamefont {B.~T.}\ \bibnamefont {K{\"a}sdorf}}, \bibinfo {author} {\bibfnamefont {B.}~\bibnamefont {Altaner}}, \bibinfo {author} {\bibfnamefont {A.}~\bibnamefont {Wenzler}}, \bibinfo {author} {\bibfnamefont {U.}~\bibnamefont {Gerland}},\ and\ \bibinfo {author} {\bibfnamefont {O.}~\bibnamefont {Lieleg}},\ }\bibfield  {title} {\bibinfo {title} {Transient binding promotes molecule penetration into mucin hydrogels by enhancing molecular partitioning},\ }\href@noop {} {\bibfield  {journal} {\bibinfo  {journal} {Biomaterials Science}\ }\textbf {\bibinfo {volume} {6}},\ \bibinfo {pages} {3373} (\bibinfo {year} {2018})}\BibitemShut {NoStop}%
\bibitem [{\citenamefont {Elberskirch}\ \emph {et~al.}(2019)\citenamefont {Elberskirch}, \citenamefont {Knoll}, \citenamefont {Moosmann}, \citenamefont {Wilhelm}, \citenamefont {von Briesen},\ and\ \citenamefont {Wagner}}]{Elberskirch_2019}%
  \BibitemOpen
  \bibfield  {author} {\bibinfo {author} {\bibfnamefont {L.}~\bibnamefont {Elberskirch}}, \bibinfo {author} {\bibfnamefont {T.}~\bibnamefont {Knoll}}, \bibinfo {author} {\bibfnamefont {A.}~\bibnamefont {Moosmann}}, \bibinfo {author} {\bibfnamefont {N.}~\bibnamefont {Wilhelm}}, \bibinfo {author} {\bibfnamefont {H.}~\bibnamefont {von Briesen}},\ and\ \bibinfo {author} {\bibfnamefont {S.}~\bibnamefont {Wagner}},\ }\bibfield  {title} {\bibinfo {title} {A novel microfluidic mucus-chip for studying the permeation of compounds over the mucus barrier},\ }\href@noop {} {\bibfield  {journal} {\bibinfo  {journal} {Journal of Drug Delivery Science and Technology}\ }\textbf {\bibinfo {volume} {54}},\ \bibinfo {pages} {101248} (\bibinfo {year} {2019})}\BibitemShut {NoStop}%
\bibitem [{\citenamefont {Jia}\ \emph {et~al.}(2021)\citenamefont {Jia}, \citenamefont {Guo}, \citenamefont {Yang}, \citenamefont {Prestidge},\ and\ \citenamefont {Thierry}}]{Jia_2021}%
  \BibitemOpen
  \bibfield  {author} {\bibinfo {author} {\bibfnamefont {Z.}~\bibnamefont {Jia}}, \bibinfo {author} {\bibfnamefont {Z.}~\bibnamefont {Guo}}, \bibinfo {author} {\bibfnamefont {C.-T.}\ \bibnamefont {Yang}}, \bibinfo {author} {\bibfnamefont {C.}~\bibnamefont {Prestidge}},\ and\ \bibinfo {author} {\bibfnamefont {B.}~\bibnamefont {Thierry}},\ }\bibfield  {title} {\bibinfo {title} {``mucus-on-chip'': A new tool to study the dynamic penetration of nanoparticulate drug carriers into mucus},\ }\href@noop {} {\bibfield  {journal} {\bibinfo  {journal} {International Journal of Pharmaceutics}\ }\textbf {\bibinfo {volume} {598}},\ \bibinfo {pages} {120391} (\bibinfo {year} {2021})}\BibitemShut {NoStop}%
\bibitem [{\citenamefont {Suh}\ \emph {et~al.}(2024)\citenamefont {Suh}, \citenamefont {Li}, \citenamefont {Pandey}, \citenamefont {Nordmann}, \citenamefont {Huang},\ and\ \citenamefont {Wu}}]{Suh_2024}%
  \BibitemOpen
  \bibfield  {author} {\bibinfo {author} {\bibfnamefont {Y.~J.}\ \bibnamefont {Suh}}, \bibinfo {author} {\bibfnamefont {A.~T.}\ \bibnamefont {Li}}, \bibinfo {author} {\bibfnamefont {M.}~\bibnamefont {Pandey}}, \bibinfo {author} {\bibfnamefont {C.~S.}\ \bibnamefont {Nordmann}}, \bibinfo {author} {\bibfnamefont {Y.~L.}\ \bibnamefont {Huang}},\ and\ \bibinfo {author} {\bibfnamefont {M.}~\bibnamefont {Wu}},\ }\bibfield  {title} {\bibinfo {title} {Decoding physical principles of cell migration under controlled environment using microfluidics},\ }\href@noop {} {\bibfield  {journal} {\bibinfo  {journal} {Biophysics Reviews}\ }\textbf {\bibinfo {volume} {5}} (\bibinfo {year} {2024})}\BibitemShut {NoStop}%
\bibitem [{\citenamefont {Wang}\ \emph {et~al.}(2024)\citenamefont {Wang}, \citenamefont {Negron}, \citenamefont {Khoshnaw}, \citenamefont {Edwards}, \citenamefont {Vu}, \citenamefont {Quatela}, \citenamefont {Park}, \citenamefont {Maldonado}, \citenamefont {Demarest}, \citenamefont {Simon}, \citenamefont {Oskay},\ and\ \citenamefont {Dong}}]{Wang_2024}%
  \BibitemOpen
  \bibfield  {author} {\bibinfo {author} {\bibfnamefont {Y.}~\bibnamefont {Wang}}, \bibinfo {author} {\bibfnamefont {C.}~\bibnamefont {Negron}}, \bibinfo {author} {\bibfnamefont {A.}~\bibnamefont {Khoshnaw}}, \bibinfo {author} {\bibfnamefont {S.}~\bibnamefont {Edwards}}, \bibinfo {author} {\bibfnamefont {H.}~\bibnamefont {Vu}}, \bibinfo {author} {\bibfnamefont {J.}~\bibnamefont {Quatela}}, \bibinfo {author} {\bibfnamefont {N.}~\bibnamefont {Park}}, \bibinfo {author} {\bibfnamefont {F.}~\bibnamefont {Maldonado}}, \bibinfo {author} {\bibfnamefont {C.}~\bibnamefont {Demarest}}, \bibinfo {author} {\bibfnamefont {V.}~\bibnamefont {Simon}}, \bibinfo {author} {\bibfnamefont {C.}~\bibnamefont {Oskay}},\ and\ \bibinfo {author} {\bibfnamefont {X.}~\bibnamefont {Dong}},\ }\bibfield  {title} {\bibinfo {title} {Sensory artificial cilia for in situ monitoring of airway physiological properties},\ }\href {https://doi.org/10.1073/pnas.2412086121} {\bibfield  {journal} {\bibinfo  {journal} {Proceedings of the National Academy
  of Sciences}\ }\textbf {\bibinfo {volume} {121}},\ \bibinfo {pages} {e2412086121} (\bibinfo {year} {2024})}\BibitemShut {NoStop}%
\bibitem [{\citenamefont {Shak}\ \emph {et~al.}(1990)\citenamefont {Shak}, \citenamefont {Capon}, \citenamefont {Hellmiss}, \citenamefont {Marsters},\ and\ \citenamefont {Baker}}]{Shak_1990}%
  \BibitemOpen
  \bibfield  {author} {\bibinfo {author} {\bibfnamefont {S.}~\bibnamefont {Shak}}, \bibinfo {author} {\bibfnamefont {D.~J.}\ \bibnamefont {Capon}}, \bibinfo {author} {\bibfnamefont {R.}~\bibnamefont {Hellmiss}}, \bibinfo {author} {\bibfnamefont {S.~A.}\ \bibnamefont {Marsters}},\ and\ \bibinfo {author} {\bibfnamefont {C.~L.}\ \bibnamefont {Baker}},\ }\bibfield  {title} {\bibinfo {title} {Recombinant human dnase i reduces the viscosity of cystic fibrosis sputum},\ }\href@noop {} {\bibfield  {journal} {\bibinfo  {journal} {Proceedings of the National Academy of Sciences}\ }\textbf {\bibinfo {volume} {87}},\ \bibinfo {pages} {9188} (\bibinfo {year} {1990})}\BibitemShut {NoStop}%
\bibitem [{\citenamefont {Shak}(1995)}]{Shak_1995}%
  \BibitemOpen
  \bibfield  {author} {\bibinfo {author} {\bibfnamefont {S.}~\bibnamefont {Shak}},\ }\bibfield  {title} {\bibinfo {title} {Aerosolized recombinant human dnase i for the treatment of cystic fibrosis},\ }\href@noop {} {\bibfield  {journal} {\bibinfo  {journal} {Chest}\ }\textbf {\bibinfo {volume} {107}},\ \bibinfo {pages} {65S} (\bibinfo {year} {1995})}\BibitemShut {NoStop}%
\bibitem [{\citenamefont {Frederiksen}\ \emph {et~al.}(2006)\citenamefont {Frederiksen}, \citenamefont {Pressler}, \citenamefont {Hansen}, \citenamefont {Koch},\ and\ \citenamefont {H{\o}iby}}]{Frederiksen_2006}%
  \BibitemOpen
  \bibfield  {author} {\bibinfo {author} {\bibfnamefont {B.}~\bibnamefont {Frederiksen}}, \bibinfo {author} {\bibfnamefont {T.}~\bibnamefont {Pressler}}, \bibinfo {author} {\bibfnamefont {A.}~\bibnamefont {Hansen}}, \bibinfo {author} {\bibfnamefont {C.}~\bibnamefont {Koch}},\ and\ \bibinfo {author} {\bibfnamefont {N.}~\bibnamefont {H{\o}iby}},\ }\bibfield  {title} {\bibinfo {title} {Effect of aerosolized rhdnase (pulmozyme{\textregistered}) on pulmonary colonization in patients with cystic fibrosis},\ }\href@noop {} {\bibfield  {journal} {\bibinfo  {journal} {Acta Paediatrica}\ }\textbf {\bibinfo {volume} {95}},\ \bibinfo {pages} {1070} (\bibinfo {year} {2006})}\BibitemShut {NoStop}%
\bibitem [{\citenamefont {Wheeler}\ \emph {et~al.}(2019)\citenamefont {Wheeler}, \citenamefont {C{\'a}rcamo-Oyarce}, \citenamefont {Turner}, \citenamefont {Dellos-Nolan}, \citenamefont {Co}, \citenamefont {Lehoux}, \citenamefont {Cummings}, \citenamefont {Wozniak},\ and\ \citenamefont {Ribbeck}}]{Wheeler_2019}%
  \BibitemOpen
  \bibfield  {author} {\bibinfo {author} {\bibfnamefont {K.~M.}\ \bibnamefont {Wheeler}}, \bibinfo {author} {\bibfnamefont {G.}~\bibnamefont {C{\'a}rcamo-Oyarce}}, \bibinfo {author} {\bibfnamefont {B.~S.}\ \bibnamefont {Turner}}, \bibinfo {author} {\bibfnamefont {S.}~\bibnamefont {Dellos-Nolan}}, \bibinfo {author} {\bibfnamefont {J.~Y.}\ \bibnamefont {Co}}, \bibinfo {author} {\bibfnamefont {S.}~\bibnamefont {Lehoux}}, \bibinfo {author} {\bibfnamefont {R.~D.}\ \bibnamefont {Cummings}}, \bibinfo {author} {\bibfnamefont {D.~J.}\ \bibnamefont {Wozniak}},\ and\ \bibinfo {author} {\bibfnamefont {K.}~\bibnamefont {Ribbeck}},\ }\bibfield  {title} {\bibinfo {title} {Mucin glycans attenuate the virulence of pseudomonas aeruginosa in infection},\ }\href@noop {} {\bibfield  {journal} {\bibinfo  {journal} {Nature Microbiology}\ }\textbf {\bibinfo {volume} {4}},\ \bibinfo {pages} {2146} (\bibinfo {year} {2019})}\BibitemShut {NoStop}%
\bibitem [{\citenamefont {Chen}\ \emph {et~al.}(2010)\citenamefont {Chen}, \citenamefont {Yang}, \citenamefont {Quinton},\ and\ \citenamefont {Chin}}]{Chen_2010}%
  \BibitemOpen
  \bibfield  {author} {\bibinfo {author} {\bibfnamefont {E.~Y.~T.}\ \bibnamefont {Chen}}, \bibinfo {author} {\bibfnamefont {N.}~\bibnamefont {Yang}}, \bibinfo {author} {\bibfnamefont {P.~M.}\ \bibnamefont {Quinton}},\ and\ \bibinfo {author} {\bibfnamefont {W.-C.}\ \bibnamefont {Chin}},\ }\bibfield  {title} {\bibinfo {title} {A new role for bicarbonate in mucus formation},\ }\href@noop {} {\bibfield  {journal} {\bibinfo  {journal} {American Journal of Physiology-Lung Cellular and Molecular Physiology}\ }\textbf {\bibinfo {volume} {299}},\ \bibinfo {pages} {L542} (\bibinfo {year} {2010})}\BibitemShut {NoStop}%
\bibitem [{\citenamefont {Gustafsson}\ \emph {et~al.}(2012)\citenamefont {Gustafsson}, \citenamefont {Ermund}, \citenamefont {Ambort}, \citenamefont {Johansson}, \citenamefont {Nilsson}, \citenamefont {Thorell}, \citenamefont {Hebert}, \citenamefont {Sj{\"o}vall},\ and\ \citenamefont {Hansson}}]{Gustafsson_2012}%
  \BibitemOpen
  \bibfield  {author} {\bibinfo {author} {\bibfnamefont {J.~K.}\ \bibnamefont {Gustafsson}}, \bibinfo {author} {\bibfnamefont {A.}~\bibnamefont {Ermund}}, \bibinfo {author} {\bibfnamefont {D.}~\bibnamefont {Ambort}}, \bibinfo {author} {\bibfnamefont {M.~E.}\ \bibnamefont {Johansson}}, \bibinfo {author} {\bibfnamefont {H.~E.}\ \bibnamefont {Nilsson}}, \bibinfo {author} {\bibfnamefont {K.}~\bibnamefont {Thorell}}, \bibinfo {author} {\bibfnamefont {H.}~\bibnamefont {Hebert}}, \bibinfo {author} {\bibfnamefont {H.}~\bibnamefont {Sj{\"o}vall}},\ and\ \bibinfo {author} {\bibfnamefont {G.~C.}\ \bibnamefont {Hansson}},\ }\bibfield  {title} {\bibinfo {title} {Bicarbonate and functional cftr channel are required for proper mucin secretion and link cystic fibrosis with its mucus phenotype},\ }\href@noop {} {\bibfield  {journal} {\bibinfo  {journal} {Journal of Experimental Medicine}\ }\textbf {\bibinfo {volume} {209}},\ \bibinfo {pages} {1263} (\bibinfo {year} {2012})}\BibitemShut {NoStop}%
\bibitem [{\citenamefont {Button}\ \emph {et~al.}(2016)\citenamefont {Button}, \citenamefont {Anderson},\ and\ \citenamefont {Boucher}}]{Button_2016}%
  \BibitemOpen
  \bibfield  {author} {\bibinfo {author} {\bibfnamefont {B.}~\bibnamefont {Button}}, \bibinfo {author} {\bibfnamefont {W.~H.}\ \bibnamefont {Anderson}},\ and\ \bibinfo {author} {\bibfnamefont {R.~C.}\ \bibnamefont {Boucher}},\ }\bibfield  {title} {\bibinfo {title} {Mucus hyperconcentration as a unifying aspect of the chronic bronchitic phenotype},\ }\href@noop {} {\bibfield  {journal} {\bibinfo  {journal} {Annals of the American Thoracic Society}\ }\textbf {\bibinfo {volume} {13}},\ \bibinfo {pages} {S156} (\bibinfo {year} {2016})}\BibitemShut {NoStop}%
\bibitem [{\citenamefont {Mellnik}\ \emph {et~al.}(2016)\citenamefont {Mellnik}, \citenamefont {Lysy}, \citenamefont {Vasquez}, \citenamefont {Pillai}, \citenamefont {Hill}, \citenamefont {Cribb}, \citenamefont {McKinley},\ and\ \citenamefont {Forest}}]{Mellnik_2016}%
  \BibitemOpen
  \bibfield  {author} {\bibinfo {author} {\bibfnamefont {J.~W.~R.}\ \bibnamefont {Mellnik}}, \bibinfo {author} {\bibfnamefont {M.}~\bibnamefont {Lysy}}, \bibinfo {author} {\bibfnamefont {P.~A.}\ \bibnamefont {Vasquez}}, \bibinfo {author} {\bibfnamefont {N.~S.}\ \bibnamefont {Pillai}}, \bibinfo {author} {\bibfnamefont {D.~B.}\ \bibnamefont {Hill}}, \bibinfo {author} {\bibfnamefont {J.}~\bibnamefont {Cribb}}, \bibinfo {author} {\bibfnamefont {S.~A.}\ \bibnamefont {McKinley}},\ and\ \bibinfo {author} {\bibfnamefont {M.~G.}\ \bibnamefont {Forest}},\ }\bibfield  {title} {\bibinfo {title} {Maximum likelihood estimation for single particle, passive microrheology data with drift},\ }\href@noop {} {\bibfield  {journal} {\bibinfo  {journal} {Journal of Rheology}\ }\textbf {\bibinfo {volume} {60}},\ \bibinfo {pages} {379} (\bibinfo {year} {2016})}\BibitemShut {NoStop}%
\bibitem [{\citenamefont {Hsu}\ \emph {et~al.}(1994)\citenamefont {Hsu}, \citenamefont {Strohl},\ and\ \citenamefont {Jamieson}}]{Hsu_1994}%
  \BibitemOpen
  \bibfield  {author} {\bibinfo {author} {\bibfnamefont {S.~H.}\ \bibnamefont {Hsu}}, \bibinfo {author} {\bibfnamefont {K.~P.}\ \bibnamefont {Strohl}},\ and\ \bibinfo {author} {\bibfnamefont {A.~M.}\ \bibnamefont {Jamieson}},\ }\bibfield  {title} {\bibinfo {title} {Role of viscoelasticity in tube model of airway reopening. i. nonnewtonian sols},\ }\href@noop {} {\bibfield  {journal} {\bibinfo  {journal} {Journal of Applied Physiology}\ }\textbf {\bibinfo {volume} {76}},\ \bibinfo {pages} {2481} (\bibinfo {year} {1994})}\BibitemShut {NoStop}%
\bibitem [{\citenamefont {Hsu}\ \emph {et~al.}(1996)\citenamefont {Hsu}, \citenamefont {Strohl}, \citenamefont {Haxhiu},\ and\ \citenamefont {Jamieson}}]{Hsu_1996}%
  \BibitemOpen
  \bibfield  {author} {\bibinfo {author} {\bibfnamefont {S.~H.}\ \bibnamefont {Hsu}}, \bibinfo {author} {\bibfnamefont {K.~P.}\ \bibnamefont {Strohl}}, \bibinfo {author} {\bibfnamefont {M.~A.}\ \bibnamefont {Haxhiu}},\ and\ \bibinfo {author} {\bibfnamefont {A.~M.}\ \bibnamefont {Jamieson}},\ }\bibfield  {title} {\bibinfo {title} {Role of viscoelasticity in the tube model of airway reopening. ii. non-newtonian gels and airway simulation},\ }\href@noop {} {\bibfield  {journal} {\bibinfo  {journal} {Journal of Applied Physiology}\ }\textbf {\bibinfo {volume} {80}},\ \bibinfo {pages} {1649} (\bibinfo {year} {1996})}\BibitemShut {NoStop}%
\bibitem [{\citenamefont {Bansil}\ and\ \citenamefont {Turner}(2018)}]{Bansil_2018}%
  \BibitemOpen
  \bibfield  {author} {\bibinfo {author} {\bibfnamefont {R.}~\bibnamefont {Bansil}}\ and\ \bibinfo {author} {\bibfnamefont {B.~S.}\ \bibnamefont {Turner}},\ }\bibfield  {title} {\bibinfo {title} {The biology of mucus: Composition, synthesis and organization},\ }\href@noop {} {\bibfield  {journal} {\bibinfo  {journal} {Advanced Drug Delivery Reviews}\ }\textbf {\bibinfo {volume} {124}},\ \bibinfo {pages} {3} (\bibinfo {year} {2018})}\BibitemShut {NoStop}%
\bibitem [{\citenamefont {Lock}\ \emph {et~al.}(2018)\citenamefont {Lock}, \citenamefont {Carlson},\ and\ \citenamefont {Carrier}}]{Lock_2018}%
  \BibitemOpen
  \bibfield  {author} {\bibinfo {author} {\bibfnamefont {J.~Y.}\ \bibnamefont {Lock}}, \bibinfo {author} {\bibfnamefont {T.~L.}\ \bibnamefont {Carlson}},\ and\ \bibinfo {author} {\bibfnamefont {R.~L.}\ \bibnamefont {Carrier}},\ }\bibfield  {title} {\bibinfo {title} {Mucus models to evaluate the diffusion of drugs and particles},\ }\href@noop {} {\bibfield  {journal} {\bibinfo  {journal} {Advanced Drug Delivery Reviews}\ }\textbf {\bibinfo {volume} {124}},\ \bibinfo {pages} {34} (\bibinfo {year} {2018})}\BibitemShut {NoStop}%
\bibitem [{\citenamefont {Taherali}\ \emph {et~al.}(2018)\citenamefont {Taherali}, \citenamefont {Varum},\ and\ \citenamefont {Basit}}]{Taherali_2018}%
  \BibitemOpen
  \bibfield  {author} {\bibinfo {author} {\bibfnamefont {F.}~\bibnamefont {Taherali}}, \bibinfo {author} {\bibfnamefont {F.}~\bibnamefont {Varum}},\ and\ \bibinfo {author} {\bibfnamefont {A.~W.}\ \bibnamefont {Basit}},\ }\bibfield  {title} {\bibinfo {title} {A slippery slope: On the origin, role and physiology of mucus},\ }\href@noop {} {\bibfield  {journal} {\bibinfo  {journal} {Advanced Drug Delivery Reviews}\ }\textbf {\bibinfo {volume} {124}},\ \bibinfo {pages} {16} (\bibinfo {year} {2018})}\BibitemShut {NoStop}%
\bibitem [{\citenamefont {Wu}\ \emph {et~al.}(2018)\citenamefont {Wu}, \citenamefont {Shan}, \citenamefont {Zhang},\ and\ \citenamefont {Huang}}]{Wu_2018}%
  \BibitemOpen
  \bibfield  {author} {\bibinfo {author} {\bibfnamefont {L.}~\bibnamefont {Wu}}, \bibinfo {author} {\bibfnamefont {W.}~\bibnamefont {Shan}}, \bibinfo {author} {\bibfnamefont {Z.}~\bibnamefont {Zhang}},\ and\ \bibinfo {author} {\bibfnamefont {Y.}~\bibnamefont {Huang}},\ }\bibfield  {title} {\bibinfo {title} {Engineering nanomaterials to overcome the mucosal barrier by modulating surface properties},\ }\href@noop {} {\bibfield  {journal} {\bibinfo  {journal} {Advanced Drug Delivery Reviews}\ }\textbf {\bibinfo {volume} {124}},\ \bibinfo {pages} {150} (\bibinfo {year} {2018})}\BibitemShut {NoStop}%
\bibitem [{\citenamefont {Sedaghat}\ \emph {et~al.}(2016{\natexlab{b}})\citenamefont {Sedaghat}, \citenamefont {Shahmardan}, \citenamefont {Norouzi}, \citenamefont {Jayathilake},\ and\ \citenamefont {Nazari}}]{Sedaghat_2016a}%
  \BibitemOpen
  \bibfield  {author} {\bibinfo {author} {\bibfnamefont {M.~H.}\ \bibnamefont {Sedaghat}}, \bibinfo {author} {\bibfnamefont {M.~M.}\ \bibnamefont {Shahmardan}}, \bibinfo {author} {\bibfnamefont {M.}~\bibnamefont {Norouzi}}, \bibinfo {author} {\bibfnamefont {P.~G.}\ \bibnamefont {Jayathilake}},\ and\ \bibinfo {author} {\bibfnamefont {M.}~\bibnamefont {Nazari}},\ }\bibfield  {title} {\bibinfo {title} {Numerical simulation of muco-ciliary clearance: immersed boundary lattice boltzmann method},\ }\href@noop {} {\bibfield  {journal} {\bibinfo  {journal} {Computers {\&} Fluids}\ }\textbf {\bibinfo {volume} {131}},\ \bibinfo {pages} {91} (\bibinfo {year} {2016}{\natexlab{b}})}\BibitemShut {NoStop}%
\bibitem [{\citenamefont {Sedaghat}\ \emph {et~al.}(2016{\natexlab{c}})\citenamefont {Sedaghat}, \citenamefont {Shahmardan}, \citenamefont {Norouzi}, \citenamefont {Nazari},\ and\ \citenamefont {Jayathilake}}]{Sedaghat_2016b}%
  \BibitemOpen
  \bibfield  {author} {\bibinfo {author} {\bibfnamefont {M.~H.}\ \bibnamefont {Sedaghat}}, \bibinfo {author} {\bibfnamefont {M.~M.}\ \bibnamefont {Shahmardan}}, \bibinfo {author} {\bibfnamefont {M.}~\bibnamefont {Norouzi}}, \bibinfo {author} {\bibfnamefont {M.}~\bibnamefont {Nazari}},\ and\ \bibinfo {author} {\bibfnamefont {P.~G.}\ \bibnamefont {Jayathilake}},\ }\bibfield  {title} {\bibinfo {title} {On the effect of mucus rheology on the muco-ciliary transport},\ }\href@noop {} {\bibfield  {journal} {\bibinfo  {journal} {Mathematical Biosciences}\ }\textbf {\bibinfo {volume} {272}},\ \bibinfo {pages} {44} (\bibinfo {year} {2016}{\natexlab{c}})}\BibitemShut {NoStop}%
\end{thebibliography}%

\end{document}